\documentclass[preprint,review,12pt]{elsarticle}

\usepackage{epsfig}         
\usepackage{cuted}
\usepackage{amsmath}
\usepackage{amssymb}
\usepackage{float}
\usepackage{xcolor}         
\usepackage[normalem]{ulem} 
\usepackage{multirow}
\usepackage{multicol}
\usepackage{subcaption}
\usepackage{lscape}
\usepackage{bm}
\usepackage{verbatim}
\usepackage{graphicx}
\usepackage{psfrag}
\usepackage{hyperref} 
\usepackage[export]{adjustbox}

\psfrag{Frequency [Hz]}{\footnotesize \hspace{-0.5cm} Frequency [Hz]}
\psfrag{Reduced wavenumber k}{\footnotesize \hspace{-0.5cm} Reduced wavenumber $k$}
\psfrag{x [m]}{\scriptsize $x$ [m]}
\psfrag{y [m]}{\scriptsize $y$ [m]}
\psfrag{z [m]}{\scriptsize $z$ [m]}
\psfrag{dmin}{\scriptsize \vspace{0.5cm} $d_{\min}$}
\psfrag{dmax}{\scriptsize \vspace{0.5cm} $d_{\max}$}
\psfrag{dmax [mm]}{\footnotesize \raisebox{-0.2cm}{$d_{\max}$ [mm]}}
\psfrag{dx}{\scriptsize \hspace{-0.2cm} $\Delta x$}
\psfrag{dtheta}{\scriptsize \hspace{-0.0cm} $\Delta \theta$}
\psfrag{lc/2}{\scriptsize \hspace{-0.35cm} $l_c/2$}
\psfrag{ls}{\scriptsize $l_s$}
\psfrag{L}{\scriptsize $L$}
\psfrag{T}{\scriptsize $\Gamma$}
\psfrag{X}{\scriptsize X}
\psfrag{M}{\scriptsize M}
\psfrag{l}{$l$}
\psfrag{Displacements [mm]}{\scriptsize \hspace{-0.75cm} Displacements [mm]}
\psfrag{Rotations [rad]}{\scriptsize \hspace{-0.5cm} Rotations [rad]}
\psfrag{Load [\%]}{\scriptsize Load [\%]}
\psfrag{20 log |FRF| [dB]}{\scriptsize $20 \log|\text{FRF}|$ [dB]}
\psfrag{20 log |FRFx| [dB]}{\scriptsize $20 \log|\text{FRF}_x|$ [dB]}
\psfrag{20 log |FRFy| [dB]}{\scriptsize $20 \log|\text{FRF}_y|$ [dB]}
\psfrag{20 log |FRFz| [dB]}{\scriptsize $20 \log|\text{FRF}_z|$ [dB]}
\psfrag{zero}{$0$}
\psfrag{one}{$1$}
\psfrag{input}{\scriptsize input}
\psfrag{output}{\scriptsize output}
\psfrag{w1}{$\overline{\omega}_1$}
\psfrag{w2s}{$\overline{\omega}_{2, \, \text{sq.}}$}
\psfrag{w2h}{$\overline{\omega}_{2, \, \text{hx.}}$}
\psfrag{w2hi}{$\overline{\omega}_{2, \, \text{hi.}}$}

\journal{International Journal of Solids and Structures}

\begin{document}

\begin{frontmatter}

\title{Band gap enhancement in periodic frames using hierarchical structures}

\author{Vin\'icius F. Dal Poggetto\fnref{label1}}
\author{Federico Bosia\fnref{label2}}
\author{Marco Miniaci\fnref{label3}}
\author{Nicola M. Pugno\fnref{label1,label4}}
\ead{nicola.pugno@unitn.it}

\fntext[label1]{Laboratory of Bio-inspired, Bionic, Nano, Meta Materials \& Mechanics, Department of Civil, Environmental and Mechanical Engineering, University of Trento, 38123 Trento, Italy}
\fntext[label2]{DISAT, Politecnico di Torino, 10129 Torino, Italy}
\fntext[label3]{CNRS, Centrale Lille, ISEN, Univ. Lille, Univ. Valenciennes, UMR 8520 - IEMN, F-59000 Lille, France}
\fntext[label4]{School of Engineering and Materials Science, Queen Mary University of London, Mile End Road, London E1 4NS, United Kingdom}

\begin{abstract}
\textit{The quest for novel designs for lightweight phononic crystals and elastic metamaterials with wide low-frequency band gaps has proven to be a significant challenge in recent years. In this context, lattice-type materials represent a promising solution, providing both lightweight properties and significant possibilities of tailoring mechanical and dynamic properties. Additionally, lattice structures also enable the generation of hierarchical architectures, in which basic constitutive elements with different characteristic length scales can be combined. In this work, we propose 1D- and 2D-periodic phononic crystals made of spatial frames inspired by a spider web-based architecture. Specifically, hierarchical plane structures based on a combination of frames with a variable cross-section are proposed and exploited to open and enhance band gaps with respect to their non-hierarchical counterparts. Our results show that hierarchy is effective in broadening existing band gaps as well as opening new full band gaps in non-hierarchical periodic structures.}
\end{abstract}


\end{frontmatter}

\section{Introduction} \label{introduction}

Phononic crystals (PCs) and elastic metamaterials (MMs) offer interesting opportunities for unprecedented wave manipulation \citep{brillouin1953wave,sigalas1993band,martinez1995sound,liu2000locally,khelif2006complete, craster2012acoustic,deymier2013acoustic,laude2015phononic,ma2016acoustic}, and can be defined as composite structures designed to yield specific wave dispersion characteristics, exploiting Bragg scattering and/or local resonance effects. Among the plethora of unconventional dynamic properties that PCs present, their capability to open frequency band gaps (BGs) is one of the most investigated. A phononic BG is a frequency region where wave propagation is not allowed, since only evanescent waves are present in the corresponding dispersion diagram \citep{lee2009spectral,laude2009evanescent}. These effects can be achieved by properly choosing specific material and geometrical configurations that allow an impedance mismatch in the system, responsible for constructive/destructive interferences associated with: (i) the lattice dimensions (Bragg scattering) \citep{khelif2006complete,deymier2013acoustic,laude2015phononic} or (ii) local resonances \citep{liu2000locally,craster2012acoustic,deymier2013acoustic}. This opened up new perspectives in many fields, ranging from microelectromechanical systems to nondestructive evaluation \citep{pennec2010two,miniaci2017proof,gliozzi2019proof}, including but not limited to wave filters and waveguiding \citep{miniaci2018experimental}, beam and wave splitting \citep{li2015acoustic,miniaci2019valley}, large scale building vibration shielding \citep{miniaci2016large}, and subwavelength imaging \citep{sukhovich2009experimental,moleron2015acoustic}.

So far, the main approach to obtain wave attenuation has involved combining materials to form the necessary impedance mismatch \citep{miranda2017flexural} or adding local resonators \citep{xiao2013flexural,nobrega2016vibration,poggetto2019optimization}, but recent advances have shown that variations in the cross-section of specific regions of the unit cell can also lead to effective vibration reduction in one- \citep{sorokin2016effects,pelat2019control} and two-dimensional \citep{tang2019periodic,bibi2019manipulation,poggetto2020widening} systems. In this context, designing simultaneously strong and lightweight PCs/MMs with desirable dynamic properties has been the quest of many researchers for decades. Among the various approaches (continuous single-phase materials, composite structures, etc.), lattice-type materials, given their porous structure and well-defined unit cell geometries that allow deviating from the properties of bulk materials, are excellent candidates to obtain lightweight structures with precisely tailored dynamic and mechanical properties. In addition, lattice- or frame-like structures allow the easy integration of hierarchical architectures in the design.

A hierarchical architecture is here understood as a structure characterized by multiple nested levels of unit cells repeated at different size scales. The structuring of these materials can be obtained by using self-similarity between different hierarchical levels \citep{mousanezhad2015hierarchical,meza2015resilient} or exploiting different multi-scale structuring \citep{chen2016hierarchical}. The use of hierarchical structures in the quasi-static regime, e.g., in architectural design and civil engineering has been largely explored, leading to considerable weight reduction without loss of mechanical performance, whereas in the dynamic regime the properties of such materials have been somewhat less investigated.

In terms of wave dynamics and attenuation, \citet{movchan2004split} have shown that localized modes can be used to tune and create low-frequency BGs, which was further reinforced by \citet{huang2010band}, showing that BGs can be shifted by the proper selection of internal stiffness and masses of the system. Hierarchical structures have been proposed by \citet{chen2016harnessing} using honeycomb architectures to design stiff and lightweight PCs, and by \citet{lim2015wave} using self-similar beam structures to improve wave propagation characteristics in hexagonal lattices. Among the various types of hierarchical metamaterials, bioinspired hierarchical PCs have shown interesting characteristics in terms of broadband wave filtering, as demonstrated by \citet{zhang2013broadband} using multiscale periodicity in one-dimensional PCs, and \citet{chen2014tunable,chen2015bio,chen2015multiband}, who also demonstrated how bioinspired composites can have their geometry tailored for broadband vibration filtering. \citet{miniaci2016spider} showed how a specific hierarchical arrangement of lattice-type structures inspired by a spider web organization (i.e., lattice-type structure with radial and circular threads linked by nodes of variable stiffness) are able to control wave propagation. The effect of bioinspired hierarchical organization on wave attenuation properties has also been experimentally investigated for the case of continuous elastic metamaterials made of single-phase continuous structures formed by self-similar unit cells with different hierarchical levels and types of hierarchy \citep{miniaci2018design}. However, practically oriented metamaterials with multi-scale wave attenuation are yet to be fully explored. One of these possibilities is the use of frame-like bioinspired structures to harness the conceptual advantages of structures with varying cross-sections.

In this paper, we propose the use of a hierarchical spider web-based lightweight solution to broaden the low- and mid-frequency BGs in a periodic metamaterial frame. The proposed hierarchical structures are constructed by replacing the regular frame elements of two-dimensional lattices by frame elements forming a 1D periodic PC. The paper is organized as follows: Section \ref{models_methods} presents the models and methods, Section \ref{results} illustrates the obtained results, and Section \ref{conclusions} presents some concluding remarks.

\section{Models and methods} \label{models_methods}

\subsection{Hierarchical structure} \label{hierarchical_structure}

The hierarchical order associated with a structure is usually defined as the number $n$ of length scales at which a recognizable sub-structure occurs \citep{lakes1993materials}: $n=0$ corresponds to a continuous (1D, 2D or 3D) element, $n=1$ (first order) represents a structure formed by elements occurring at a single length scale, $n=2$ (second-order) an arrangement of first-order structures, and so on. 

\subsubsection{First hierarchical level}

\begin{figure}[h!]
  \centering
  \makebox[\textwidth]{
  \begin{subfigure}[h]{.2\textwidth}
    \centering
    \includegraphics[width=2.25cm,height=2.25cm]{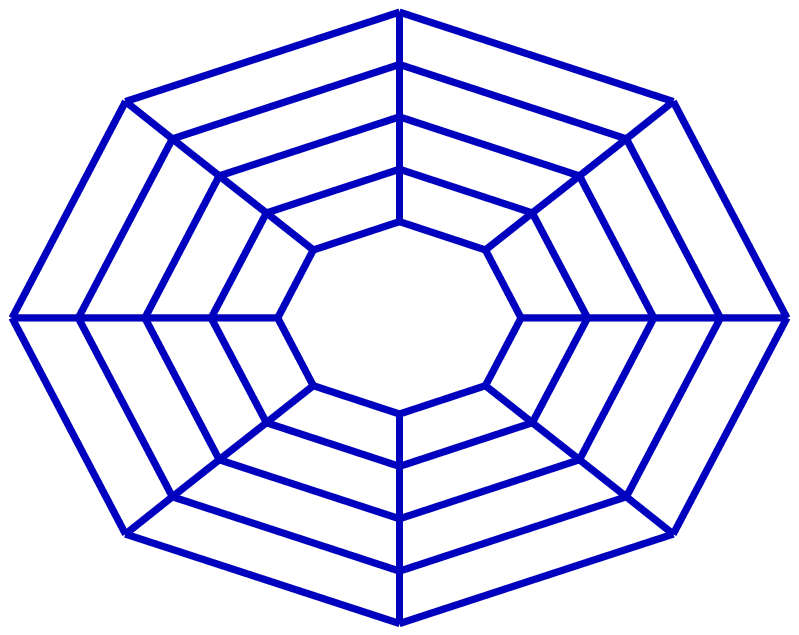}
  \end{subfigure}
  \begin{subfigure}[h]{.25\textwidth}
    \centering
    \includegraphics[height=1.5cm]{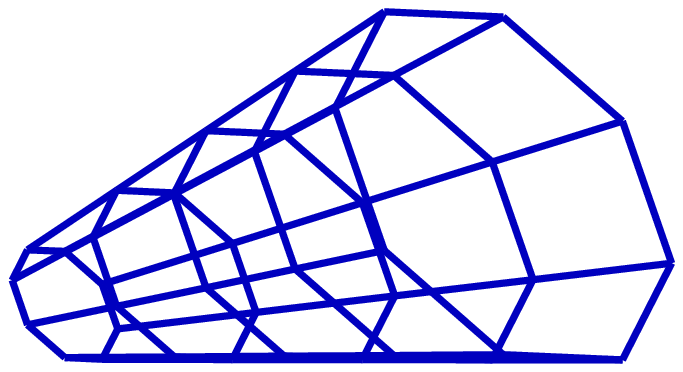}
  \end{subfigure}
  \begin{subfigure}[h]{.30\textwidth}
    \centering
    \includegraphics[height=1.5cm]{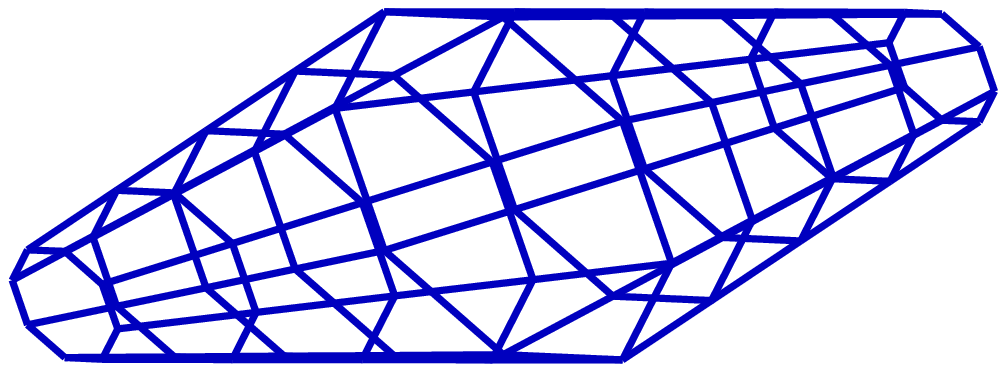}
  \end{subfigure}
  \begin{subfigure}[h]{.30\textwidth}
    \centering
    \includegraphics[height=1.5cm]{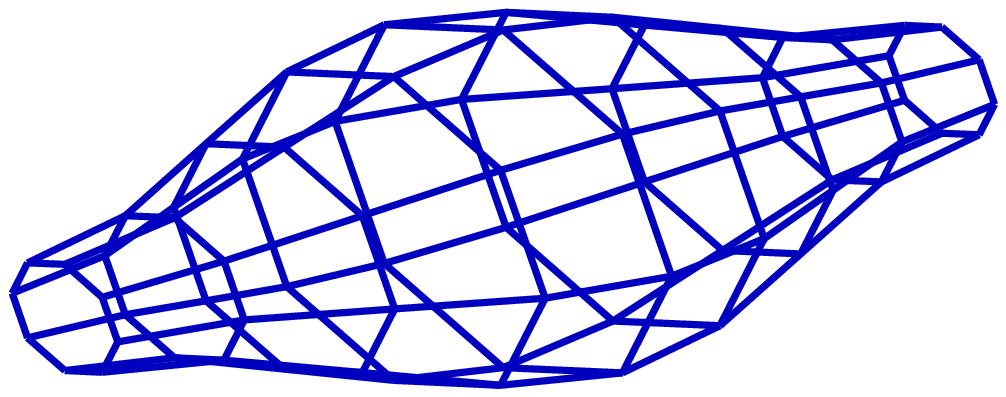}
  \end{subfigure}
  }
  \makebox[\textwidth]{
  \begin{subfigure}[h]{.2\textwidth}
    \caption{}
    \label{rationale1}
  \end{subfigure}
  \begin{subfigure}[h]{.25\textwidth}
    \caption{}
    \label{rationale2}
  \end{subfigure}
  \begin{subfigure}[h]{.30\textwidth}
    \caption{}
    \label{rationale3}
  \end{subfigure}
  \begin{subfigure}[h]{.30\textwidth}
    \caption{}
    \label{rationale4}
  \end{subfigure}
  }
  \makebox[\textwidth]{
  \begin{subfigure}[h]{.3\textwidth}
    \centering
    \includegraphics[height=1.5cm]{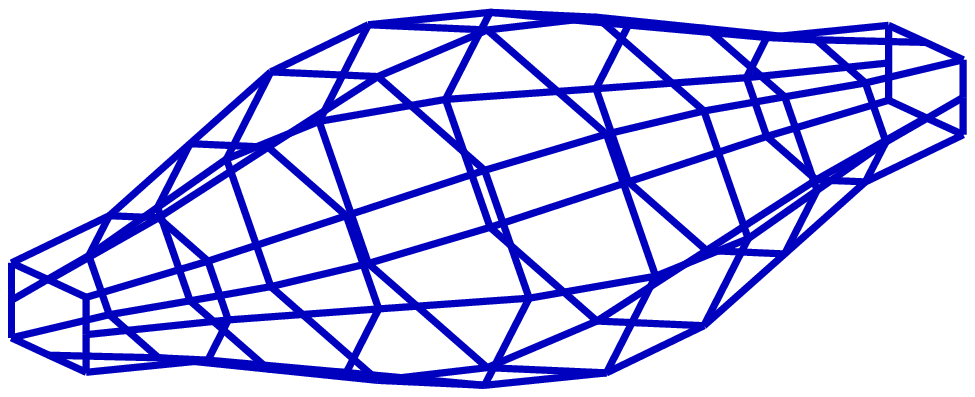}
  \end{subfigure}
  \begin{subfigure}[h]{.4\textwidth}
    \centering
    \includegraphics[width=5.0cm]{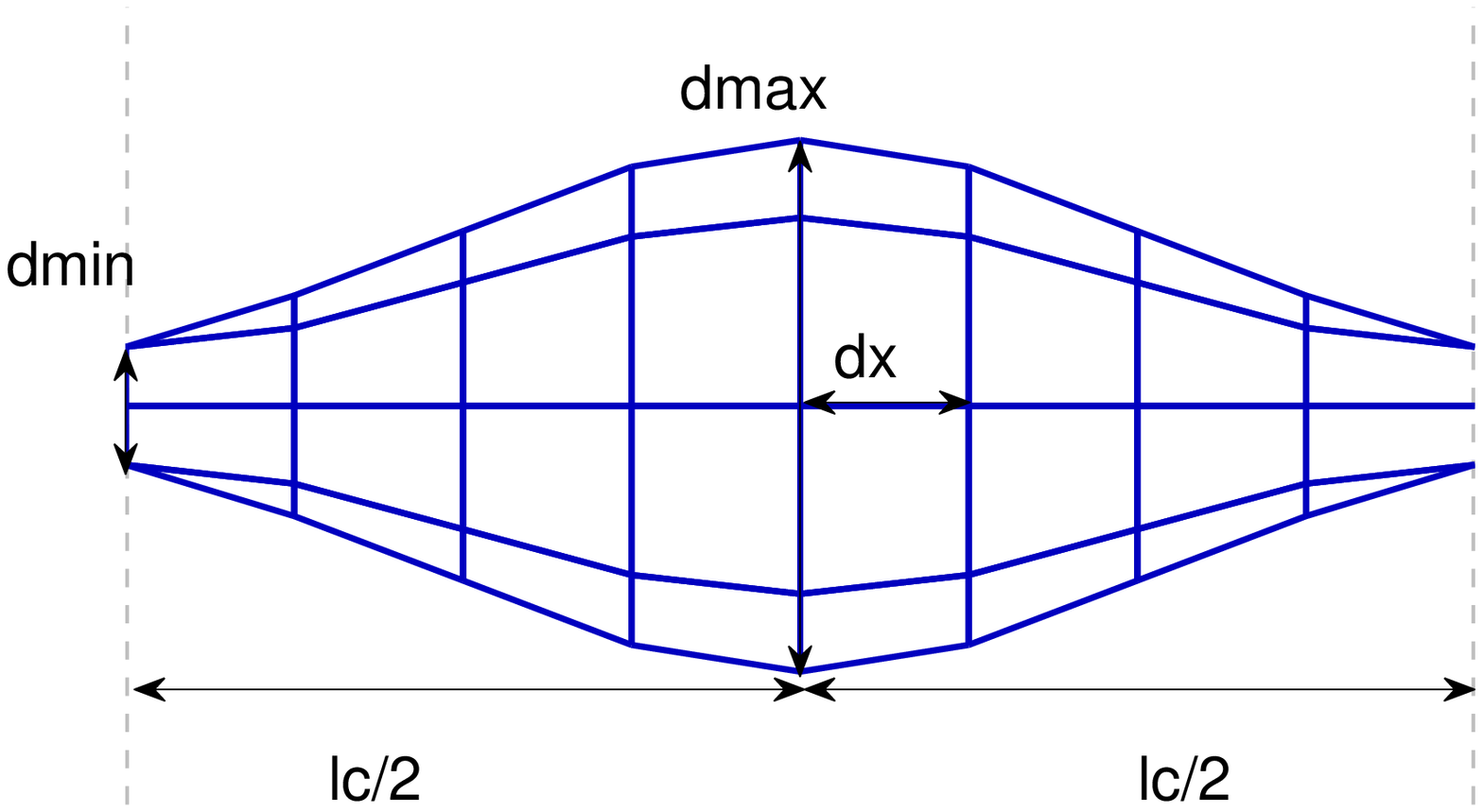}
  \end{subfigure}
  \begin{subfigure}[h]{.35\textwidth}
    \centering
    \includegraphics[width=3.0cm,height=3.0cm]{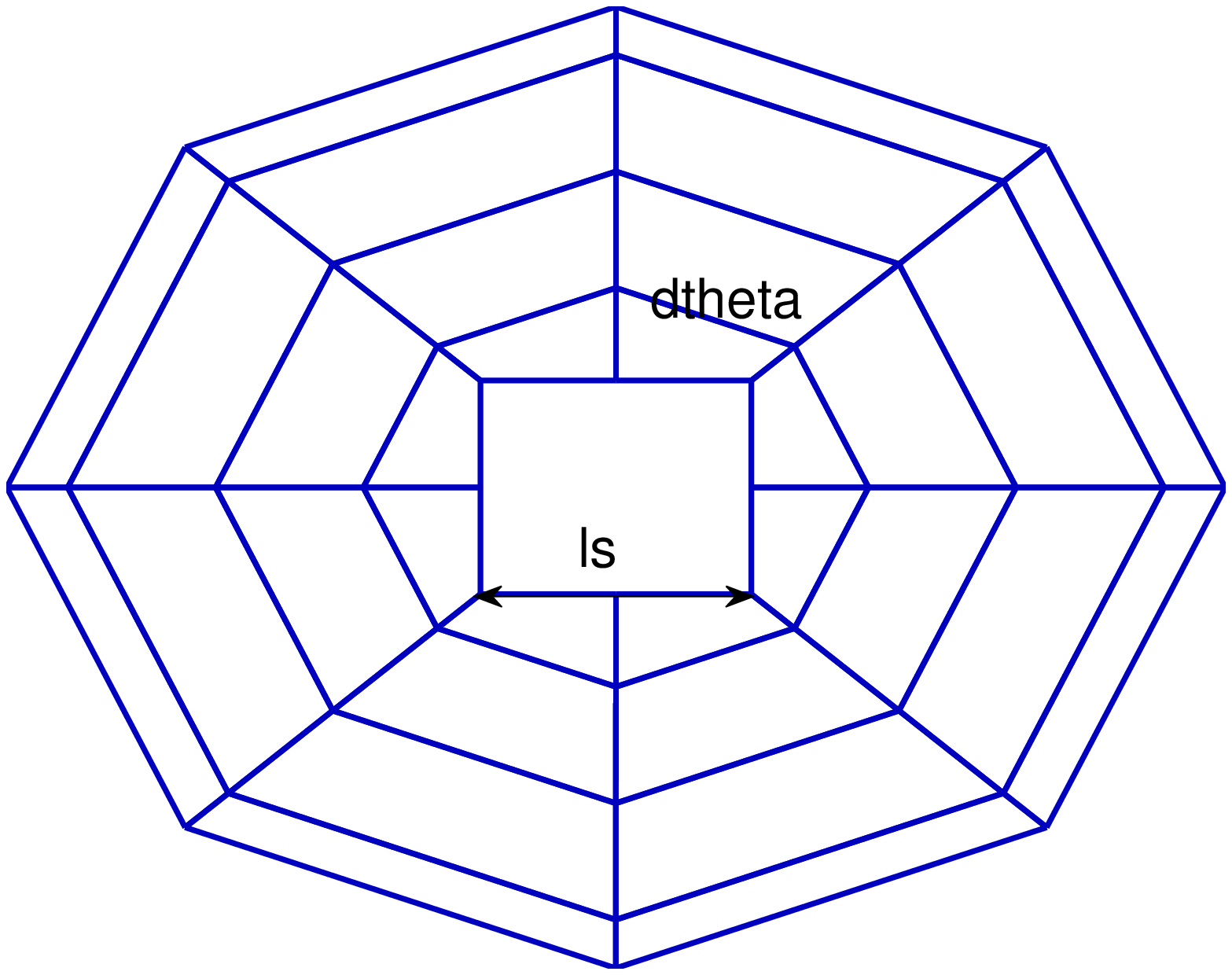}
  \end{subfigure}
  }
  \makebox[\textwidth]{
  \begin{subfigure}[h]{.3\textwidth}
    \caption{}
    \label{rationale5}
  \end{subfigure}
  \begin{subfigure}[h]{.4\textwidth}
    \caption{}
    \label{rationale6}
  \end{subfigure}
  \begin{subfigure}[h]{.35\textwidth}
    \caption{}
    \label{rationale7}
  \end{subfigure}
  }
  \caption{Procedure for the development of a one-dimensional PC using a spider web-based design: (\subref{rationale1}) initial shape with radial and viscid threads, (\subref{rationale2}) extended structure, (\subref{rationale3}) reflected extended structure, (\subref{rationale4}) sinusoidal shape applied to viscid threads, (\subref{rationale5}) square endings which enable connection between elements at orthogonal angles, (\subref{rationale6}) side view: the structure has a varying circular cross-section, with a maximum diameter $d_{\max}$ at its center and minimum diameter $d_{\min}$ at the edges with equally spaced sections, (\subref{rationale7}) section view: equally spaced radial nodes ending in a square cross-section.}
  \label{geometry_rationale}
\end{figure}

We begin by defining a structure ($n=1$) obtained by arranging one-dimensional solid frame elements ($n=0$), so that it can be used as a one-dimensional PC, controlling waves in its longitudinal direction. The rationale for the development of this structure is depicted in Figure \ref{geometry_rationale}.

The initial cross-sectional shape resembles a spider web structure (Figure \ref{rationale1}), which contains radial threads (connecting the center and the outer regions) and viscid threads (connecting adjacent radial threads) \citep{cranford2012nonlinear,miniaci2016spider}. This structure can then be deformed in the direction orthogonal to the web plane to form a three-dimensional structure (Figure \ref{rationale2}). Further details relative to this deformation are provided in Section \ref{band_diagrams}. The latter, in turn, can be reflected with respect to the original plane of the structure to form a symmetrical frame structure (Figure \ref{rationale3}). Then, the distribution of the diameter of viscid threads is adjusted into a sinusoidal shape (Figure \ref{rationale4}), which provides a smoother transition between cross-sections. The connection between elements at an orthogonal angle is further enabled by using a square terminal cross-section (Figure \ref{rationale5}).

The proposed overall design of the one-dimensional PC using the three-dimensional structure shown in Figure \ref{rationale6} is constructed using frame elements with circular cross-sections of radius $r$. The total length of the structure is $l_c$, its diameter is maximum ($d_{\max}$) at $x=0$ and minimum ($d_{\min}$) at $x = \pm l_c/2$, with nodes equally spaced in the longitudinal direction, using $n_x$ cross-sections ($\Delta x = l_c/(n_x-1)$). The diameter $d$ of the structure can be can be described at each $x$-coordinate as $d = d(x)$, given by
\begin{equation}
 d(x) = \frac{d_{\max} + d_{\min}}{2} + \frac{d_{\max} - d_{\min}}{2} \cos \bigg( \frac{2\pi}{l_c} x \bigg) \, .
\end{equation}

The structure is terminated using a square cross-section with side $l_s$ at $x = \pm l_c/2$ (Figure \ref{rationale7}), which is inscribed in a circle of diameter $d_{\min} = l_s\sqrt{2}$. All nodes are also equally spaced with respect to the longitudinal centerline using $n_r$ subdivisions ($\Delta \theta = 2\pi /n_r$), yielding a total of $n_x \cdot n_r$ nodes.

For the corresponding band diagram computations, the periodic cell depicted in Figure \ref{rationale6} will be used, with contours represented using dashed gray lines. Notice that the square cross-section is removed at the right edge, so that the stiffnesses of elements connecting corresponding nodes in the periodic structure are not accounted for twice. This one-dimensional PC will be used to construct the upper hierarchical levels, leading to two-dimensional PCs, as described in the following section.

\subsubsection{Second hierarchical level}

\begin{figure}[h!]
  \centering
  \makebox[\textwidth]{
  \begin{subfigure}[h]{.3\textwidth}
    \centering
    \includegraphics[height=3.5cm]{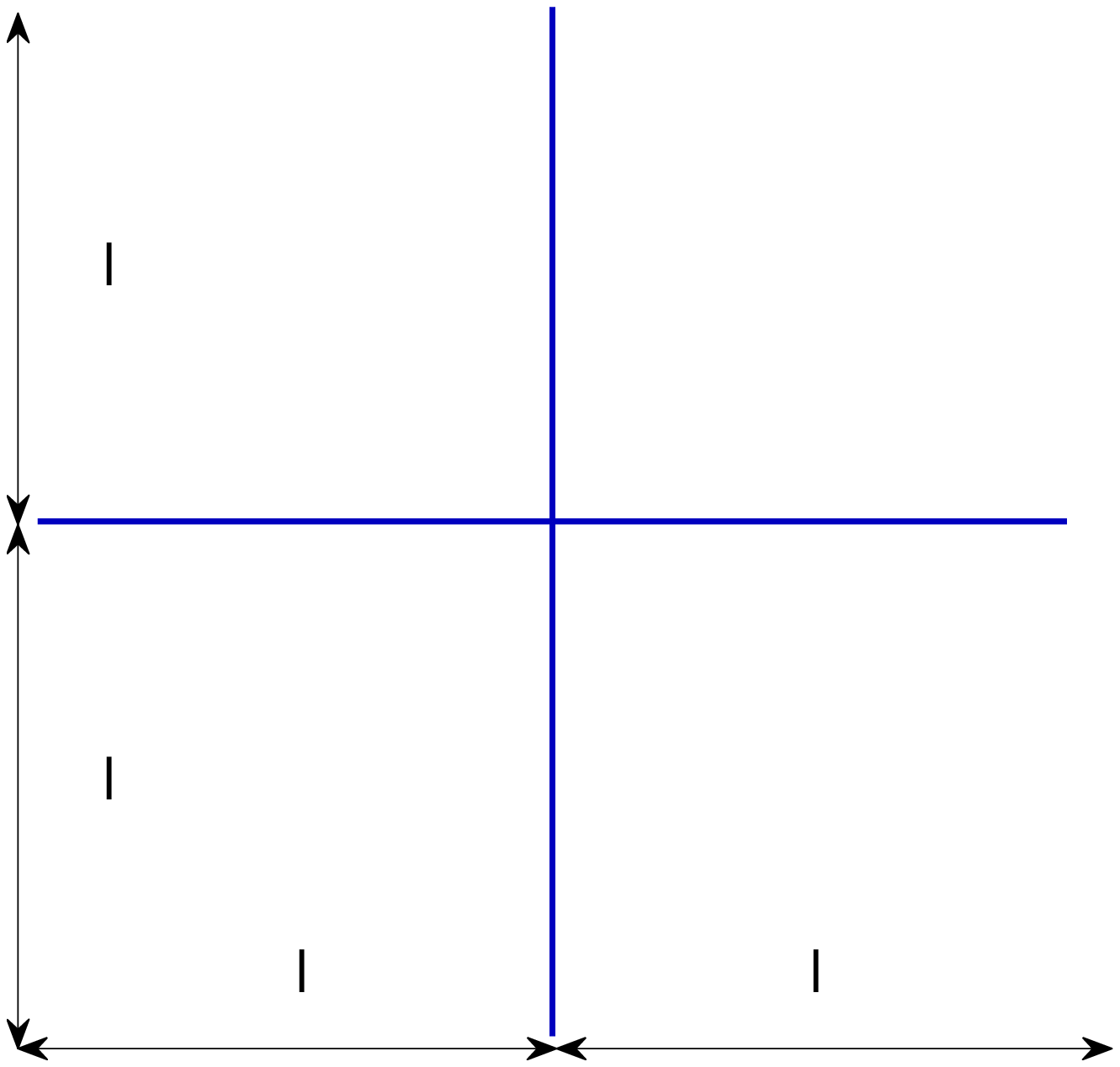}
    \caption{}
    \label{level2_square_top}
  \end{subfigure}
  \begin{subfigure}[h]{.3\textwidth}
    \centering
    \includegraphics[height=3.5cm]{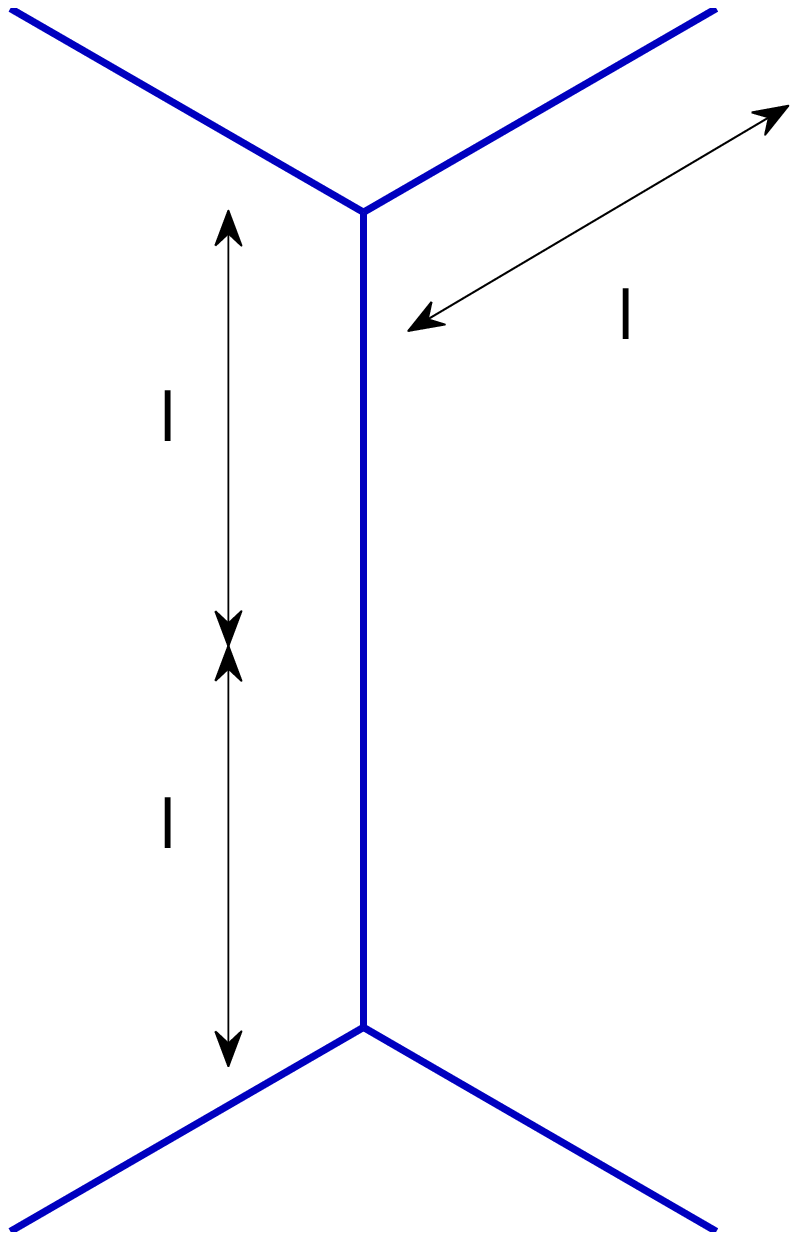}
    \caption{}
    \label{level2_hexagonal_top}
  \end{subfigure}
  \begin{subfigure}[h]{.3\textwidth}
    \centering
    \includegraphics[height=3.5cm]{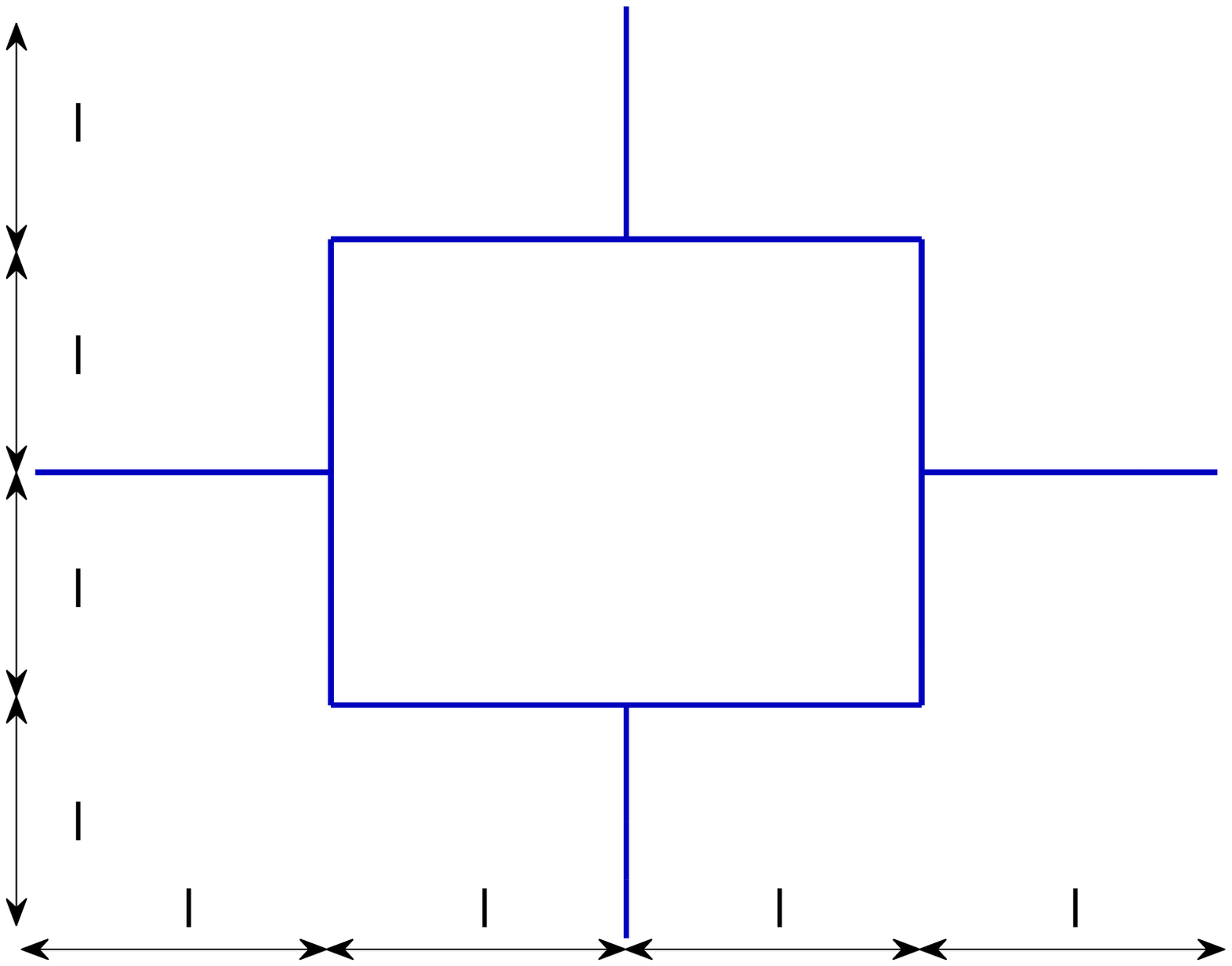}
    \caption{}
    \label{level2_square_inclusion_top}
  \end{subfigure}
  }
  \vspace{0.5cm}
  \makebox[\textwidth]{
  \begin{subfigure}[h]{.3\textwidth}
    \centering
    \includegraphics[height=3.5cm]{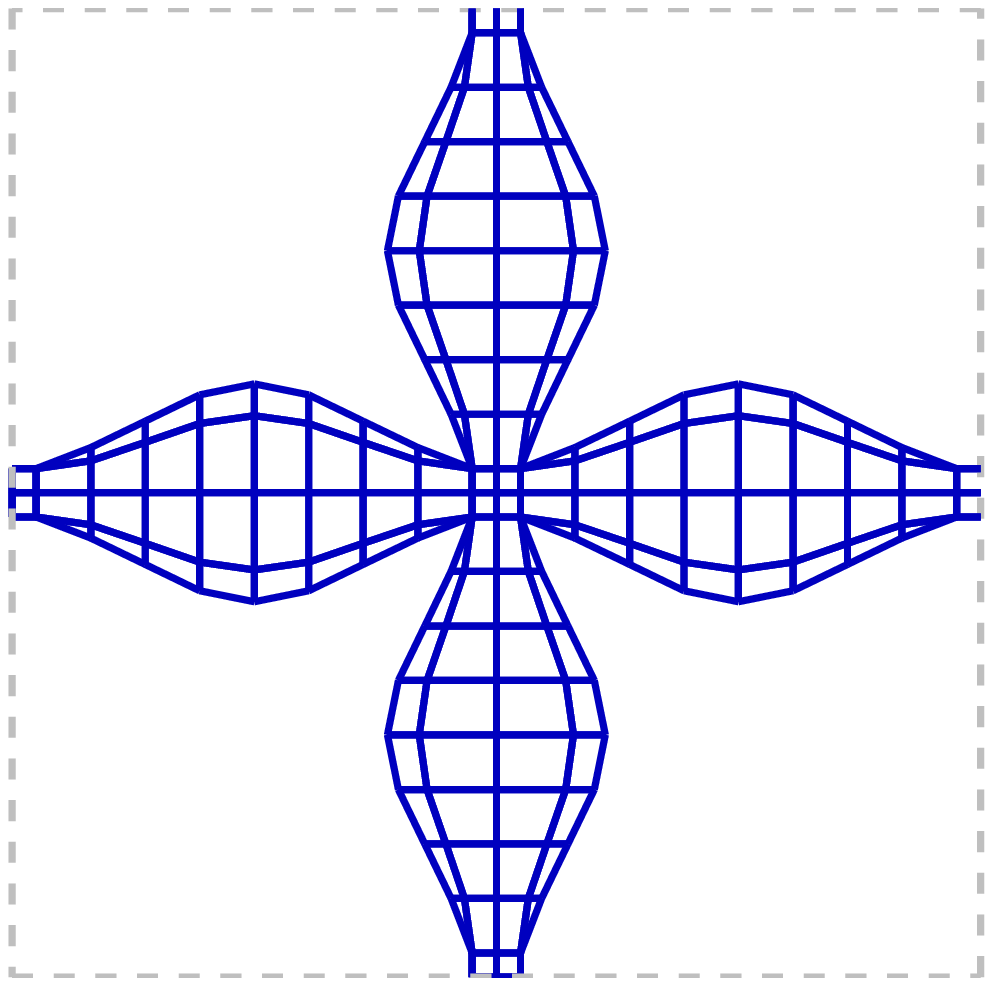}
    \caption{}
    \label{level2_spiderweb_square_top}
  \end{subfigure}
  \begin{subfigure}[h]{.3\textwidth}
    \centering
    \includegraphics[height=3.5cm]{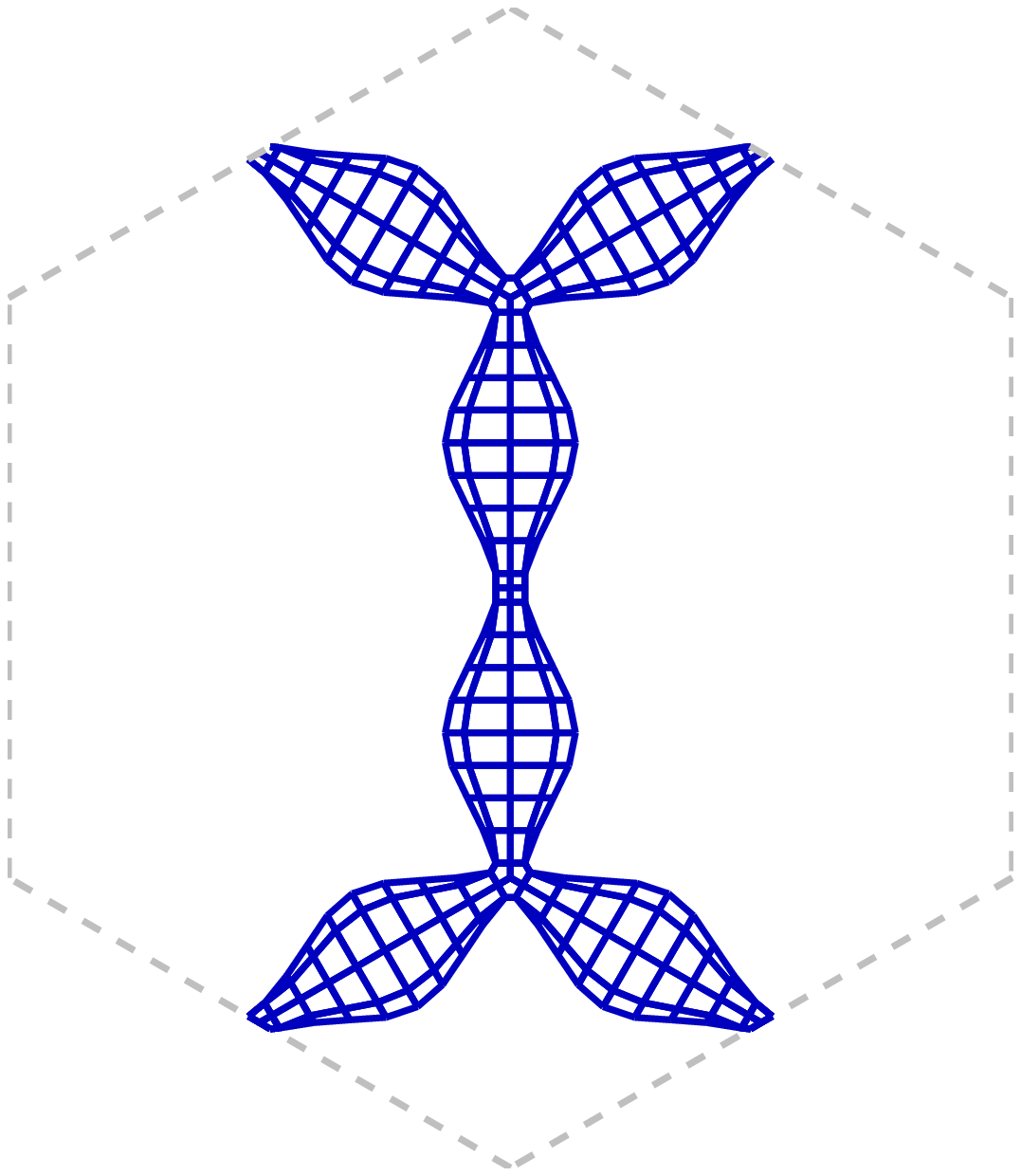}
    \caption{}
    \label{level2_spiderweb_hexagonal_top}
  \end{subfigure}
  \begin{subfigure}[h]{.3\textwidth}
    \centering
    \includegraphics[height=3.5cm]{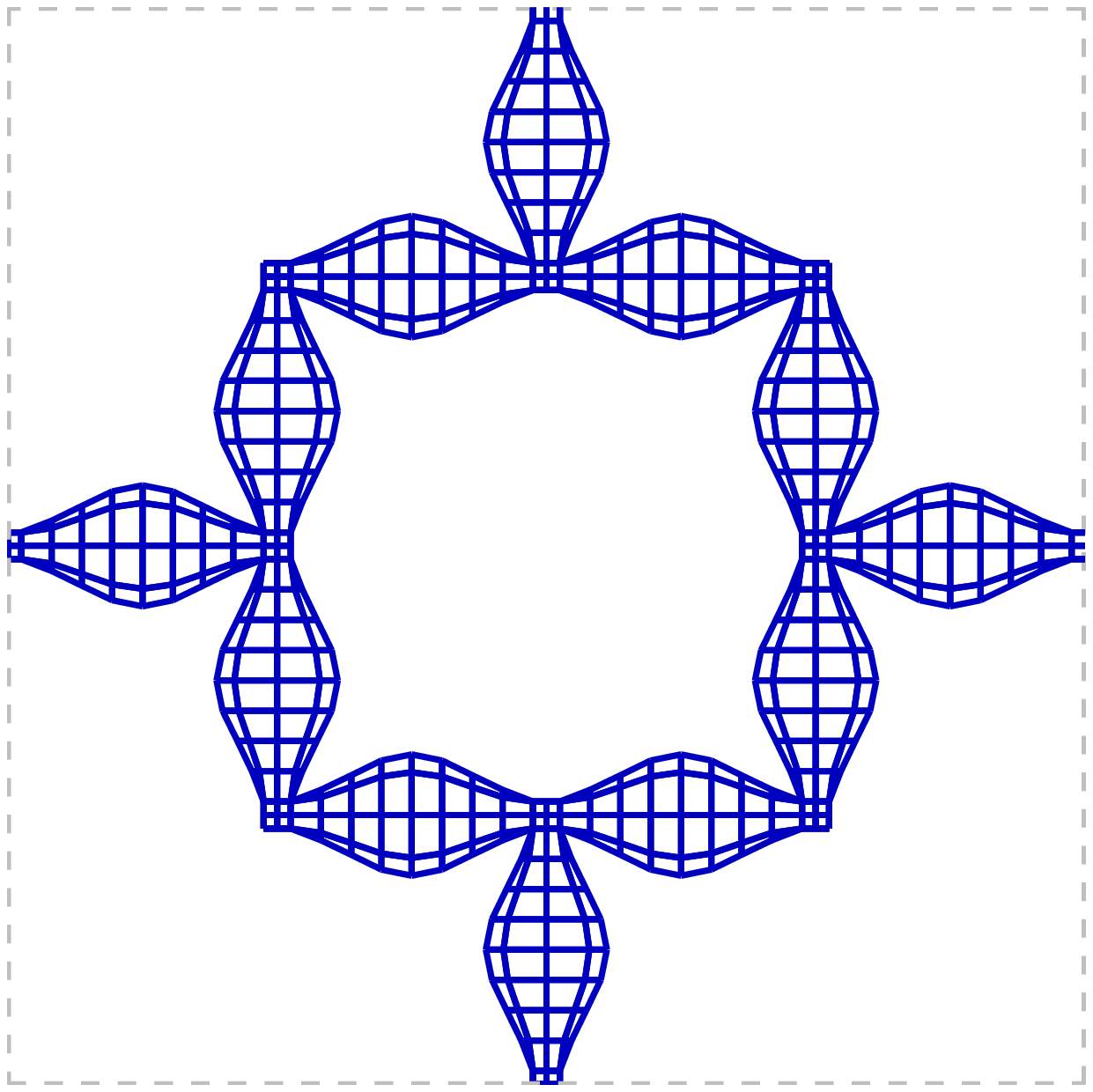}
    \caption{}
    \label{level2_spiderweb_square_inclusion_top}
  \end{subfigure}
  }
  
  \begin{subfigure}[h]{.4\textwidth}
    \centering
    \includegraphics[height=3.0cm]{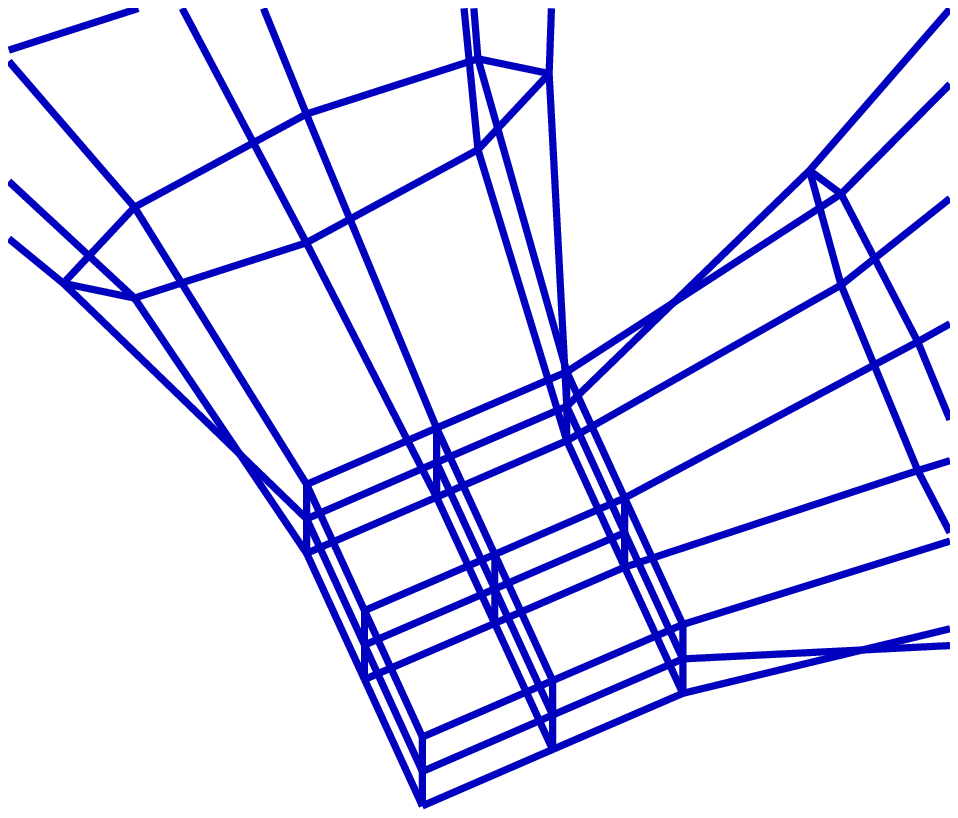}
    \caption{}
    \label{level2_spiderweb_square_inclusion_cube}
  \end{subfigure}
  \begin{subfigure}[h]{.4\textwidth}
    \centering
    \includegraphics[height=3.0cm]{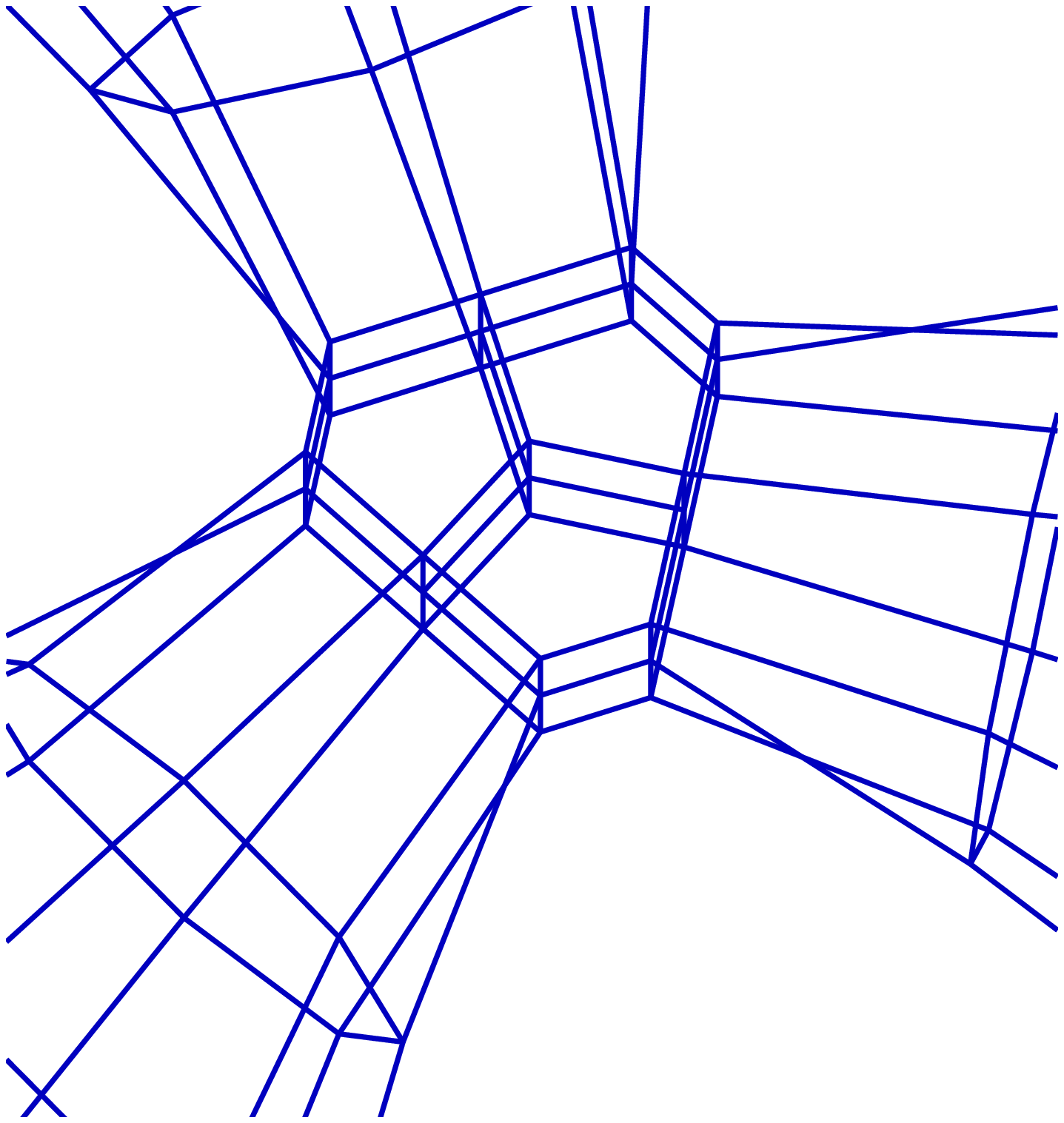}
    \caption{}
    \label{level2_spiderweb_hexagonal_triangle}
  \end{subfigure}

  \caption{Proposed two-dimensional structures. The simple frame structures depicted in (\subref{level2_square_top}), (\subref{level2_hexagonal_top}), and (\subref{level2_square_inclusion_top}) are built using one-dimensional frame elements of length $l$ to form square, hexagonal, and square lattices, respectively. The resulting hierarchical counterparts are shown in (\subref{level2_spiderweb_square_top}), (\subref{level2_spiderweb_hexagonal_top}), and (\subref{level2_spiderweb_square_inclusion_top}), respectively. The junctions of one-dimensional PCs at $90^o$ and $120^o$ angles are depicted in (\subref{level2_spiderweb_square_inclusion_cube}) and (\subref{level2_spiderweb_hexagonal_triangle}), respectively.}
  \label{level2_square_vs_spiderweb_square}
\end{figure}

Next, we describe how the previously developed first-order hierarchical structure can be used to build the two-dimensional PCs. In Figure \ref{level2_square_vs_spiderweb_square}, three types of two-dimensional lattices are proposed, together with their corresponding hierarchical counterparts.

Geometries presented in Figures \ref{level2_square_top} and \ref{level2_hexagonal_top} are used as simple examples of square and hexagonal lattices, respectively, obtained by coupling several one-dimensional frame elements of length $l$: the former is obtained by connecting $4$ frame elements at $90^o$ angles, and the latter is obtained by connecting $3$ frame elements at $120^o$ angles (similar to a ``Y'') and reflecting it with respect to its base to obtain a vertical symmetry. The two-dimensional lattice presented in Figure \ref{level2_square_inclusion_top} is a square resonator of side length $2l$ connected using frame elements of side $l$. These structures can be modified by considering them to be second-order hierarchical structures, constituted by first-order PCs, as depicted in Figures \ref{level2_spiderweb_square_top} --  \ref{level2_spiderweb_square_inclusion_top}. The assembly of one-dimensional PCs oriented at $90^o$ and $120^o$ angles is made using the connection elements formed by their square endings (Figures \ref{level2_spiderweb_square_inclusion_cube} and \ref{level2_spiderweb_hexagonal_triangle}).

The first-order and second-order hierarchical structures are investigated using the modeling approach presented in the next section.

\subsection{Dynamic models} \label{dynamic_models}

Frame elements can be modeled by superposing the effects of rod, shaft, and beam elements, considering forces and displacements in $x$ for the longitudinal direction, $y$ and $z$ for transverse directions with respect to the frame element local axis. The degrees-of-freedom (DOFs) for the $i$-th node are represented by the displacement vector $\mathbf{q}_i$, and the forces and moments in the same node are represented using vector $\mathbf{f}_i$, respectively given by
\begin{subequations}
\begin{align}
 \mathbf{q}_i &= \left\{
 \begin{array}{cccccc}
  u_{xi} & u_{yi} & u_{zi} & \phi_{xi} & \phi_{yi} & \phi_{zi}
 \end{array} \right\}^T \, , \\
 \mathbf{f}_i &= \left\{
 \begin{array}{cccccc}
  F_{xi} & F_{yi} & F_{zi} & M_{xi} & M_{yi} & M_{zi}
 \end{array} \right\}^T \, ,
\end{align}
\end{subequations}
where $u_{xi}$, $u_{yi}$, and $u_{zi}$ represent displacements, $\phi_{xi}$, $\phi_{yi}$, and $\phi_{zi}$ rotations, $F_{xi}$, $F_{yi}$, and $F_{zi}$ forces, and $M_{xi}$, $M_{yi}$, and $M_{zi}$ moments, all given at the $i$-th node for the $x$, $y$, and $z$ directions, respectively.

The relations between the displacements and forces at each node for the rod, shaft, and beam elements are thoroughly described in the literature in the case of linear elastic behavior using stiffness and mass matrices, which were here obtained using the procedure outlined in \citep{cook2001concepts}. After the process of assembly to obtain the global stiffness matrix $\mathbf{K}$ and the mass matrix $\mathbf{M}$, if one considers harmonic excitations and displacements, it is possible to write
\begin{equation} \label{DqF}
 \mathbf{D} \mathbf{q} = \mathbf{f} \, ,
\end{equation}
where $\mathbf{D} = \mathbf{K} - \omega^2 \mathbf{M}$ is the dynamic stiffness matrix, $\omega$ is the circular frequency, $\mathbf{q}$ is the nodal displacement vector, and $\mathbf{f}$ is the nodal force vector for all DOFs in the structure, respectively.

Equation (\ref{DqF}) is suited for the application of periodic conditions and the determination of the propagation characteristics of the considered structure.

\subsection{Band structure computation}

\begin{figure}[h!]
  \centering
  \makebox[\textwidth]{
  \begin{subfigure}[h]{.35\textwidth}
    \centering
    \includegraphics[height=1.25cm]{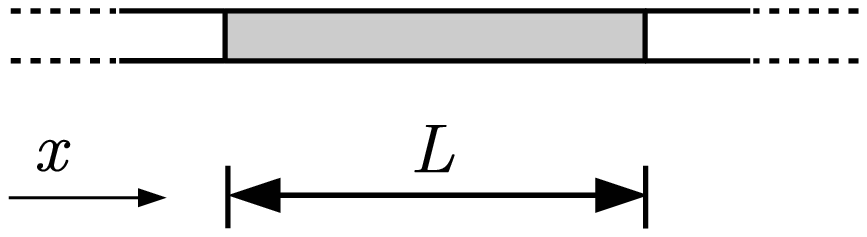}
  \end{subfigure}
  \begin{subfigure}[h]{.35\textwidth}
    \centering
    \includegraphics[height=4.5cm]{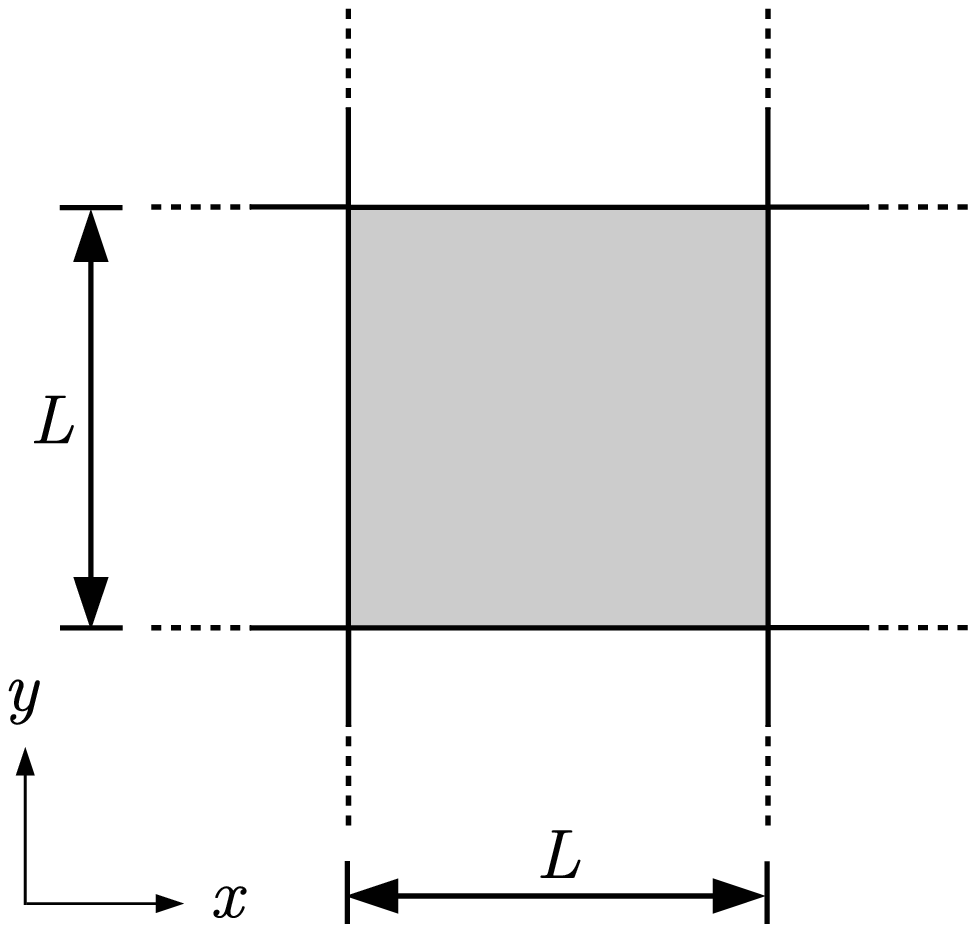}
  \end{subfigure}
  \begin{subfigure}[h]{.35\textwidth}
    \centering
    \includegraphics[height=4.5cm]{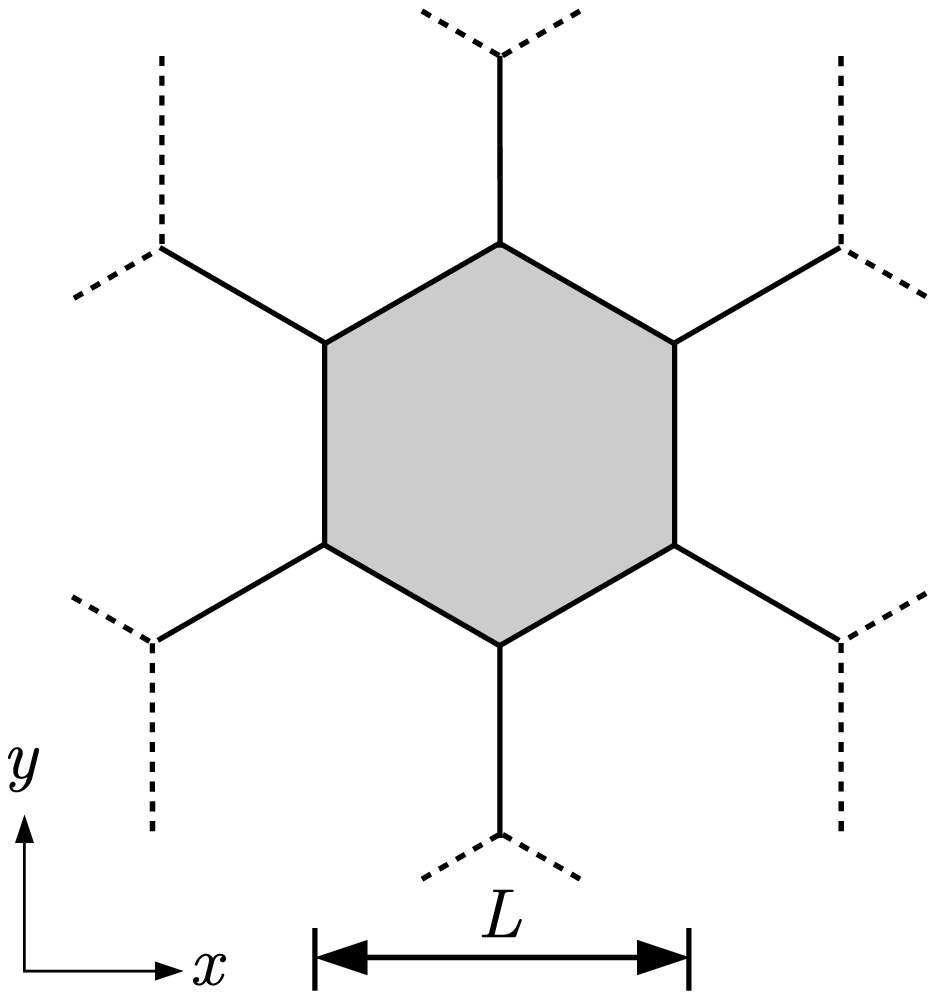}
  \end{subfigure}
  }
  \makebox[\textwidth]{
  \begin{subfigure}[h]{.35\textwidth}
    \caption{}
    \label{fe_1d}
  \end{subfigure}
  \begin{subfigure}[h]{.35\textwidth}
    \caption{}
    \label{fe_2d_square}
  \end{subfigure}
  \begin{subfigure}[h]{.35\textwidth}
    \caption{}
    \label{fe_2d_hexagonal}
  \end{subfigure}
  }
  \makebox[\textwidth]{
  \begin{subfigure}[h]{.35\textwidth}
    \centering
    \includegraphics[height=1.3cm]{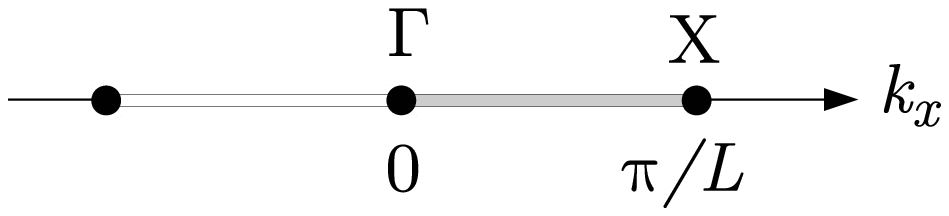}
  \end{subfigure}
  \begin{subfigure}[h]{.35\textwidth}
    \centering
    \includegraphics[height=3.5cm]{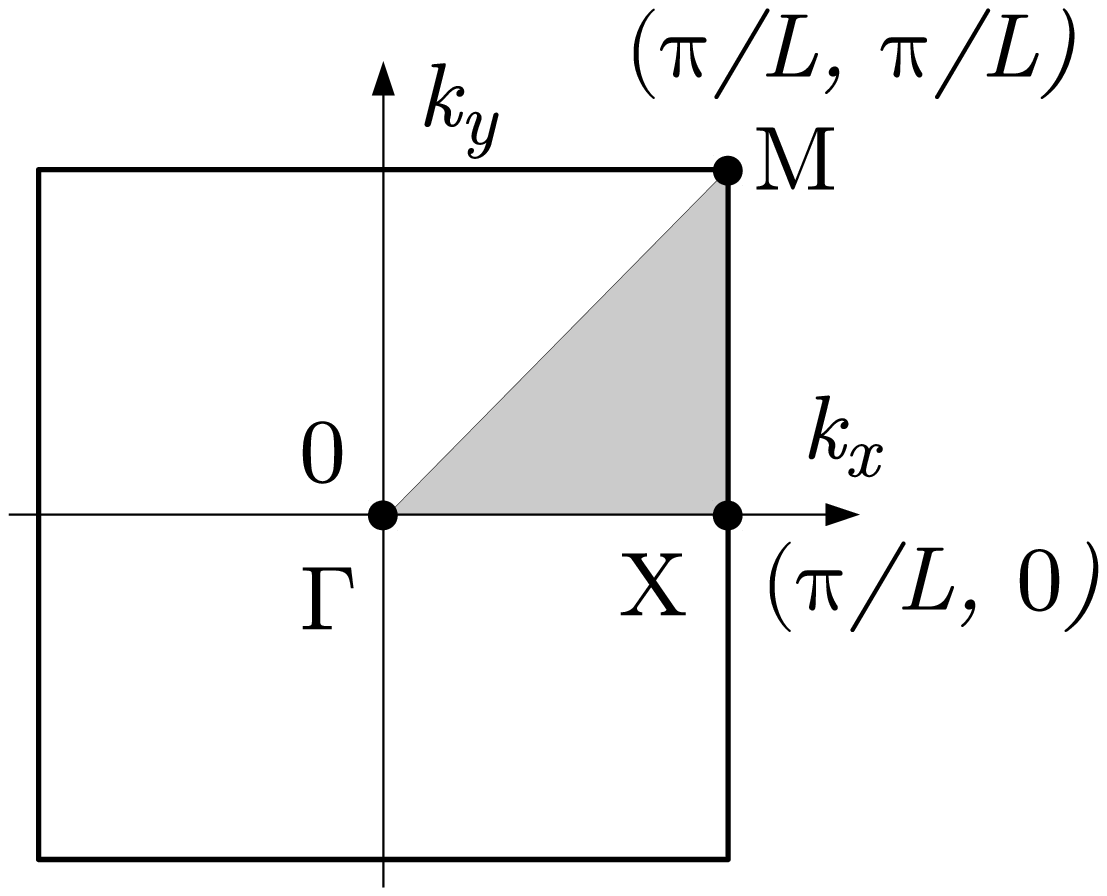}
  \end{subfigure}
  \begin{subfigure}[h]{.35\textwidth}
    \centering
    \includegraphics[height=3.5cm]{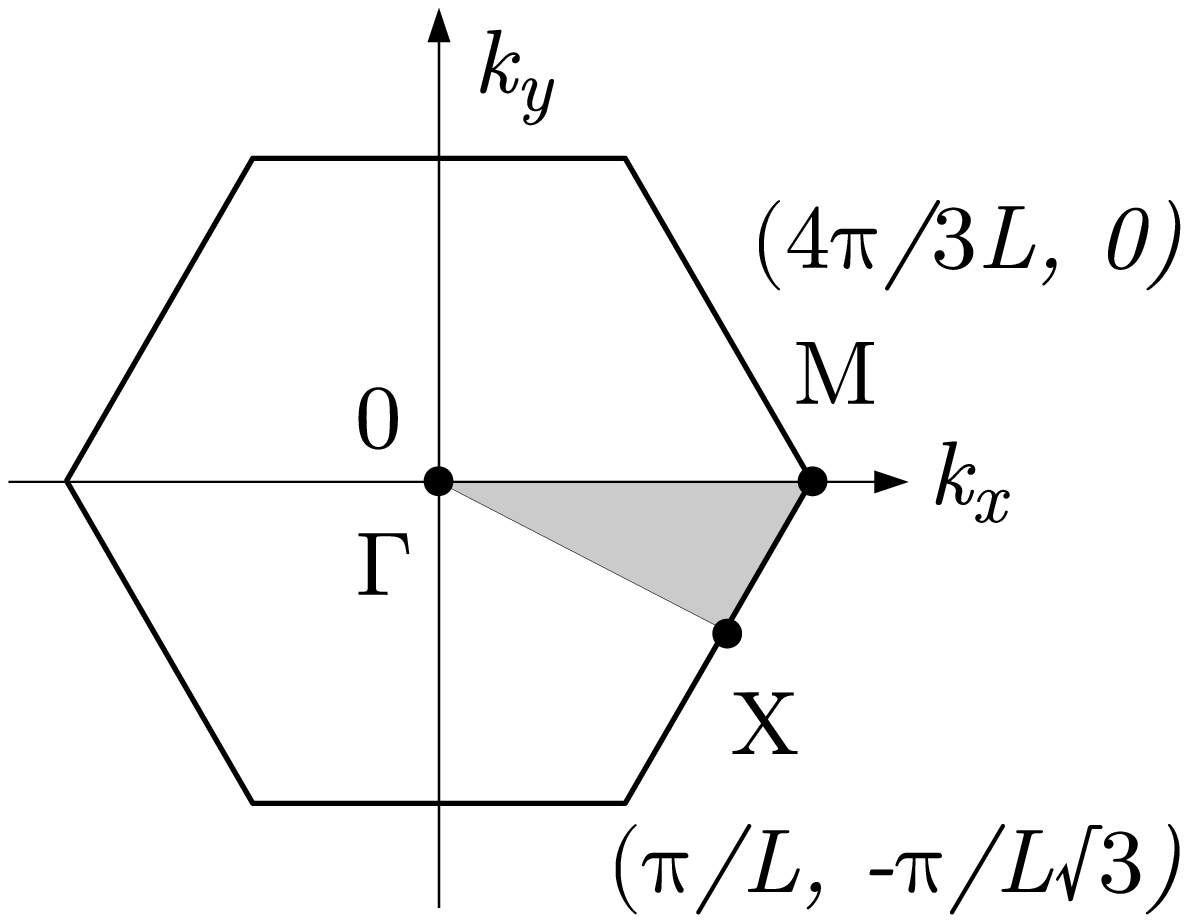}
  \end{subfigure}
  }
  \makebox[\textwidth]{
  \begin{subfigure}[h]{.35\textwidth}
    \caption{}
    \label{fbz1d}
  \end{subfigure}
  \begin{subfigure}[h]{.35\textwidth}
    \caption{}
    \label{fbz2d_square}
  \end{subfigure}
  \begin{subfigure}[h]{.35\textwidth}
    \caption{}
    \label{fbz2d_hexagonal}
  \end{subfigure}
  }
  \caption{Periodic media with respective periodic cells (marked using gray areas) for (\subref{fe_1d}) one-dimensional and two-dimensional (\subref{fe_2d_square}) square and (\subref{fe_2d_hexagonal}) hexagonal lattices. First Brillouin zones and their respective irreducible zones (marked using gray areas) for (\subref{fbz1d}) one-dimensional and two-dimensional (\subref{fbz2d_square}) square and (\subref{fbz2d_hexagonal}) hexagonal lattices.}
  \label{periodic_cells}
\end{figure}

Band diagrams report the relation between wavenumbers and wave propagation frequencies, and can be obtained using FE-based methods. Henceforth, we consider the periodic one-dimensional lattice of length $L$ and two-dimensional square and hexagonal lattices of size $L$, as depicted in Figures \ref{fe_1d} -- \ref{fe_2d_hexagonal},with periodic cells highlighted in gray.

For the computation of band diagrams, we consider the approach described by \citet{mace2008modelling}, leading to a $\omega = \omega(\mathbf{k})$ problem \citep{miranda2019flexural}, where $\omega$ is the propagating wave frequency and $\mathbf{k}$ is the wave vector. Further details on the implementation of this method can be found in \ref{app_wfem}. The corresponding first Brillouin zone (FBZ) \citep{brillouin1953wave} regions are depicted in Figures \ref{fbz1d} -- \ref{fbz2d_hexagonal} with their irreducible Brillouin zone (IBZ) regions marked using gray areas, and denoting the components of wave vector $\mathbf{k}$ as $k_x$ and $k_y$. These definitions follow the work of \citet{maurin2018probability}.

In the case of two-dimensional periodic lattices, the usual procedure is to scan the contour of the first IBZ, which is sufficient for the determination of the BGs, as long as symmetry is maintained. This band diagram computation procedure can now be used with the proposed PCs with periodicity in one and two dimensions for the investigation of elastic wave BGs.

\section{Results} \label{results}

For the computation of results, we consider the material properties of Digital ABS Plus (PolyJet 3D printing material \citep{digitalabsplus}), in view of a potential experimental realization: Young's modulus $E = 2800$ MPa, Poisson's ratio $\nu = 0.35$ (assumed), and mass density $\rho = 1175$ kg/m$^3$. This is a commonly used material in 3D printing. We now investigate the dispersion relations for the one- and two-dimensional structures, determining the weight reduction and stiffness variation when using the hierarchical structures, and the corresponding BGs. To determine the stability of the proposed structures, a brief analysis of their quasistatic properties is presented in \ref{app_quasistatic}.

\subsection{Band diagrams} \label{band_diagrams}

\subsubsection{First hierarchical level}

\begin{figure}[h!]
  \makebox[\textwidth]{
  \centering
  \begin{minipage}[l]{.5\textwidth}
    \begin{subfigure}[h]{\textwidth}
      \centering
      \includegraphics[height=5cm]{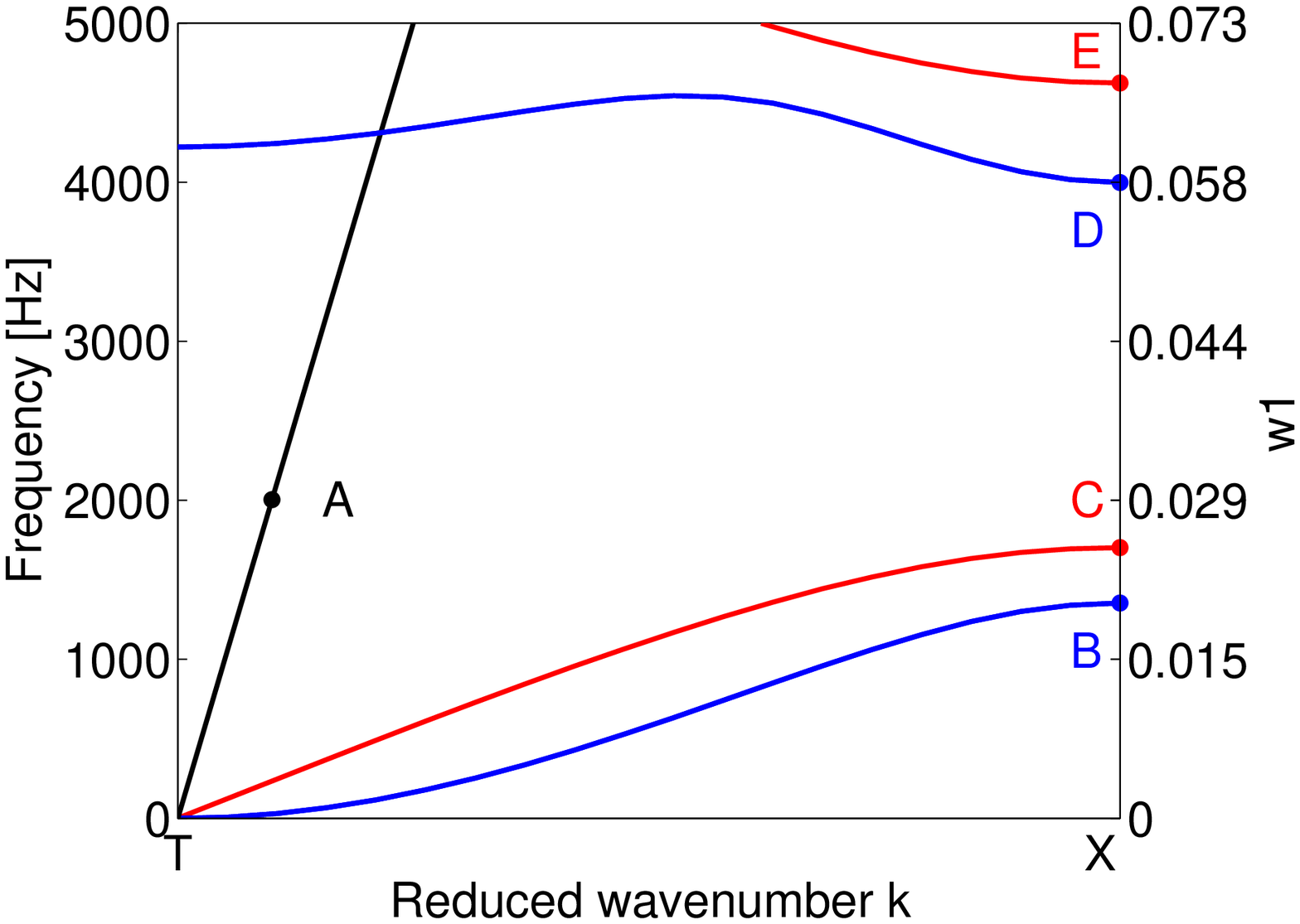}
    \end{subfigure}
    \end{minipage}
    \begin{minipage}[c]{.65\textwidth}
    \begin{subfigure}[h]{.3\textwidth}
      \centering
      \includegraphics[height=1.75cm,trim={3cm 3.5cm 3cm 3.5cm},clip]{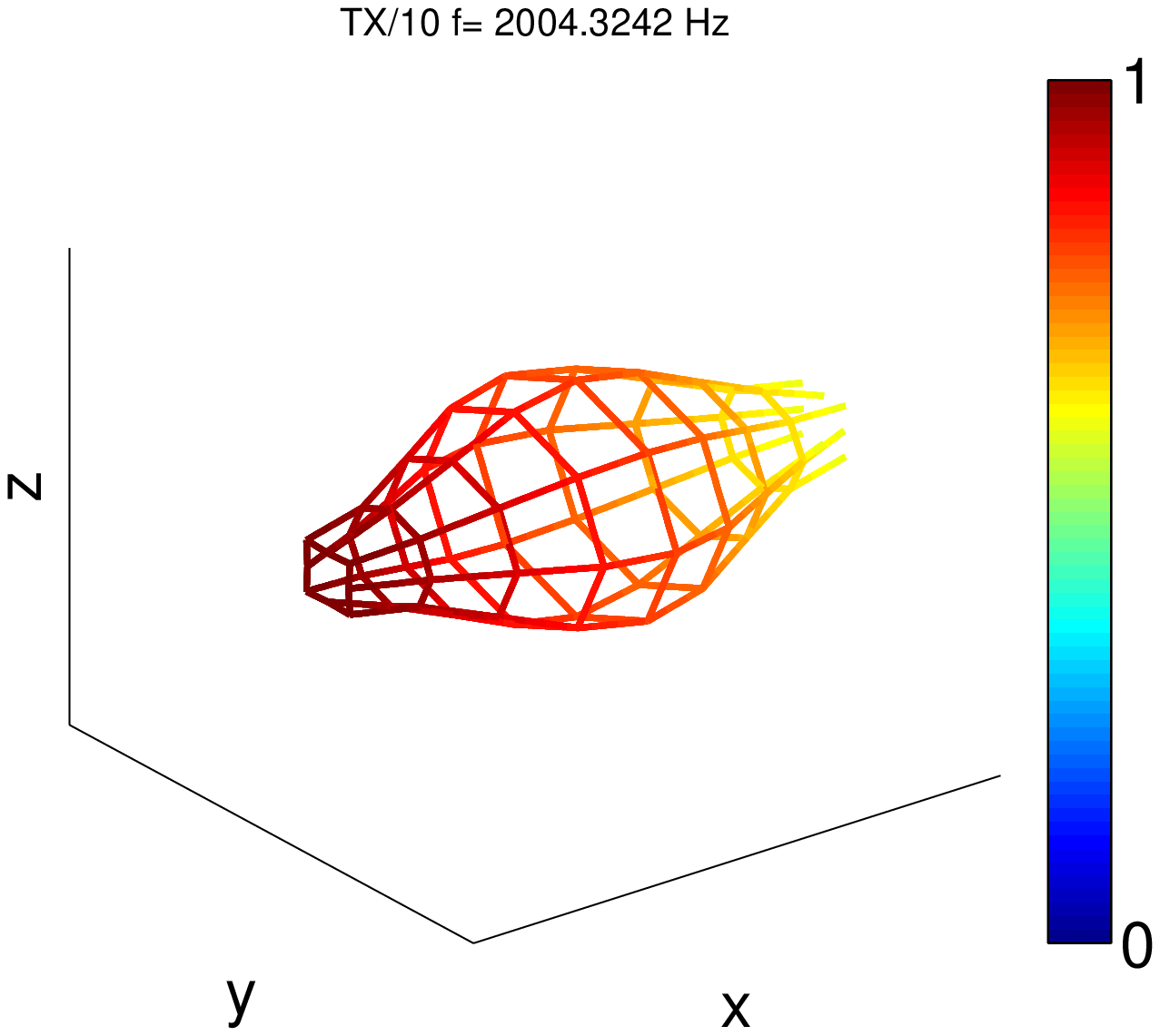}
      \\\vspace{-0.5cm}A
    \end{subfigure}
    \begin{subfigure}[h]{.3\textwidth}
      \centering
      \includegraphics[height=1.75cm,trim={3cm 3.5cm 3cm 3.5cm},clip]{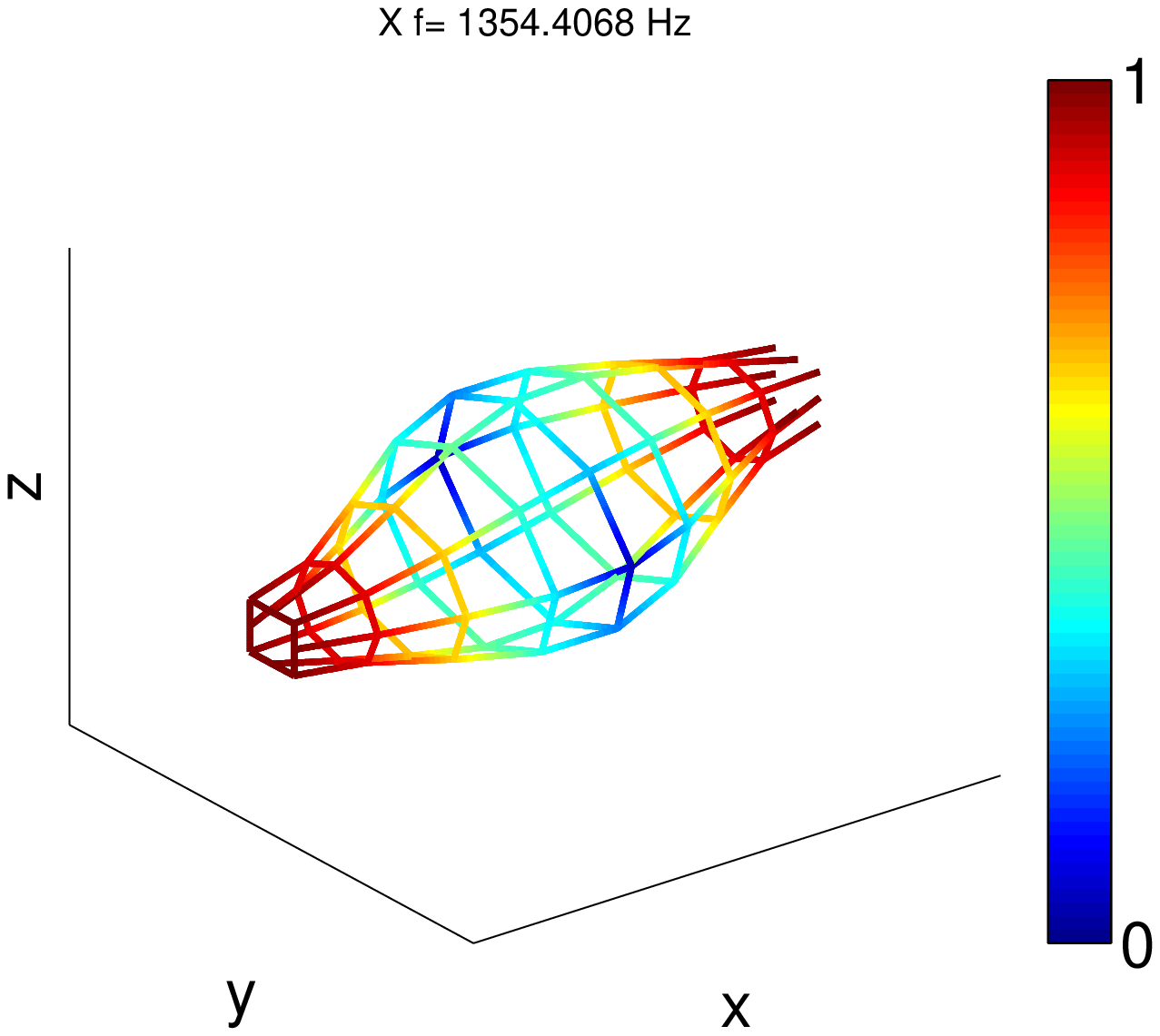}
      \\\vspace{-0.5cm}B
    \end{subfigure}
    \begin{subfigure}[h]{.3\textwidth}
      \centering
      \includegraphics[height=1.75cm,trim={3cm 3.5cm 3cm 3.5cm},clip]{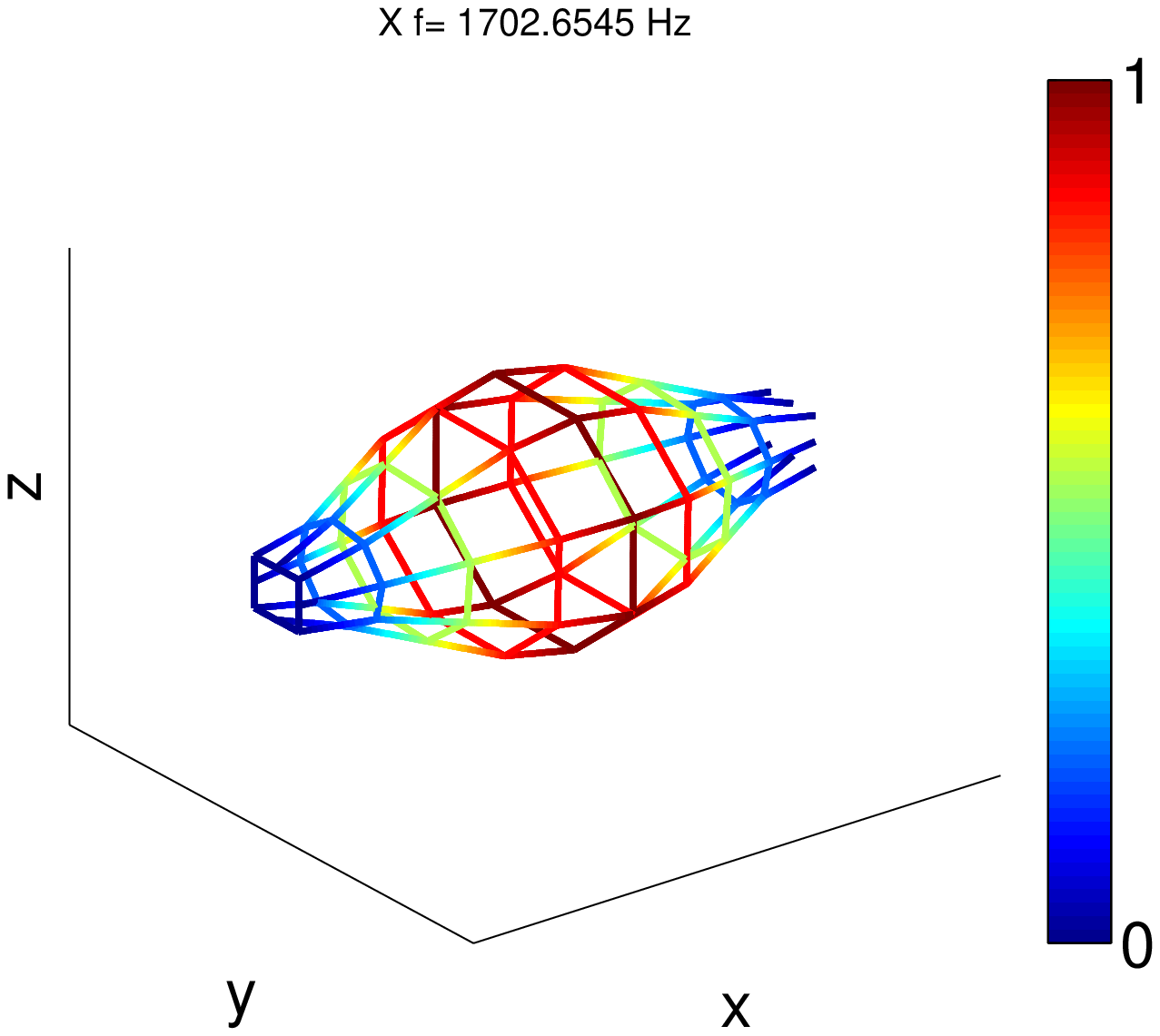}
      \\\vspace{-0.5cm}C
    \end{subfigure}
    
    \begin{subfigure}[h]{.3\textwidth}
      \centering
      \includegraphics[height=1.75cm,trim={3cm 3.5cm 3cm 3.5cm},clip]{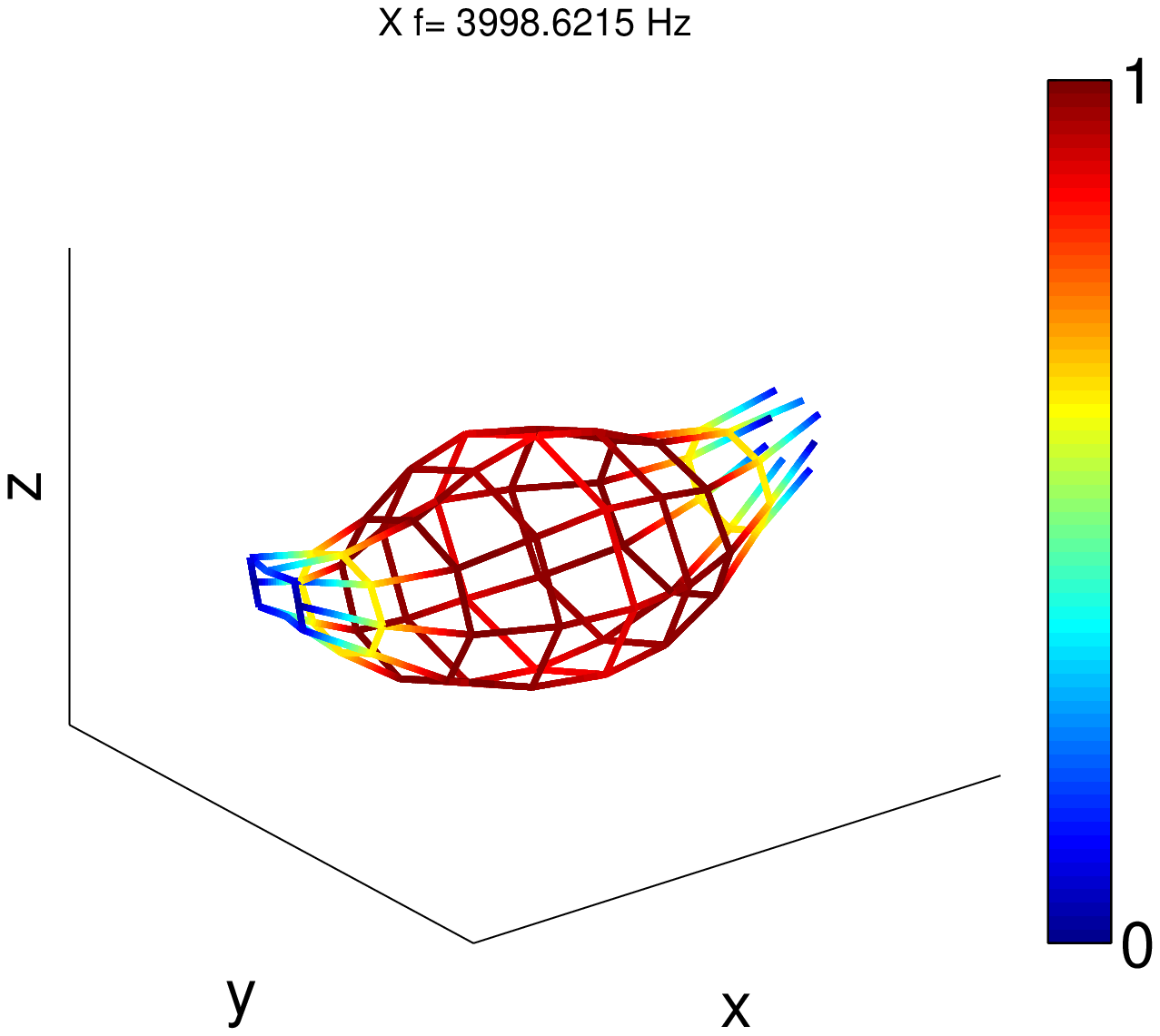}
      \\\vspace{-0.5cm}D
    \end{subfigure}
    \begin{subfigure}[h]{.3\textwidth}
      \centering
      \includegraphics[height=1.75cm,trim={3cm 3.5cm 3cm 3.5cm},clip]{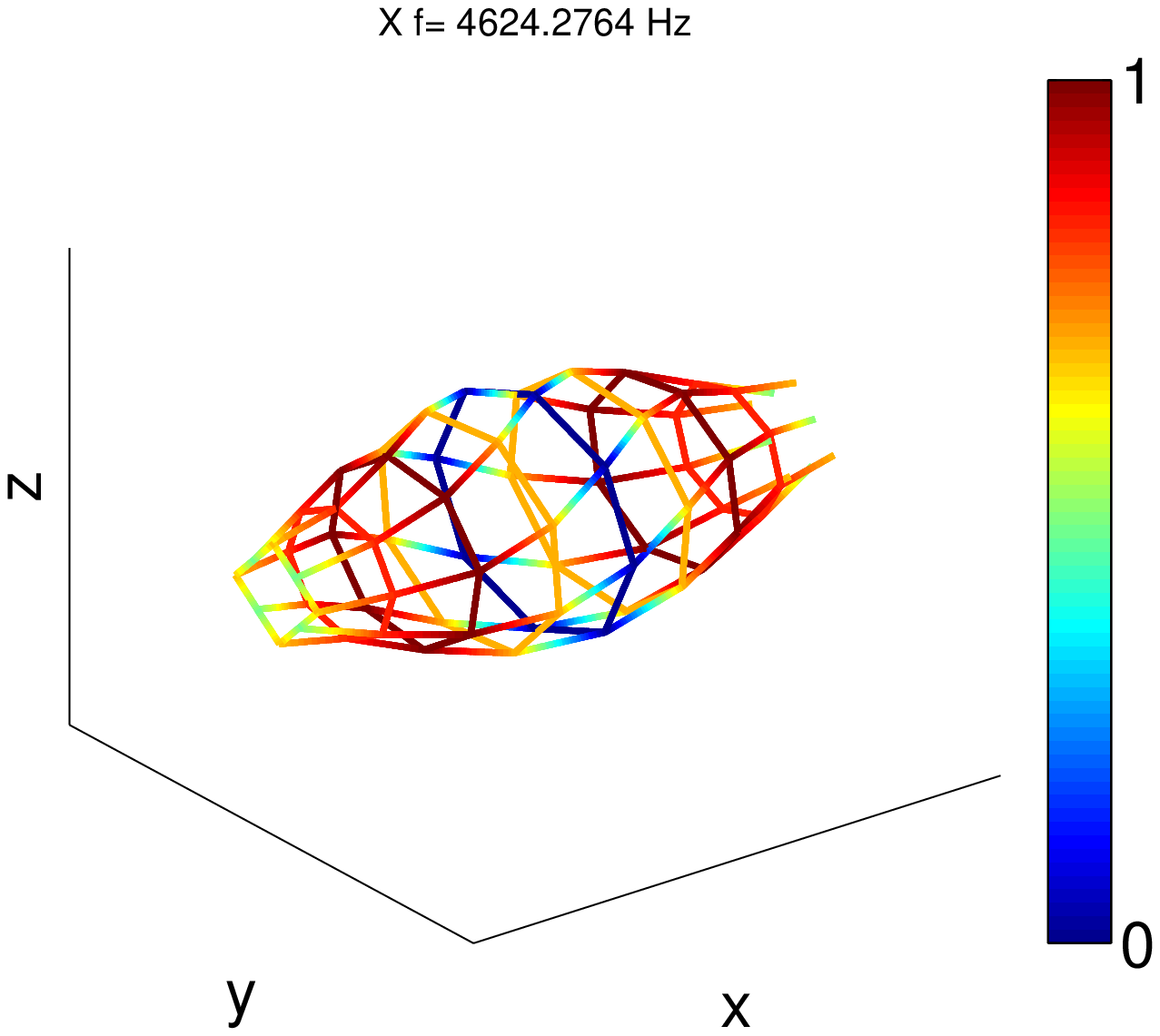}
      \\\vspace{-0.5cm}E
    \end{subfigure}
    \begin{subfigure}[h]{.3\textwidth}
      \centering
      \includegraphics[height=2.5cm,trim={12cm 0cm 0cm 0cm},clip]{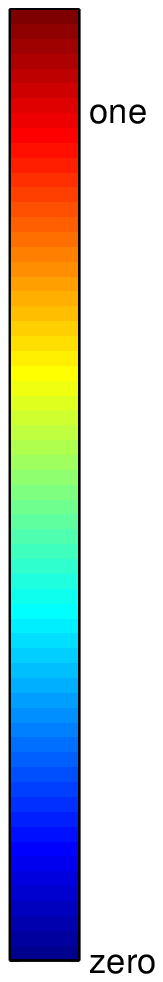}\\
    \end{subfigure}

  \end{minipage}
  }
  \makebox[\textwidth]{
  \begin{minipage}[l]{.5\textwidth}
      \begin{subfigure}[h]{\textwidth}
        \caption{}
        \label{level1_spiderweb_band_diagram}
      \end{subfigure}
  \end{minipage}
  \begin{minipage}[c]{.65\textwidth}
      \begin{subfigure}[h]{\textwidth}
        \caption{}
        \label{level1_spiderweb_modes}
      \end{subfigure}
  \end{minipage}
  }
  
  \caption{(\subref{level1_spiderweb_band_diagram}) Band diagram and the (\subref{level1_spiderweb_modes}) corresponding mode shapes of the first hierarchical level, colored according to their normalized displacement amplitudes (maximum absolute displacement is $1$, no displacement is $0$). Bands are colored according to their dominant behavior: bending (\textcolor{blue}{\textbf{--}}, wave mode shapes B and D), torsional (\textcolor{red}{\textbf{--}}, wave mode shapes C and E), and longitudinal (\textbf{--}, wave mode A). A bending BG is observed between $1354$ Hz and $3999$ Hz, and a torsional BG between $1703$ Hz and $4624$ Hz.}
  \label{level1_spiderweb_band_modes}
\end{figure}

The first investigation is made using a length of $l_c = 22.5$ mm for the one-dimensional PC shown in Figure \ref{geometry_rationale}, with $n_x = 9$ cross-sections, $n_r = 8$ radial threads, elements of circular cross-section of radius $r = 300$ $\mu$m, and fixed values of $d_{\max} = l_c/2 = 11.25$ mm, $l_s = 2.5$ mm, and $d_{\min} = l_s \sqrt{2} = 3.54$ mm. This structure occupies a volume of approximately $970$ mm$^3$, with a mass of $0.124 \times 10^{-3}$ kg, thus yielding an apparent mass density of approximately $128$ kg/m$^3$, i.e., $11 \%$ of the homogeneous material mass density. This low apparent mass density is due to the porous structure of the rather sparse periodic cell. The corresponding band diagram and the displacements of the wave mode shapes are shown in Figure \ref{level1_spiderweb_band_modes} with the dimensionless angular frequency given by $\overline{\omega}_1 = \omega \, l_c / 2 \pi c_L$, where $c_L = \sqrt{E/\rho}$ is the longitudinal wave speed in the homogeneous material.

Figure \ref{level1_spiderweb_band_diagram} shows that there are no complete BGs in this frequency range, i.e., at any given frequency, there is always at least one propagating mode. However, when the wave modes are observed separately (Figure \ref{level1_spiderweb_modes}), it is possible to see that only the longitudinal mode A inhibits the nucleation of a complete BG. As a consequence, the structure presents bending as well as torsional BGs between $1354$ Hz and $3999$ Hz, and between $1703$ Hz and $4624$ Hz, respectively.

\begin{figure}[h!]
  \makebox[\textwidth]{
    \begin{subfigure}[h]{.425\textwidth}
      \centering
      \includegraphics[height=4cm,width=5.5cm]{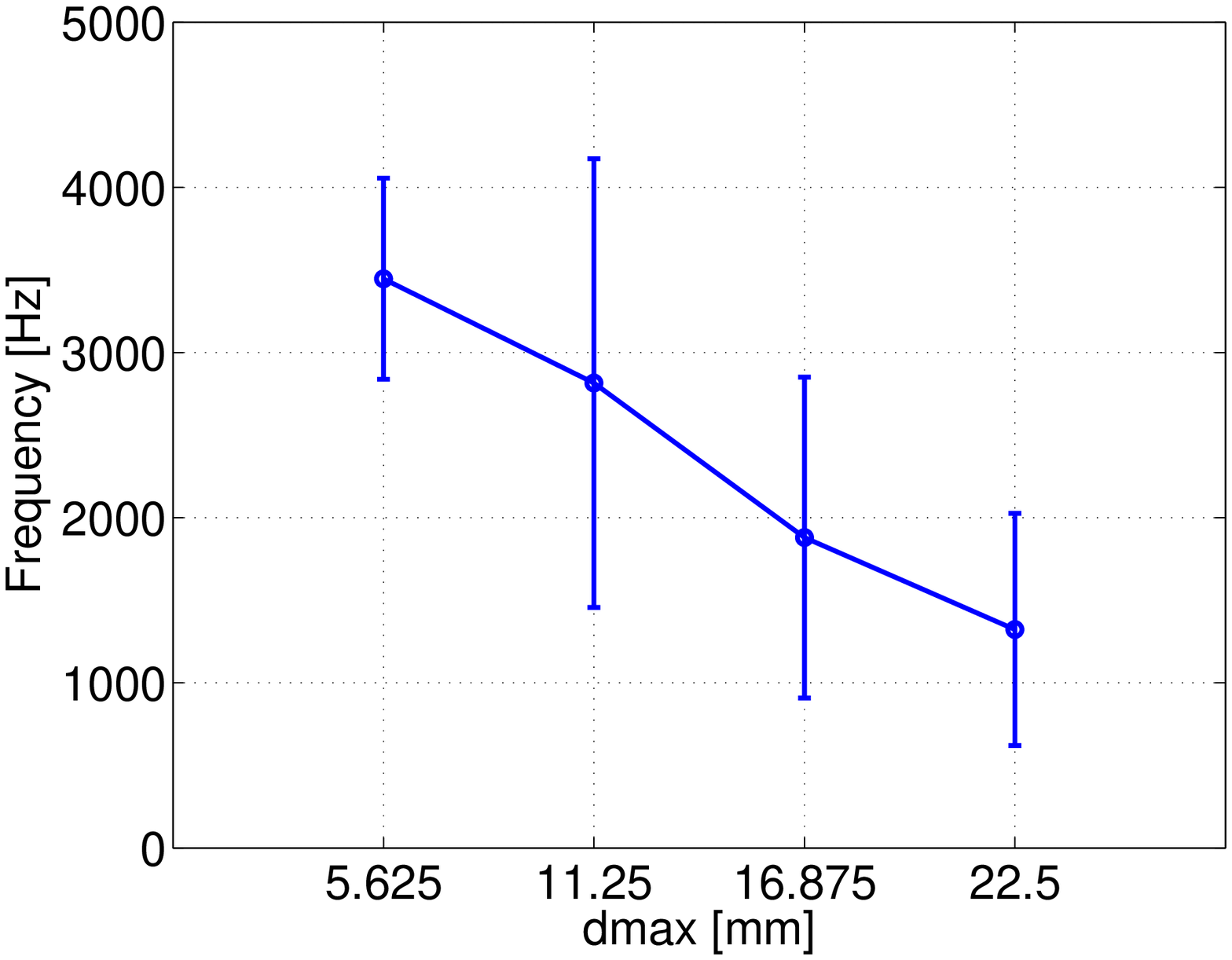}
      \caption{}
      \label{level1_spiderweb_investigation_ls1p5}
    \end{subfigure}
    \begin{subfigure}[h]{.425\textwidth}
      \centering
      \includegraphics[height=4cm,width=5.5cm]{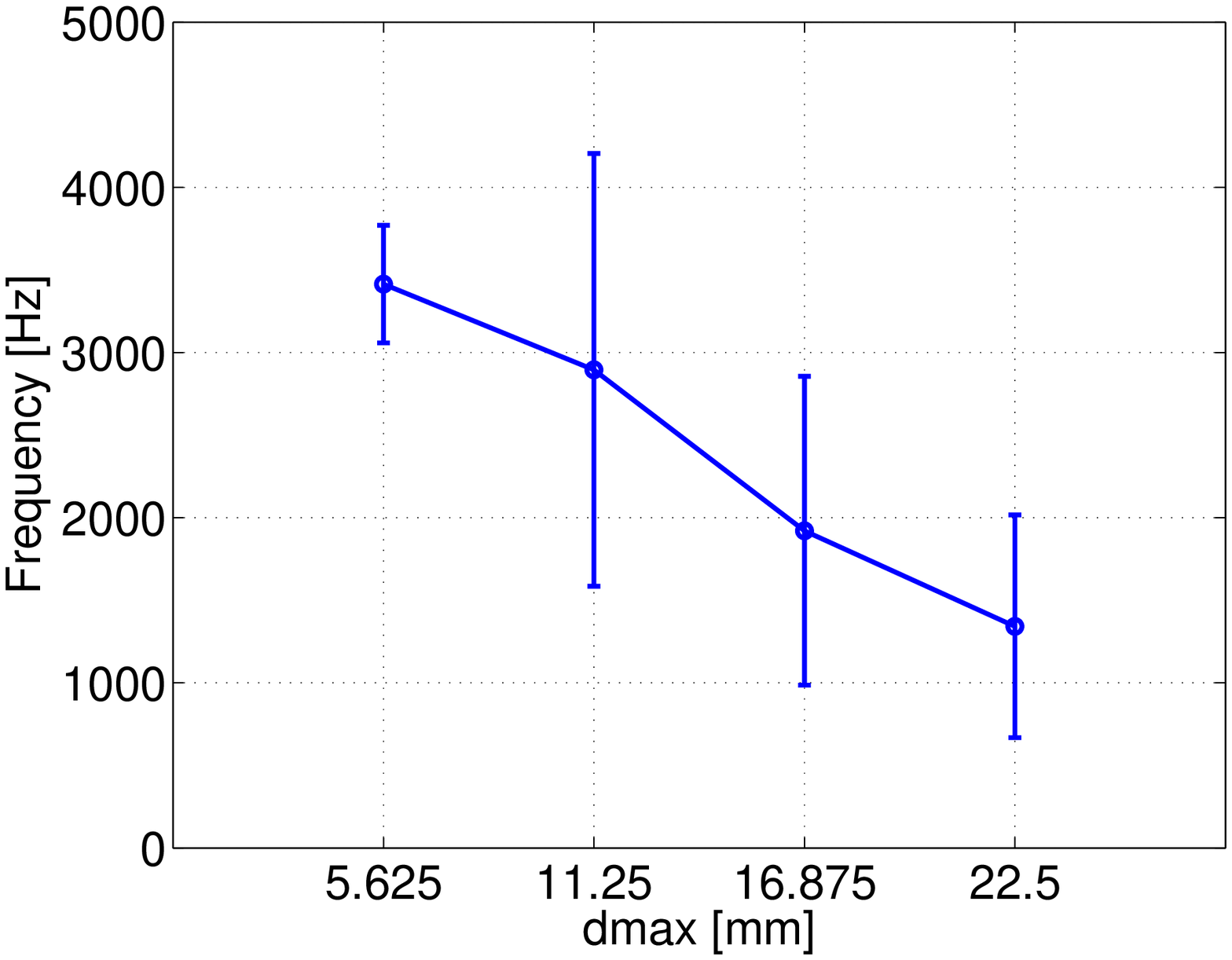}
      \caption{}
      \label{level1_spiderweb_investigation_ls2p0}
    \end{subfigure}
    \begin{subfigure}[h]{.425\textwidth}
      \centering
      \includegraphics[height=4cm,width=5.5cm]{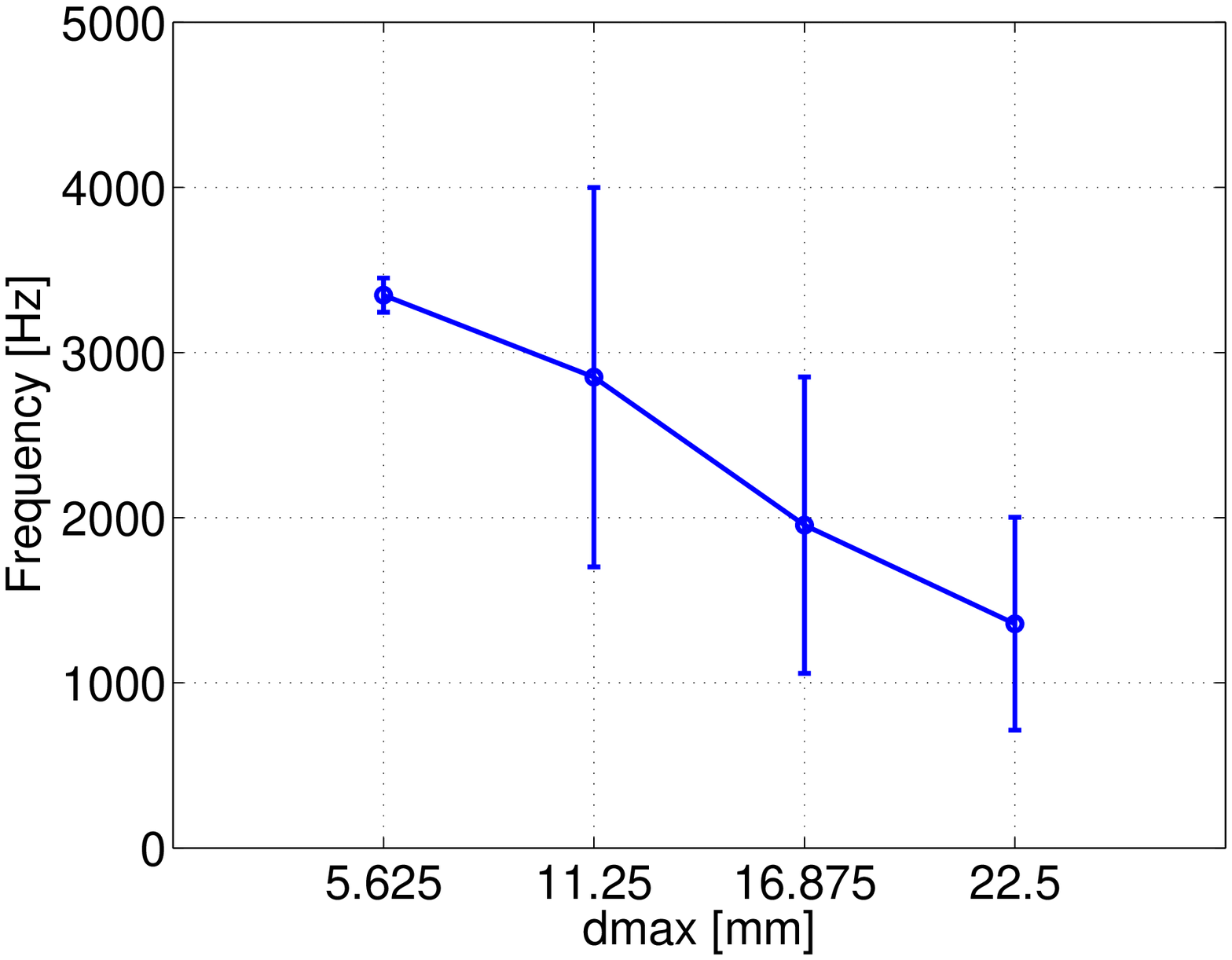}
      \caption{}
      \label{level1_spiderweb_investigation_ls2p5}
    \end{subfigure}
  }

  \makebox[\textwidth]{  
    \begin{subfigure}[h]{.425\textwidth}
      \centering
      \includegraphics[height=4cm,width=5.5cm]{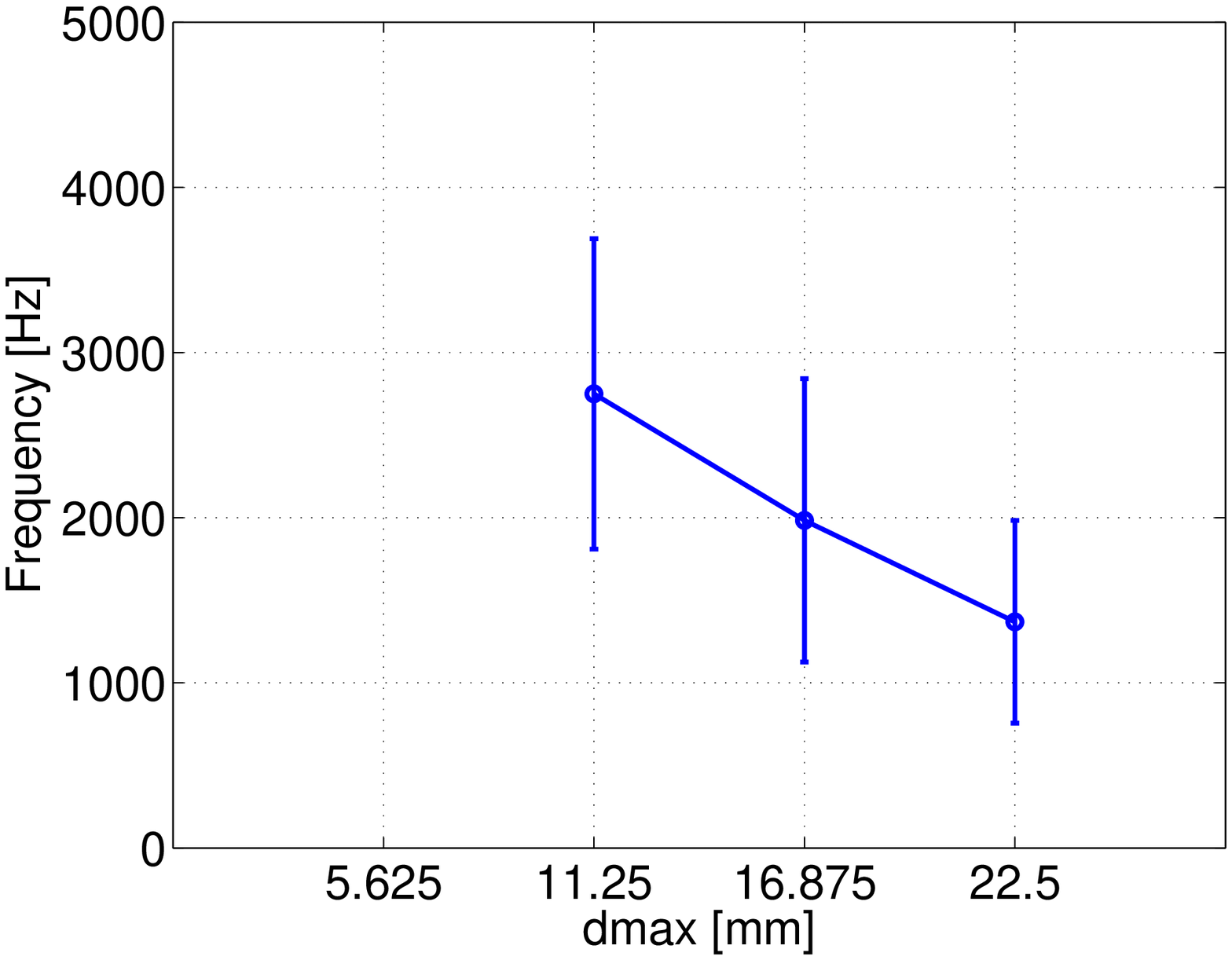}
      \caption{}
      \label{level1_spiderweb_investigation_ls3p0}
    \end{subfigure}
    \begin{subfigure}[h]{.425\textwidth}
      \centering
      \includegraphics[height=4cm,width=5.5cm]{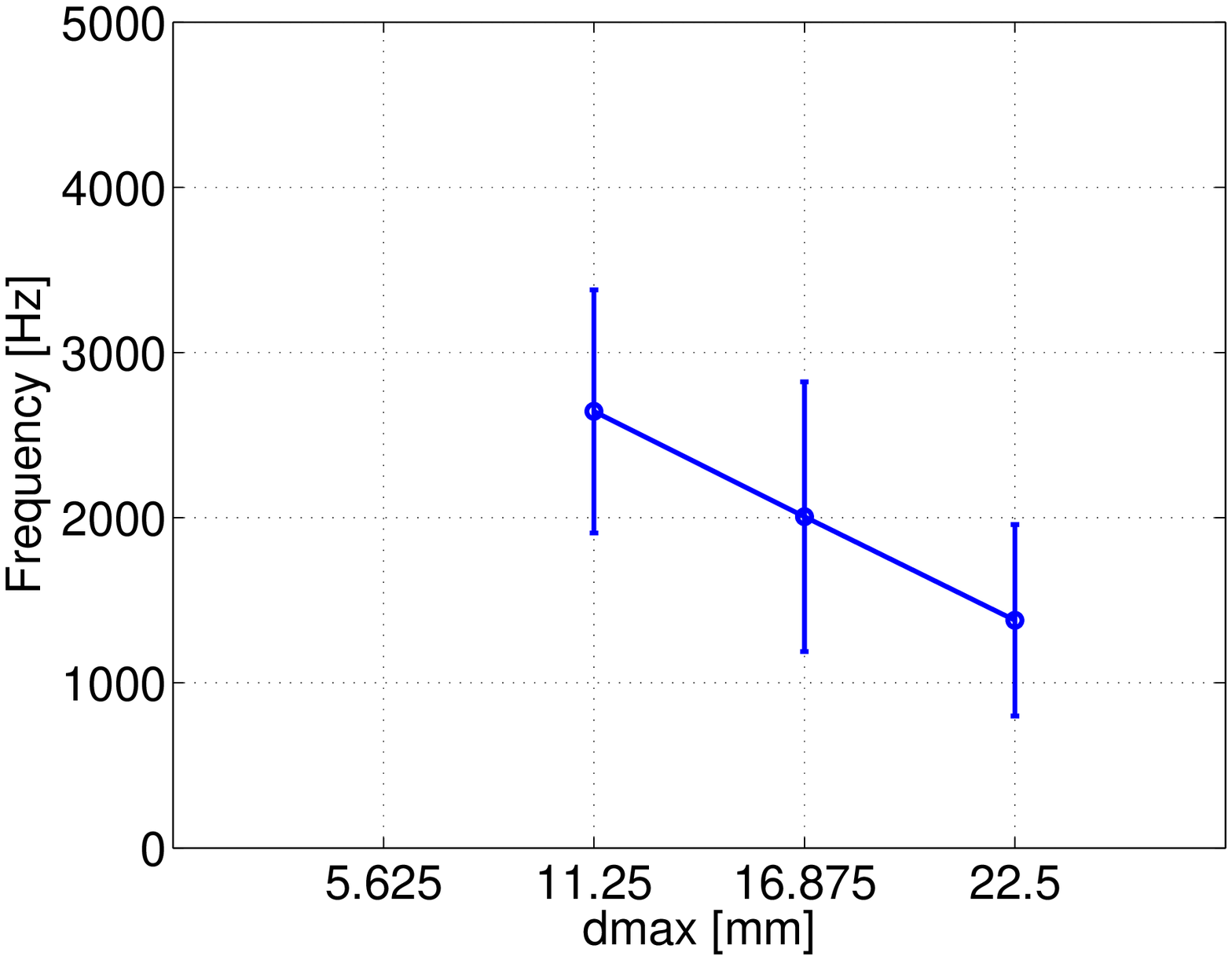}
      \caption{}
      \label{level1_spiderweb_investigation_ls3p5}
    \end{subfigure}
  }
  \caption{Mean frequency (dots) and width (vertical bars) of the lowest BG using several values for $l_s$: (\subref{level1_spiderweb_investigation_ls1p5}) $1.5$ mm, (\subref{level1_spiderweb_investigation_ls2p0}) $2.0$ mm, (\subref{level1_spiderweb_investigation_ls2p5}) $2.5$ mm, (\subref{level1_spiderweb_investigation_ls3p0}) $3.0$ mm, and (\subref{level1_spiderweb_investigation_ls3p5}) $3.5$ mm.}
  \label{level1_spiderweb_investigation}
\end{figure}

Next, we investigate the BGs computed for the values of $l_s = \{ 1.5, \, 2.0, \, 2.5, \, 3.0, \, 3.5 \}$ mm and varying $d_{\max}$ between $0$ and $22.5$ mm. These results are shown in Figure \ref{level1_spiderweb_investigation}.

Figure \ref{level1_spiderweb_investigation} shows the first BG mean frequency and width deriving from torsional and bending behavior. No longitudinal wave mode (tension/compression) BGs are present in these configurations. Band gaps were not computed for $d_{\max} < d_{\min}$. The BGs mean frequencies show a monotonic decreasing behavior for increasing values of $d_{\max}$. For small values of $l_s$ (Figures \ref{level1_spiderweb_investigation_ls1p5} -- \ref{level1_spiderweb_investigation_ls2p5}), the BG width shows a large variation with respect to $d_{\max}$. This relation is less sensitive for larger values of $l_s$ (Figures \ref{level1_spiderweb_investigation_ls3p0} -- \ref{level1_spiderweb_investigation_ls3p5}, for instance). For the investigated values of $l_s$ (and corresponding $d_{\min}$), the value $d_{\max} = 11.25$ mm seems to be the more suitable for BG nucleation. However, since values of $d_{\max}$ were not further discretized, it is not possible to state that other values of $d_{\max}$ would not yield larger BGs. This may be a topic for future investigation using optimization techniques.

When dealing with elastic wave attenuation, one usually aims to obtain wide and low-frequency BGs. Since BG frequencies are associated with the dimensions of PCs (Bragg scattering), it is usual to try to achieve a compromise between the BG mean frequency and the dimensions of the corresponding structure. Since Figure \ref{level1_spiderweb_investigation_ls2p5} shows a good compromise between BG mean frequency, width, and a smooth geometry (with respect to the relation between $l_s$ and $l_c$), we further investigate the hierarchical two-dimensional structures using $l_s = 2.5$ mm and $d_{\max} = 11.25$ mm.

\subsubsection{Second hierarchical level} \label{second_level}

\begin{figure}[h!]
  \centering
  \makebox[\textwidth]{
  \begin{minipage}[c]{.35\textwidth}
    \begin{subfigure}[h]{\textwidth}
      \centering
      \includegraphics[width=5cm]{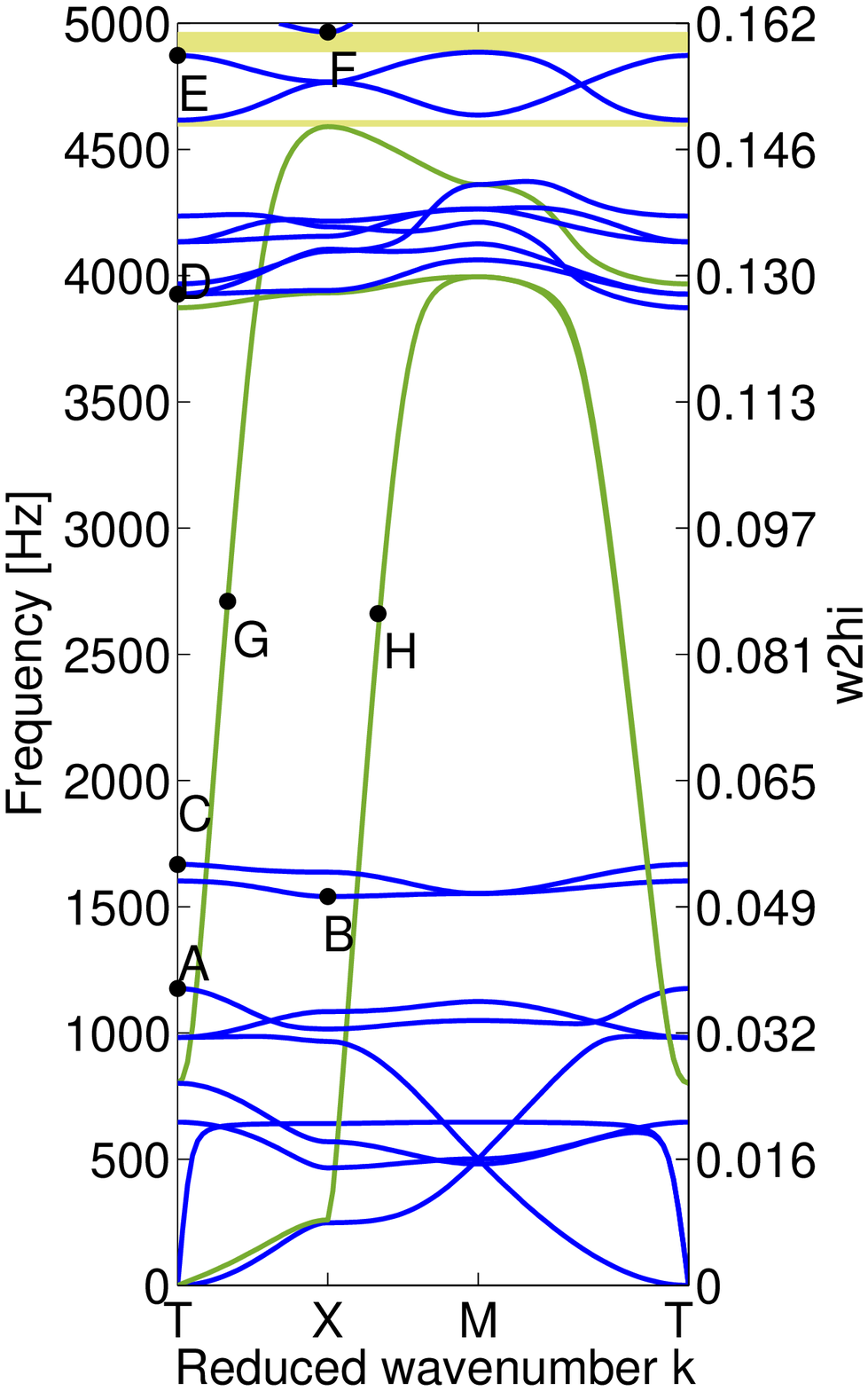}
    \end{subfigure}
  \end{minipage}
  
  \begin{minipage}[c]{.5\textwidth}
  \centering
  \begin{subfigure}[h]{.45\textwidth}
    \centering
    \includegraphics[height=2.0cm,trim={3.5cm 3.5cm 3.5cm 3.5cm},clip]{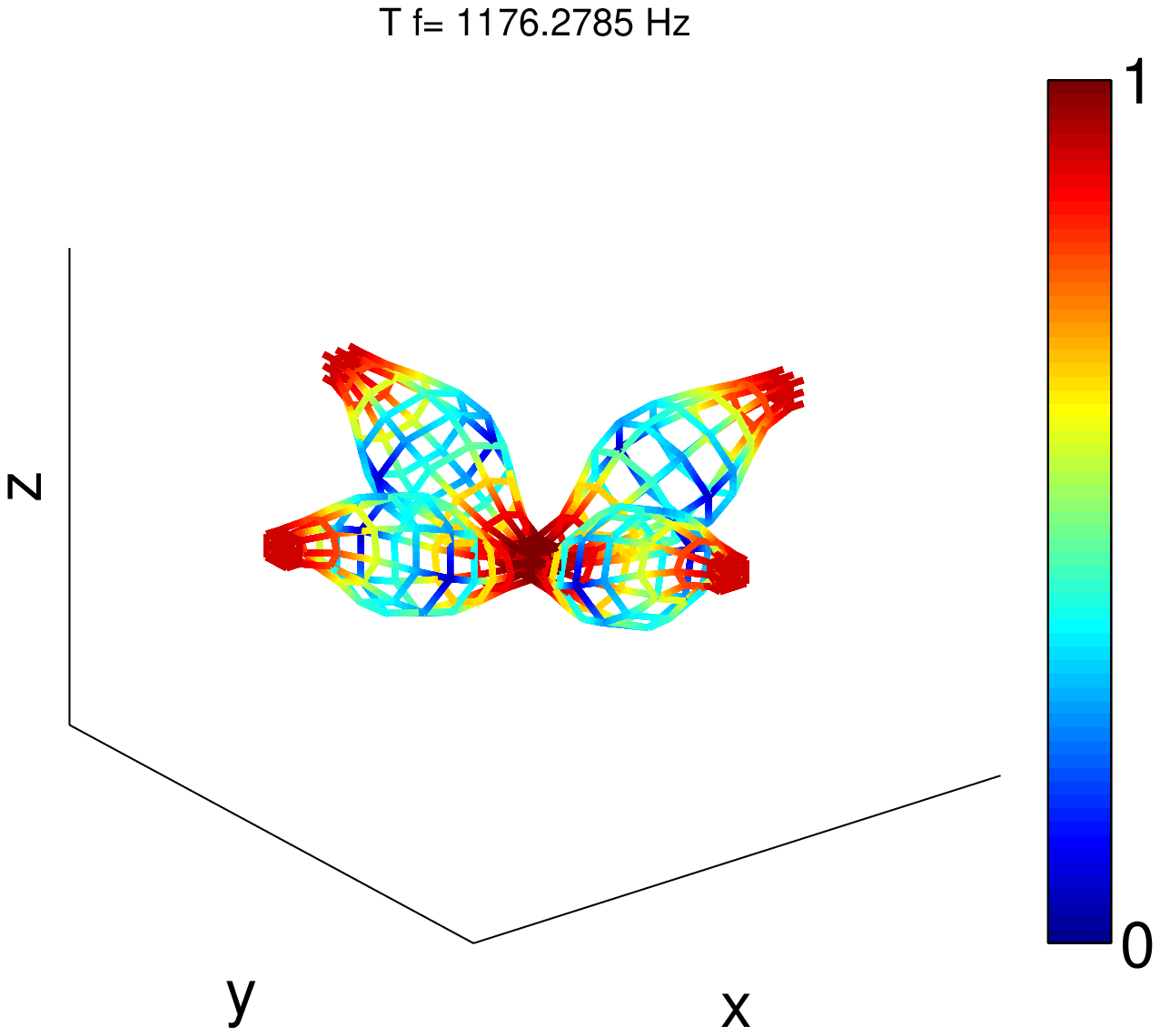}
    \\\vspace{-1cm}A
  \end{subfigure}
  \begin{subfigure}[h]{.45\textwidth}
    \centering
    \includegraphics[height=2.0cm,trim={3.5cm 3.5cm 3.5cm 3.5cm},clip]{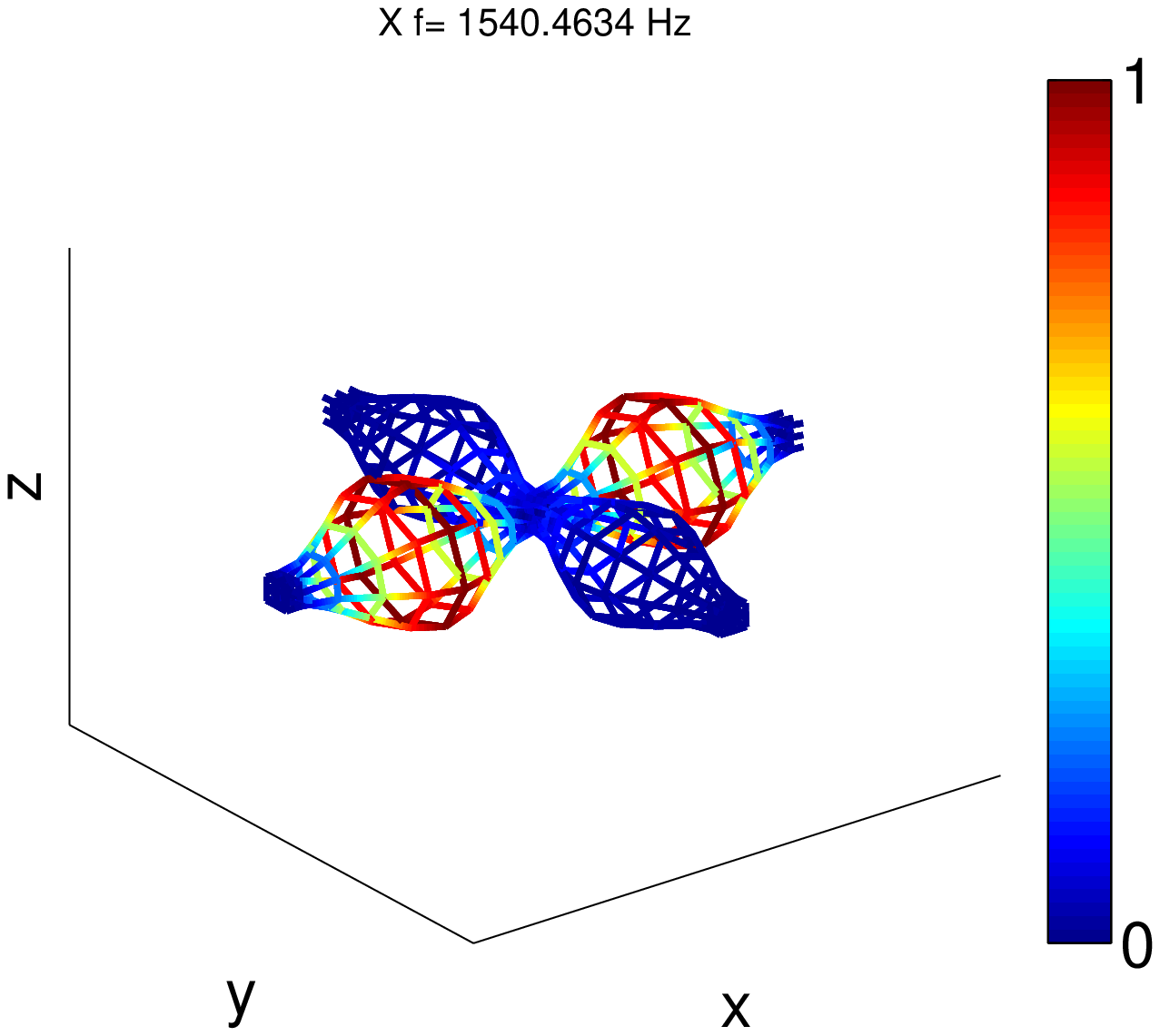}
    \\\vspace{-1cm}B
  \end{subfigure}
  \\\vspace{0.25cm}
  \begin{subfigure}[h]{.45\textwidth}
    \centering
    \includegraphics[height=2.0cm,trim={3.5cm 3.5cm 3.5cm 3.5cm},clip]{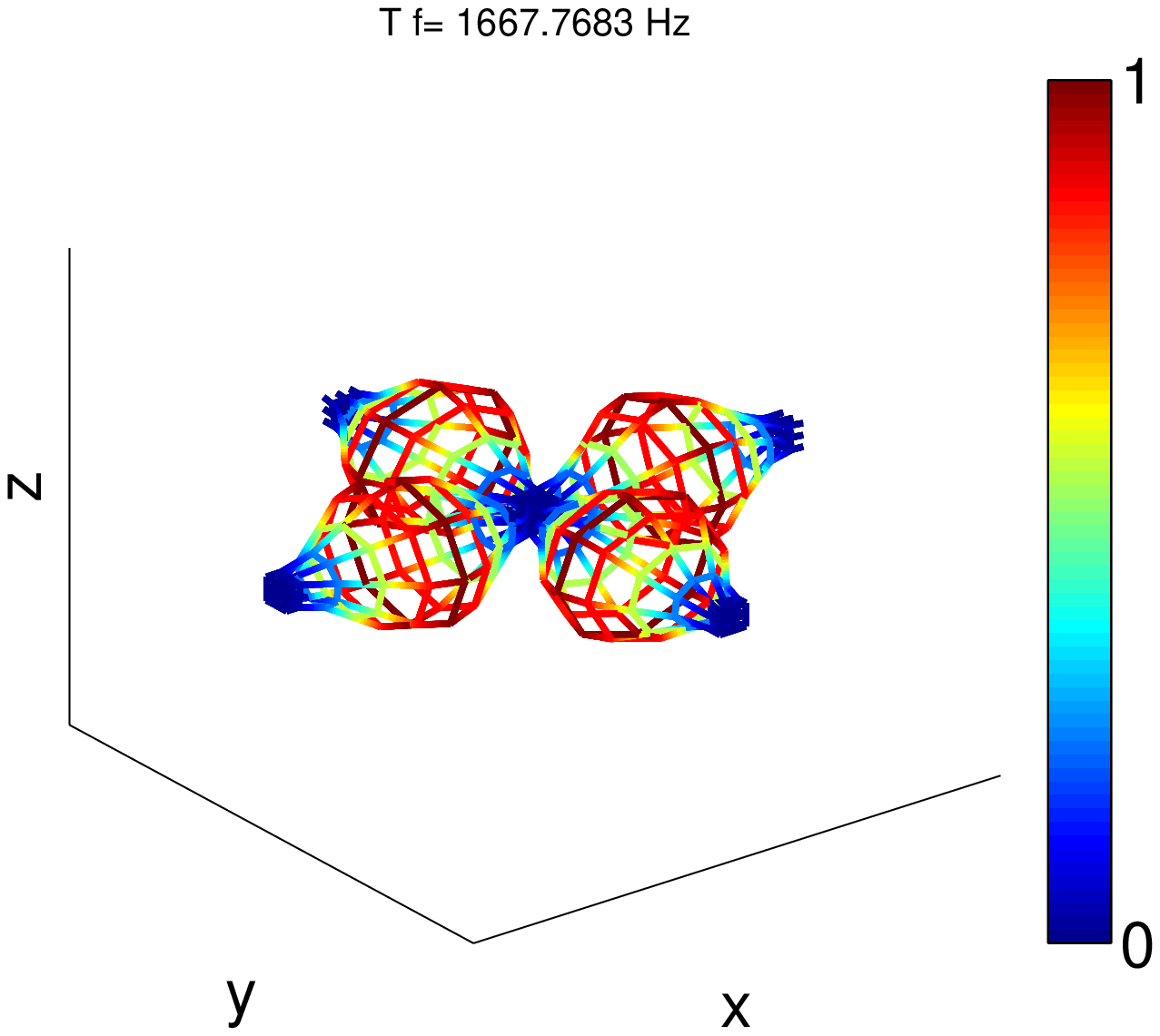}
    \\\vspace{-1cm}C
  \end{subfigure}
  \begin{subfigure}[h]{.45\textwidth}
    \centering
    \includegraphics[height=2.0cm,trim={3.5cm 3.5cm 3.5cm 3.5cm},clip]{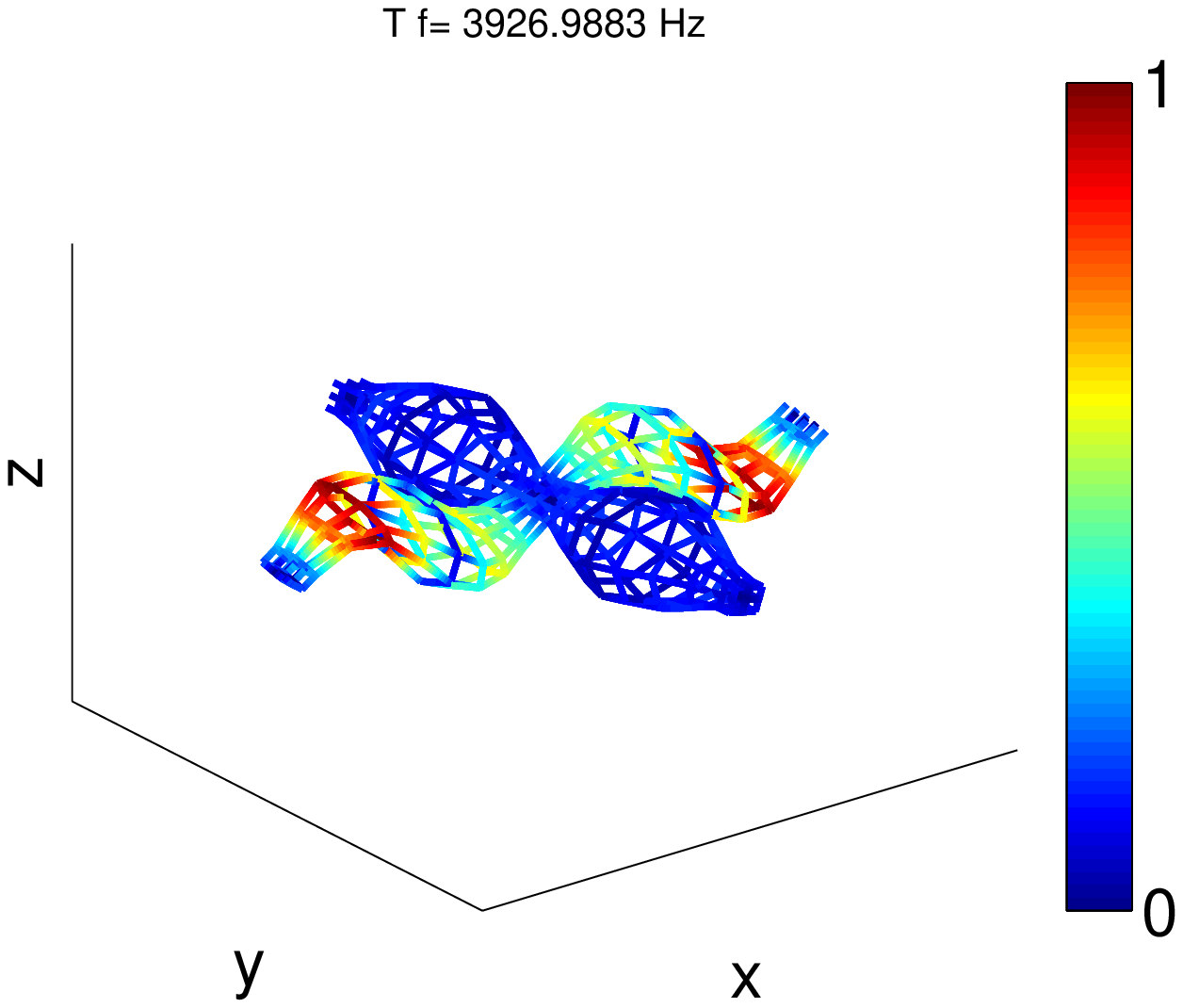}
    \\\vspace{-1cm}D
  \end{subfigure}
  \\\vspace{0.25cm}
  \begin{subfigure}[h]{.45\textwidth}
    \centering
    \includegraphics[height=2.0cm,trim={3.5cm 3.5cm 3.5cm 3.5cm},clip]{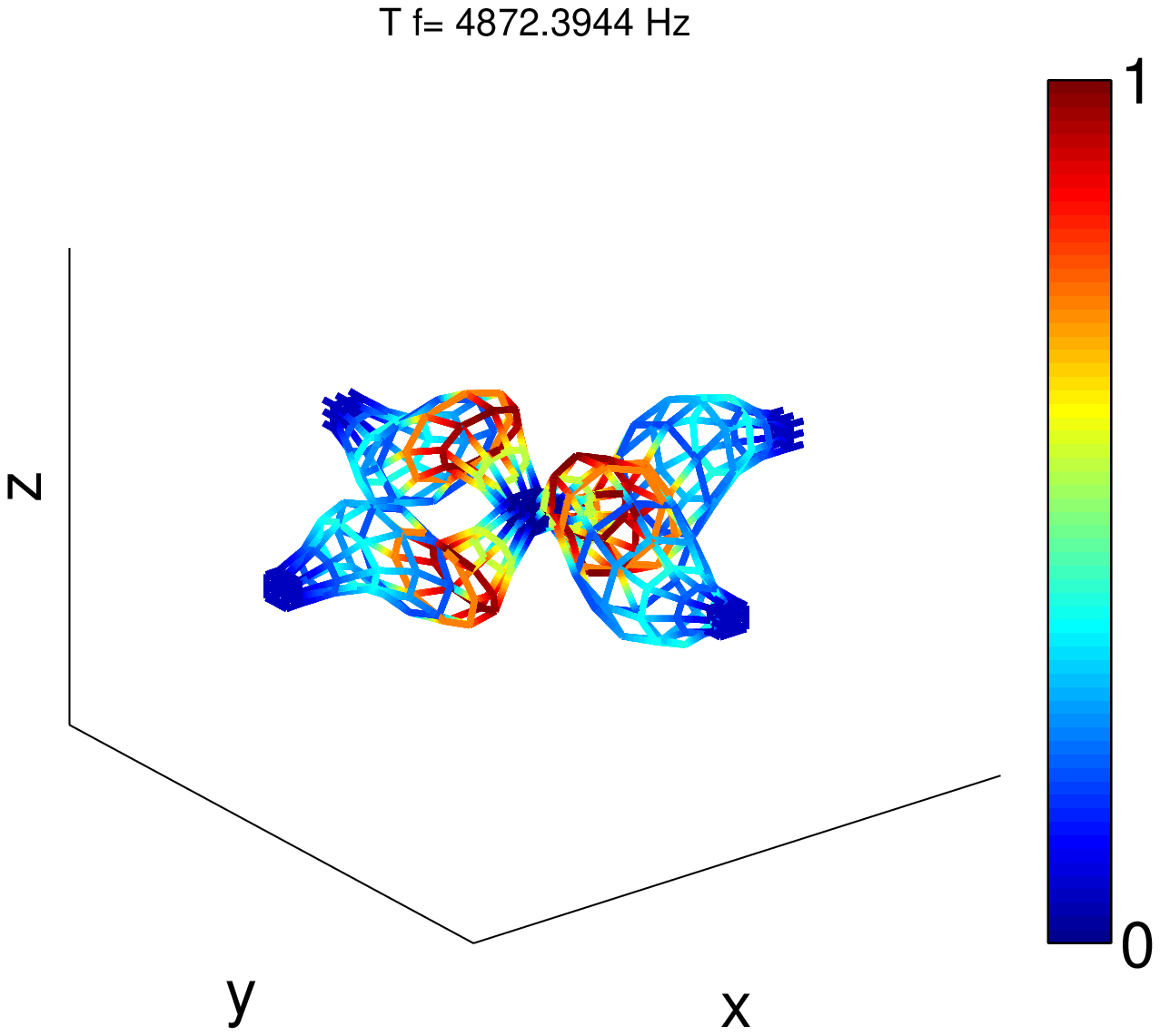}
    \\\vspace{-1cm}E
  \end{subfigure}
  \begin{subfigure}[h]{.45\textwidth}
    \centering
    \includegraphics[height=2.0cm,trim={3.5cm 3.5cm 3.5cm 3.5cm},clip]{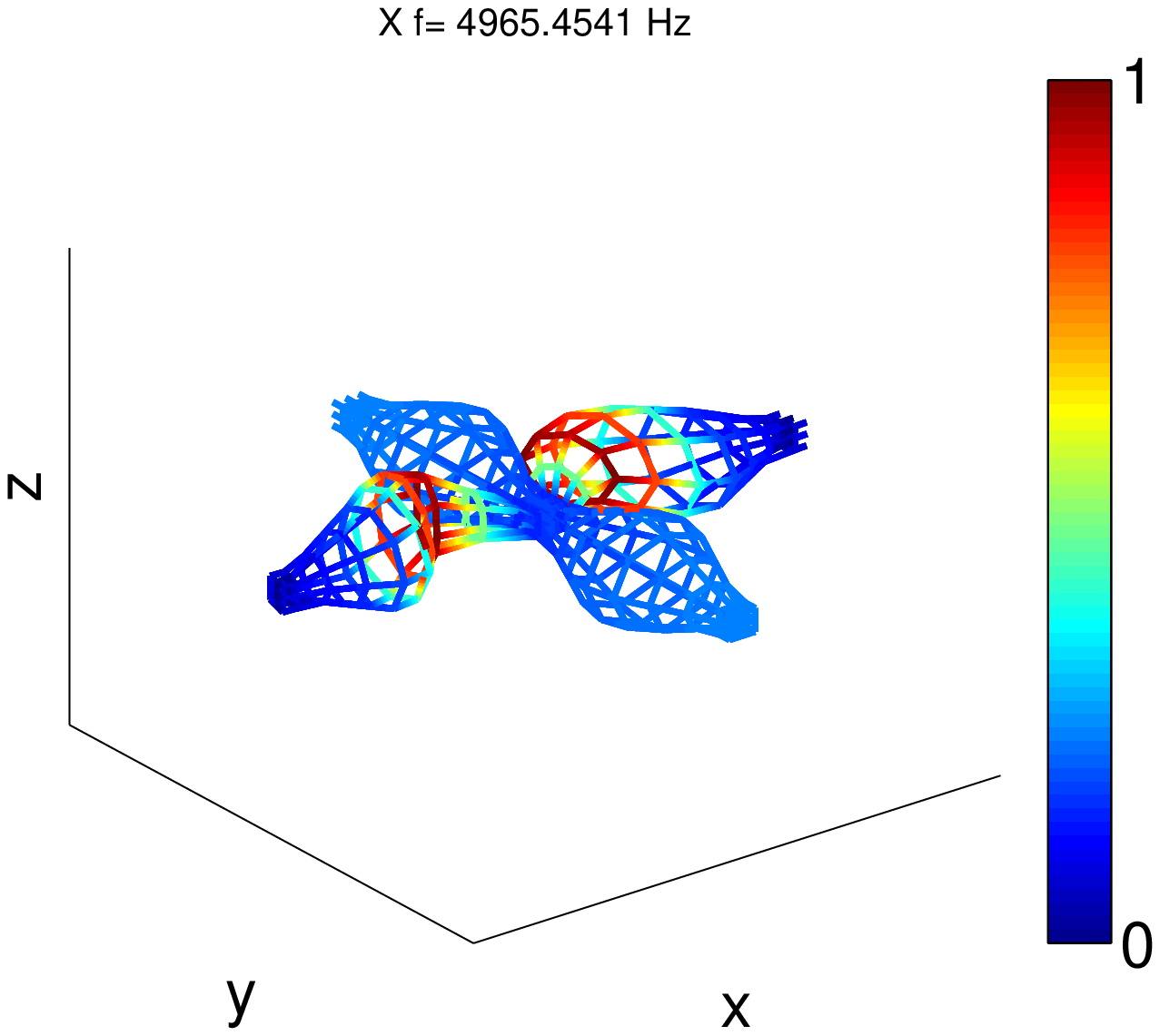}
    \\\vspace{-1cm}F
  \end{subfigure}
  \\\vspace{0.25cm}
  \begin{subfigure}[h]{.45\textwidth}
    \centering
    \includegraphics[height=2.0cm,trim={3.5cm 3.5cm 3.5cm 3.5cm},clip]{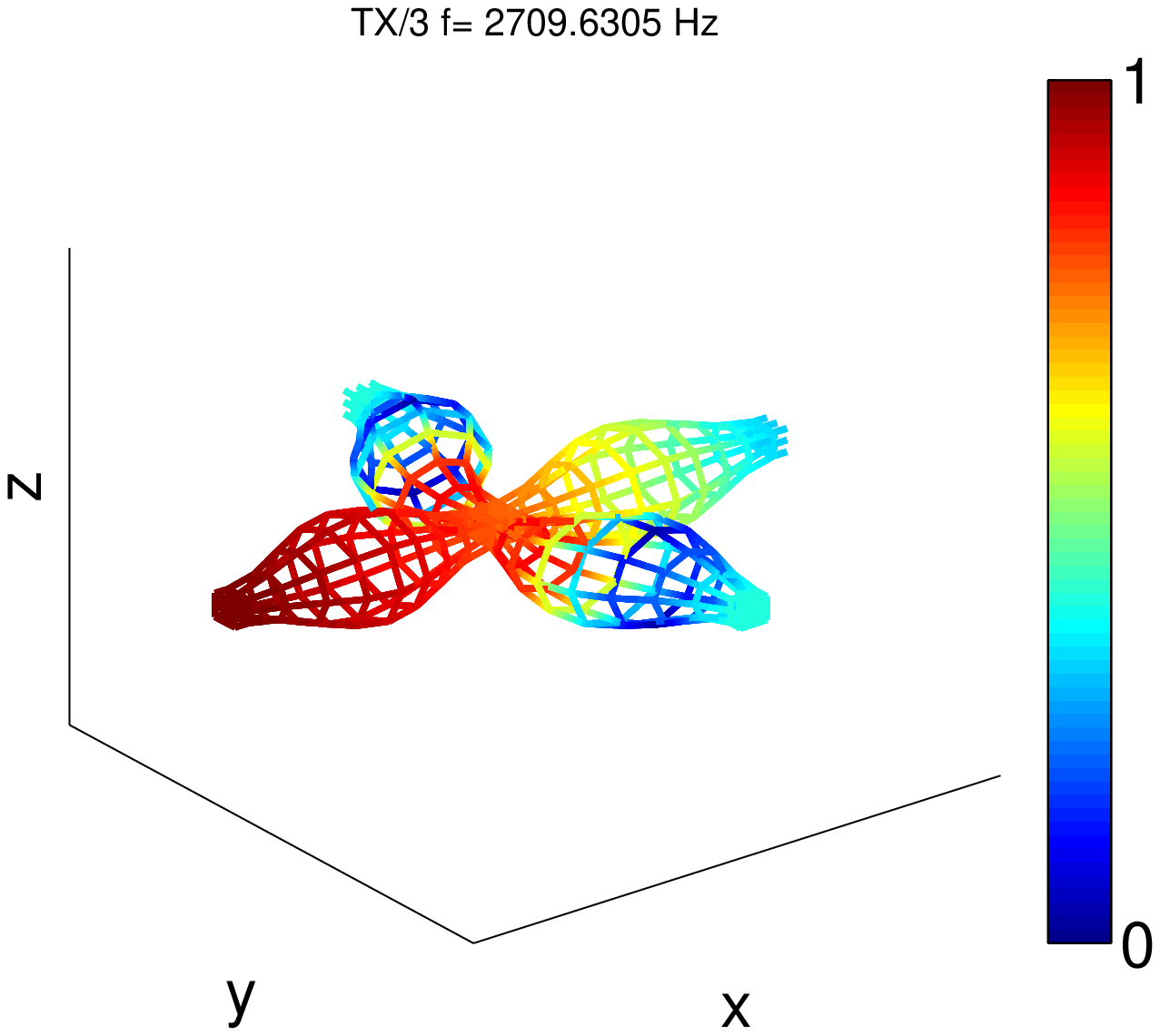}
    \\\vspace{-1cm}G
  \end{subfigure}
  \begin{subfigure}[h]{.45\textwidth}
    \centering
    \includegraphics[height=2.0cm,trim={3.5cm 3.5cm 3.5cm 3.5cm},clip]{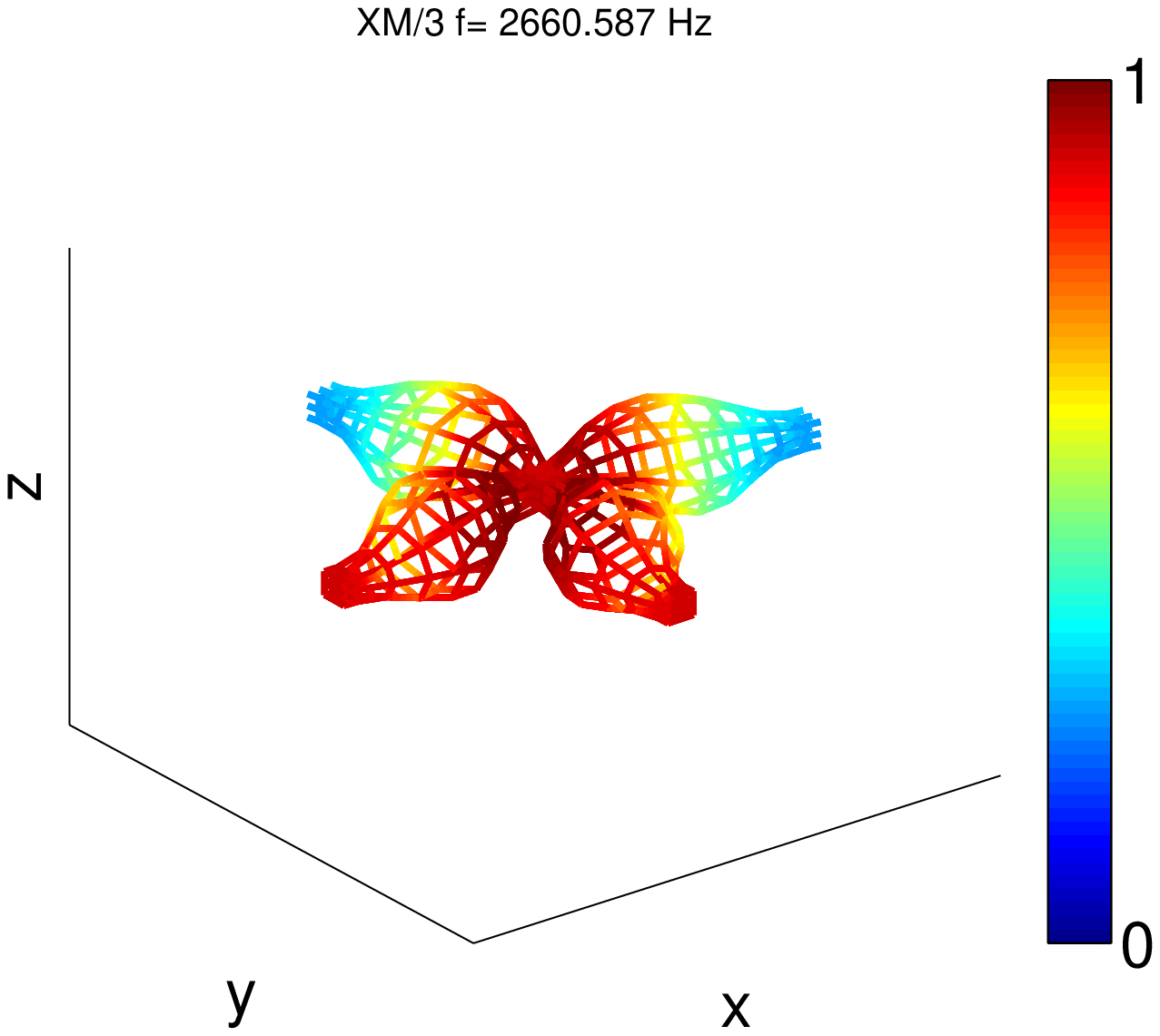}
    \\\vspace{-1cm}H
  \end{subfigure}
  \end{minipage}
  
  \begin{minipage}[c]{.1\textwidth}
      \begin{subfigure}[h]{.15\textwidth}
      \centering
      \includegraphics[height=5cm,trim={12cm 0cm 0cm 0cm},clip]{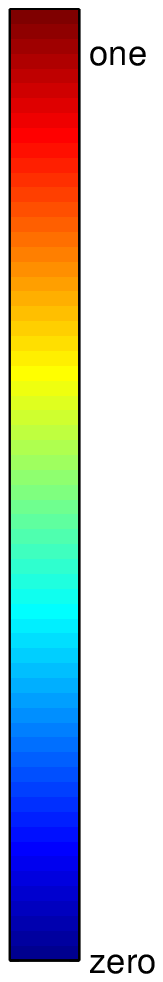}\\
    \end{subfigure}
  \end{minipage}
  }
  
  \makebox[\textwidth]{
  \begin{minipage}{.35\textwidth}
    \begin{subfigure}[h]{\textwidth}
      \caption{}
      \label{level2_spiderweb_square_band_diagram}
    \end{subfigure}
  \end{minipage}
  \begin{minipage}{.5\textwidth}
    \begin{subfigure}[h]{\textwidth}
      \caption{}
      \label{level2_spiderweb_square_modes}
    \end{subfigure}
  \end{minipage}
  }
  \caption{Band diagram and mode shapes (colored according to the magnitude of normalized displacements) from the two-dimensional square lattice obtained using the structure from Figure \ref{level2_spiderweb_square_top}. (\subref{level2_spiderweb_square_band_diagram}) Band diagram with interesting wave mode shapes indicated using letters A -- H. (\subref{level2_spiderweb_square_modes}) Mode shapes A -- D are located at the boundaries of potential BGs if modes highlighted in green (such as G and H) are not excited; wave mode shapes E and F denote full BG edges between $4886$ Hz and $4966$ Hz.}
  \label{level2_spiderweb_square_band_modes}
\end{figure}

Since the structures presented in Figures \ref{level2_square_top} and \ref{level2_hexagonal_top} have no special features such as inclusions or local resonators, they are not notable as periodic structures, and their band diagrams are presented in \ref{app_simple_band_diagrams}. However, the periodic lattices obtained by the repetition of the structures presented in Figures \ref{level2_spiderweb_square_top} and \ref{level2_spiderweb_hexagonal_top} may present interesting wave propagation characteristics. Thus, we investigate these structures, where each frame element has a length of $l = l_c + l_s = 25.0$ mm, which yields a  square periodic cell with lattice constant $2l = 50.0$ mm and a hexagonal periodic cell with lattice constant $2 l \sqrt{3} = 86.6$ mm. These periodic cells have the respective masses of $0.550 \times 10^{-3}$ kg and $0.836 \times 10^{-3}$ kg, corresponding to a more than $50 \%$ mass reduction compared to their non-hierarchical counterparts (see \ref{app_simple_band_diagrams}). The resulting band diagram and wave mode displacement shapes using the structure from Figure \ref{level2_spiderweb_square_top} are depicted in Figure \ref{level2_spiderweb_square_band_modes}, with full BGs (for all wave vectors) marked as yellow rectangles. Particularly interesting mode shapes (further discussed below) are marked using green lines. The dimensionless angular frequency is given by $\overline{\omega}_{2, \, \text{sq.}} = \omega 2l / 2 \pi c_L$.


Figure \ref{level2_spiderweb_square_band_diagram} shows the resulting band diagram of the two-dimensional PC. Mode shapes depicted using letters A -- H (see Figure \ref{level2_spiderweb_square_modes}) help us to gain further insight about the deformation mechanisms of the structure. Wave modes A and D represent out-of-plane bending modes, and modes B and C represent torsional modes. The frequency ranges between these modes ($1176$ Hz and $1540$ Hz for A -- B, and $1668$ Hz and $3927$ Hz for C -- D) present a very low mode density. Mode shapes that inhibit these regions from presenting complete BGs are marked using green lines, and examples of their deformations are denoted using G and H, showing the interaction between in-plane bending and longitudinal modes. In practice, this means that since wave modes G and H present only in-plane displacements, out-of-plane waves do not excite them, and we can expect strong attenuation in these regions. Also, a complete BG is found in the region with edges E and F (between $4886$ Hz and $4966$ Hz) and a very small BG between $4591$ Hz and $4616$ Hz.

\begin{figure}[h!]
  \centering
  \makebox[\textwidth]{
  \begin{minipage}{.35\textwidth}
    \begin{subfigure}[h]{\textwidth}
      \centering
      \includegraphics[width=5cm]{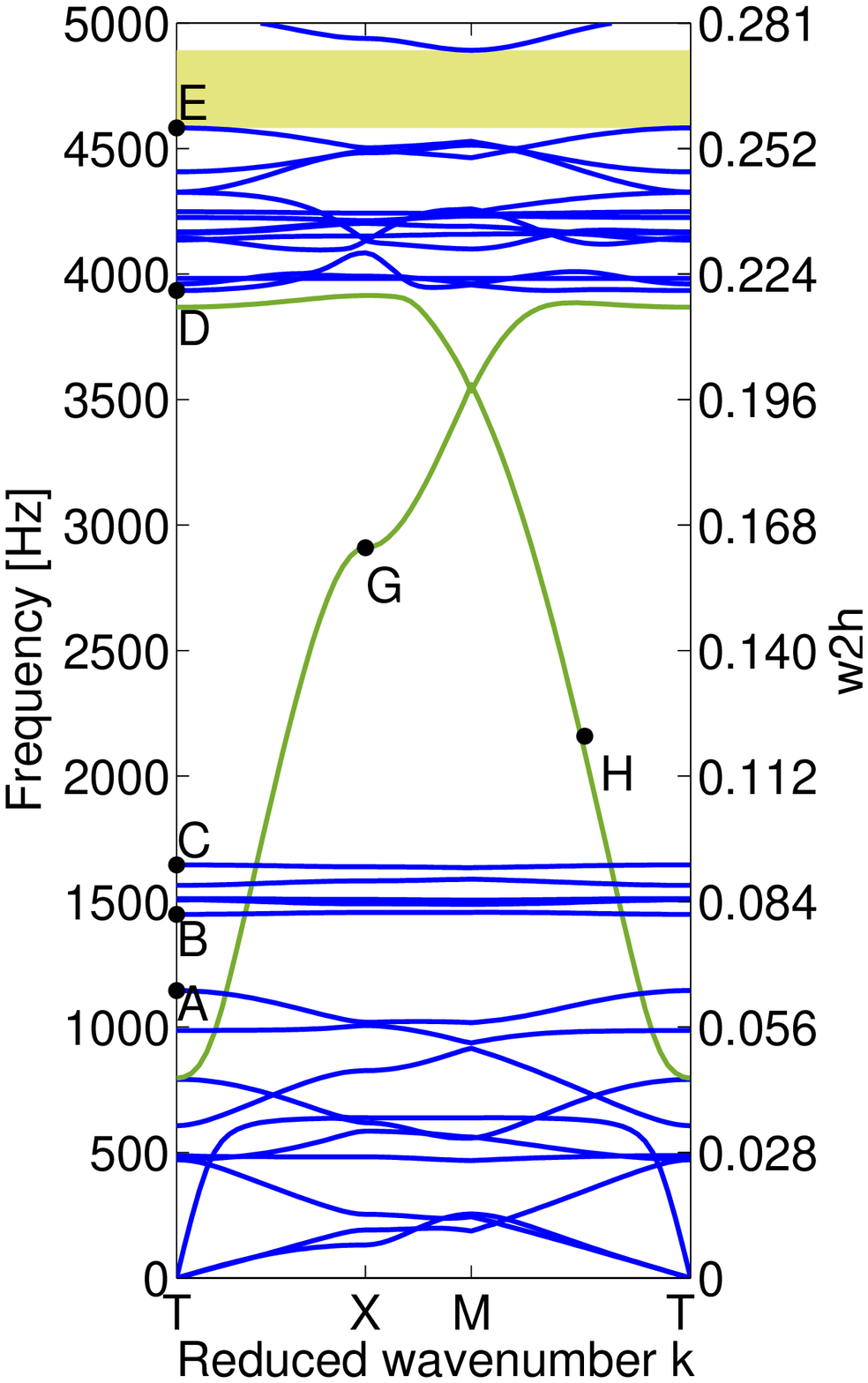}
  \end{subfigure}
  \end{minipage}
  
  \begin{minipage}{.5\textwidth}
  \centering
  \begin{subfigure}[h]{.45\textwidth}
    \centering
    \includegraphics[height=2.0cm,trim={3.5cm 3.5cm 3.5cm 3.5cm},clip]{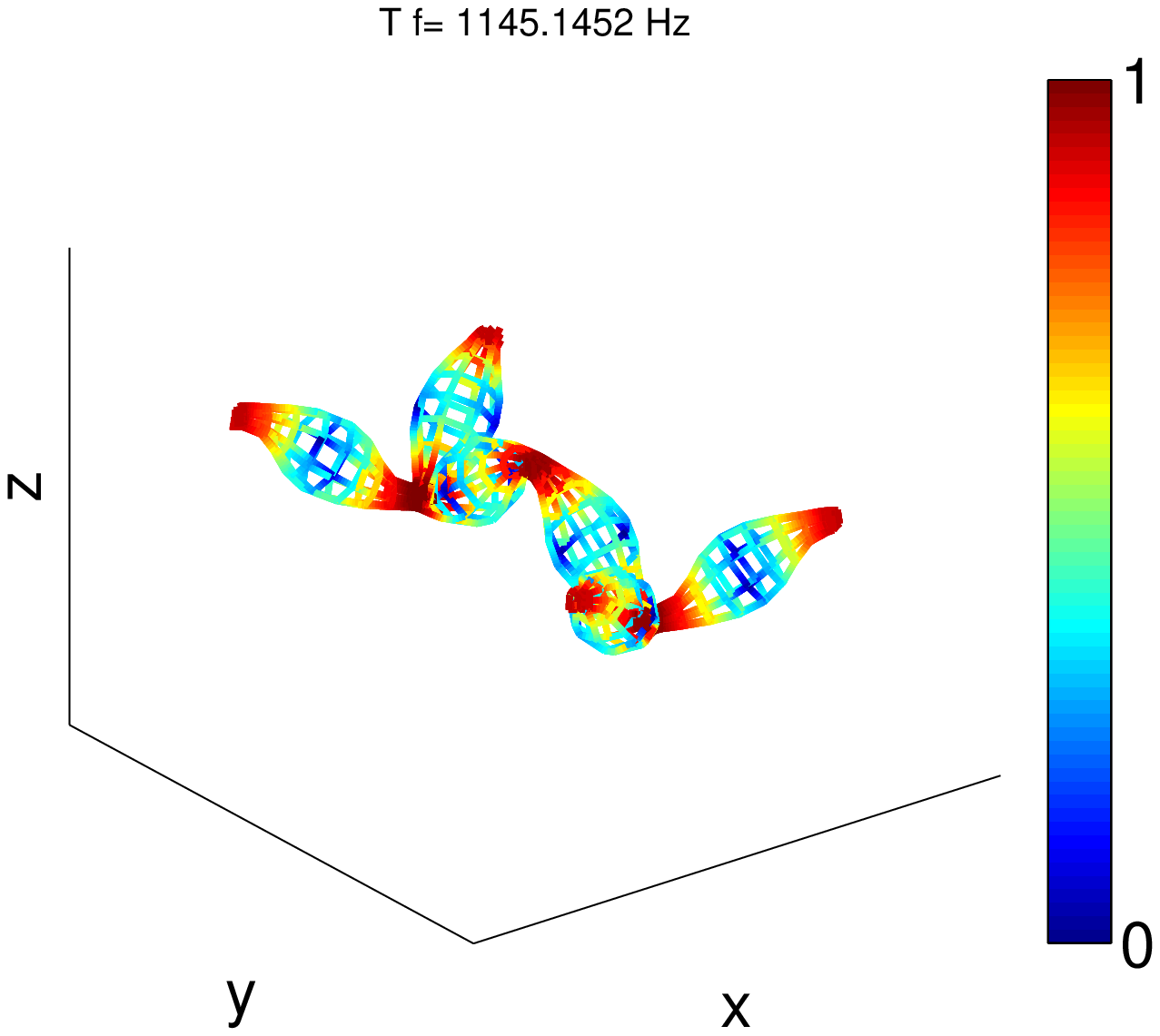}
    \\\vspace{-0.75cm}A
  \end{subfigure}
  \begin{subfigure}[h]{.45\textwidth}
    \centering
    \includegraphics[height=2.0cm,trim={3.5cm 3.5cm 3.5cm 3.5cm},clip]{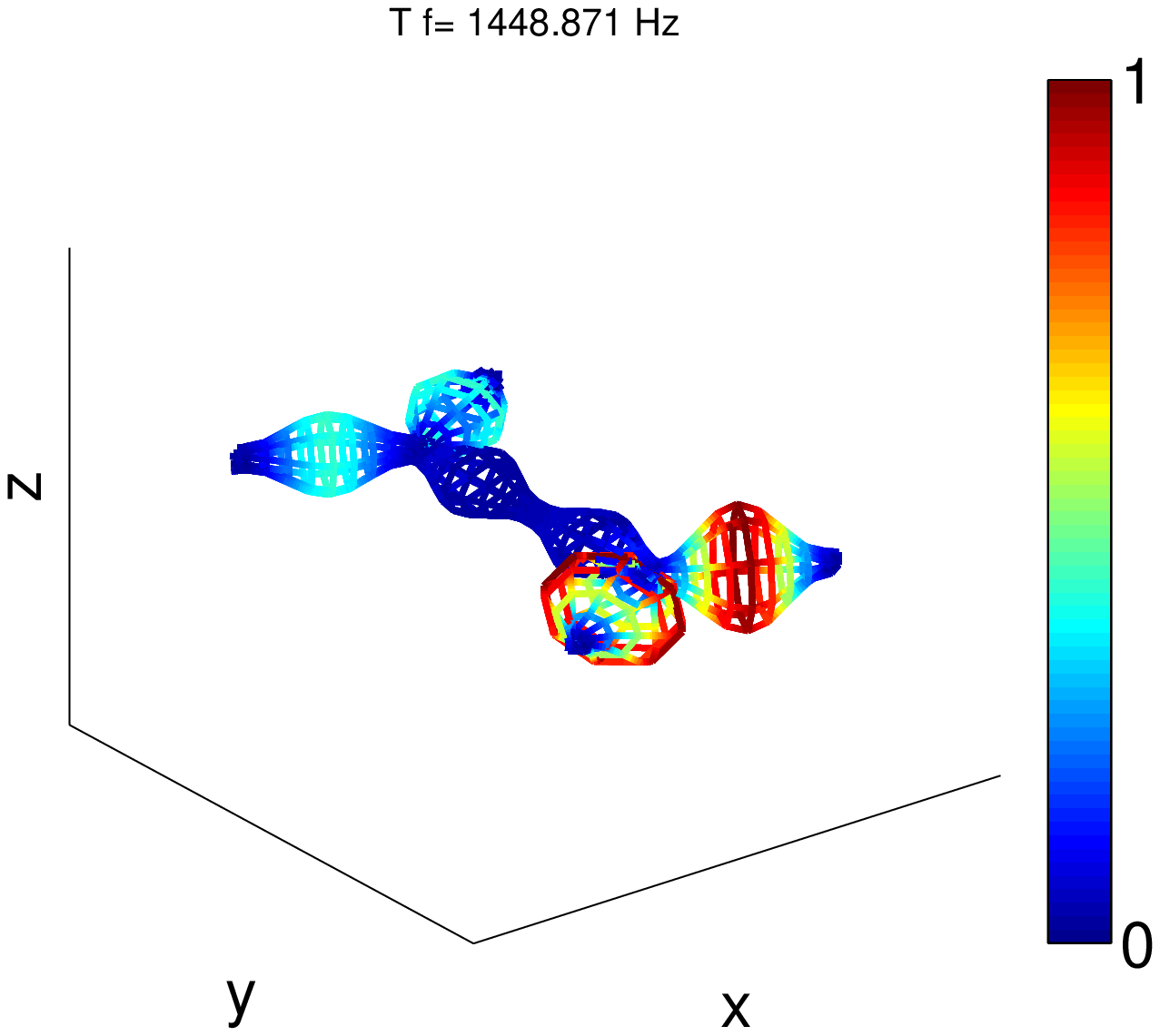}
    \\\vspace{-0.75cm}B
  \end{subfigure}
  \\\vspace{0.25cm}
  \begin{subfigure}[h]{.45\textwidth}
    \centering
    \includegraphics[height=2.0cm,trim={3.5cm 3.5cm 3.5cm 3.5cm},clip]{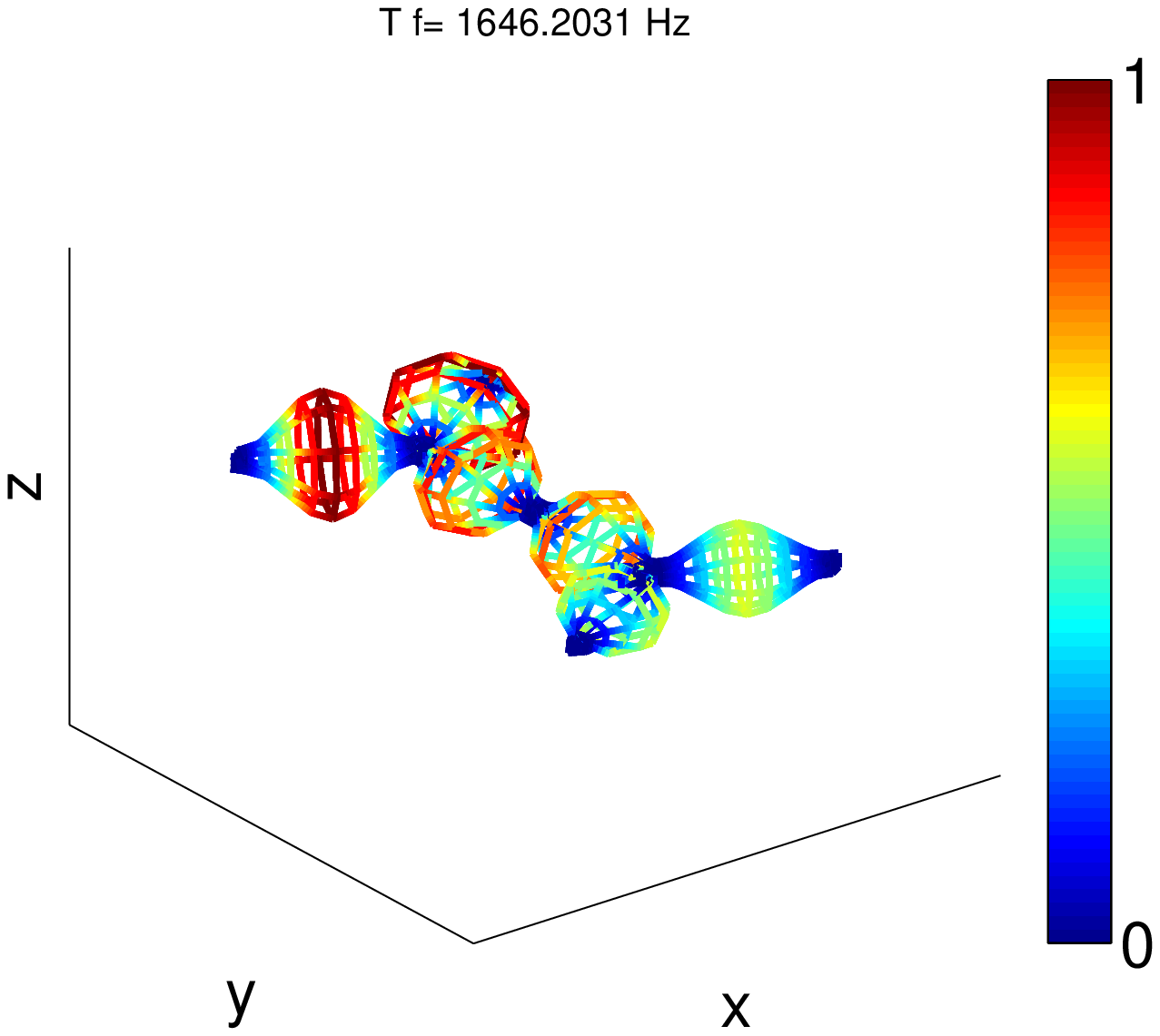}
    \\\vspace{-0.75cm}C
  \end{subfigure}
  \begin{subfigure}[h]{.45\textwidth}
    \centering
    \includegraphics[height=2.0cm,trim={3.5cm 3.5cm 3.5cm 3.5cm},clip]{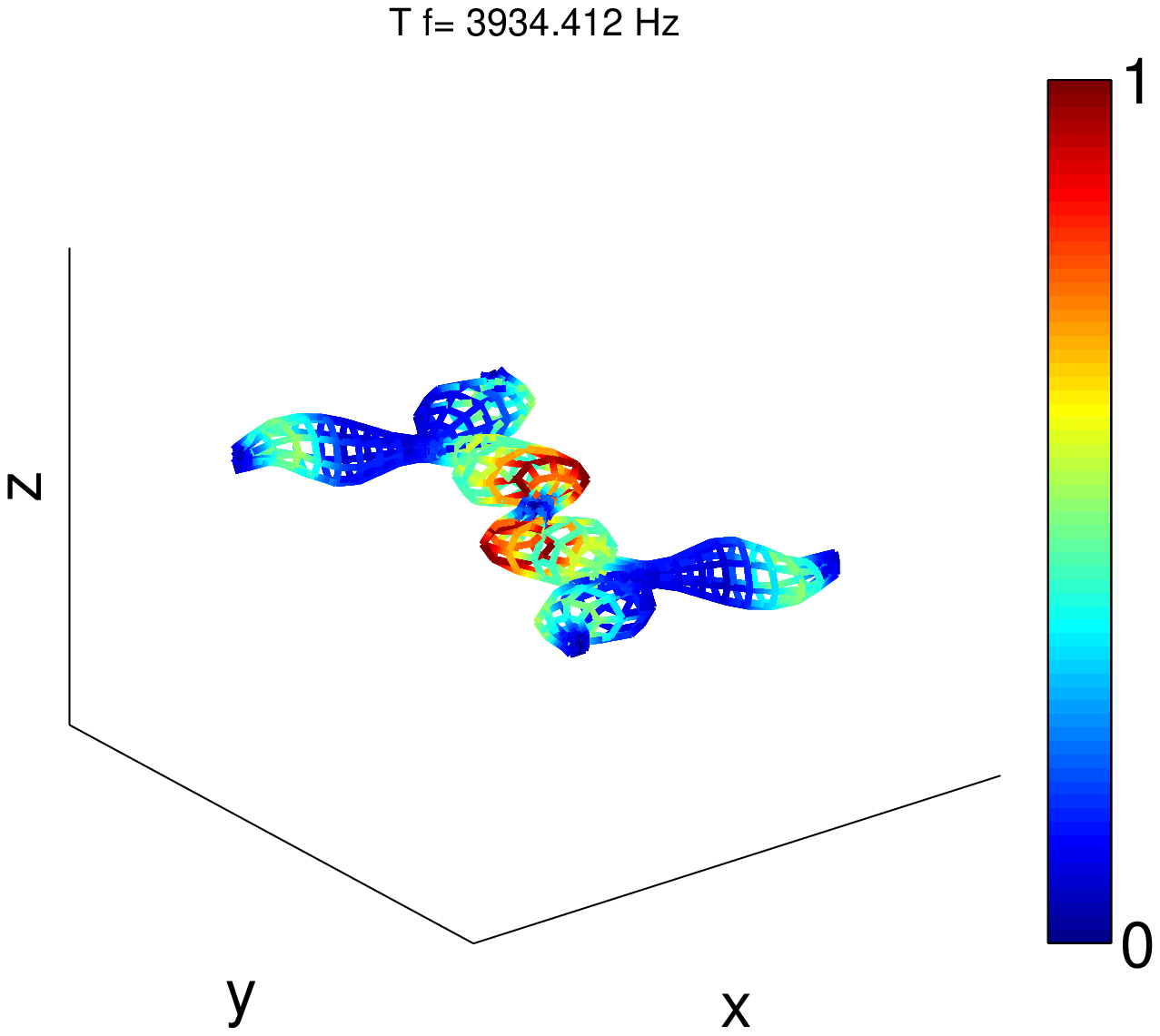}
    \\\vspace{-0.75cm}D
  \end{subfigure}
  \\\vspace{0.25cm}
  \begin{subfigure}[h]{.45\textwidth}
    \centering
    \includegraphics[height=2.0cm,trim={3.5cm 3.5cm 3.5cm 3.5cm},clip]{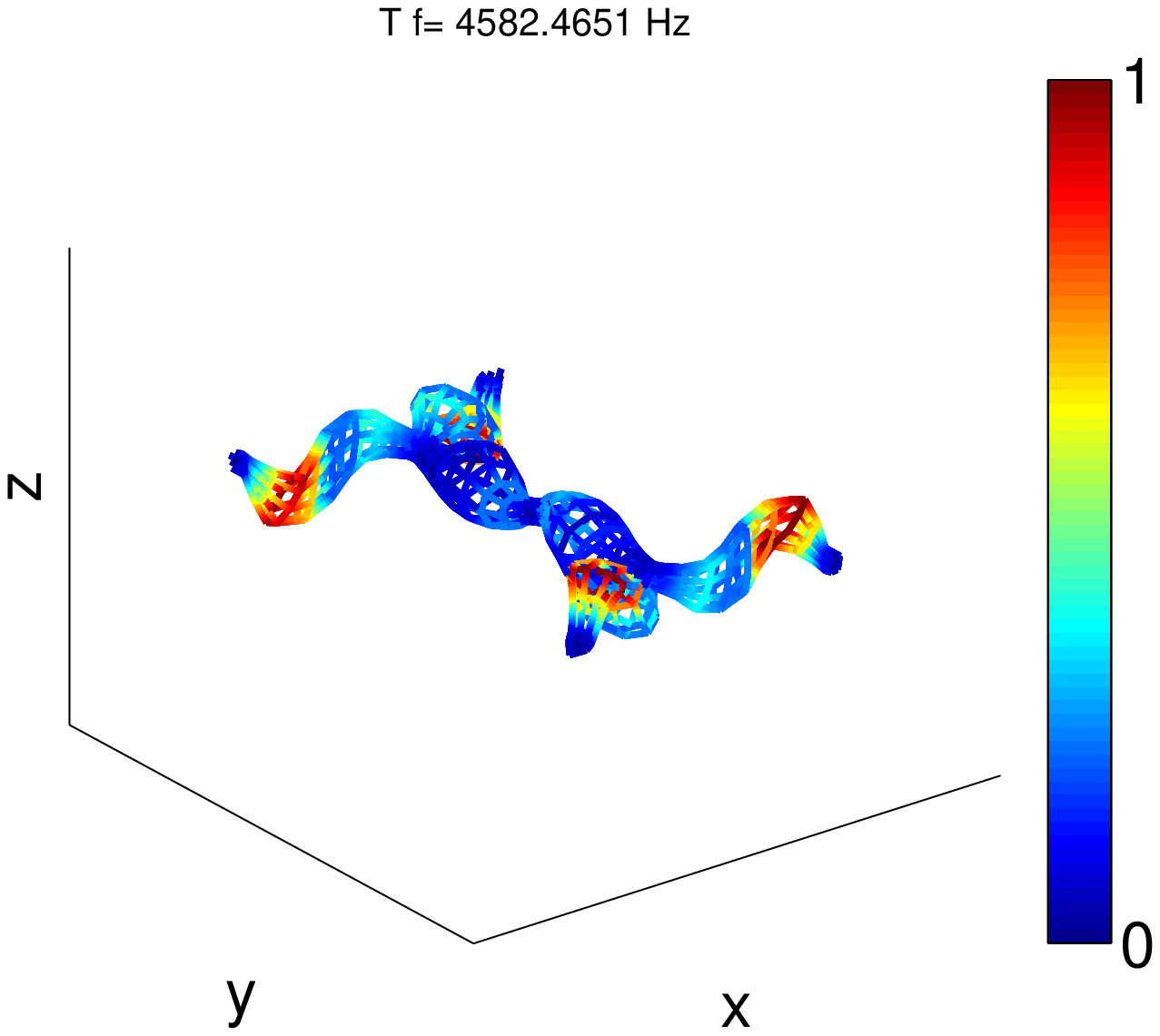}
    \\\vspace{-0.75cm}E
  \end{subfigure}
  \begin{subfigure}[h]{.45\textwidth}
    \centering
    \includegraphics[height=2.0cm,trim={3.5cm 3.5cm 3.5cm 3.5cm},clip]{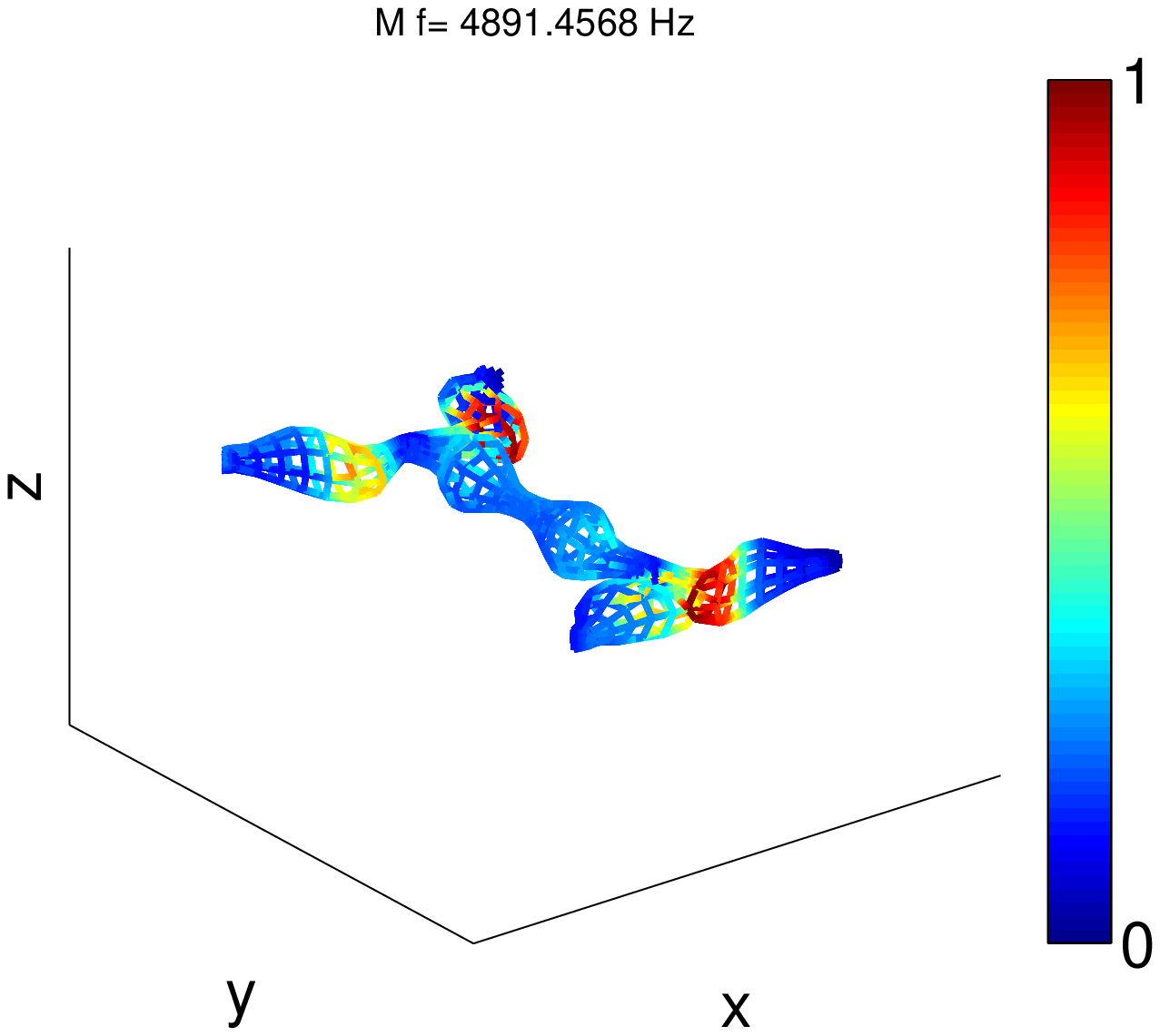}
    \\\vspace{-0.75cm}F
  \end{subfigure}
  \\\vspace{0.25cm}
  \begin{subfigure}[h]{.45\textwidth}
    \centering
    \includegraphics[height=2.0cm,trim={3.5cm 3.5cm 3.5cm 3.5cm},clip]{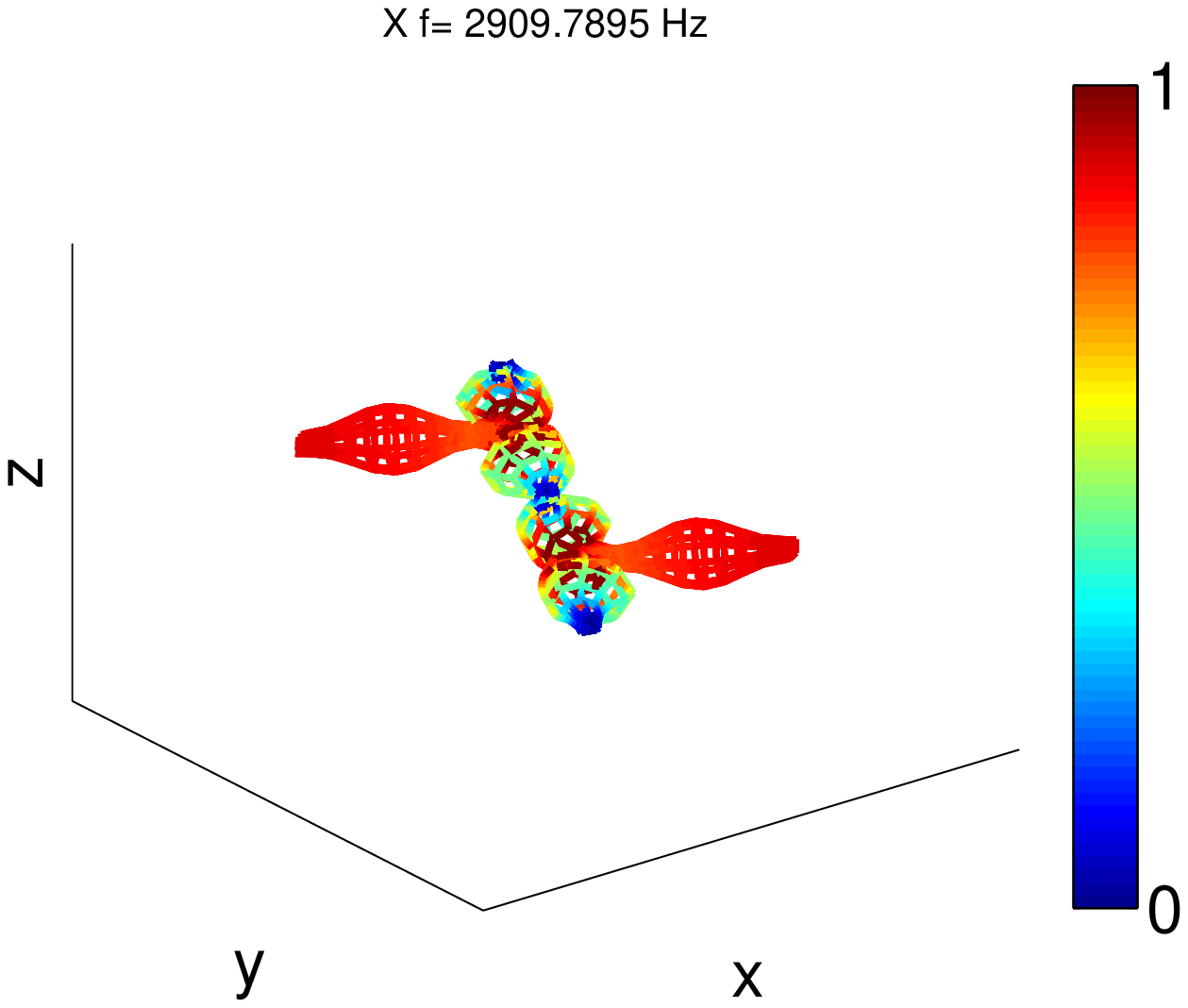}
    \\\vspace{-0.75cm}G
  \end{subfigure}
  \begin{subfigure}[h]{.45\textwidth}
    \centering
    \includegraphics[height=2.0cm,trim={3.5cm 3.5cm 3.5cm 3.5cm},clip]{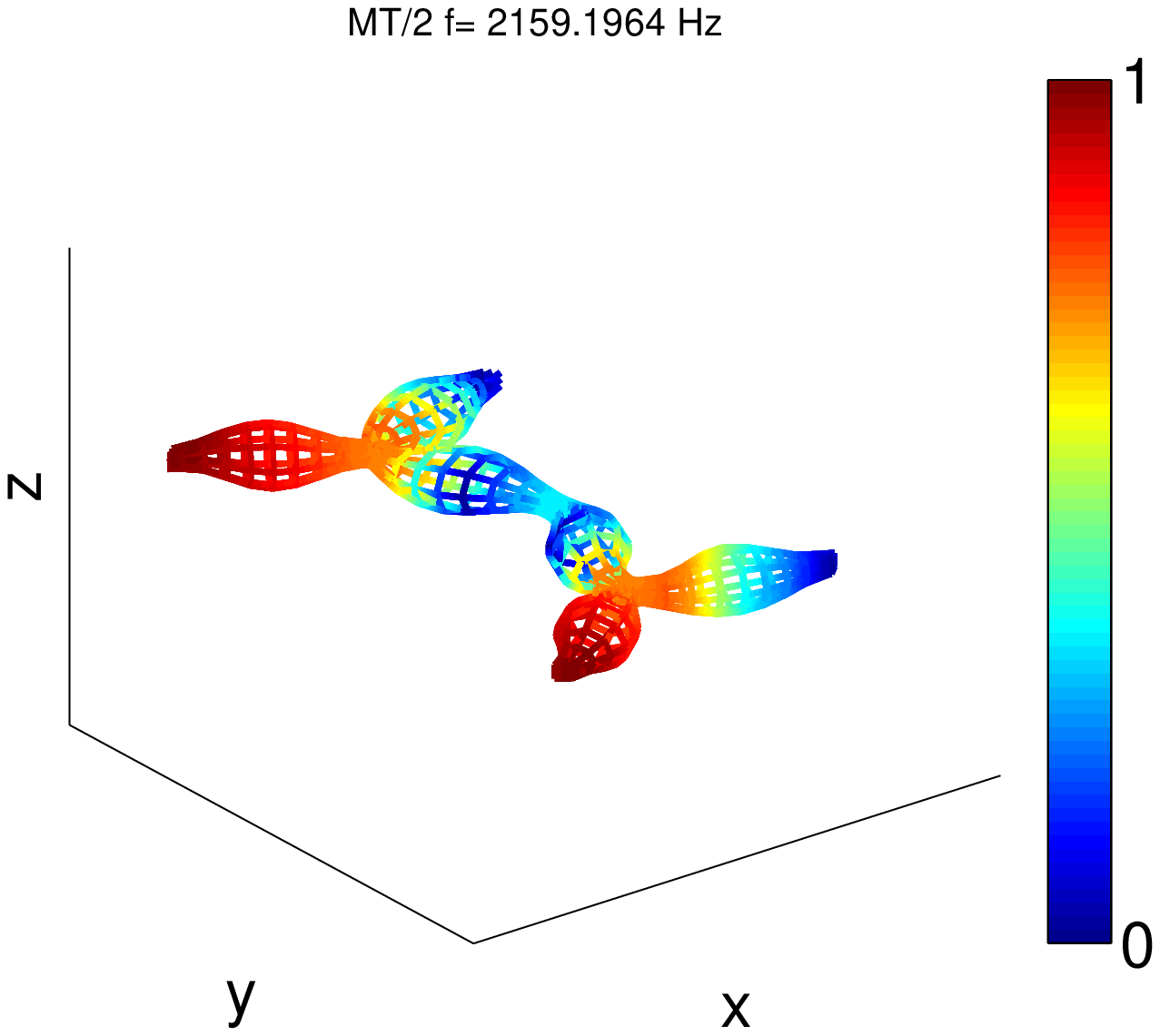}
    \\\vspace{-0.75cm}H
  \end{subfigure}
  \end{minipage}
  
  \begin{minipage}{.1\textwidth}
      \begin{subfigure}[h]{.15\textwidth}
      \centering
      \includegraphics[height=5cm,trim={12cm 0cm 0cm 0cm},clip]{./colorbar_normalized_large.eps}\\
    \end{subfigure}
  \end{minipage}
  }
  
  \makebox[\textwidth]{
  \begin{minipage}{.35\textwidth}
    \begin{subfigure}[h]{\textwidth}
      \caption{}
      \label{level2_spiderweb_hexagonal_band_diagram}
    \end{subfigure}
  \end{minipage}
  \begin{minipage}{.5\textwidth}
    \begin{subfigure}[h]{\textwidth}
      \caption{}
      \label{level2_spiderweb_hexagonal_modes}
    \end{subfigure}
  \end{minipage}  
  }

  \caption{Band diagram and wave mode shapes (colored according to the magnitude of normalized displacements) from the two-dimensional hexagonal lattice obtained using the structure from Figure \ref{level2_spiderweb_hexagonal_top}. (\subref{level2_spiderweb_hexagonal_band_diagram}) Band diagram with interesting wave mode shapes indicated using letters A -- H. (\subref{level2_spiderweb_hexagonal_modes}) Wave mode shapes A, B, C, and D are at the contour of a region that has the potential to attenuate waves, even though it is not a complete BG, and wave modes G and H are used to illustrate typical modes; wave mode shapes E and F denote full BG edges between $4582$ Hz and $4891$ Hz.}
  \label{level2_spiderweb_hexagonal_band_modes}
\end{figure}

The same analysis is now made using the structure from Figure \ref{level2_spiderweb_hexagonal_top}, and is presented in Figure \ref{level2_spiderweb_hexagonal_band_modes}. The dimensionless angular frequency is given by $\overline{\omega}_{2, \, \text{hx.}} = \omega 2l \sqrt{3} / 2 \pi c_L$.


Figure \ref{level2_spiderweb_hexagonal_band_diagram} presents a similar interaction between wave modes of the one-dimensional PC as that presented for the square case (Figure \ref{level2_spiderweb_square_band_diagram}). 
As in the previouse case, in Figure \ref{level2_spiderweb_hexagonal_modes} we examine the mode shapes of the structure (using the letters A -- H). Wave mode A represents an out-of-plane bending mode, B and C torsional modes, and D an in-plane bending mode. The frequency ranges between these modes ($1145$ Hz and $1449$ Hz for A -- B, and $1646$ Hz and $3934$ Hz for C -- D) show a very low mode density. Examples of wave modes that inhibit these regions from displaying complete BGs are denoted using G and H, showing the interaction between in-plane bending and longitudinal modes. Once again, strong wave attenuation is expected for out-of-plane waves, since they do not excite wave modes G and H. This also shows that what inhibits a complete BG is not an in-plane bending mode itself (as mode D), but the coupling of such a mode with longitudinal modes, which have no BGs in the one-dimensional PC. A complete BG is found in the region with edges E and F (between $4582$ Hz and $4891$ Hz).

\begin{figure}[h!]
  \makebox[\textwidth]{
  \begin{minipage}[l]{.35\textwidth}
    \begin{subfigure}[h]{\textwidth}
      \centering
      \includegraphics[width=5cm]{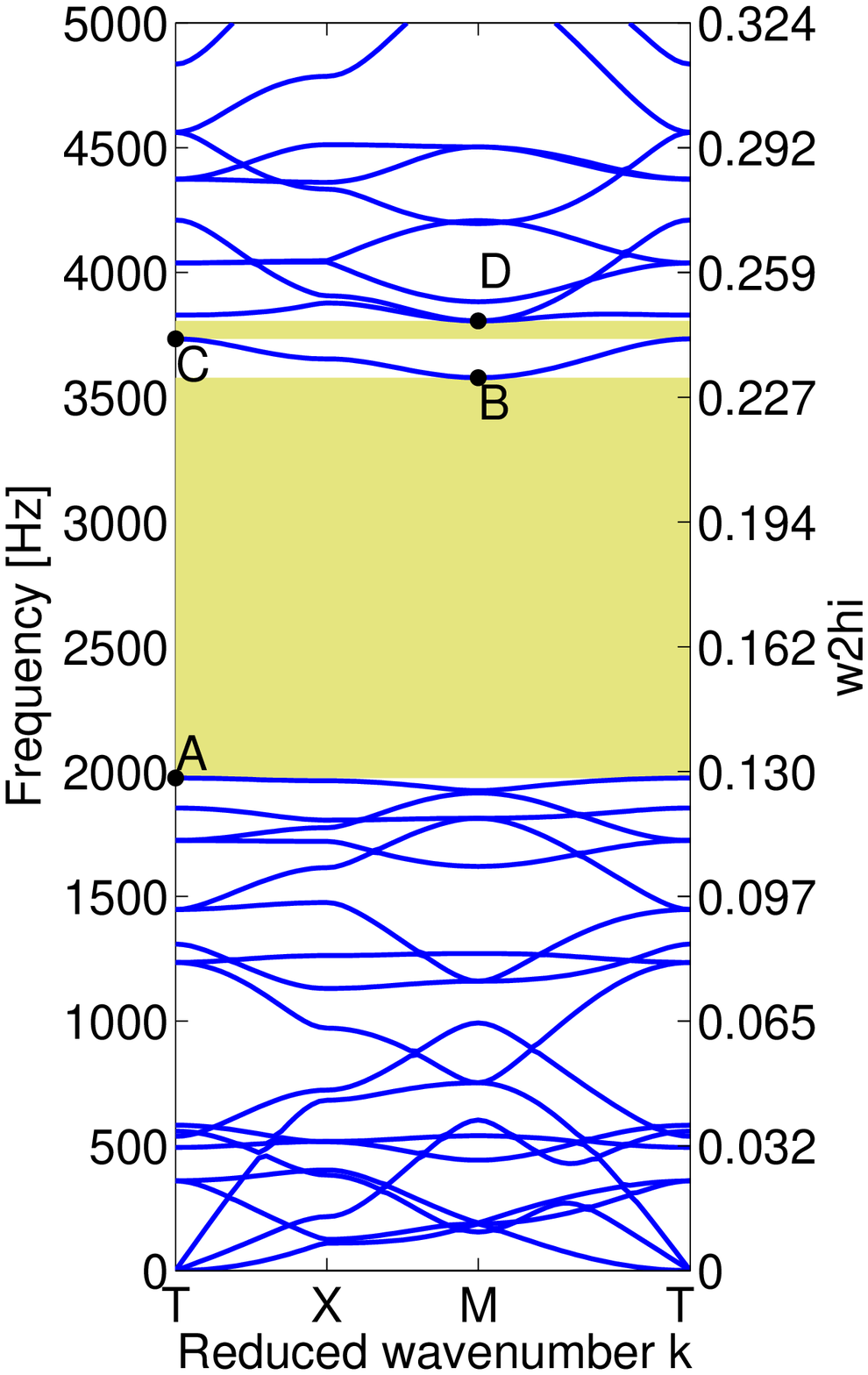}
    \end{subfigure}
  \end{minipage}
  
  \begin{minipage}[c]{.35\textwidth}
  \centering
    \begin{subfigure}[h]{\textwidth}
      \centering
      \includegraphics[height=2.0cm,trim={3.5cm 3.5cm 3.5cm 3.5cm},clip]{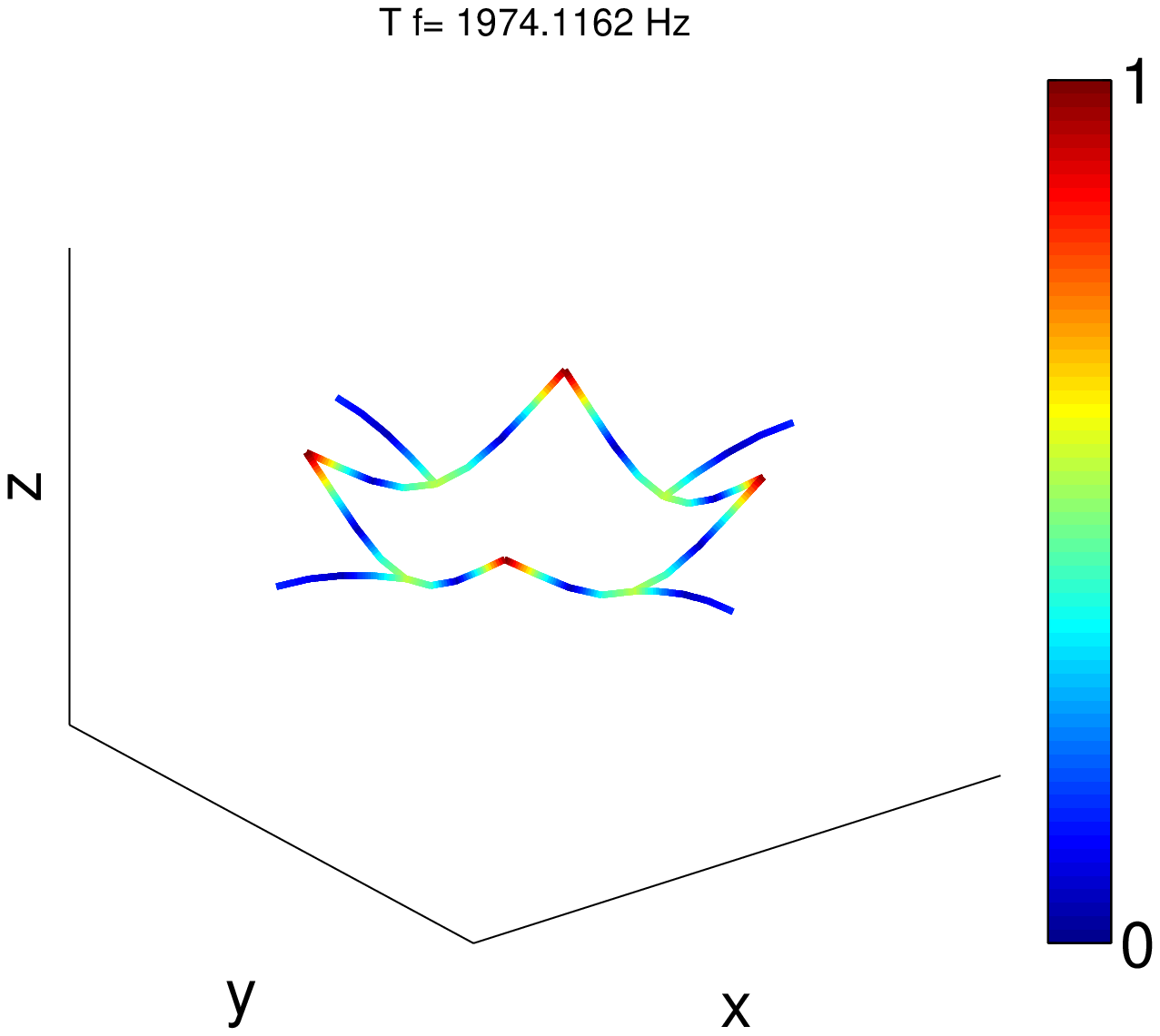}
      \\\vspace{-0.75cm}A
    \end{subfigure}
    \\\vspace{0.25cm}
    \begin{subfigure}[h]{\textwidth}
      \centering
      \includegraphics[height=2.0cm,trim={3.5cm 3.5cm 3.5cm 3.5cm},clip]{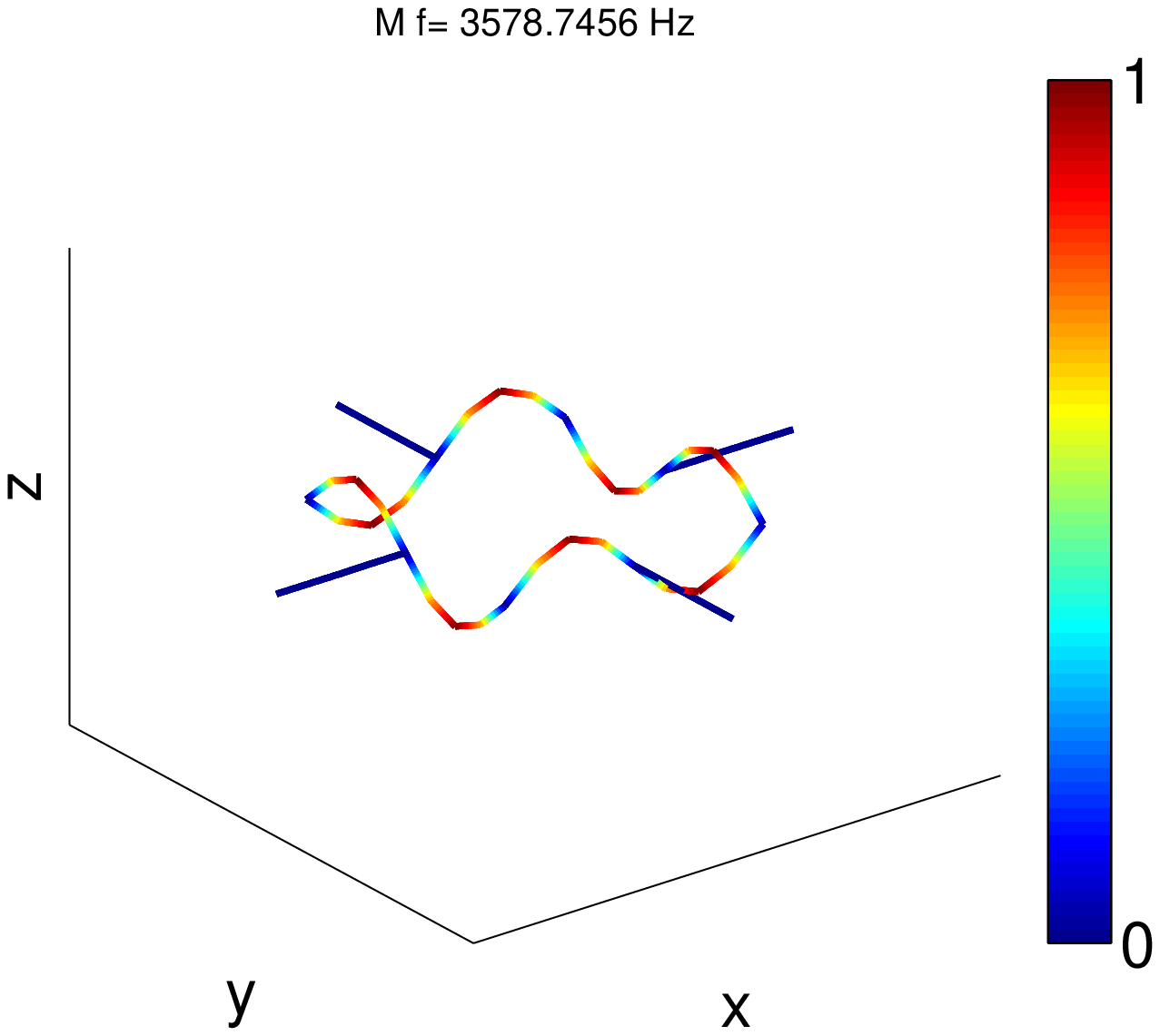}
      \\\vspace{-0.75cm}B
    \end{subfigure}
    \\\vspace{0.25cm}
    \begin{subfigure}[h]{\textwidth}
      \centering
      \includegraphics[height=2.0cm,trim={3.5cm 3.5cm 3.5cm 3.5cm},clip]{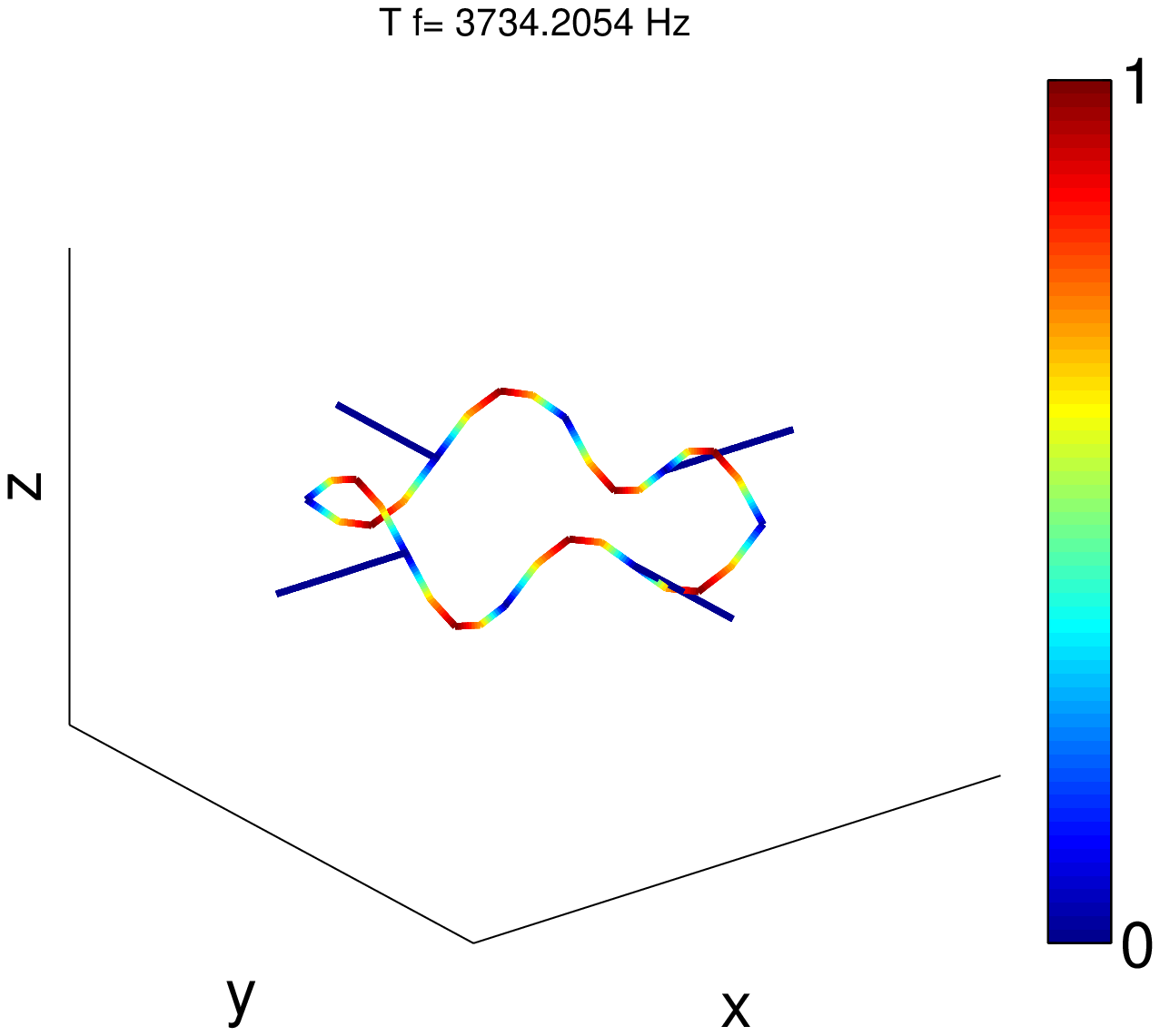}
      \\\vspace{-0.75cm}C
    \end{subfigure}
    \\\vspace{0.25cm}
    \begin{subfigure}[h]{\textwidth}
      \centering
      \includegraphics[height=2.0cm,trim={3.5cm 3.5cm 3.5cm 3.5cm},clip]{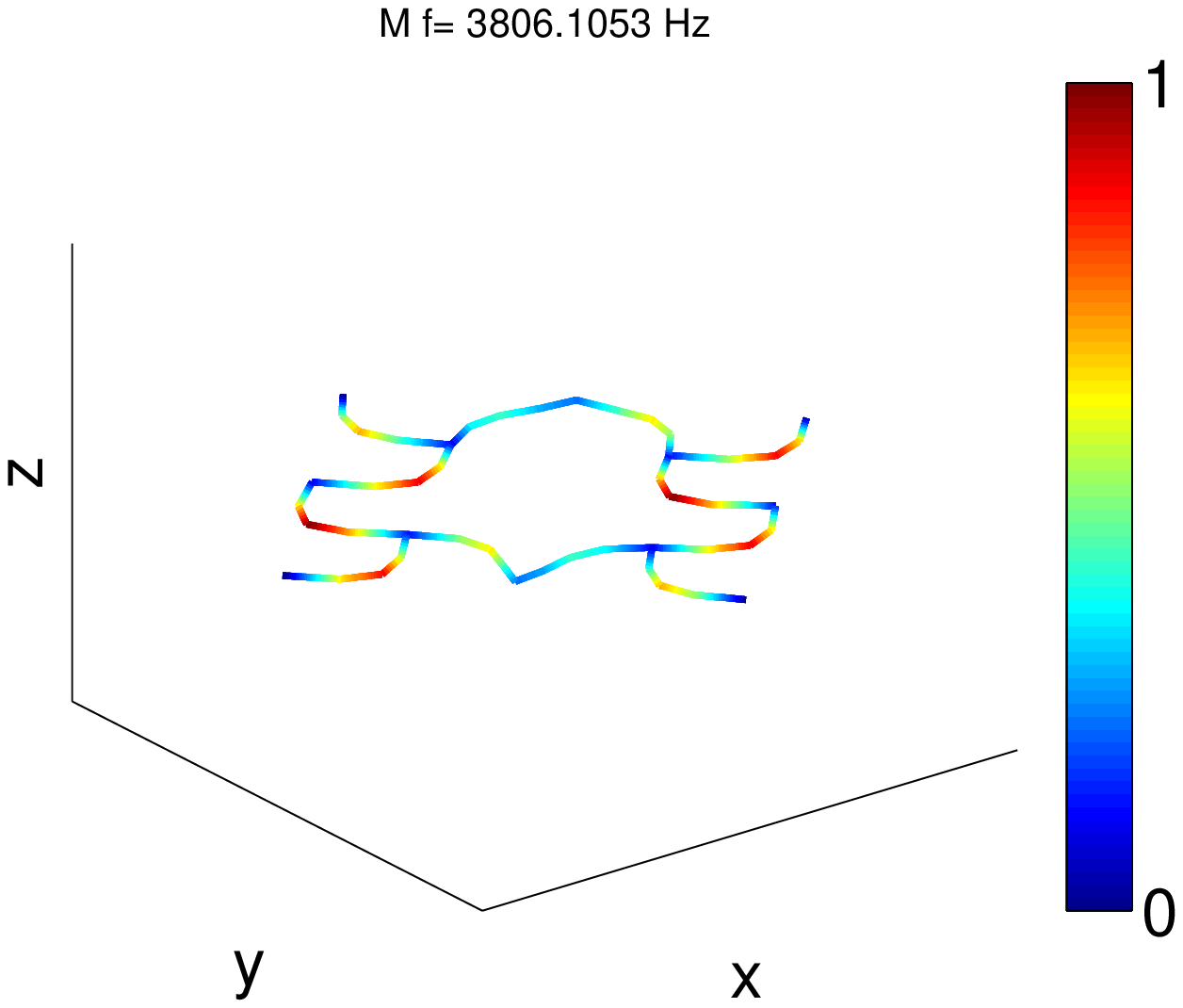}
      \\\vspace{-0.75cm}D
    \end{subfigure}
  \end{minipage}

  \begin{minipage}[c]{.1\textwidth}
    \begin{subfigure}[h]{.15\textwidth}
      \centering
      \includegraphics[height=5cm,trim={12cm 0cm 0cm 0cm},clip]{./colorbar_normalized_large.eps}\\
    \end{subfigure}
  \end{minipage}
  }
  
  \makebox[\textwidth]{
  \begin{minipage}[c]{.35\textwidth}
    \begin{subfigure}[h]{\textwidth}
      \centering
      \caption{}
      \label{level2_square_inclusion_band_diagram}
    \end{subfigure}
  \end{minipage}    
  \begin{minipage}[c]{.35\textwidth}
    \begin{subfigure}[h]{\textwidth}
      \centering
      \caption{}
      \label{level2_square_inclusion_modes}
    \end{subfigure}
  \end{minipage}
  }
  \caption{Band diagram of the structure presented in Figure \ref{level2_square_inclusion_top}. (\subref{level2_square_inclusion_band_diagram}) Several complete BGs are found, and the most pronounced are opened in the $1974$ Hz and $3579$ Hz, and $3734$ Hz and $3806$ Hz ranges, and BG edges are marked using letters A--D. (\subref{level2_square_inclusion_modes}) Corresponding displacement modes are shown and wave mode shapes (colored according to the magnitude of normalized displacements): modes A, B, and C are dominated by an out-of-plane bending, and mode D presents an in-plane bending behavior.}
  \label{level2_square_inclusion_band_modes}
\end{figure}

The structure shown in Figure \ref{level2_square_inclusion_top} presents distinctive features which characterize it as a square inclusion of dimension $2l \times 2l$ connected to adjacent cells by elements of length $l$. For this reason, we first analyze the square lattice obtained by the repetition of this periodic cell, with a total side length of $4l = 100.0$ mm, and consider elements with a circular cross-section with a diameter of $l_s \sqrt{2} = 3.54$ mm, yielding a total mass of $3.46 \times 10^{-3}$ kg. The band diagram and corresponding wave mode shapes are shown in Figure \ref{level2_square_inclusion_band_modes}, with the dimensionless angular frequency given by $\overline{\omega}_{2, \, \text{hi.}} = \omega 4l / 2 \pi c_L$.

Figure \ref{level2_square_inclusion_band_diagram} shows the opening of a wide BG between $1974$ Hz and $3579$ Hz, and a small BG between $3734$ Hz and $3806$ Hz ranges. These BG edges are marked using letters A--D and the corresponding mode shapes are shown in Figure \ref{level2_square_inclusion_modes}. It is possible to see that bending behavior dominates BG formation, which suggests that a hierarchical structure that presents bending BGs could benefit from this geometrical configuration.

\begin{figure}[h!]
  
  \makebox[\textwidth]{
  \begin{minipage}[c]{.35\textwidth}
    \begin{subfigure}[h]{\textwidth}
      \centering
      \includegraphics[width=5cm]{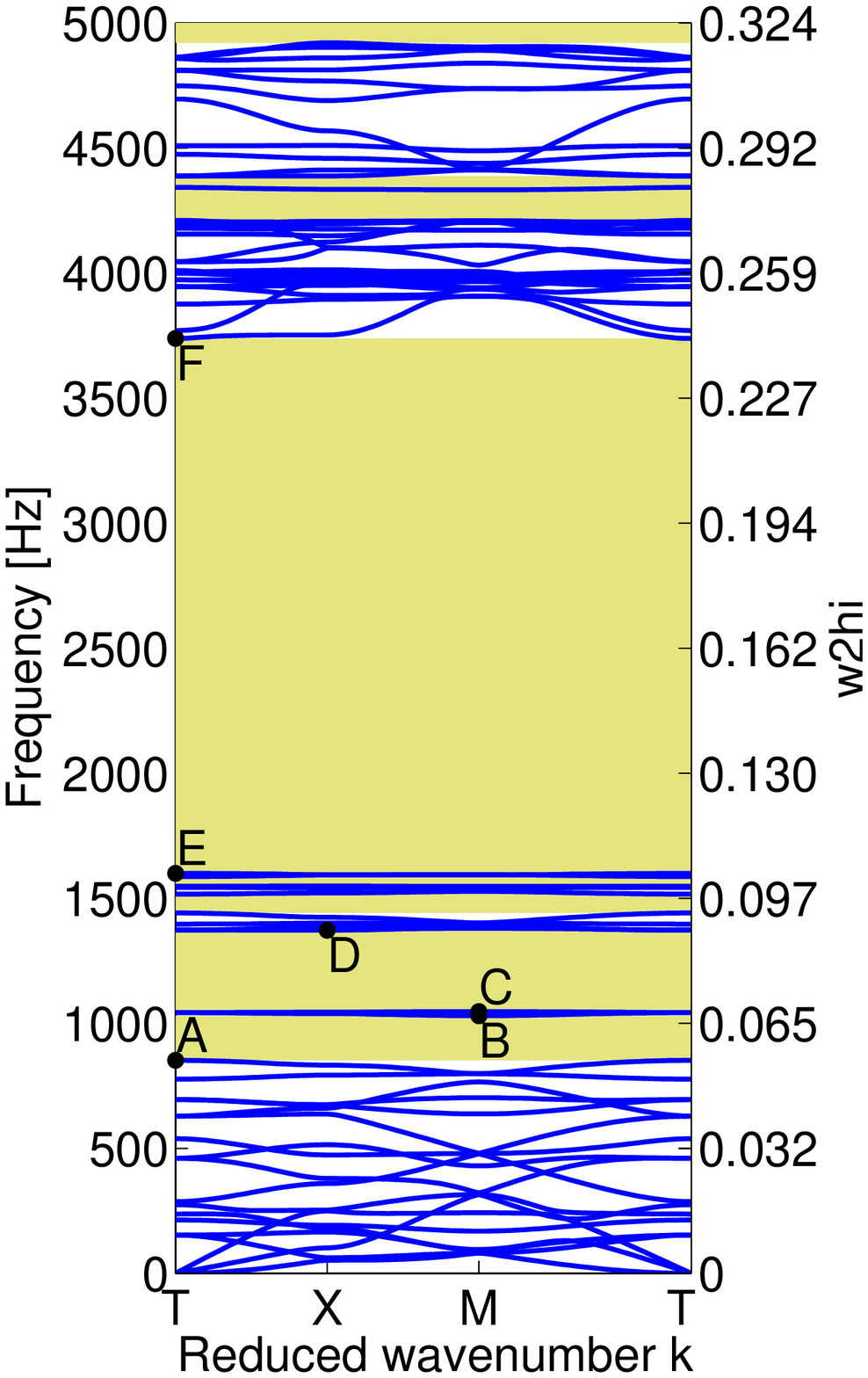}
    \end{subfigure}
  \end{minipage}
  
  \begin{minipage}[c]{.5\textwidth}
  \centering
    \begin{subfigure}[h]{.45\textwidth}
      \centering
      \includegraphics[width=4.0cm,trim={3.5cm 3.5cm 3.5cm 3.5cm},clip]{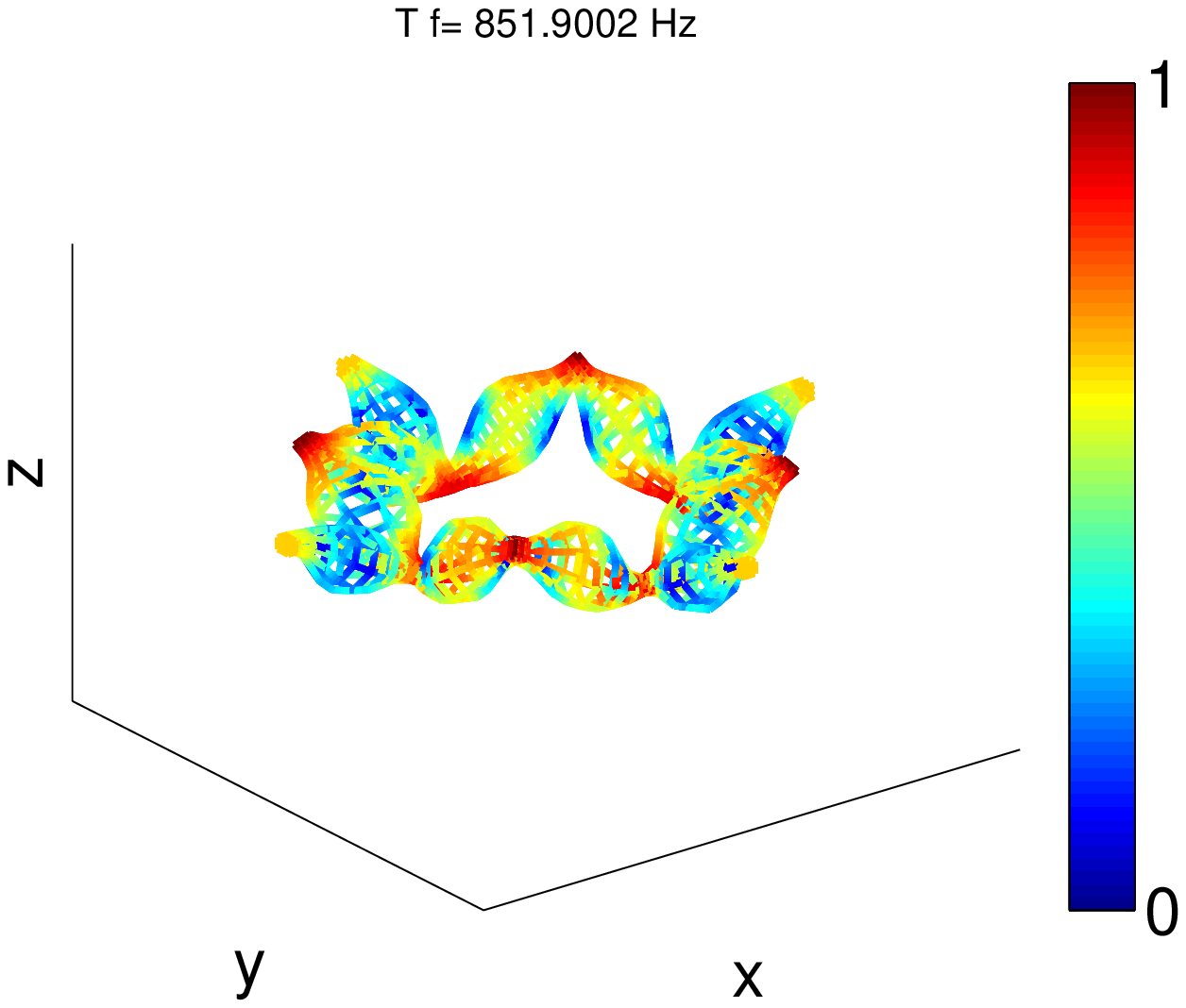}
      \\\vspace{-0.5cm}A
    \end{subfigure}
    \begin{subfigure}[h]{.45\textwidth}
      \centering
      \includegraphics[width=4.0cm,trim={3.5cm 3.5cm 3.5cm 3.5cm},clip]{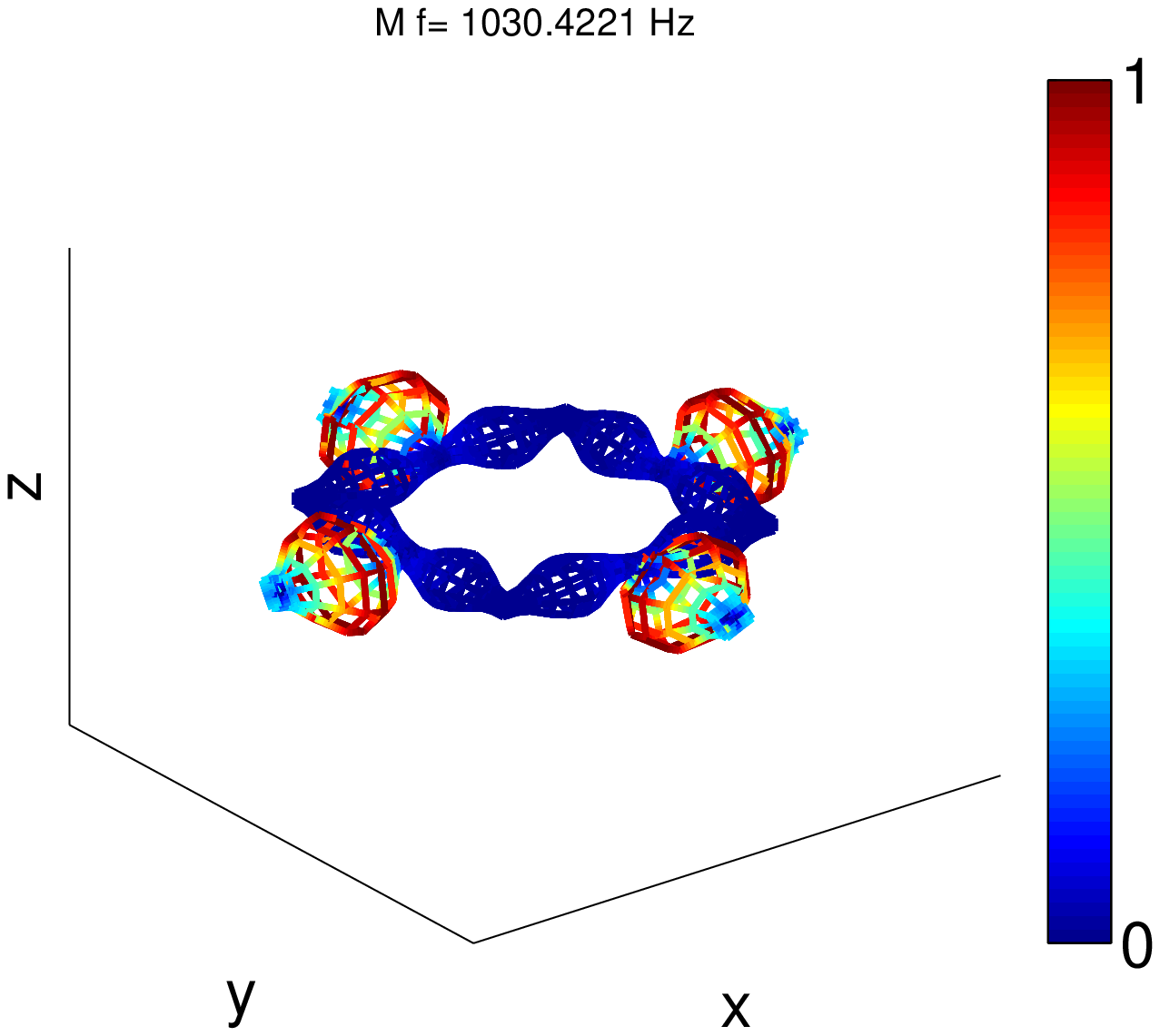}
      \\\vspace{-0.5cm}B
    \end{subfigure}
    \\\vspace{0.5cm}
    \begin{subfigure}[h]{.45\textwidth}
      \centering
      \includegraphics[width=4.0cm,trim={3.5cm 3.5cm 3.5cm 3.5cm},clip]{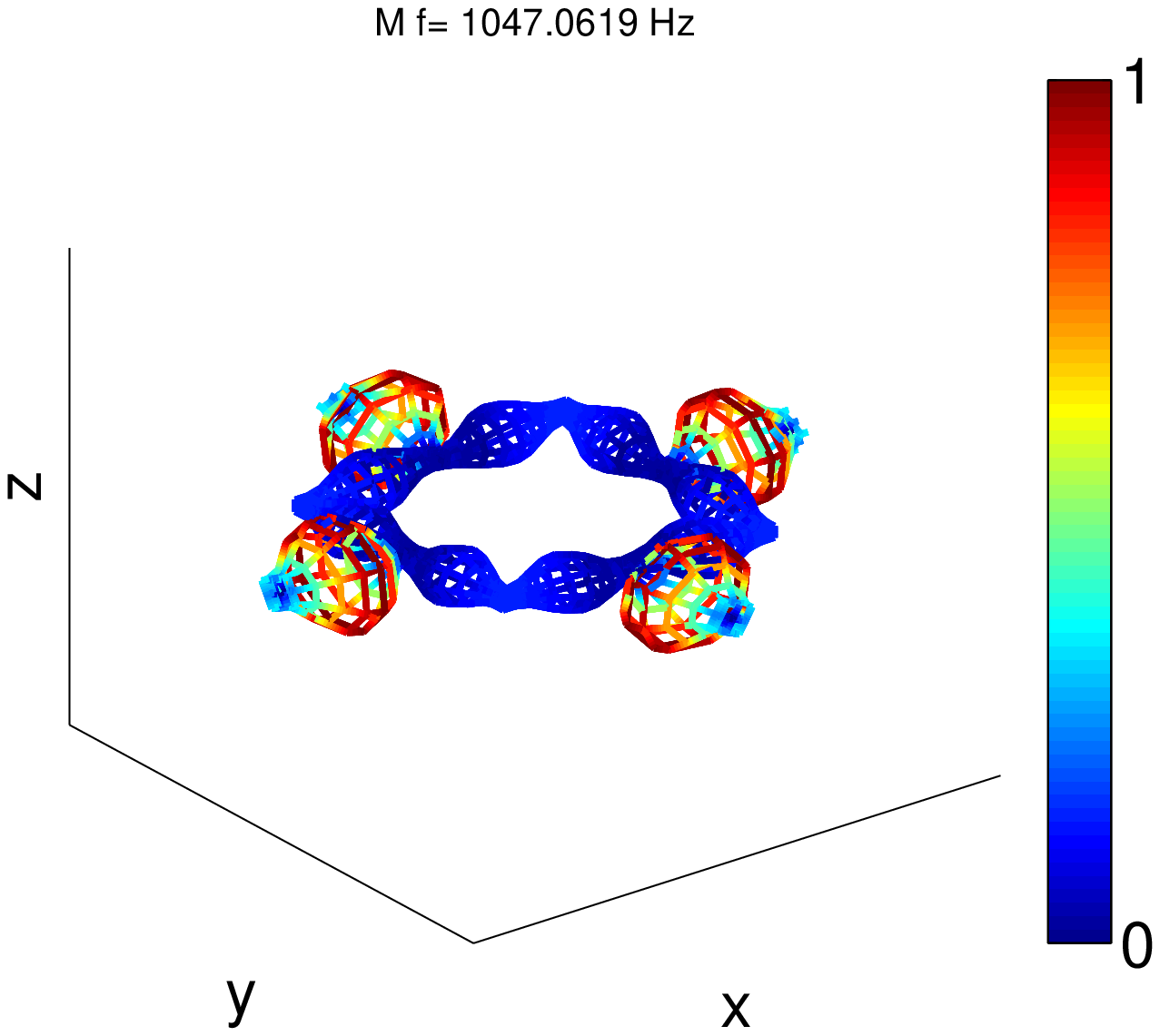}
      \\\vspace{-0.5cm}C
    \end{subfigure}
    \begin{subfigure}[h]{.45\textwidth}
      \centering
      \includegraphics[width=4.0cm,trim={3.5cm 3.5cm 3.5cm 3.5cm},clip]{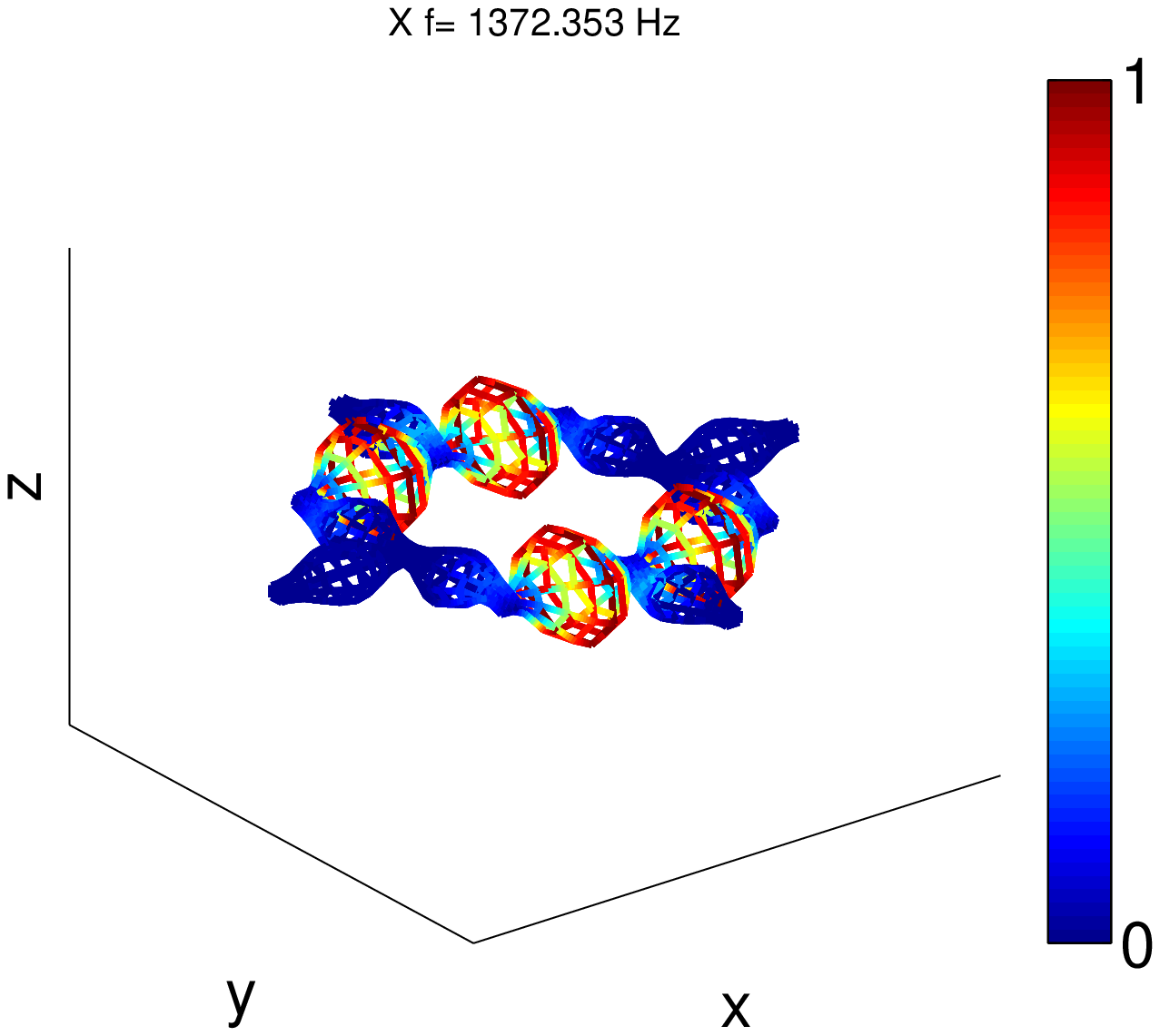}
      \\\vspace{-0.5cm}D
    \end{subfigure}
    \\\vspace{0.5cm}
    \begin{subfigure}[h]{.45\textwidth}
      \centering      
      \includegraphics[width=4.0cm,trim={3.5cm 3.5cm 3.5cm 3.5cm},clip]{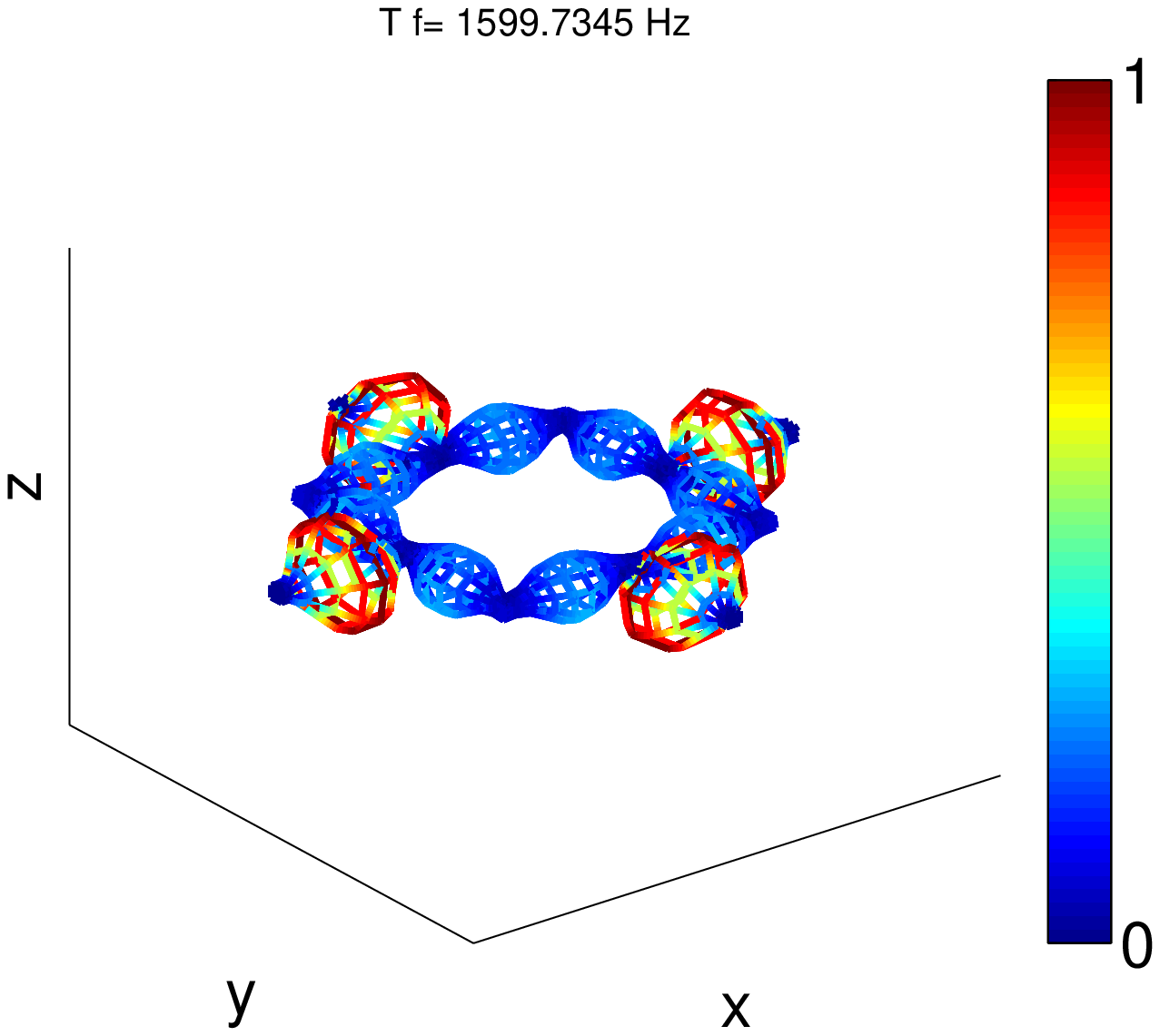}
      \\\vspace{-0.5cm}E
    \end{subfigure}
    \begin{subfigure}[h]{.45\textwidth}
      \centering
      \includegraphics[width=4.0cm,trim={3.5cm 3.5cm 3.5cm 3.5cm},clip]{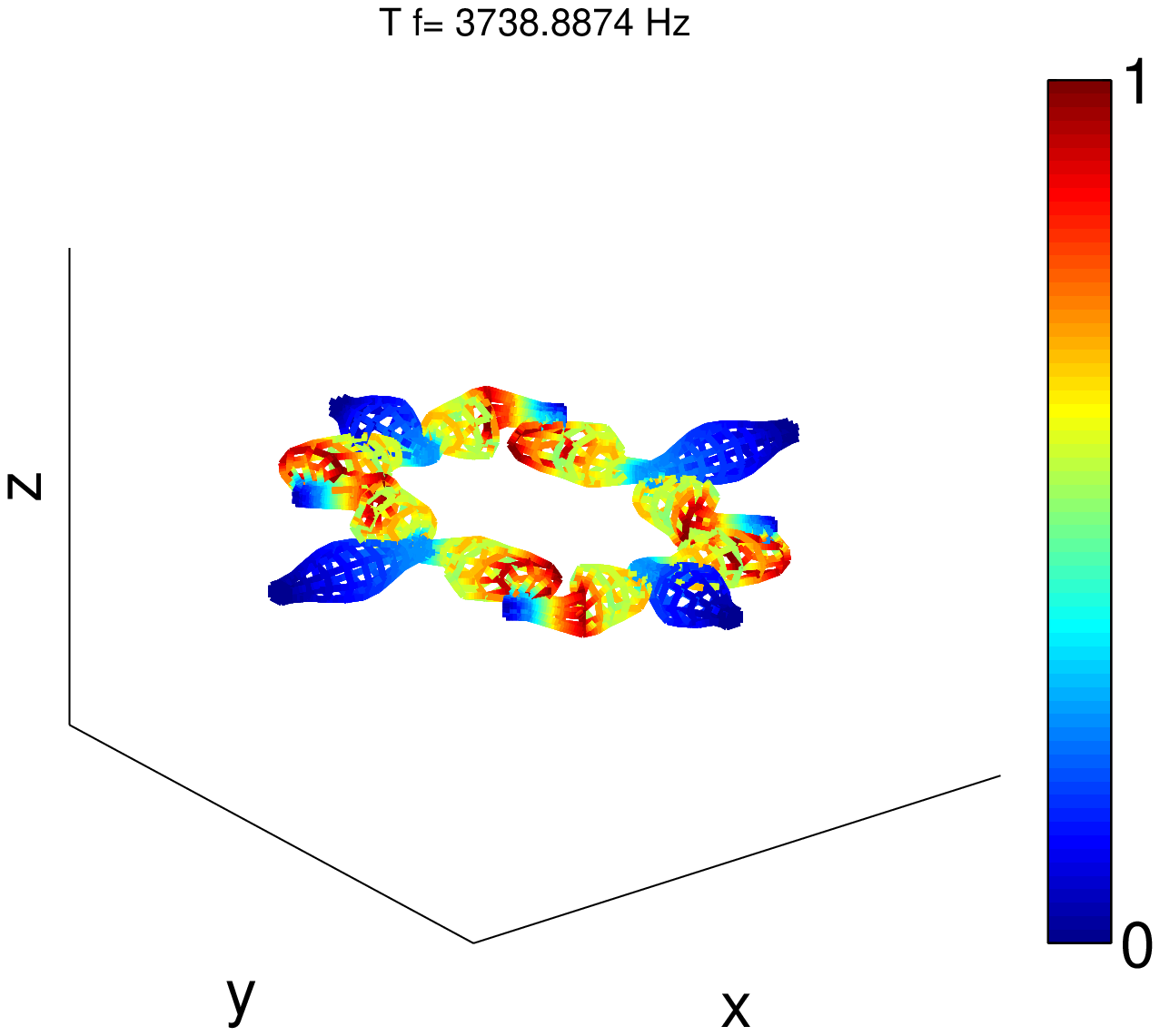}
      \\\vspace{-0.5cm}F
    \end{subfigure}

  \end{minipage}
  
  \begin{minipage}[c]{.1\textwidth}
    \begin{subfigure}[h]{\textwidth}
      \centering
      \includegraphics[height=5cm,trim={12cm 0cm 0cm 0cm},clip]{./colorbar_normalized_large.eps}\\
    \end{subfigure}
  \end{minipage}
  }
  
  \makebox[\textwidth]{
    \begin{minipage}[c]{.35\textwidth}
      \begin{subfigure}[h]{\textwidth}
        \caption{}
        \label{level2_spiderweb_square_inclusion_band_diagram}
      \end{subfigure}
    \end{minipage}
    \begin{minipage}[c]{.5\textwidth}
      \begin{subfigure}[h]{\textwidth}
        \caption{}
        \label{level2_spiderweb_square_inclusion_modes}       
      \end{subfigure}
    \end{minipage}    
  }
  
  \caption{Band diagram of the structure presented in \ref{level2_spiderweb_square_inclusion_top}. (\subref{level2_spiderweb_square_inclusion_band_diagram}) Several BGs are verified, in the $851.9$ Hz -- $1030$ Hz, $1047$ Hz -- $1372$ Hz, and $1600$ Hz -- $3739$ Hz ranges; BG edges are marked using letters A--F and the corresponding displacement modes are shown in (\subref{level2_spiderweb_square_inclusion_modes}): modes B, C, D, and E are dominated by torsional behavior and modes A and F display bending behavior.}
  \label{level2_spiderweb_square_inclusion_band_modes}
\end{figure}

Therefore, we now assess the wave propagation characteristics of the hierarchical structure presented in Figure \ref{level2_spiderweb_square_inclusion_top}, with a periodic cell with a side length of $4l = 100.0$ mm and a total mass of $1.68 \times 10^{-3}$ kg mass (over $50 \%$ mass reduction with respect to its non-hierarchical counterpart). The resulting band diagram and wave mode shapes are presented in Figure \ref{level2_spiderweb_square_inclusion_band_modes}.

Figure \ref{level2_spiderweb_square_inclusion_band_diagram} shows several BGs, which are opened in the $851.9$ Hz -- $1030$ Hz, $1047$ Hz -- $1372$ Hz, and $1600$ Hz -- $3739$ Hz regions. The wave mode shapes corresponding to the BG edges are shown in Figure \ref{level2_spiderweb_square_inclusion_modes}, and it is possible to see that the first BG opens due to bending behavior and closes due to torsional behavior, while the second BG is completely controlled by torsional behavior. The third BG behaves as the opposite of the first one (opens due to torsional behavior and closes due to bending).

The presence of the lower-frequency BGs may be explained due to the behavior of the one-dimensional PC (Figure \ref{level1_spiderweb_band_diagram}), which indicates BGs at close frequencies to those observed in the two-dimensional lattice. Deviations are expected when comparing the BG edges in the one-dimensional PC and the two-dimensional structure due to the cubic connection elements (not modeled in the one-dimensional PC).

Also, unlike previous cases (see Figures \ref{level2_spiderweb_square_band_modes} and \ref{level2_spiderweb_hexagonal_band_modes}), longitudinal wave modes do not inhibit BG formation, which further reinforces the conclusion that multi-level hierarchical structuring is able to harness additional advantageous features compared to the simple combination of the components which form the whole structure.

\begin{figure}[h!]
  \centering
  \includegraphics[width=14cm]{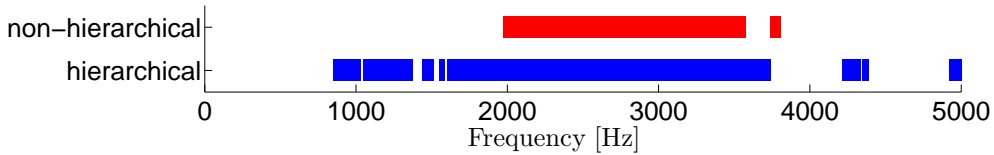}
  \caption{Comparison between BGs presented in Figures \ref{level2_square_inclusion_band_diagram} and \ref{level2_spiderweb_square_inclusion_band_diagram}. The BGs obtained using the constant circular cross-section are marked in red (\textcolor{red}{$\blacksquare$}), and BGs obtained using the spider web-like design are marked in blue (\textcolor{blue}{$\blacksquare$}).}
  \label{band_gap_comparison}
\end{figure}

Apart from the reduction in terms of mass, there is a noticeable advantage in achieving lower and wider BGs using a hierarchical structure over its non-hierarchical counterpart, as demonstrated in Figure \ref{band_gap_comparison}.

The reduced use of material associated with the hierarchical design may be beneficial in terms of wave controlling characteristics, but detrimental in terms of reduction in stiffness. To assess this potential disadvantage, a brief comparative analysis of the quasistatic properties of the proposed hierarchical structures is shown in \ref{app_quasistatic} using structures with an equivalent mass.

It is also important to notice that although additive manufacturing leads to numerous possibilities in design, variability in the fabrication process may be detrimental in terms of attenuation performance \citep{beli2016influence,fabro2020uncertainties}. Even though modern 3D printers can achieve fine resolutions down to $14$ $\mu$m \citep{digitalabsplus}, the fabrication process of hierarchical structures can be more complex than their homogeneous counterparts.

Since the hierarchical structure presented in Figure \ref{level2_spiderweb_square_inclusion_top} presents noticeable complete and full BGs (whereas more simple geometries such as those presented in Figures \ref{level2_spiderweb_square_top} and \ref{level2_spiderweb_hexagonal_top} do not), it is interesting to further explore the influence of hierarchical structuring on the robustness of the obtained BGs.

\subsection{Frequency response functions}

\subsubsection{Correlation between frequency response functions and band diagrams}

To further characterize the dynamic response of the proposed structures, we present here a correlation between complete full BGs, frequency regions with a low density of wave modes, and frequency response functions (FRFs) computed using a finite repetition of periodic cells by applying the input on one side and obtaining the response on the other side of the structure \citep{jia2018designing}.

For the computation of the FRFs, we apply a sinusoidal force in three different directions ($x$, $y$, and $z$) on the leftmost central node of a structure obtained from the repetition ($3 \times 3$) of the periodic cell, and take the displacement outputs at the center node of the other side of the structure considering the same direction as the applied force, thus obtaining $3$ separate FRFs, computed as output displacement/input force, calculated in the same direction. This is done essentially to separately assess in-plane longitudinal, in-plane transverse, and out-of-plane transverse behaviors.

When considering hierarchical structures, more than one node can be located at the leftmost central position (see, for instance, Figure \ref{level2_spiderweb_square_top}, where $9$ nodes are located on the left face of the periodic cell), in which case, a unitary force is equally divided into all of the input nodes. Accordingly, the output displacements are computed using an equal number of output nodes (in Figure \ref{level2_spiderweb_square_top}, there are also $9$ nodes located on the right face of the periodic cell), and combined to yield a RMS displacement ($\sqrt{ \frac{1}{n} \sum_{i=1}^n u_i^2}$, for every $i$-th node from the $n$ nodes at the output).

\begin{figure}[H]
  \makebox[\textwidth]{
  \centering
  \begin{subfigure}[t]{0.075\textwidth}
    \includegraphics[scale=0.5]{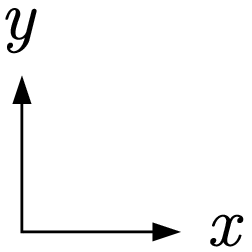}
  \end{subfigure}
  \begin{subfigure}[t]{.45\textwidth}
    \centering
    \includegraphics[height=5cm,width=6cm]{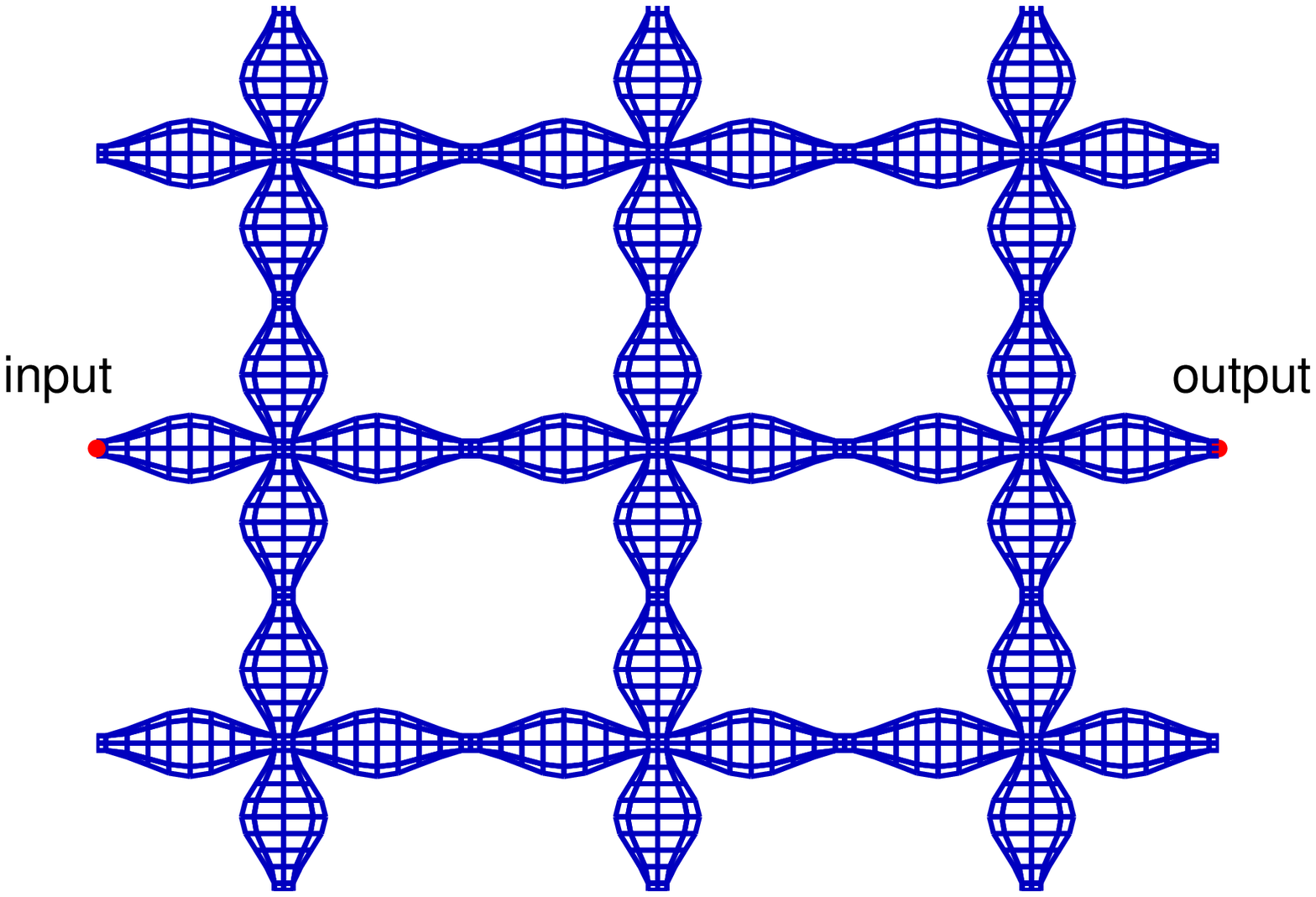}
    \caption{}
    \label{level2_spiderweb_square_frf_top}
  \end{subfigure}
  \begin{subfigure}[t]{.5\textwidth}
    \centering
    \raisebox{1cm}{
    \includegraphics[width=10cm]{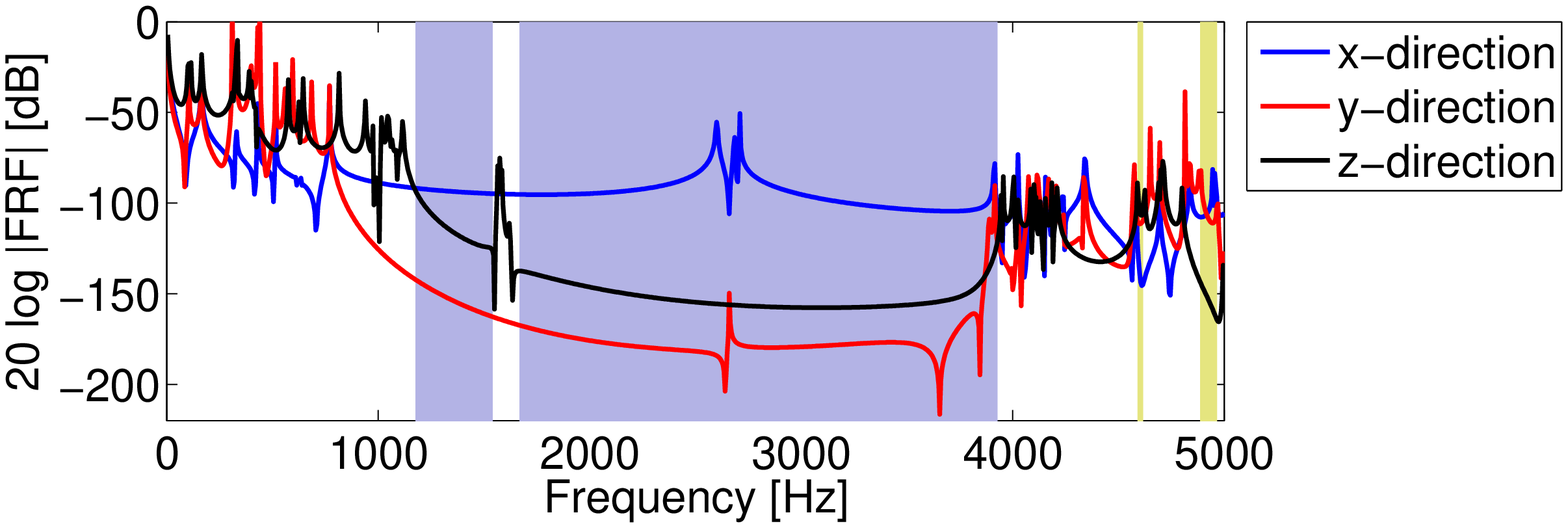}
    }
    \caption{}
    \label{level2_spiderweb_square_frf_xyz}
  \end{subfigure}  
  }
  \caption{Computed FRFs considering (\subref{level2_spiderweb_square_frf_top}) a force input at the center of the left side and (\subref{level2_spiderweb_square_frf_xyz}) outputs averaged at the center nodes of the right side. }
  \label{level2_spiderweb_square_frf}
\end{figure}

\begin{figure}[H]
  \makebox[\textwidth]{
  \centering
  \begin{subfigure}[t]{0.075\textwidth}
    \includegraphics[scale=0.5]{./xy_coordinates.eps}
  \end{subfigure}
  \begin{subfigure}[t]{.45\textwidth}
      \centering
      \includegraphics[height=3cm,width=6cm]{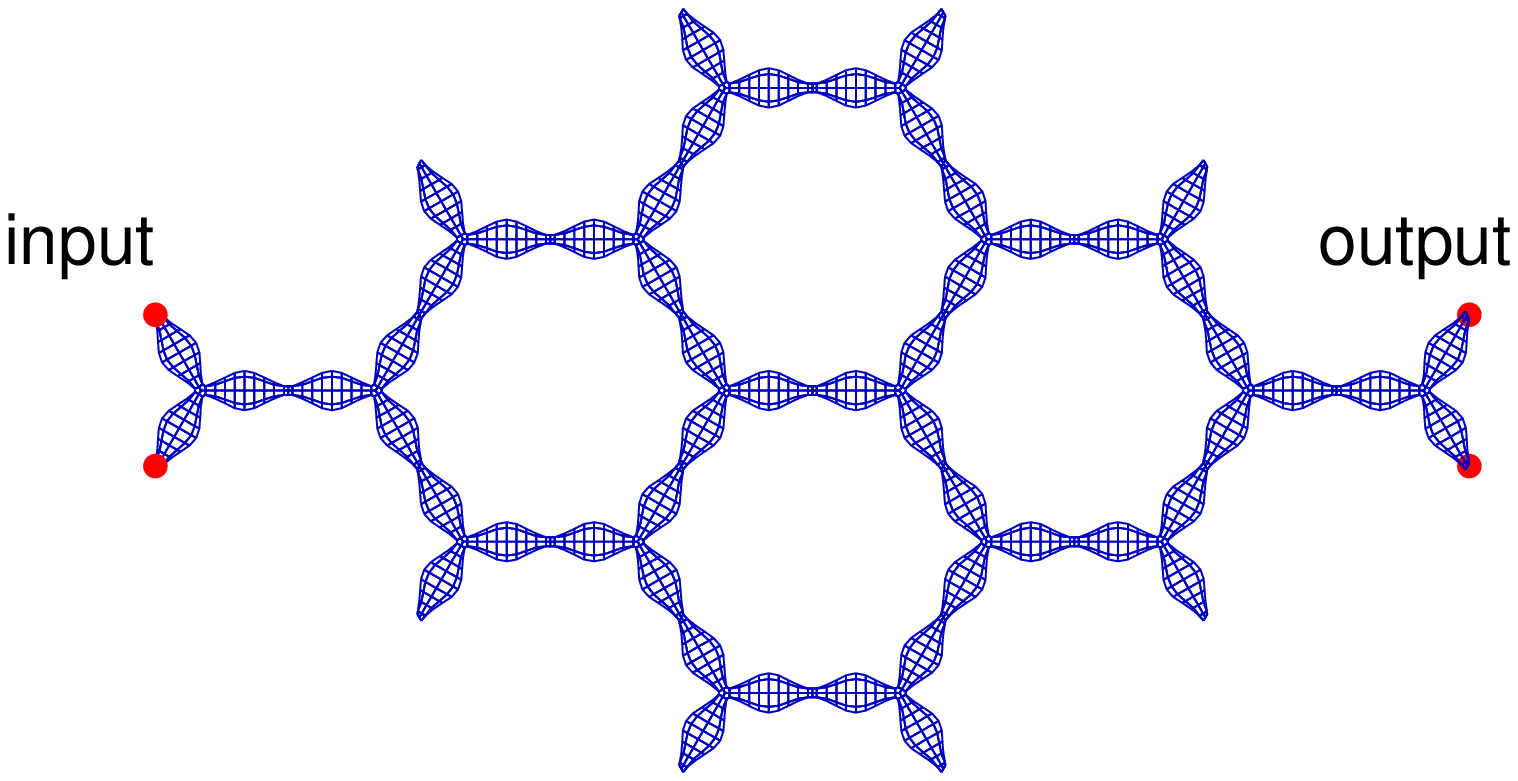}
      \caption{}
      \label{level2_spiderweb_hexagonal_frf_top}
  \end{subfigure}    
  \begin{subfigure}[t]{.5\textwidth}
      \centering
      \includegraphics[width=10cm]{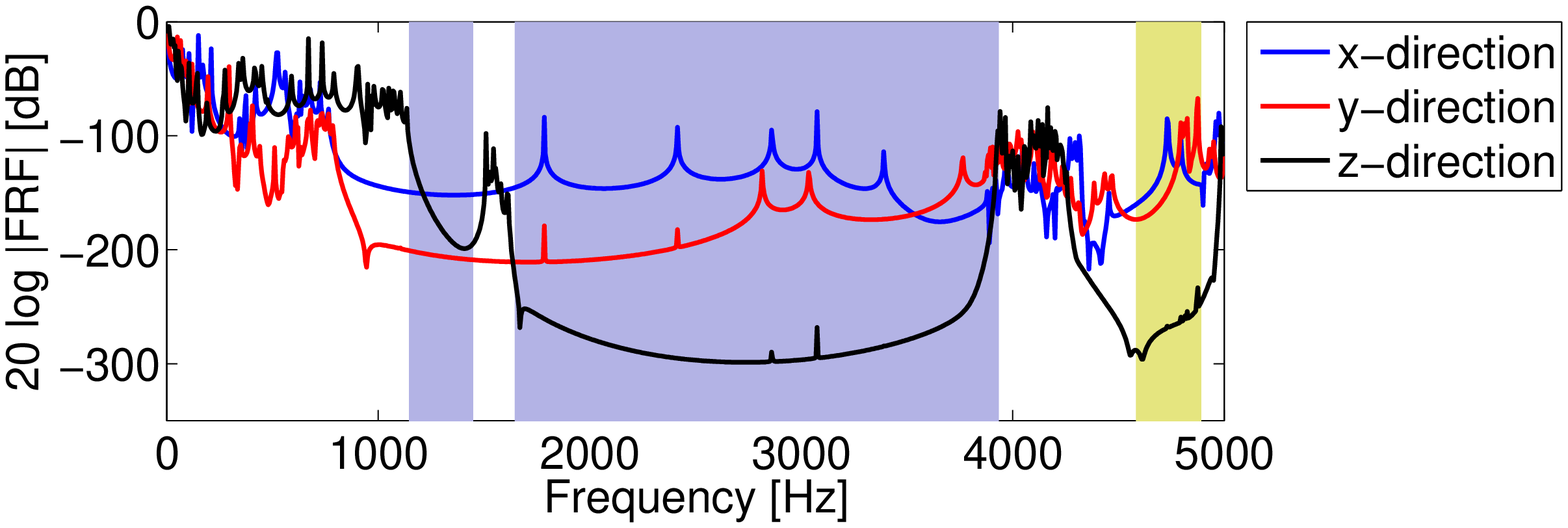}
      \caption{}
      \label{level2_spiderweb_hexagonal_frf_xyz}
  \end{subfigure}  
  }
  \caption{Computed FRFs considering (\subref{level2_spiderweb_hexagonal_frf_top}) a force input at the leftmost side nodes and (\subref{level2_spiderweb_hexagonal_frf_xyz}) outputs averaged at rightmost side nodes.}
  \label{level2_spiderweb_hexagonal_frf}
\end{figure}

We will use the examples of structures depicted in Figures \ref{level2_spiderweb_square_top} ($9$ nodes at the left and right sides) and \ref{level2_spiderweb_hexagonal_top} ($18$ nodes at the left and right sides) -- with band diagrams shown in Figures \ref{level2_spiderweb_square_band_diagram} and \ref{level2_spiderweb_hexagonal_band_diagram}, respectively -- for this first illustration, since they present narrow BGs, which in this case serve to show that wide frequency regions with low mode density (which may not be proper BGs) may be preferable over narrow BGs. Results are shown in Figures \ref{level2_spiderweb_square_frf} and \ref{level2_spiderweb_hexagonal_frf}. The previously computed regions with a low density of wave modes are marked in blue, and full BGs are marked in yellow.

Figure \ref{level2_spiderweb_square_frf} is a good illustration of the effects observed in Figure \ref{level2_spiderweb_square_band_modes}: longitudinal wave modes hinder the frequency regions marked in blue from showing complete BGs. For these regions, the $x$-direction FRF shows the smallest attenuation. On the other hand, when bending and torsional modes are excited, waves have large attenuation for a broad frequency range, as shown in the $y$- and $z$-directions. Also, the full BGs computed at high frequencies are somewhat difficult to be associated with specific attenuation ranges due to their narrow widths. Another interesting observation is that attenuation at higher frequencies (above approximately $3900$ Hz) is considerably stronger than that observed at lower frequencies (below approximately $1200$ Hz). This can also be explained using the corresponding band diagram: for lower frequencies, several modes with non-negligible group velocity can be noticed ($\frac{\partial \omega}{\partial k}$, see \citet{brillouin1953wave}), which indicates that wave energy can be carried further; however, for higher frequencies, wave modes are contained in a narrow frequency range with low group velocities, thus indicating localized behavior and limited energy propagation. These observations are in agreement with those reported by \citet{miniaci2018design}.

Similar relations can be observed between Figures \ref{level2_spiderweb_hexagonal_frf} and \ref{level2_spiderweb_hexagonal_band_modes}: the coupling between longitudinal and in-plane bending wave modes hinders larger attenuation in the $x$- and $y$- direction, while for the most part of the analyzed frequency ranges, the $z$-direction has large attenuation due to the fact that out-of-plane bending is not coupled with longitudinal modes. The same reasoning regarding greater attenuation at higher frequencies due to wave modes with low group velocity applies.

Results presented here illustrate how, even when no complete BGs are detected, there may still be wide frequency ranges with large attenuation. Additionally, for certain types of excitation (especially out-of-plane), the discussed structures behave effectively as filters in large frequency ranges.

\subsubsection{Robustness of band gaps}

Since the existence of BGs is mainly due to the periodicity of the structure (Bragg scattering), geometrical imperfections that occur in fabrication processes may hinder periodicity and affect the BGs adversely. For instance, variations in the radii of circular elements that form the frame structures may lead to non-periodicity.

To evaluate this possibility and its effect on BG properties, we assume that the radius of every circular element used to construct the frame structures follows a Gaussian distribution \citep{abramowits1972handbook}, with the mean equal to its nominal value, and the standard deviation of $33.33$ $\mu$m (in which case, $99\%$ of radii are within the $\pm 100$ $\mu$m tolerance). This standard deviation value was chosen to reflect a typical resolution of a 3D printer, which represents an uncertainty of about $2 \%$ for the radii of elements in the non-hierarchical two-dimensional periodic cell (Figure \ref{level2_square_inclusion_top}), and about $11 \%$ for elements in the hierarchical version (Figure \ref{level2_spiderweb_square_inclusion_top}).

We follow the same procedure of applying forces in the central leftmost nodes, and computing RMS displacement values at the central rightmost nodes of a $3 \times 3$ periodic cell finite structure. The results obtained using the periodic structure shown in Figure \ref{level2_square_inclusion_top} are described in \ref{app_square_inclusion_frf}, and show no significant BG edges variations under the assumed radii variation. 

\begin{figure}[h!]
  \makebox[\textwidth]{
  \begin{minipage}[l]{.45\textwidth}
  \begin{subfigure}[t]{\textwidth}
    {\centering
    \includegraphics[height=5cm,width=6cm]{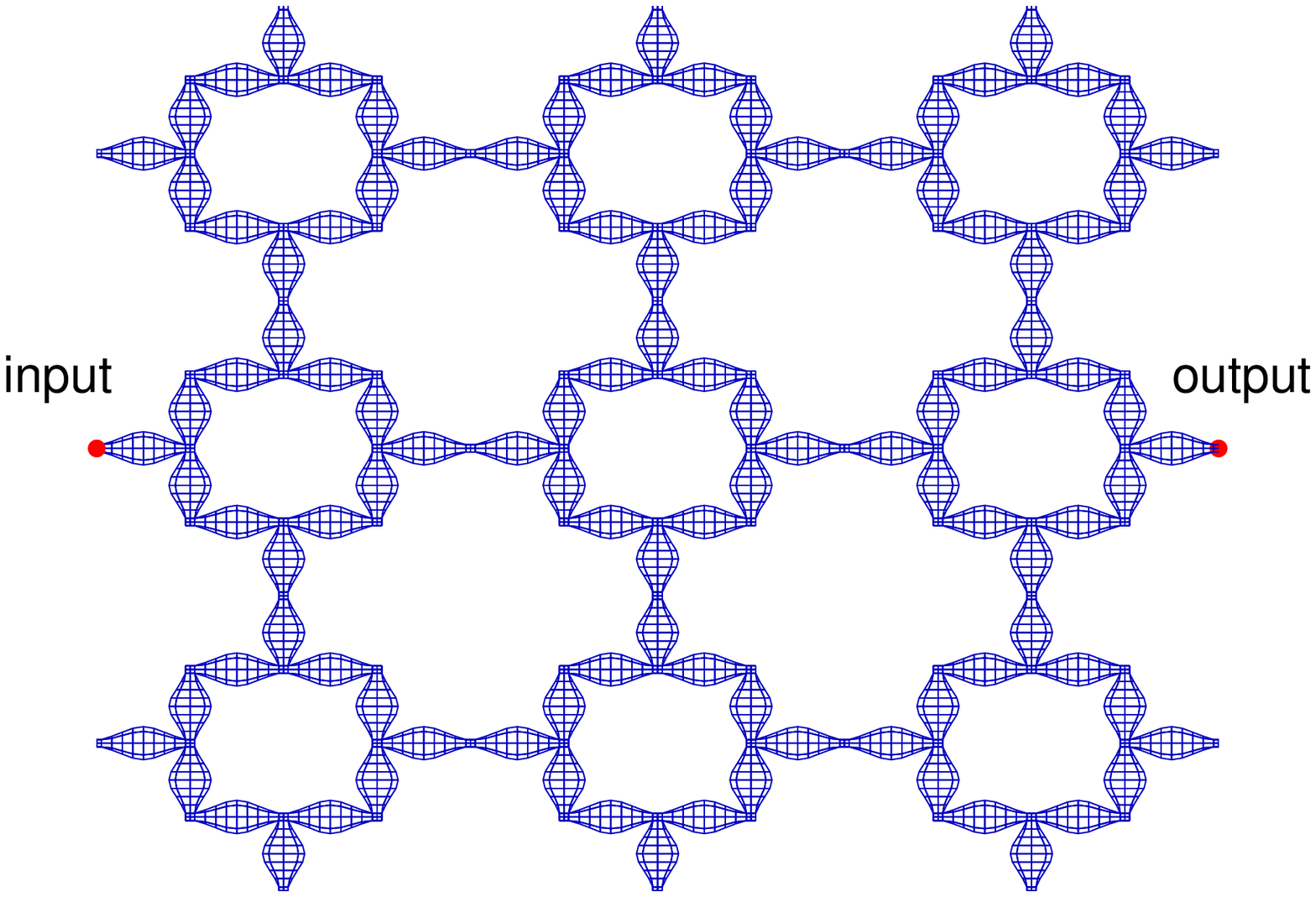}}
    \includegraphics[scale=0.5]{./xy_coordinates.eps}
    \caption{}
    \label{level2_spiderweb_square_inclusion_frf_top}
  \end{subfigure}
  \end{minipage}
    \begin{minipage}[c]{.6\textwidth}
    
    \begin{subfigure}[h]{\textwidth}
      \centering
      \includegraphics[width=10cm]{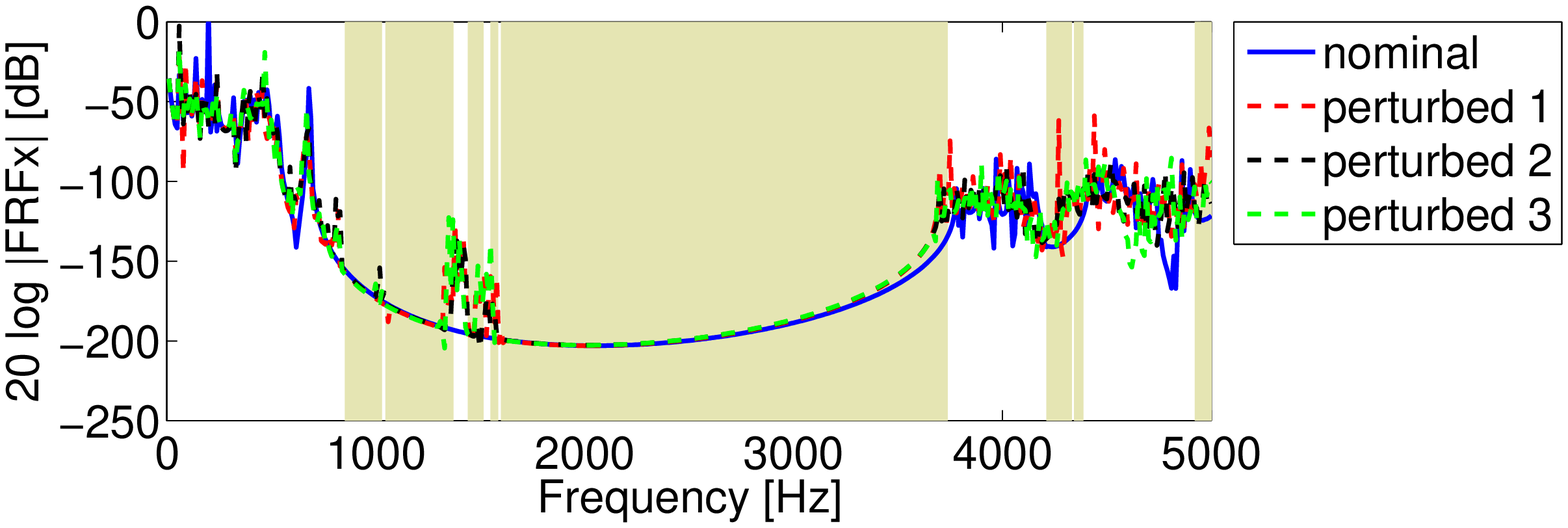}
      \caption{}
      \label{level2_spiderweb_square_inclusion_frf_x}
    \end{subfigure}
    
    \begin{subfigure}[h]{\textwidth}
      \centering
      \includegraphics[width=10cm]{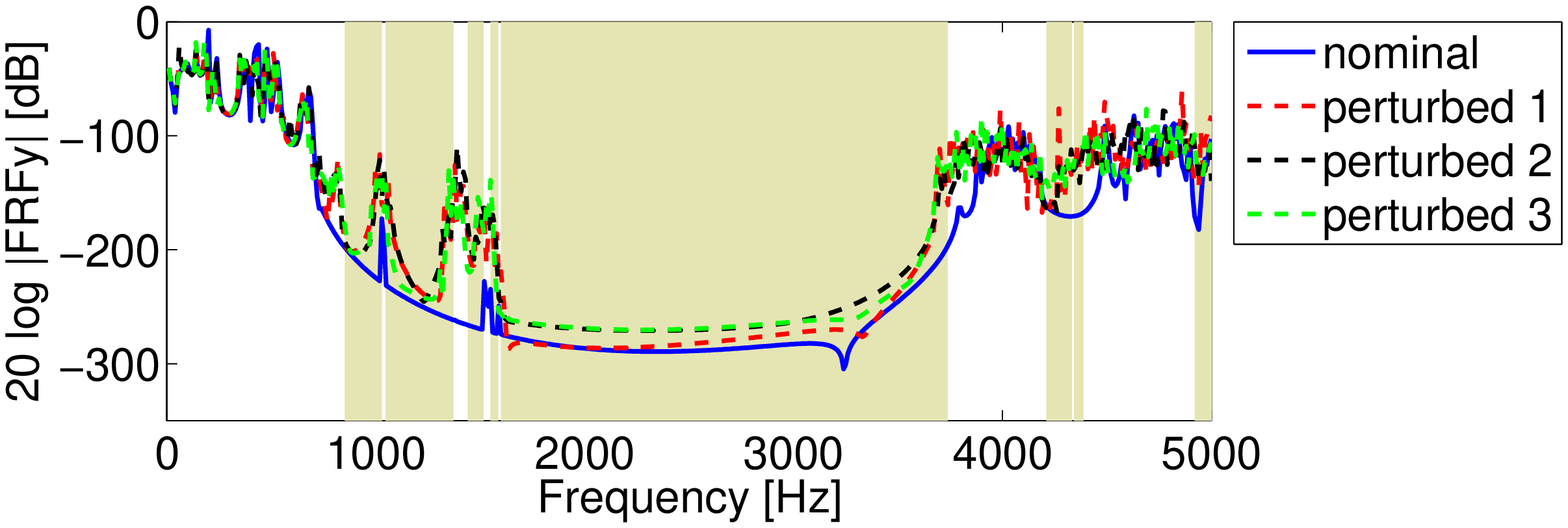}
      \caption{}
      \label{level2_spiderweb_square_inclusion_frf_y}
    \end{subfigure}
    
    \begin{subfigure}[h]{\textwidth}
      \centering
      \includegraphics[width=10cm]{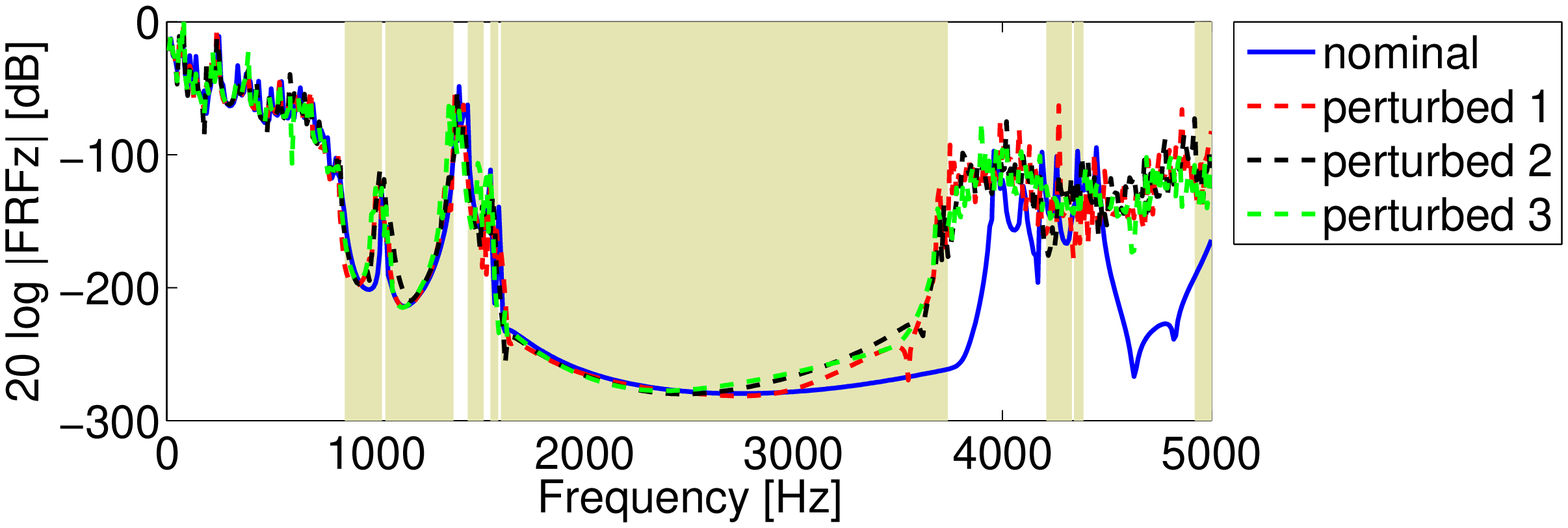}
      \caption{}
      \label{level2_spiderweb_square_inclusion_frf_z}
    \end{subfigure}
  \end{minipage}
  }
  \caption{Variations in FRFs considering (\subref{level2_spiderweb_square_inclusion_frf_top}) a force input at the center of the left side and outputs averaged at the center nodes of the right side with (\subref{level2_spiderweb_square_inclusion_frf_x}) $x$-, (\subref{level2_spiderweb_square_inclusion_frf_y}) $y$-, and (\subref{level2_spiderweb_square_inclusion_frf_z}) $z$-direction FRFs.}
  \label{level2_spiderweb_square_inclusion_frf}
\end{figure}

The results for the hierarchical structure are shown in Figure \ref{level2_spiderweb_square_inclusion_frf}. Frequency response functions computed using nominal radii values are marked in blue (\textcolor{blue}{--}, nominal), and those obtained using perturbed samples are marked in red (\textcolor{red}{- -}, perturbed 1), black (- -, perturbed 2), and green (\textcolor{green}{- -}, perturbed 3). The previously computed BGs are marked in yellow. 

The FRFs computed using the structure depicted in Figure \ref{level2_spiderweb_square_inclusion_frf_top} are shown in Figures \ref{level2_spiderweb_square_inclusion_frf_x} -- \ref{level2_spiderweb_square_inclusion_frf_z} and provide more insight with respect to the effects of a more considerable standard deviation in terms of relative dimensions of the elements radii. In all FRFs, there is a very good correlation between the BGs lower limits and the attenuation regions, even when the variation of elements radii are considered. The same cannot be stated regarding the upper limits, i.e., in all three cases the attenuation regions presented by the perturbed samples are lower than the limit stated by the band diagrams. This suggests that the lower limits of Bragg scattering BGs are more robust regarding periodicity than the corresponding upper limits. This also motivates future research in terms of the physical mechanisms responsible for BG formation and their robustness with respect to variations in geometrical and material configurations, since geometric perturbations may strongly alter specific modes responsible for certain BGs (as in the cases above $4000$ Hz).

\section{Concluding remarks} \label{conclusions}

In this work, we investigated hierarchical structures designed using a spider web-inspired design to broaden low- and mid-frequency range BGs using periodic frames. The definition of a first-order hierarchical structure was presented combining the geometry features of a deformed spider web-based shape to harness the promising wave-controlling characteristics of shape-varying cross-section phononic crystals. The resulting one-dimensional phononic crystal has symmetric sinusoidal-shape varying cross-section and a fixed number of radial and viscid threads, with square cross-section terminations. The band diagram of the resulting structure was computed and shows BGs deriving from torsional and bending behaviors between $1703$ Hz and $4624$ Hz.

Two-dimensional frame structures were proposed to assess the effects of replacing regular one-dimensional frame elements by the previously defined phononic crystal. Three resulting hierarchical structures were designed: a square lattice, a hexagonal lattice, and another square lattice based on a resonator-like two-dimensional structure. The resulting band diagrams for the simple square and hexagonal lattices show narrow complete BGs. However, they also show a very low mode density for large frequency ranges (widths of $2569$ Hz for the square lattice and $2592$ Hz for the hexagonal lattice), where the only computed wave modes are associated with longitudinal wave behavior. This results in wide attenuation frequency ranges, especially when considering out-of-plane excitations. This was also confirmed by examining the FRFs of a $3 \times 3$ finite structure obtained using the proposed periodic cells.

Band diagrams computed for the resonator-like structure show a large initial BG ($1974$ Hz -- $3579$ Hz), which is further broadened in the hierarchical structure counterpart ($1600$ Hz -- $3739$ Hz), while lower frequency BGs are also opened ($851.9$ Hz -- $1030$ Hz and $1047$ Hz -- $1372$ Hz, for instance). The hierarchical structure also presents a $50 \%$ weight reduction when compared to its non-hierarchical counterpart. The robustness of the resulting BGs was verified using FRF analyses, considering a normal distribution for the radii of its constituting elements, which implies a deviation from the structure periodicity. Results show that these BGs have a lower edge with little sensitivity to the lack of periodicity, while their corresponding upper edges may suffer degradation. In practice, this implies in less effective narrow BGs, while wide BGs are expected to be observed.

In general, our results suggest that hierarchical frame-like structures provide ideal lightweight systems for the generation of wide low-frequency band gaps. Hierarchical structuring adds further degrees of freedom to tune the structural properties in order to obtain attenuation of various modes over several frequency ranges. The possibilities for the design of such structures is virtually limitless, so that a bioinspired approach (e.g. spider-web inspired frames) could provide a useful starting point.

\section{Acknowledgments}

VDP, FB, MM, and NMP are  supported  by  the EU H2020  FET  Open ``Boheme''  grant  No. 863179.

\appendix
\section{Band structure computation} \label{app_wfem}

The method used in this work is based on \citep{mace2008modelling}. First, the dynamic stiffness matrix (see Eq. \ref{DqF}) is partitioned \citep{bathe1996finite} into active DOFs (a), related to nodes located at the borders of the periodic cell, and internal DOFs (i), using
\begin{equation} \label{D_partitioned}
 \left[
 \begin{array}{cc}
  \mathbf{D}_{\text{aa}} & \mathbf{D}_{\text{ai}} \\
  \mathbf{D}_{\text{ia}} & \mathbf{D}_{\text{ii}}
 \end{array}
 \right]
 \left\{
 \begin{array}{cc}
  \mathbf{q}_{\text{a}} \\
  \mathbf{q}_{\text{i}} \\
 \end{array}
 \right\} = 
 \left\{
 \begin{array}{cc}
  \mathbf{f}_{\text{a}} \\
  \mathbf{f}_{\text{i}} \\
 \end{array}
 \right\} \, ,
\end{equation}
where $\mathbf{D}_{\text{mn}} = \mathbf{K}_{\text{mn}} - \omega^2 \mathbf{M}_{\text{mn}}$ for $\{ \text{m}, \, \text{n}\} = \{\text{a}, \, \text{i}\}$.

\begin{figure}[h!]
  \centering
  \makebox[\textwidth]{
  \begin{subfigure}[h]{.35\textwidth}
    \centering
    \includegraphics[height=2.35cm]{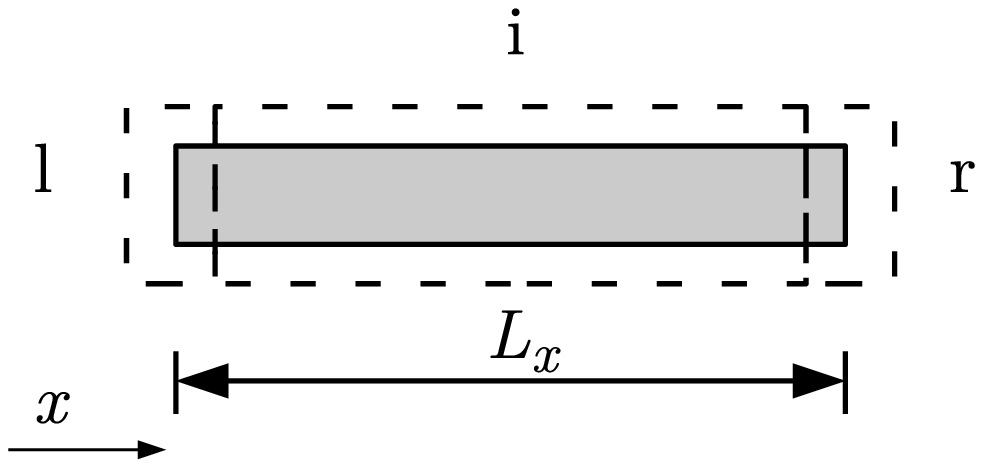}
  \end{subfigure}
  \begin{subfigure}[h]{.40\textwidth}
    \centering
    \includegraphics[height=4.5cm]{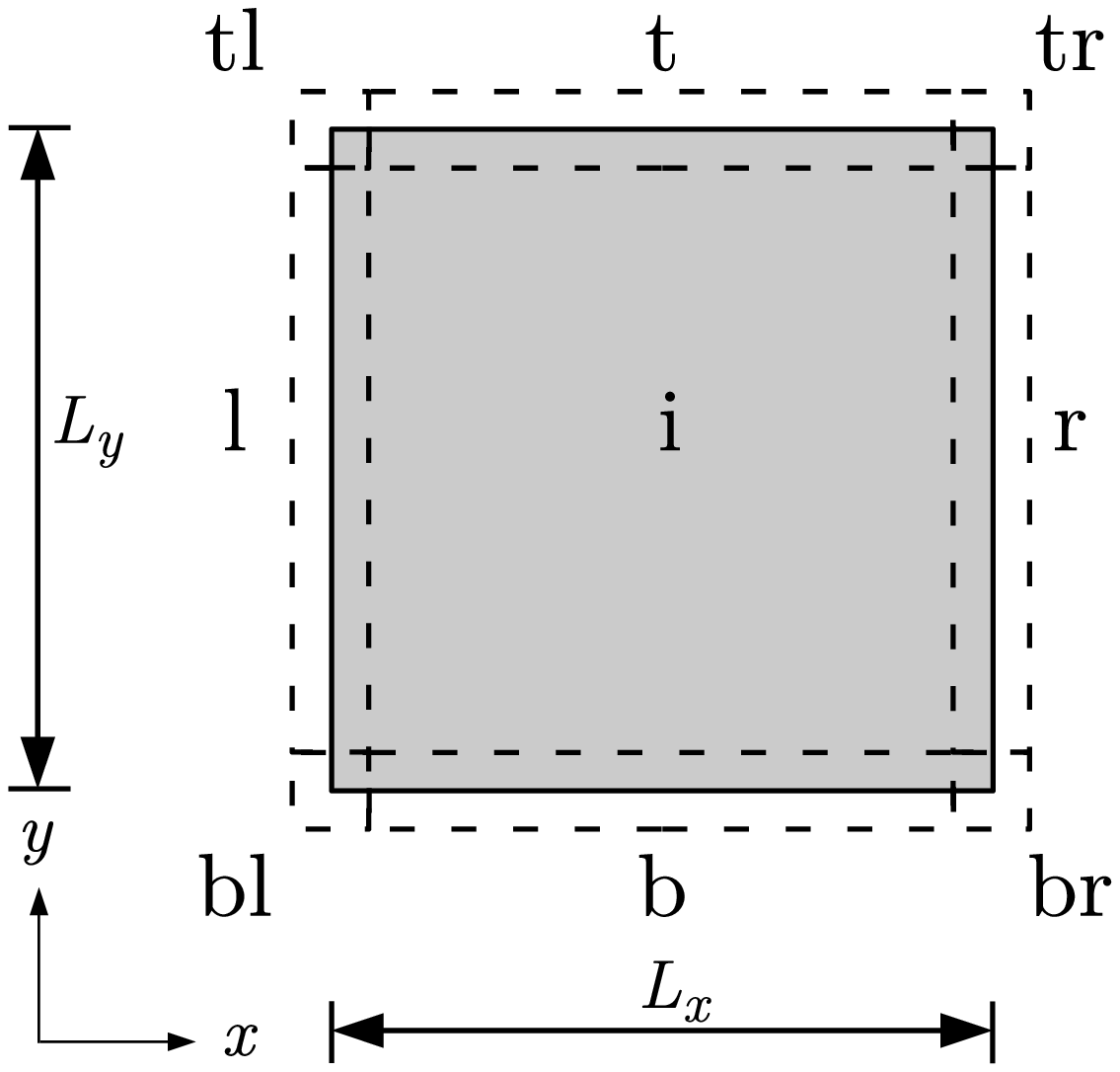}
  \end{subfigure}  
  \begin{subfigure}[h]{.35\textwidth}
    \centering
    \includegraphics[height=4.5cm]{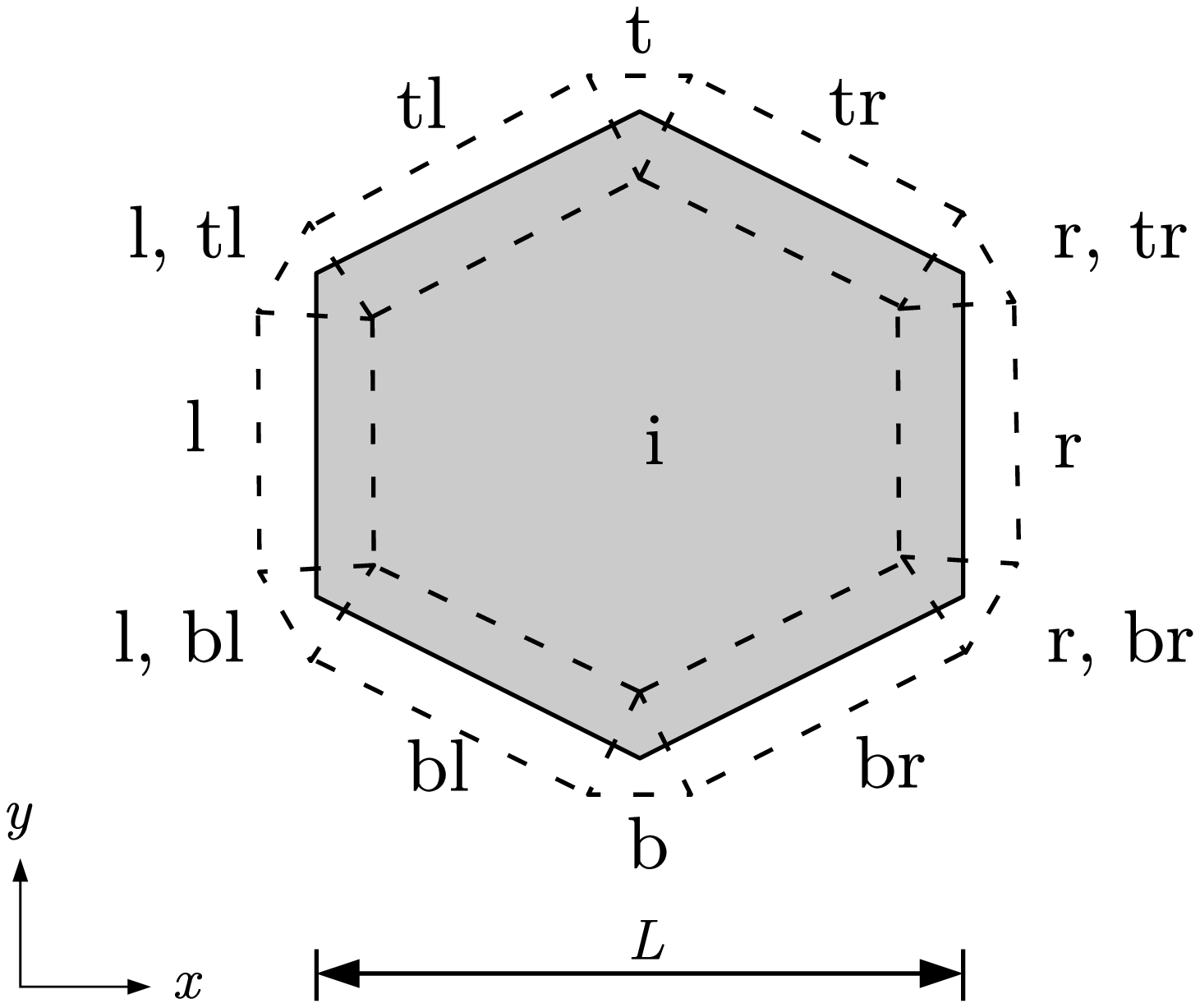}
  \end{subfigure}
  }
  \makebox[\textwidth]{
  \begin{subfigure}[h]{.35\textwidth}
    \caption{}
    \label{partitioned_fe_1d}
  \end{subfigure}
  \begin{subfigure}[h]{.40\textwidth}
    \caption{}
    \label{partitioned_fe_2d_square}
  \end{subfigure}  
  \begin{subfigure}[h]{.35\textwidth}
    \caption{}
    \label{partitioned_fe_2d_hexagonal}
  \end{subfigure}
  }
  \caption{Partitioned nodes for the (\subref{partitioned_fe_1d}) one- and two-dimensional (\subref{partitioned_fe_2d_square}) square and (\subref{partitioned_fe_2d_hexagonal}) hexagonal periodic cells. Every node which is not internal (i) belongs to the active (a) set of DOFs. The periodic cells dimensions are represented by $L_x$ and $L_y$.}
  \label{partitioned_fe_nodes}
\end{figure}

The active and internal nodes for the one- and two-dimensional periodic cells (see Figure \ref{periodic_cells}) are depicted in Figure \ref{partitioned_fe_nodes}, with characteristic dimensions shown in each case. Active nodes are constituted by the set of boundary nodes, following the subindex convention: l stands for left, r for right, t for top, b for bottom; two indexes indicate regions belonging to a diagonal: tl for top-left, tr for top-right, bl for bottom-left, br for bottom right, and indexes separated by a comma indicate adjacent regions, for instance, in the hexagonal lattice, l, tl is adjacent to the l and tl regions, and so on.

According to the partitioning shown in Figure \ref{partitioned_fe_nodes}, the sets of active DOFs in one- and two-dimensional square and hexagonal periodic cells can be described, respectively, using
\begin{subequations}
\begin{align}
 \mathbf{q}_\text{a}^{\text{1D}} &= 
  \left\{
  \begin{array}{cccccccc}
   \mathbf{q}_\text{l} & \mathbf{q}_\text{r} 
  \end{array}
  \right\}^T \, , \\
 \mathbf{q}_\text{a}^{\text{2D, sq.}} &= 
  \left\{
  \begin{array}{cccccccc}
   \mathbf{q}_\text{bl} & \mathbf{q}_\text{l} & \mathbf{q}_\text{b} & \mathbf{q}_\text{tl} & \mathbf{q}_\text{br} & \mathbf{q}_\text{tr} & \mathbf{q}_\text{r} & \mathbf{q}_\text{t}
  \end{array}
  \right\}^T \, , \\
\mathbf{q}_\text{a}^{\text{2D, hx.}} &= 
  \left\{
  \begin{array}{cccccccccccc}
   \mathbf{q}_\text{l} & \mathbf{q}_\text{l, bl} & \mathbf{q}_\text{bl} & \mathbf{q}_\text{b} & \mathbf{q}_\text{br} & \mathbf{q}_\text{r} & \mathbf{q}_\text{r, br} & \mathbf{q}_\text{t} & \mathbf{q}_\text{tr} & \mathbf{q}_\text{r, tr} & \mathbf{q}_\text{l, tl} & \mathbf{q}_\text{tl}
  \end{array}
  \right\}^T \, .
\end{align}
\end{subequations}

The periodicity condition can be applied by using Bloch's theorem \citep{bloch1929quantenmechanik} on the boundaries of the periodic cells by using the relation
\begin{equation} \label{qa_qc}
 \mathbf{q}_\text{a} = \mathbf{\Lambda}_\text{R} \mathbf{q}_\text{c} \, ,
\end{equation}
where $\mathbf{\Lambda}_\text{R}$ is a linear transformation between the reduced displacements and rotations vector $\mathbf{q}_\text{c}$ and the active DOFs $\mathbf{q}_\text{a}$. In the case of one- and two-dimensional periodic cells, the reduced vectors are given by
\begin{subequations}
\begin{align}
  \mathbf{q}_\text{c}^{\text{1D}} &=  \mathbf{q}_\text{l} \, , \\
  \mathbf{q}_\text{c}^{\text{2D, sq.}} &= 
  \left\{
  \begin{array}{ccc}
   \mathbf{q}_\text{bl} & \mathbf{q}_\text{l} & \mathbf{q}_\text{b}
  \end{array}
  \right\}^T \, , \\
  \mathbf{q}_\text{c}^{\text{2D, hx.}} &= 
  \left\{
  \begin{array}{ccccc}
   \mathbf{q}_\text{l} & \mathbf{q}_\text{l, bl} & \mathbf{q}_\text{bl} & \mathbf{q}_\text{b} & \mathbf{q}_\text{br}
  \end{array}
  \right\}^T \, .
\end{align}
\end{subequations}
and the linear transformations of DOFs are given by
\begin{subequations}
\begin{align}
  \mathbf{\Lambda}_\text{R}^\text{1D} &= 
  \left[
  \begin{array}{cc}
   \mathbf{I} & \lambda_x \mathbf{I} \\
  \end{array}
  \right]^T \, , \\
  \mathbf{\Lambda}_\text{R}^\text{2D, sq.} &= 
  \left[
  \begin{array}{cccccccc}
   \mathbf{I} & \mathbf{0} & \mathbf{0} & \lambda_y \mathbf{I} & \lambda_x \mathbf{I} & \lambda_x \lambda_y \mathbf{I} & \mathbf{0} & \mathbf{0} \\
   \mathbf{0} & \mathbf{I} & \mathbf{0} & \mathbf{0} & \mathbf{0} & \mathbf{0} & \lambda_x \mathbf{I} & \mathbf{0} \\
   \mathbf{0} & \mathbf{0} & \mathbf{I} & \mathbf{0} & \mathbf{0} & \mathbf{0} & \mathbf{0} & \lambda_y \mathbf{I} 
  \end{array}
  \right]^T \, , \\
  \mathbf{\Lambda}_\text{R}^\text{2D, hx.} &= 
  \left[
  \begin{array}{cccccccccccc}
   \mathbf{I} & \mathbf{0} & \mathbf{0} & \mathbf{0} & \mathbf{0} & \lambda_x' \mathbf{I} & \mathbf{0} & \mathbf{0} & \mathbf{0} & \mathbf{0} & \mathbf{0} & \mathbf{0} \\
   \mathbf{0} & \mathbf{I} & \mathbf{0} & \mathbf{0} & \mathbf{0} & \mathbf{0} & \lambda_x' \mathbf{I} & \lambda_y' \mathbf{I} & \mathbf{0} & \mathbf{0} & \mathbf{0} & \mathbf{0} \\
   \mathbf{0} & \mathbf{0} & \mathbf{I} & \mathbf{0} & \mathbf{0} & \mathbf{0} & \mathbf{0} & \mathbf{0} & \lambda_y' \mathbf{I} & \mathbf{0} & \mathbf{0} & \mathbf{0} \\
   \mathbf{0} & \mathbf{0} & \mathbf{0} & \mathbf{I} & \mathbf{0} & \mathbf{0} & \mathbf{0} & \mathbf{0} & \mathbf{0} & \lambda_y' \mathbf{I} & \lambda_x'^{-1} \lambda_y' \mathbf{I} & \mathbf{0} \\
   \mathbf{0} & \mathbf{0} & \mathbf{0} & \mathbf{0} & \mathbf{I} & \mathbf{0} & \mathbf{0} & \mathbf{0} & \mathbf{0} & \mathbf{0} & \mathbf{0} & \lambda_x'^{-1} \lambda_y' \mathbf{I} \\
  \end{array}
  \right]^T \, ,
\end{align}
\end{subequations}
where $\lambda_x = e^{\text{i} k_x L_x}$, $\lambda_y = e^{\text{i} k_y L_y}$, $\lambda_x' = e^{\text{i} k_x L}$, $\lambda_y' = e^{\text{i}(k_x + k_y \sqrt{3})L/2}$, and $k_x$ and $k_y$ are the Cartesian components of the wave vector $\mathbf{k}$, i.e., $\mathbf{k} = k_x \hat{\mathbf{i}} + k_y \hat{\mathbf{j}}$.

The Bloch condition is also applied to forces at boundary nodes using
\begin{equation} \label{fa_zero}
 \mathbf{\Lambda}_\text{L} \mathbf{f}_\text{a} = \mathbf{0} \, ,
\end{equation}
where $\mathbf{\Lambda}_\text{L}$ can be written for the case of one- and two- dimensional periodic cells using
\begin{subequations}
\begin{align}
  \mathbf{\Lambda}_\text{L}^\text{1D} &= 
  \left[
  \begin{array}{cc}
   \mathbf{I} & \lambda_x^{-1} \mathbf{I} \\
  \end{array}
  \right] \, , \\
  \mathbf{\Lambda}_\text{L}^\text{2D, sq.} &= 
  \left[
  \begin{array}{cccccccc}
   \mathbf{I} & \mathbf{0} & \mathbf{0} & \lambda_y^{-1} \mathbf{I} & \lambda_x^{-1} \mathbf{I} & \lambda_x^{-1} \lambda_y^{-1} \mathbf{I} & \mathbf{0} & \mathbf{0} \\
   \mathbf{0} & \mathbf{I} & \mathbf{0} & \mathbf{0} & \mathbf{0} & \mathbf{0} & \lambda_x^{-1} \mathbf{I} & \mathbf{0} \\
   \mathbf{0} & \mathbf{0} & \mathbf{I} & \mathbf{0} & \mathbf{0} & \mathbf{0} & \mathbf{0} & \lambda_y^{-1} \mathbf{I} 
  \end{array}
  \right] \, , \\
  \mathbf{\Lambda}_\text{L}^\text{2D, hx.} &= 
  \left[
  \begin{array}{cccccccccccc}
   \mathbf{I} & \mathbf{0} & \mathbf{0} & \mathbf{0} & \mathbf{0} & \lambda_x'^{-1} \mathbf{I} & \mathbf{0} & \mathbf{0} & \mathbf{0} & \mathbf{0} & \mathbf{0} & \mathbf{0} \\
   \mathbf{0} & \mathbf{I} & \mathbf{0} & \mathbf{0} & \mathbf{0} & \mathbf{0} & \lambda_x'^{-1} \mathbf{I} & \lambda_y'^{-1} \mathbf{I} & \mathbf{0} & \mathbf{0} & \mathbf{0} & \mathbf{0} \\
   \mathbf{0} & \mathbf{0} & \mathbf{I} & \mathbf{0} & \mathbf{0} & \mathbf{0} & \mathbf{0} & \mathbf{0} & \lambda_y'^{-1} \mathbf{I} & \mathbf{0} & \mathbf{0} & \mathbf{0} \\
   \mathbf{0} & \mathbf{0} & \mathbf{0} & \mathbf{I} & \mathbf{0} & \mathbf{0} & \mathbf{0} & \mathbf{0} & \mathbf{0} & \lambda_y'^{-1} \mathbf{I} & \lambda_x' \lambda_y'^{-1} \mathbf{I} & \mathbf{0} \\
   \mathbf{0} & \mathbf{0} & \mathbf{0} & \mathbf{0} & \mathbf{I} & \mathbf{0} & \mathbf{0} & \mathbf{0} & \mathbf{0} & \mathbf{0} & \mathbf{0} & \lambda_x' \lambda_y'^{-1} \mathbf{I} \\
  \end{array}
  \right] \, .
\end{align}
\end{subequations}

Equations (\ref{D_partitioned}), (\ref{qa_qc}), and (\ref{fa_zero}) can be combined to yield 
\begin{equation}
 \left[
 \begin{array}{cc}
  \mathbf{\Lambda}_\text{L} & \mathbf{0} \\
  \mathbf{0} & \mathbf{I} \\
 \end{array}
 \right]
 \left[
 \begin{array}{cc}
  \mathbf{D}_\text{aa} & \mathbf{D}_\text{ai} \\
  \mathbf{D}_\text{ia} & \mathbf{D}_\text{ii} \\
 \end{array}
 \right]
 \left[
 \begin{array}{cc}
  \mathbf{\Lambda}_\text{R} & \mathbf{0} \\
  \mathbf{0} & \mathbf{I} \\
 \end{array}
 \right] 
 \left\{
 \begin{array}{c}
  \mathbf{q}_\text{c} \\
  \mathbf{q}_\text{i}
 \end{array}
 \right\}
 = 
 \left[
 \begin{array}{cc}
  \mathbf{\Lambda}_\text{L} & \mathbf{0} \\
  \mathbf{0} & \mathbf{I} \\
 \end{array}
 \right]
 \left\{
 \begin{array}{c}
  \mathbf{f}_\text{c} \\
  \mathbf{f}_\text{i}
 \end{array}
 \right\} = \mathbf{0} \, ,
\end{equation}
which, with the use of Eq. (\ref{DqF}), leads to
\begin{equation} \label{eigen_wk}
  \left[
    \begin{array}{cc}
      \mathbf{\Lambda}_\text{L} \mathbf{K}_\text{aa} \mathbf{\Lambda}_\text{R} & \mathbf{\Lambda}_\text{L} \mathbf{K}_\text{ai} \\
      \mathbf{K}_\text{ia} \mathbf{\Lambda}_\text{R}  & \mathbf{K}_\text{ii} \\
    \end{array}
 \right]
 \left\{
 \begin{array}{c}
  \mathbf{q}_\text{c} \\
  \mathbf{q}_\text{i}
 \end{array}
 \right\} =
 \omega^2
   \left[
    \begin{array}{cc}
      \mathbf{\Lambda}_\text{L} \mathbf{M}_\text{aa} \mathbf{\Lambda}_\text{R} & \mathbf{\Lambda}_\text{L} \mathbf{M}_\text{ai} \\
      \mathbf{M}_\text{ia} \mathbf{\Lambda}_\text{R}  & \mathbf{M}_\text{ii} \\
    \end{array}
 \right]
 \left\{
 \begin{array}{c}
  \mathbf{q}_\text{c} \\
  \mathbf{q}_\text{i}
 \end{array}
 \right\} \, .
\end{equation}

Equation (\ref{eigen_wk}) represents an eigenproblem with eigenvectors given by $\left\{
\begin{array}{cc} \mathbf{q}_\text{c} & \mathbf{q}_\text{i} \end{array} \right\}^T$, representing wave modes, and eigenvalues given by $\omega^2$, representing propagating wave frequencies. This is commonly referred to as a a $\omega = \omega(\mathbf{k})$ band diagram computation problem, where the propagating wave frequencies can be determined for the wave vectors restricted to the FBZ regions (depicted in Figures \ref{fbz1d} -- \ref{fbz2d_hexagonal}).

\section{Quasistatic properties} \label{app_quasistatic}

The stiffness and stress concentration of the two-dimensional structures (Figures \ref{level2_square_inclusion_top} -- \ref{level2_spiderweb_square_inclusion_top}) need to be evaluated to assess the effect of substituting full circular cross-sections with their hierarchical counterparts. For the sake of a fair comparison, the quasistatic properties are computed here using structures with the same mass, i.e., the radii of elements are not all the same. Results here illustrated are meant to be comparative, since no real measurements were made to estimate force and moments intensities.

\begin{figure}[H]
  \makebox[\textwidth]{
  \begin{subfigure}[t]{.25\textwidth}
    \centering
    \includegraphics[height=3.0cm,width=3.0cm]{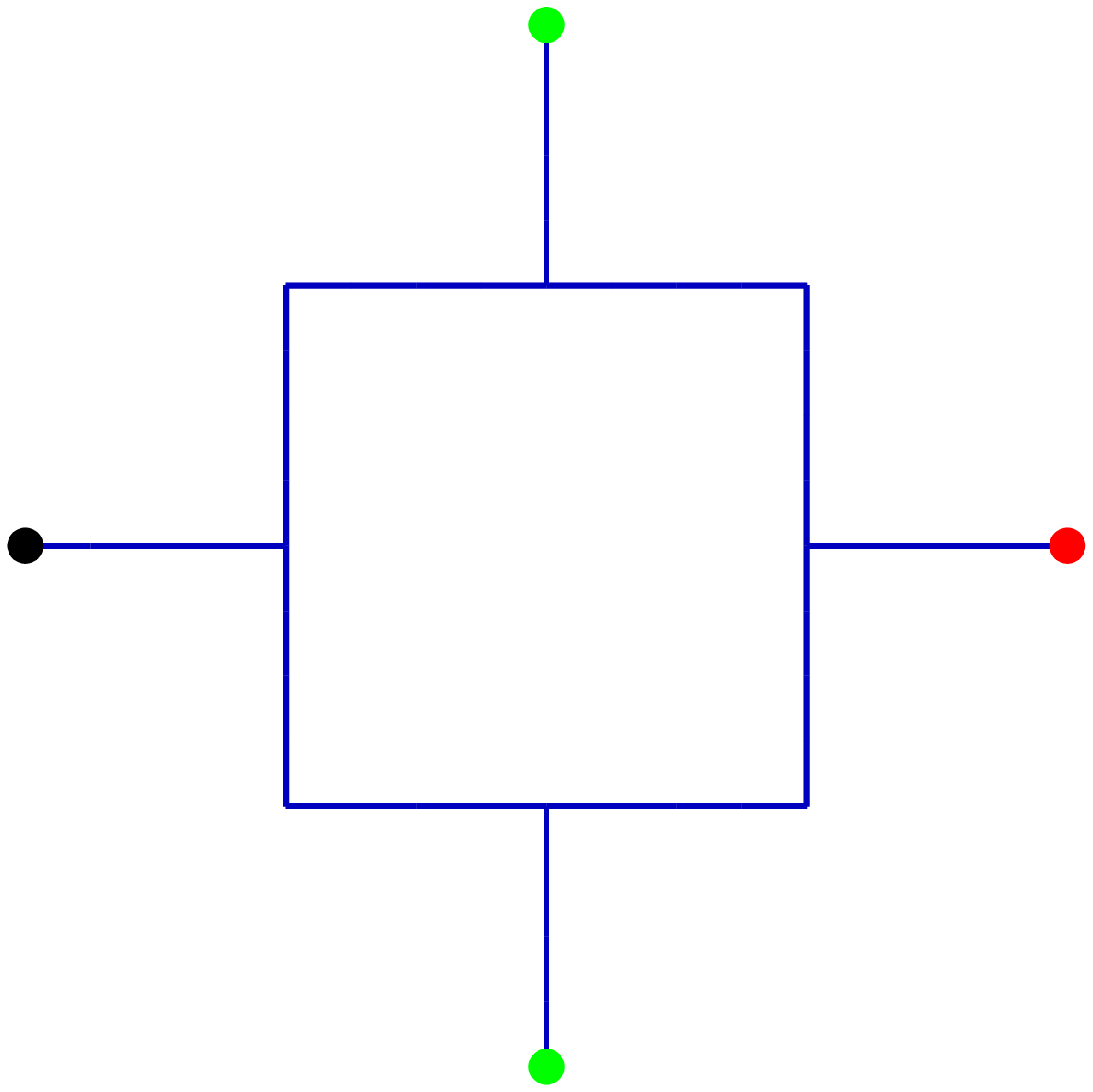}
    \caption{}
    \label{level2_square_inclusion_quasi}
  \end{subfigure}
  \begin{subfigure}[t]{.25\textwidth}
    \centering
    \includegraphics[height=3.0cm,width=3.0cm]{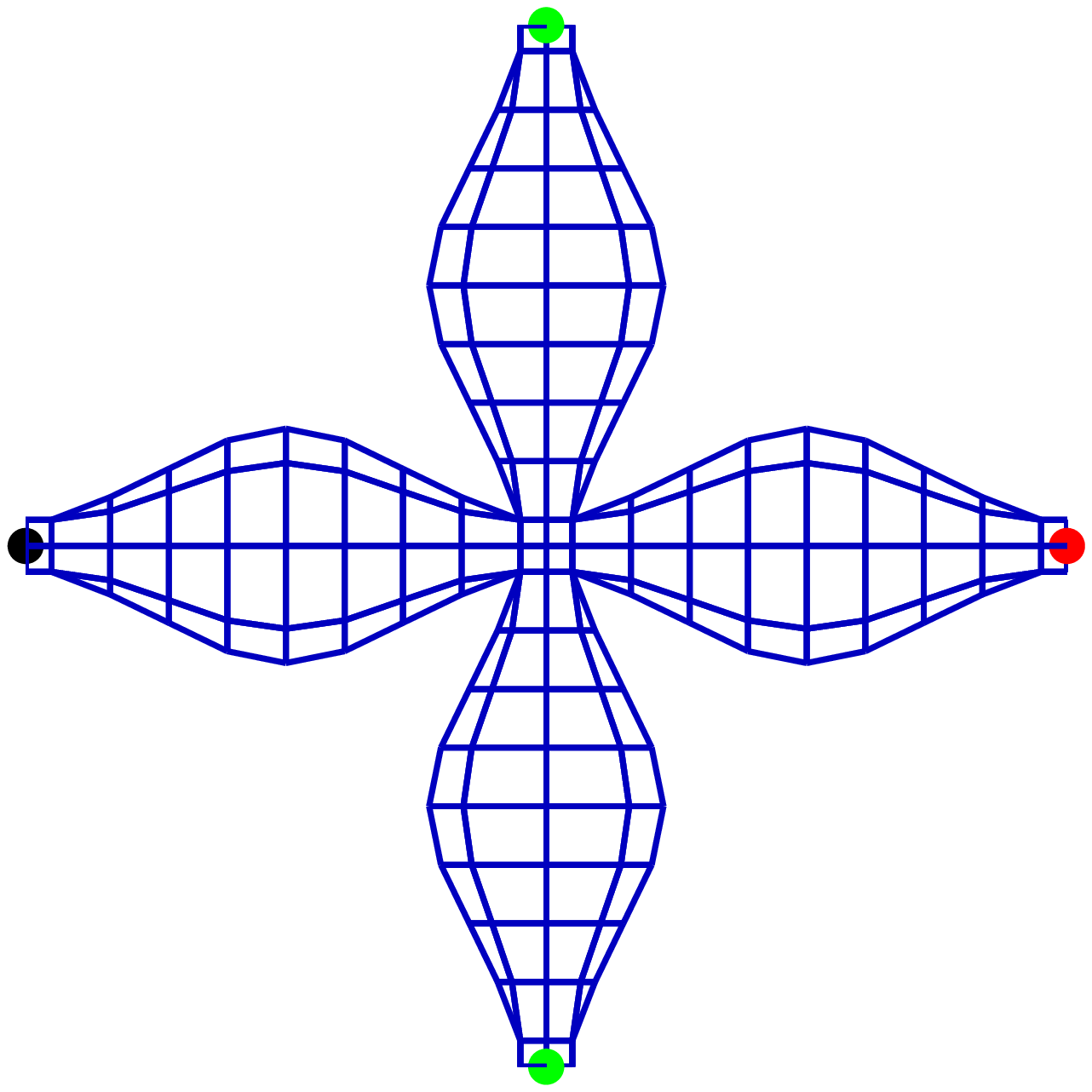}
    \caption{}
    \label{level2_spiderweb_square_quasi}
  \end{subfigure}
  \begin{subfigure}[t]{.25\textwidth}
    \centering
    \includegraphics[height=3.0cm]{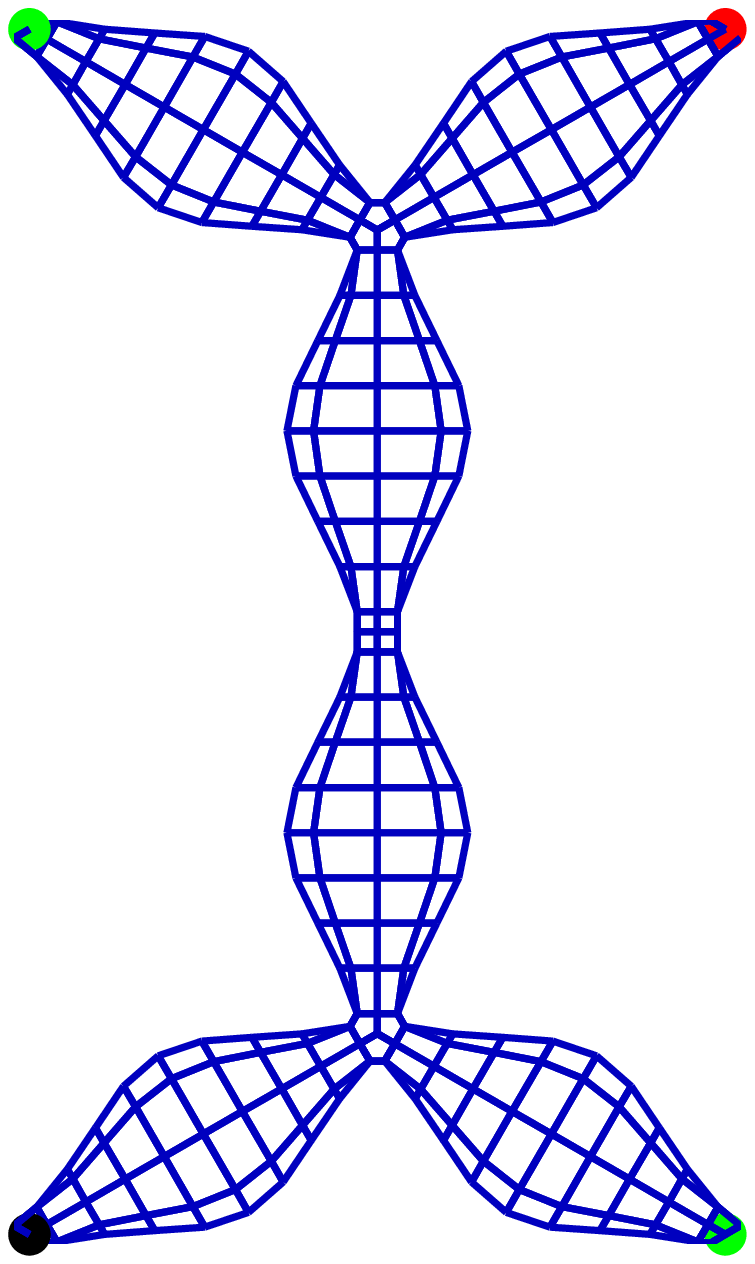}
    \caption{}
    \label{level2_spiderweb_hexagonal_quasi}
  \end{subfigure}
  \begin{subfigure}[t]{.25\textwidth}
    \centering
    \includegraphics[height=3.0cm,width=3.0cm]{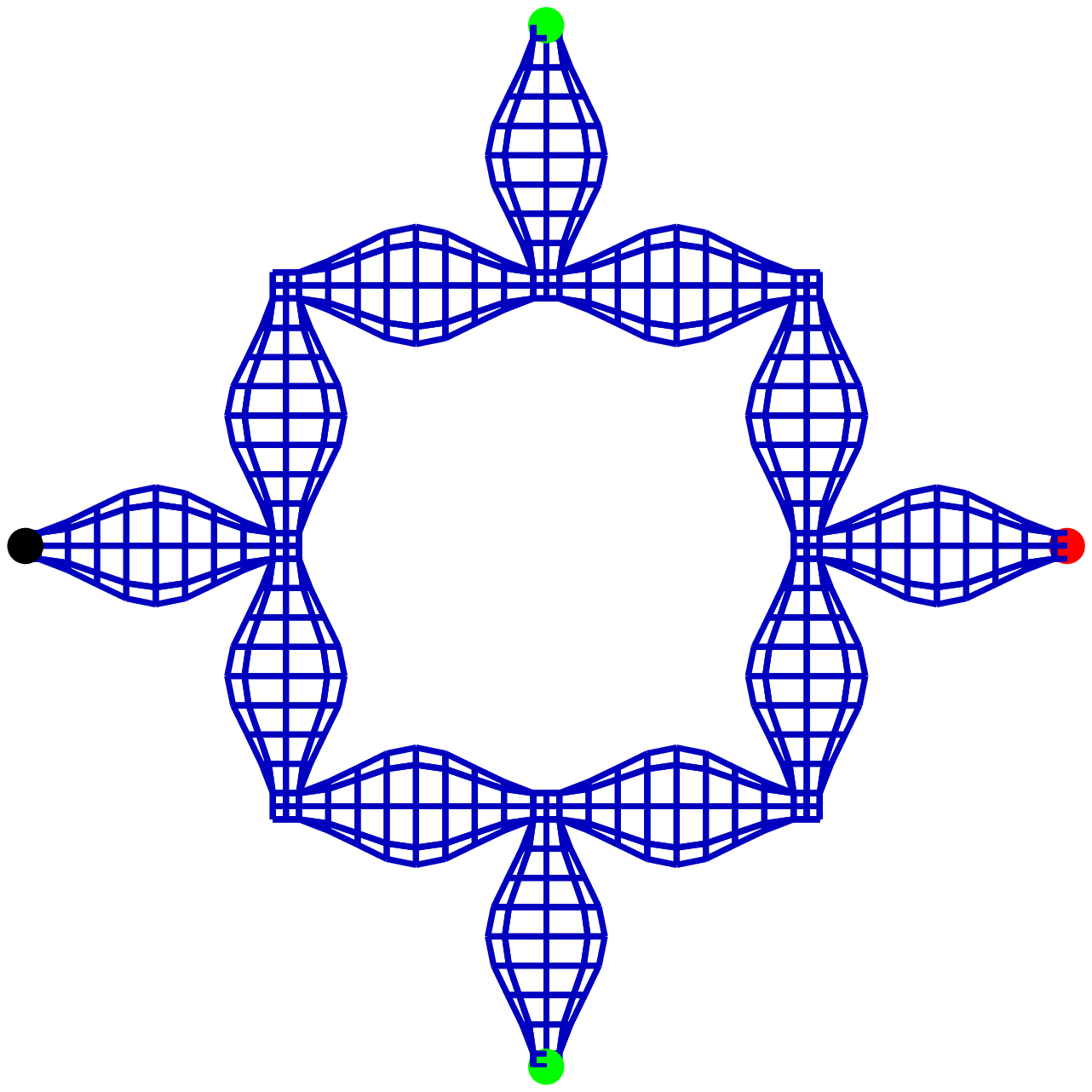}
    \caption{}
    \label{level2_spiderweb_square_inclusion_quasi}
  \end{subfigure}
  }
  \caption{Considered structures for quasistatic properties evaluation. Regions marked in black are restrained in both cases. In-plane bending behavior is assessed by applying a $1$ mN force in the in-plane direction, orthogonal to the structure. Torsion behavior is assessed by also fixing the regions marked in red, and applying forces in opposing directions using the regions marked in greed, for a total $1$ mN$\cdot$m moment.}
  \label{level2_quasi_bcs_forces}
\end{figure}

Numerical simulations are performed using the Newton-Raphson method \citep{bathe1996finite}. Quasistatic progressive loads are applied to evaluate stresses and load-displacement curves associated with in-plane bending behavior and torsion behavior. The considered structures are shown in Figure \ref{level2_quasi_bcs_forces} to illustrate the locations of displacements and restricted DOFs.

For in-plane bending displacements, the region indicated by the black dots is completely restrained, and a $1$ mN force is applied to the region indicated by the red dot, orthogonal and in the same plane as the structure. To evaluate torsional behavior, a moment of $1$ mN$\cdot$m is applied in the out-of-plane direction, using forces in opposing directions applied to the regions indicated by the green dots, while regions indicated by black and red dots are restrained.

\begin{figure}[H]
  \makebox[\textwidth]{
  \begin{subfigure}[t]{.3\textwidth}
    \centering
    \includegraphics[height=3.0cm,trim={0cm 0cm 3.5cm 0cm},clip]{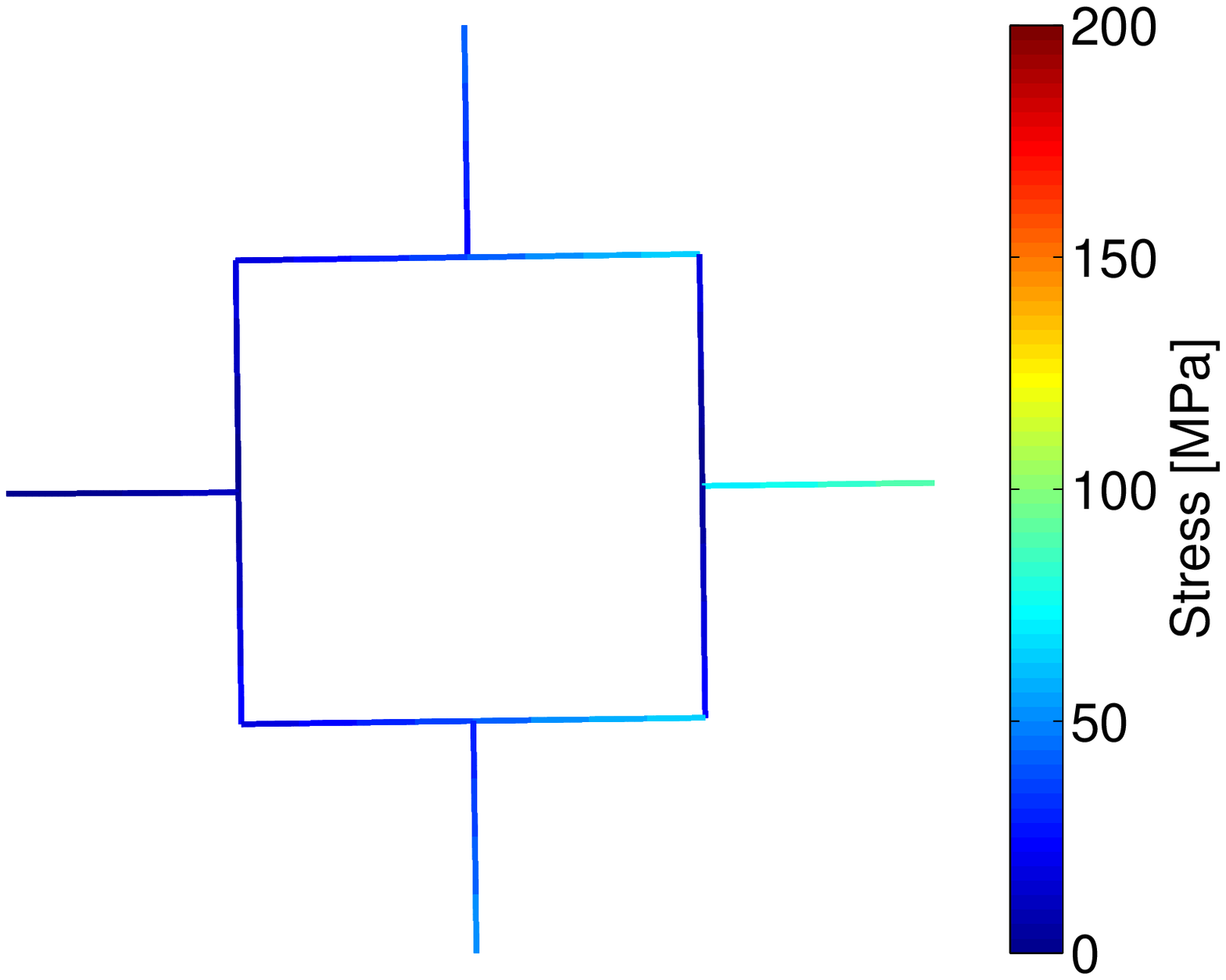}
    \caption{}
    \label{level2_square_inclusion_quasi_disp_defr}
  \end{subfigure}
  \begin{subfigure}[t]{.3\textwidth}
    \centering
    \includegraphics[height=3.0cm,trim={0cm 0cm 3.5cm 0cm},clip]{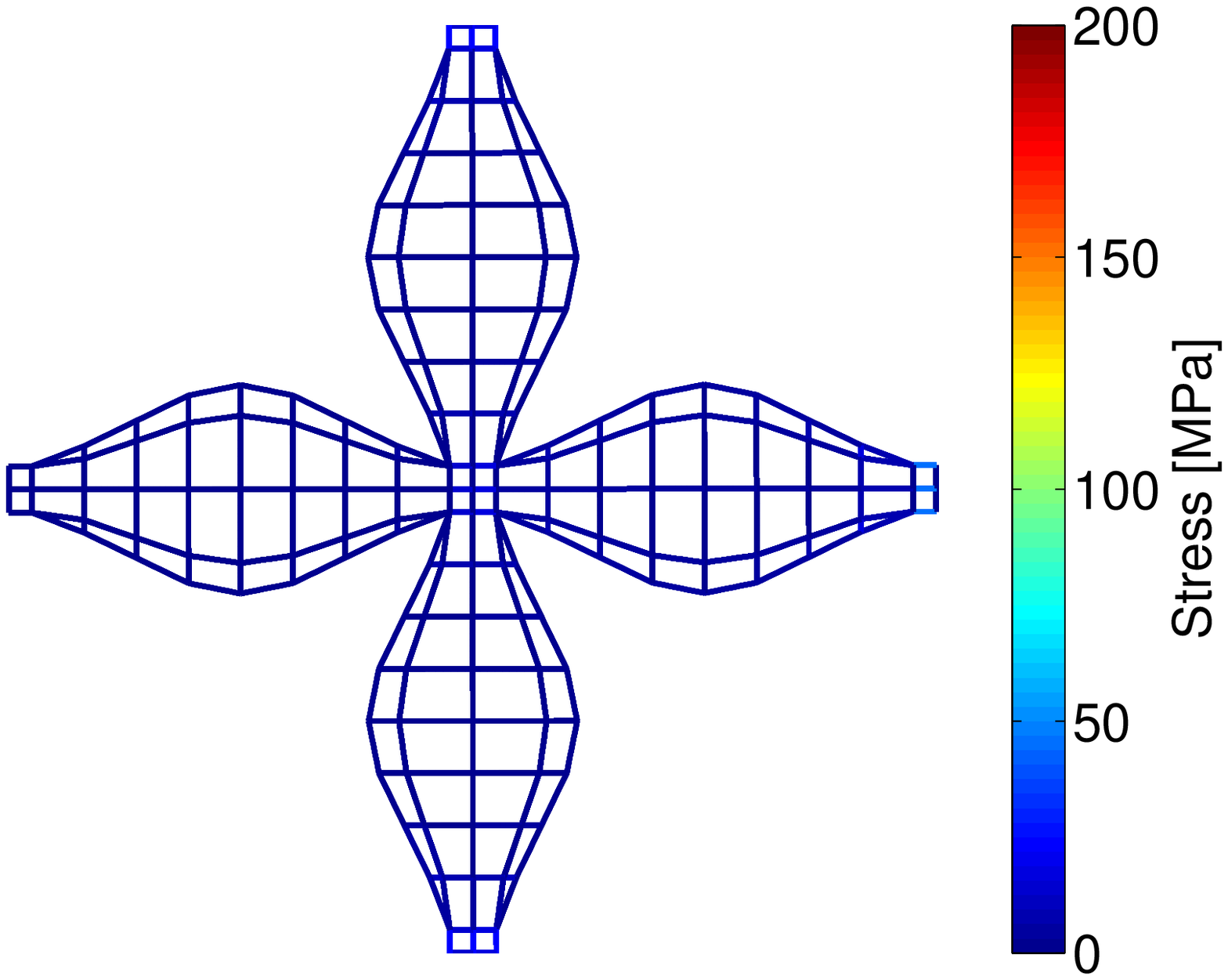}
    \caption{}
    \label{level2_spiderweb_square_quasi_disp_defr}
  \end{subfigure}
  \begin{subfigure}[t]{.3\textwidth}
    \centering
    \includegraphics[height=3.0cm,trim={0cm 0cm 3.5cm 0cm},clip]{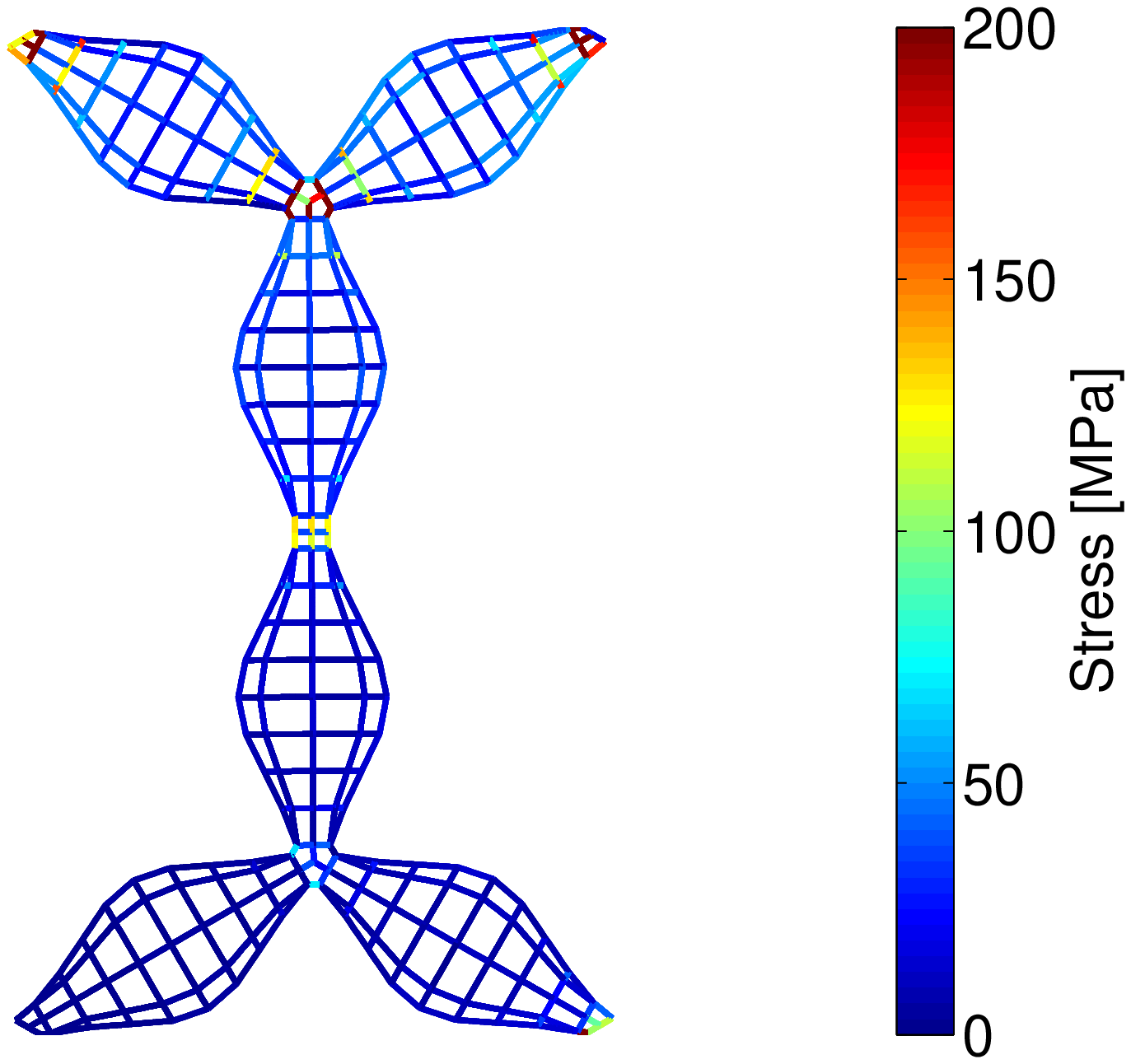}
    \caption{}
    \label{level2_spiderweb_hexagonal_quasi_disp_defr}
  \end{subfigure}
  \begin{subfigure}[t]{.3\textwidth}
    \centering
    \includegraphics[height=3.0cm]{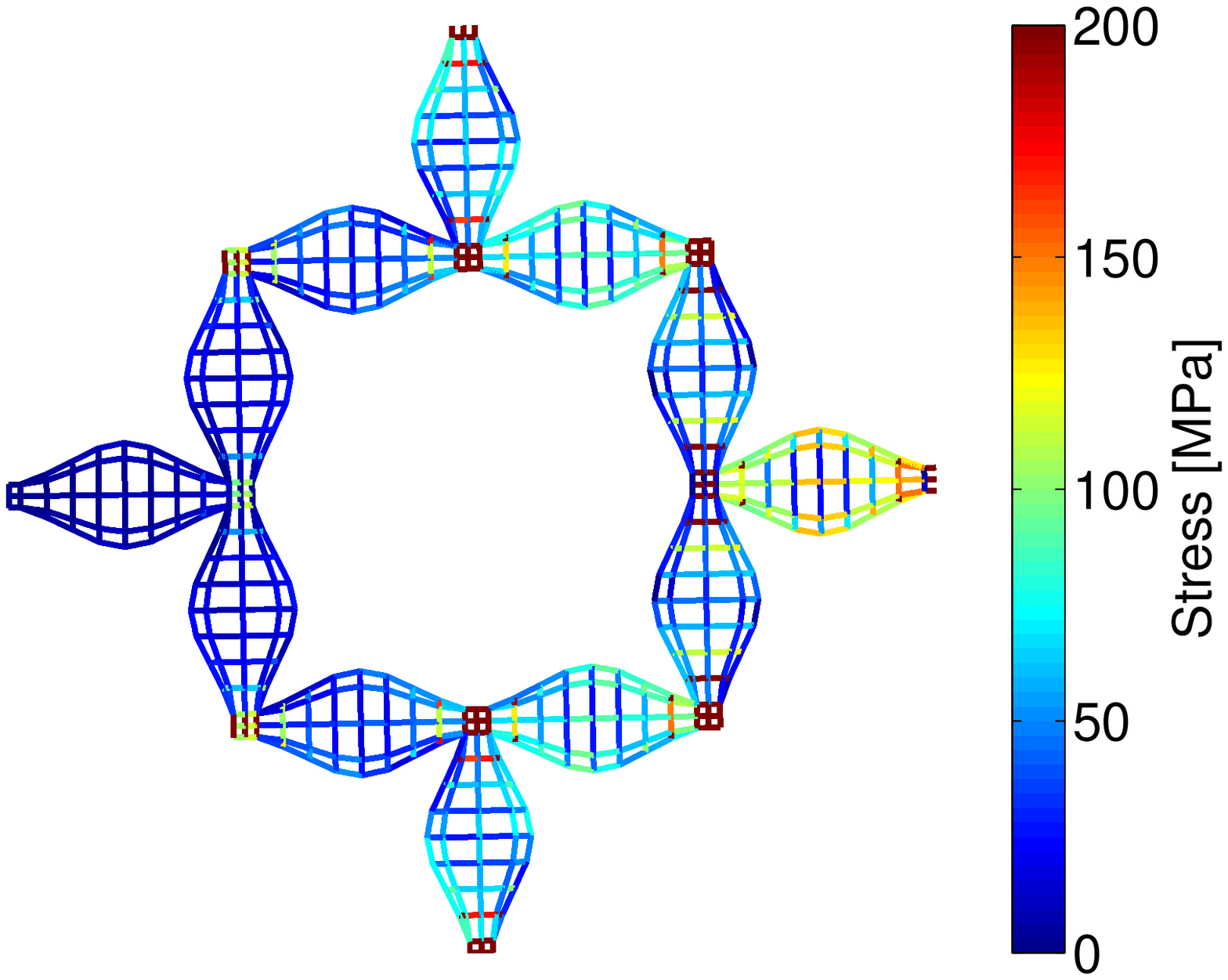}
    \caption{}
    \label{level2_spiderweb_square_inclusion_quasi_disp_defr}
  \end{subfigure}
  }
  \makebox[\textwidth]{
  \begin{subfigure}[t]{.325\textwidth}
    \centering
    \includegraphics[height=3.5cm]{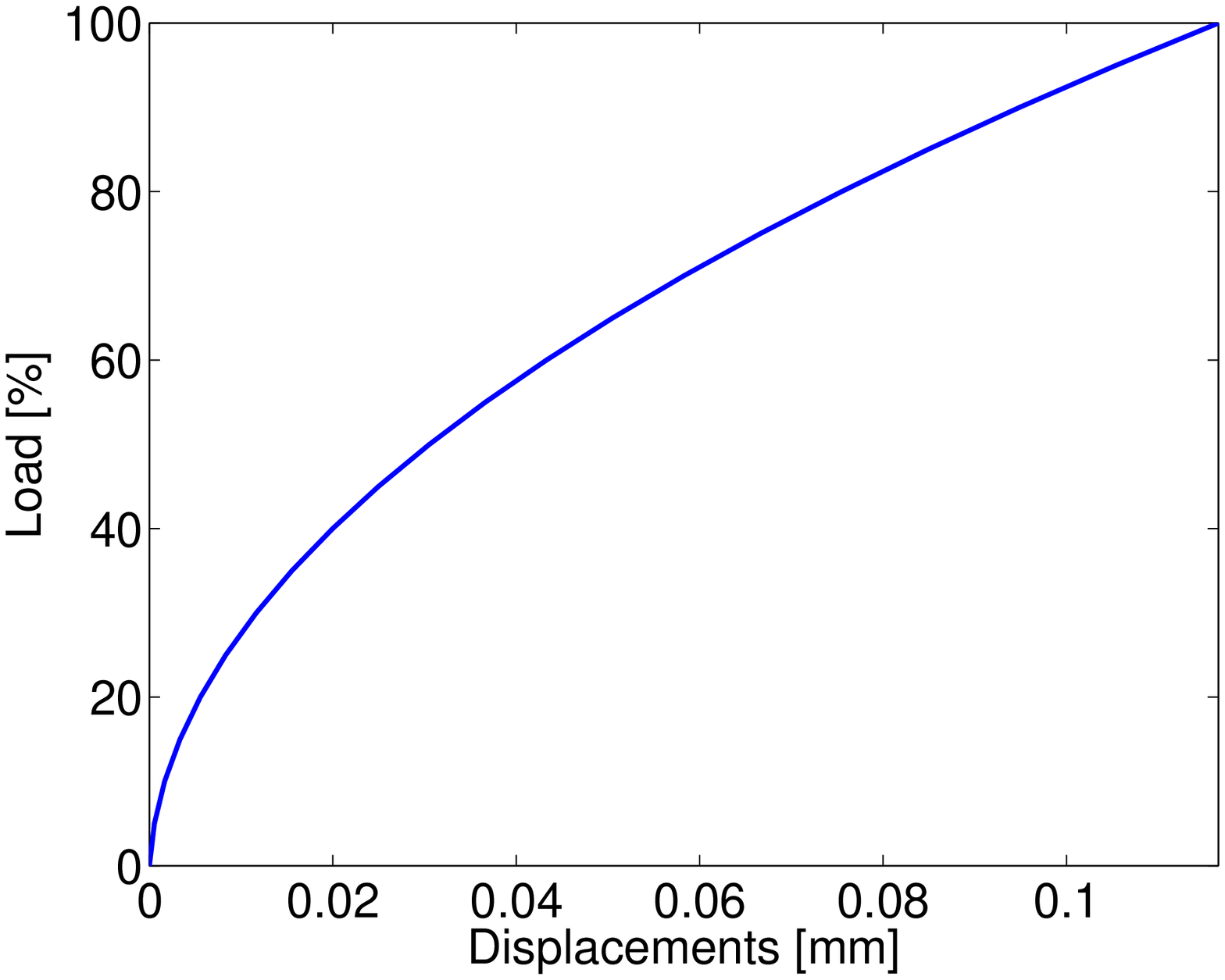}
    \caption{}
    \label{level2_square_inclusion_quasi_disp}
  \end{subfigure}
  \begin{subfigure}[t]{.325\textwidth}
    \centering
    \includegraphics[height=3.5cm]{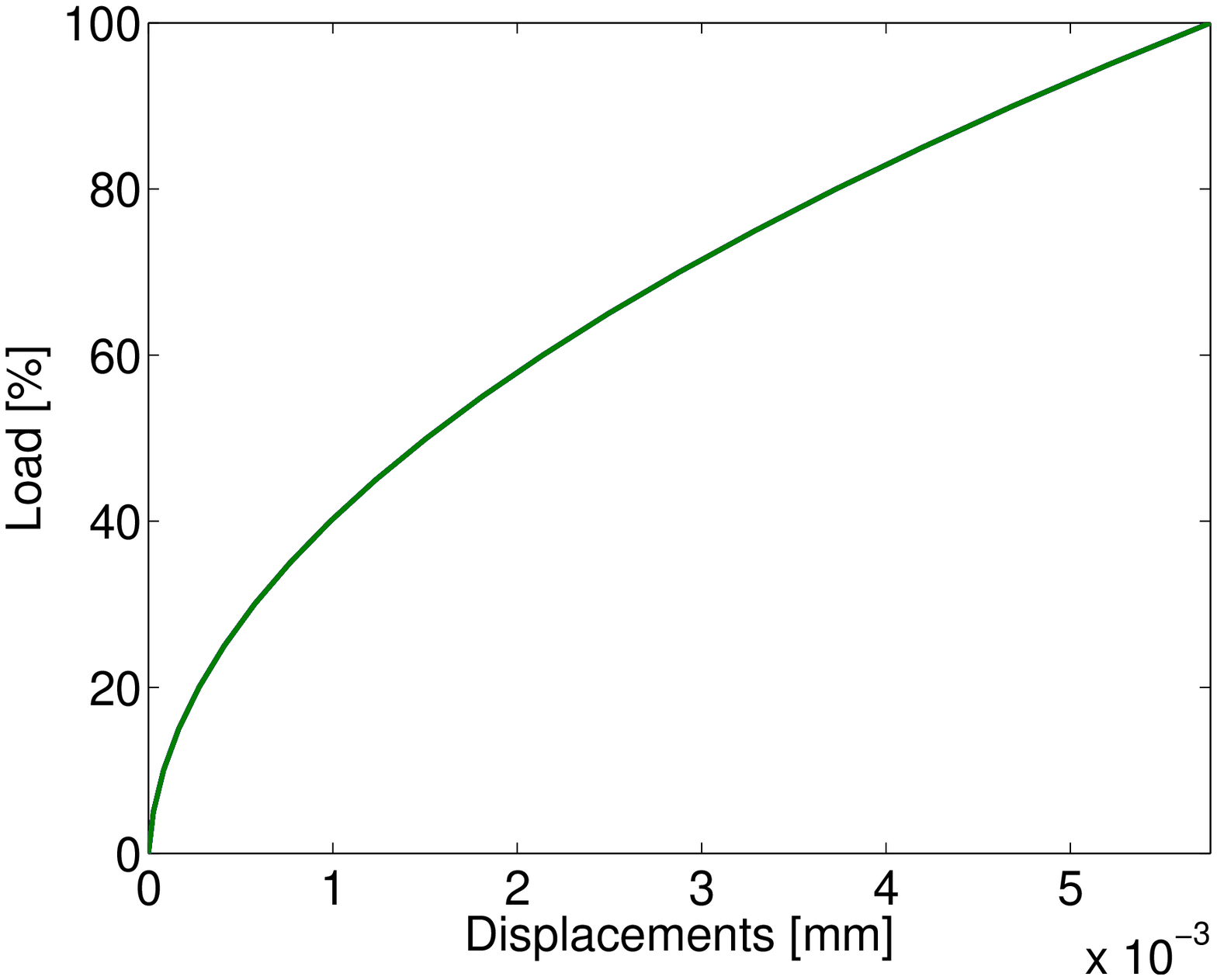}
    \caption{}
    \label{level2_spiderweb_square_quasi_disp}
  \end{subfigure}
  \begin{subfigure}[t]{.325\textwidth}
    \centering
    \includegraphics[height=3.5cm]{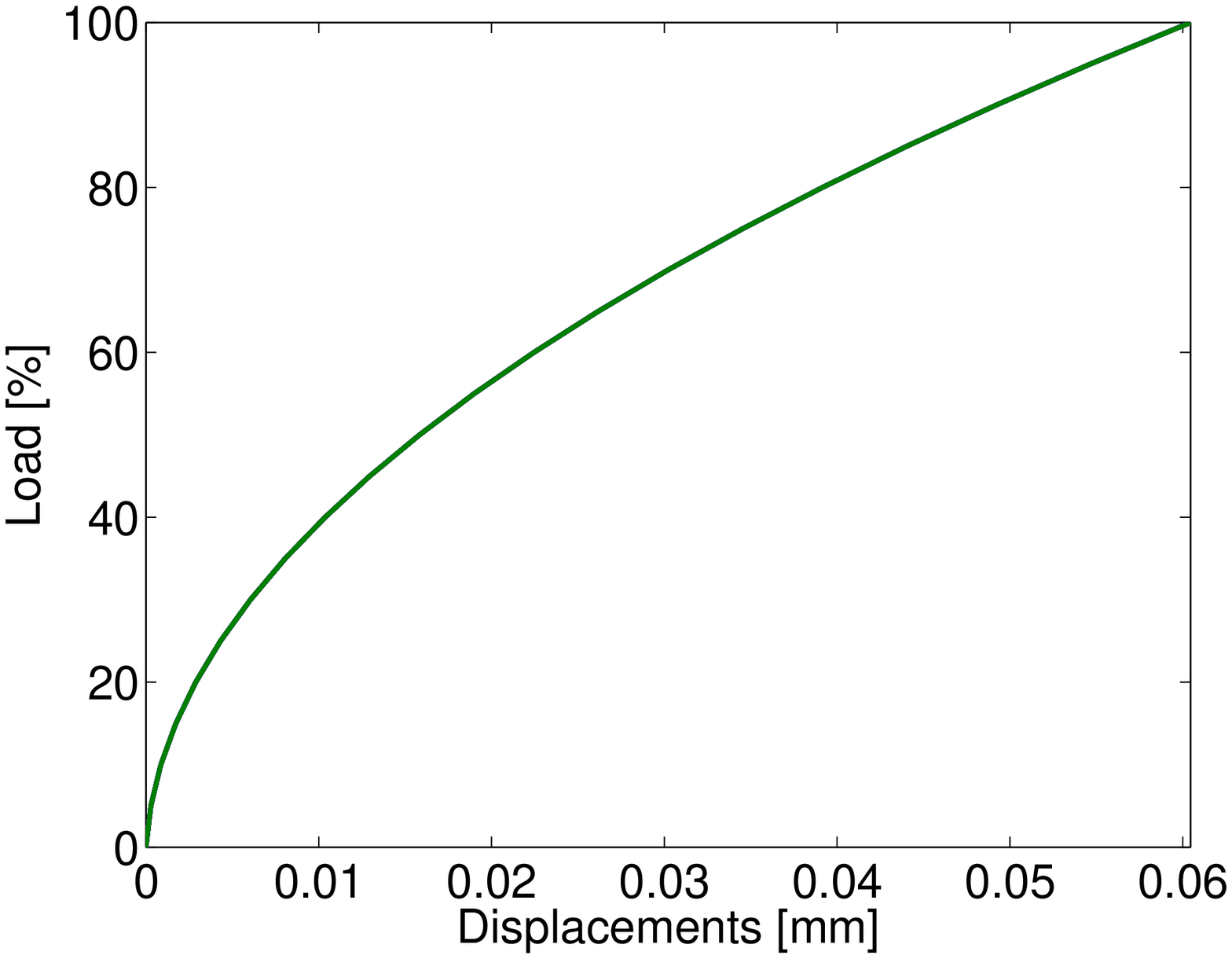}
    \caption{}
    \label{level2_spiderweb_hexagonal_quasi_disp}
  \end{subfigure}
  \begin{subfigure}[t]{.325\textwidth}
    \centering
    \includegraphics[height=3.5cm]{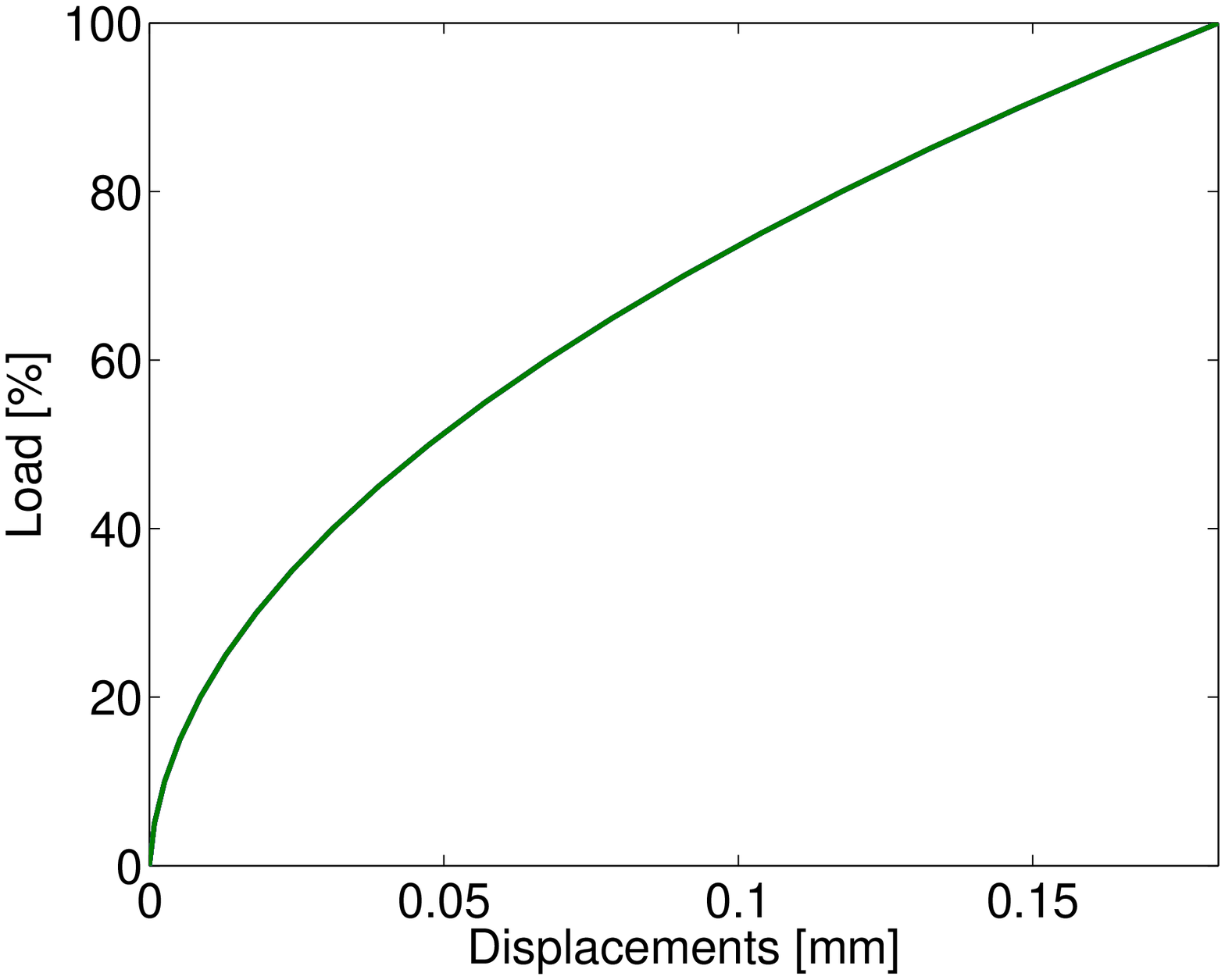}
    \caption{}
    \label{level2_spiderweb_square_inclusion_quasi_disp}
  \end{subfigure}
  }
  \caption{Stresses and load-displacement curves considering in-plane bending behavior for (\subref{level2_square_inclusion_quasi_disp_defr}, \subref{level2_square_inclusion_quasi_disp}) non-hierarchical, (\subref{level2_spiderweb_square_quasi_disp_defr}, \subref{level2_spiderweb_square_quasi_disp}) square hierarchical, (\subref{level2_spiderweb_hexagonal_quasi_disp_defr}, \subref{level2_spiderweb_hexagonal_quasi_disp}) hexagonal hierarchical, and (\subref{level2_spiderweb_square_inclusion_quasi_disp_defr}, \subref{level2_spiderweb_square_inclusion_quasi_disp}) hierarchical square structures.}
  \label{level2_quasi_shear}
\end{figure}

\begin{figure}[H]
  \makebox[\textwidth]{
  \begin{subfigure}[t]{.325\textwidth}
    \centering
    \includegraphics[height=4.0cm,trim={0cm 0cm 3.5cm 0cm},clip]{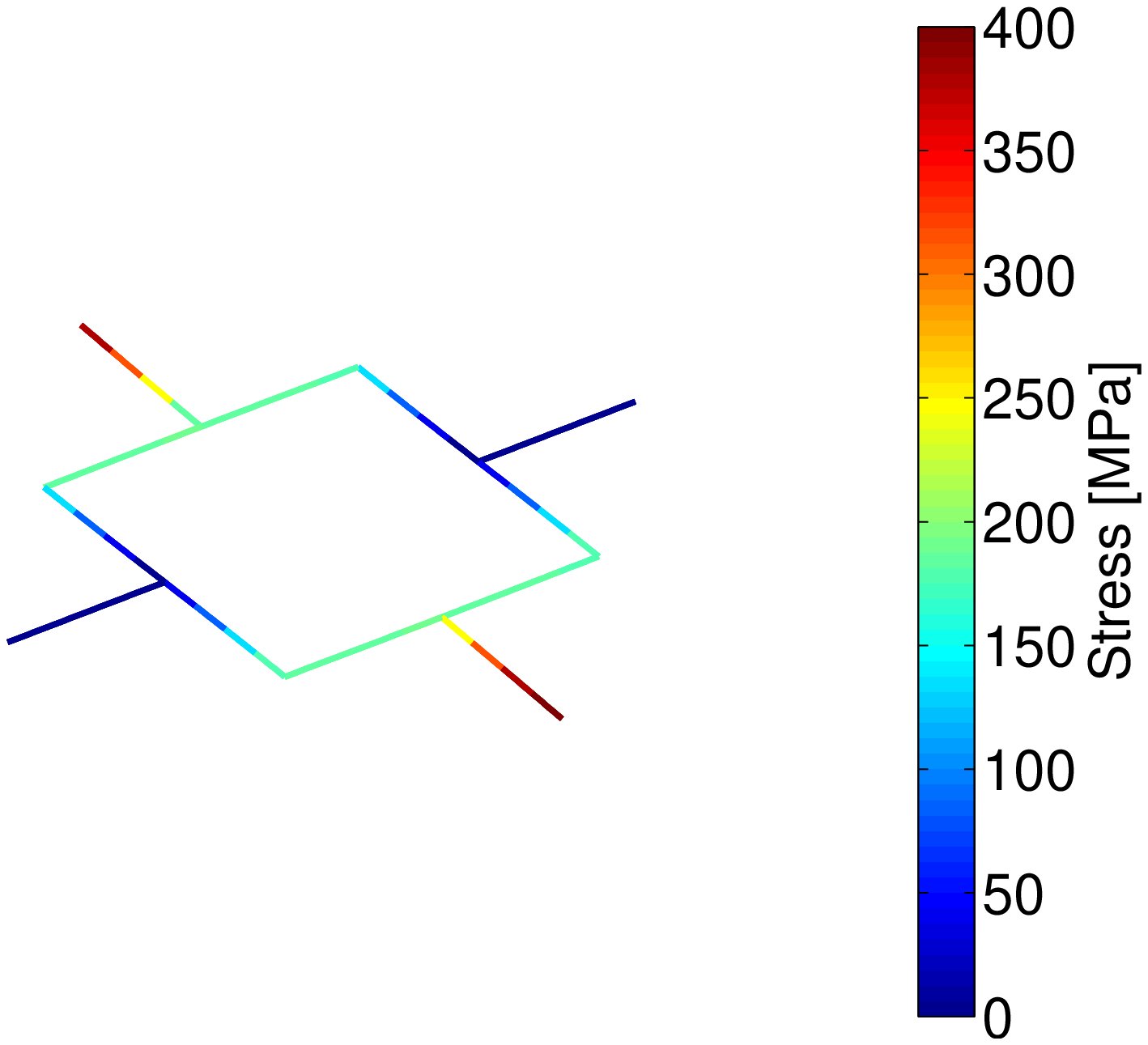}
    \caption{}
    \label{level2_square_inclusion_quasi_rot_defr}
  \end{subfigure}
  \begin{subfigure}[t]{.325\textwidth}
    \centering
    \includegraphics[height=4.0cm,trim={0cm 0cm 3.5cm 0cm},clip]{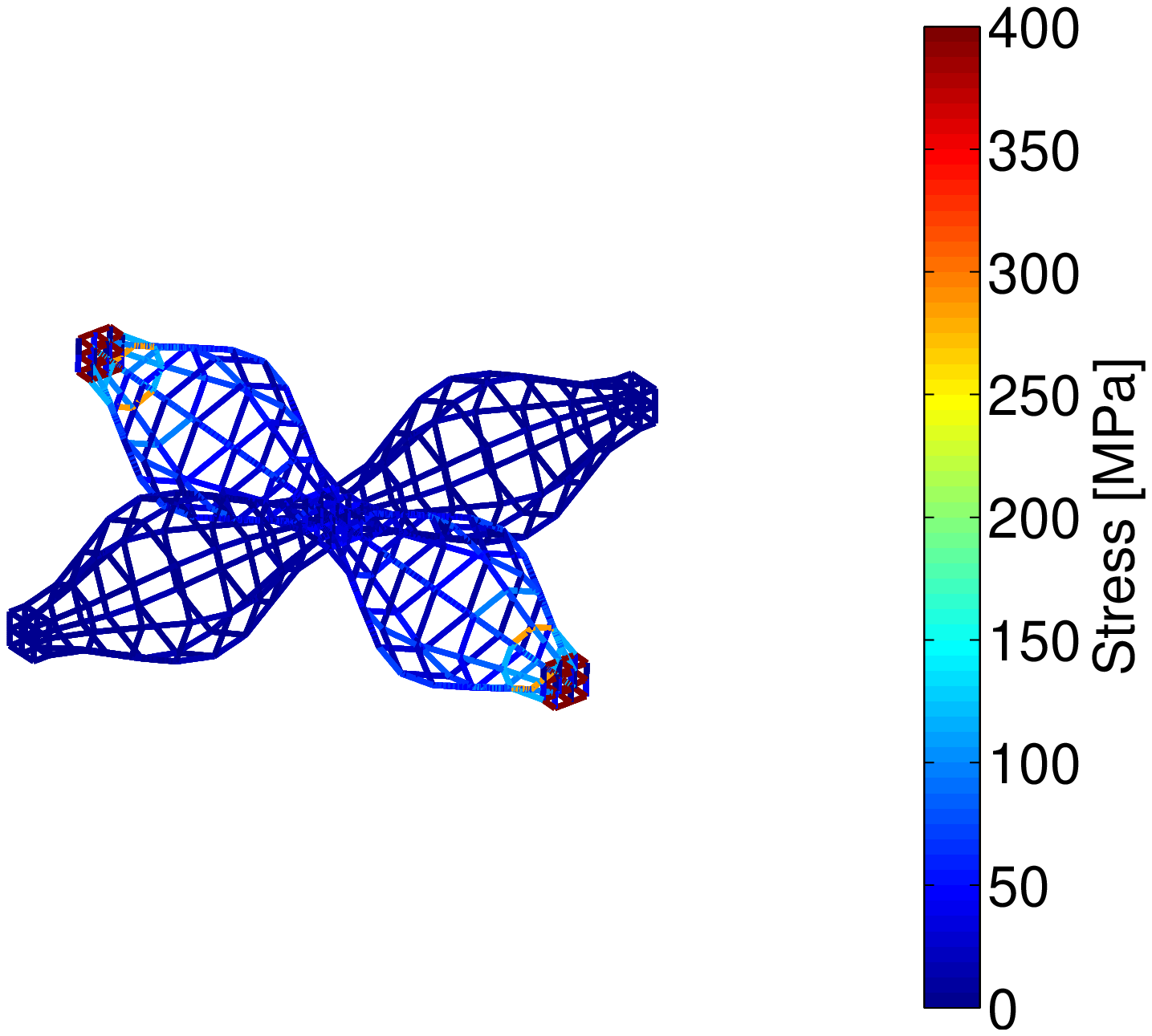}
    \caption{}
    \label{level2_spiderweb_square_quasi_rot_defr}
  \end{subfigure}
  \begin{subfigure}[t]{.325\textwidth}
    \centering
    \includegraphics[height=4.0cm,trim={0cm 0cm 3.5cm 0cm},clip]{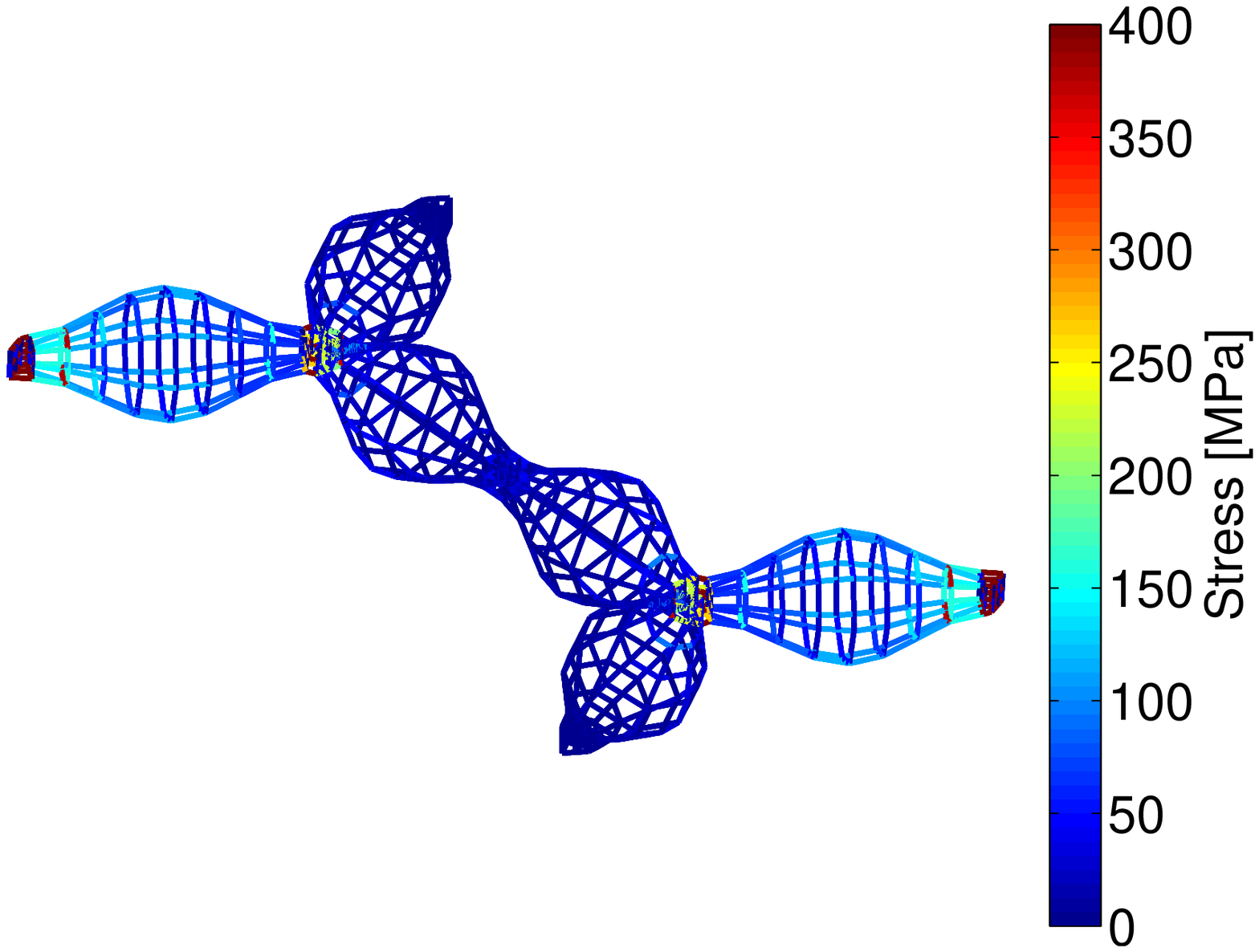}
    \caption{}
    \label{level2_spiderweb_hexagonal_quasi_rot_defr}
  \end{subfigure}
  \begin{subfigure}[t]{.325\textwidth}
    \centering
    \includegraphics[height=4.0cm]{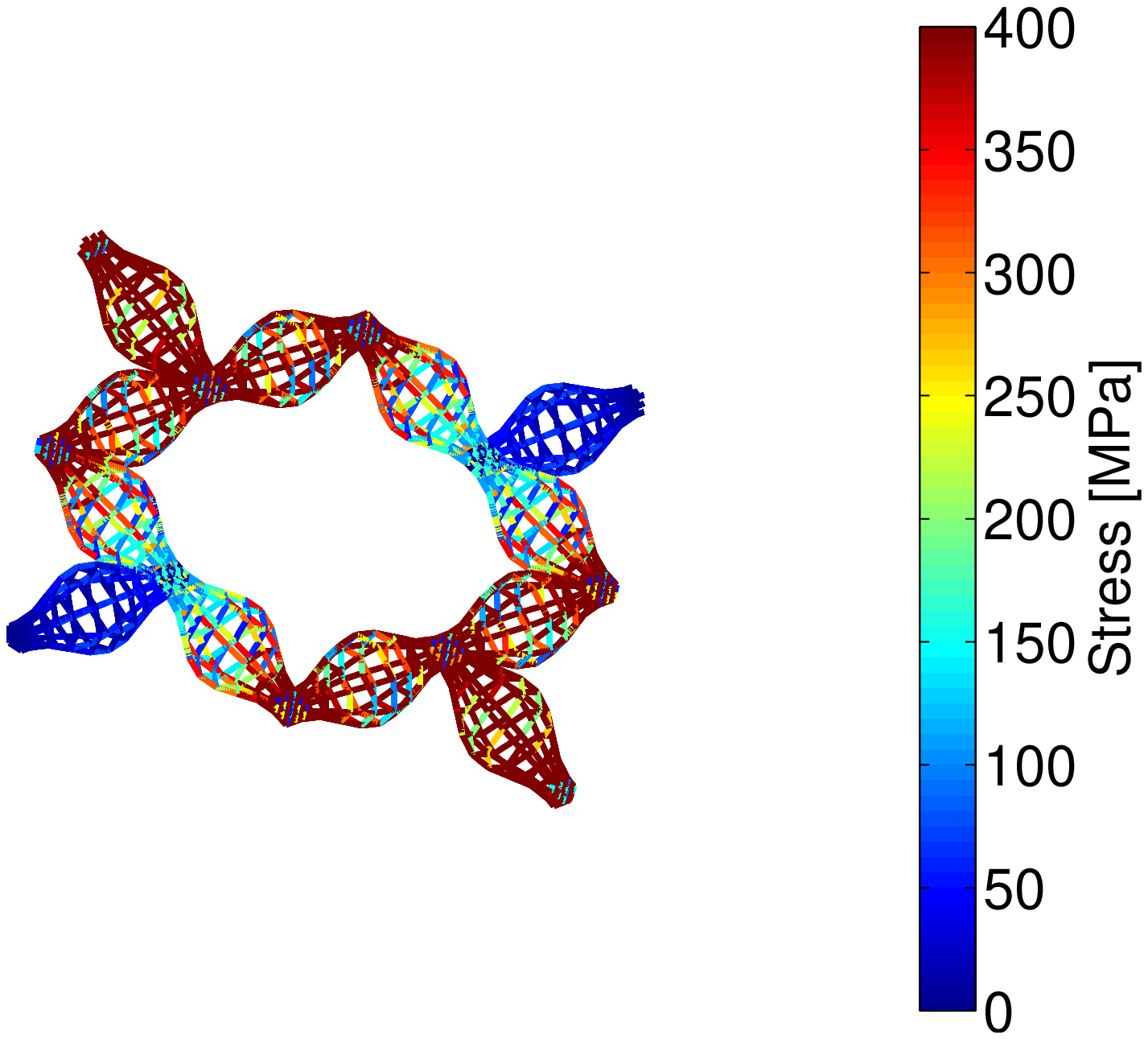}
    \caption{}
    \label{level2_spiderweb_square_inclusion_quasi_rot_defr}
  \end{subfigure}
  }
  \makebox[\textwidth]{
  \begin{subfigure}[t]{.325\textwidth}
    \centering
    \includegraphics[height=3.5cm]{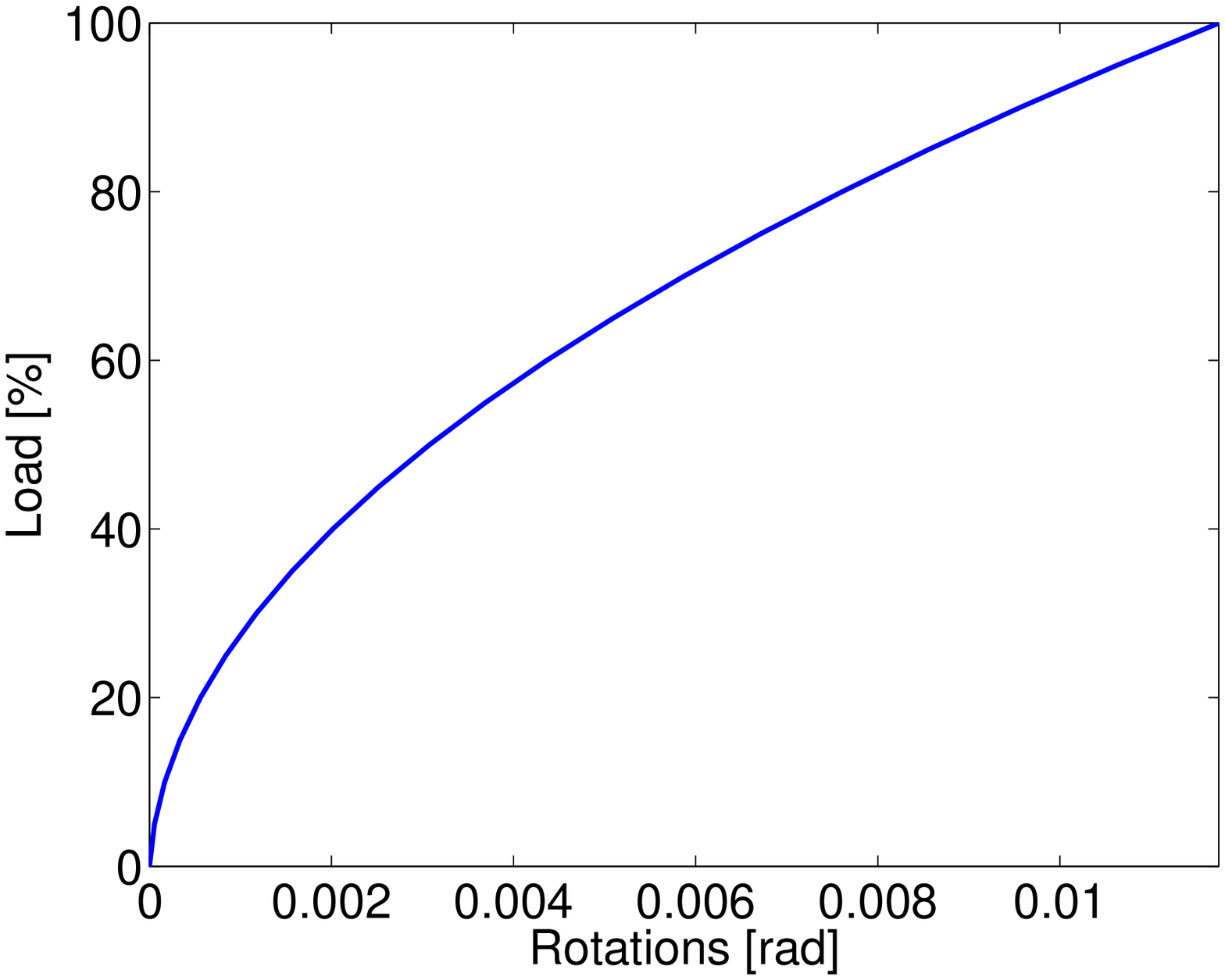}
    \caption{}
    \label{level2_square_inclusion_quasi_rot}
  \end{subfigure}
  \begin{subfigure}[t]{.325\textwidth}
    \centering
    \includegraphics[height=3.5cm]{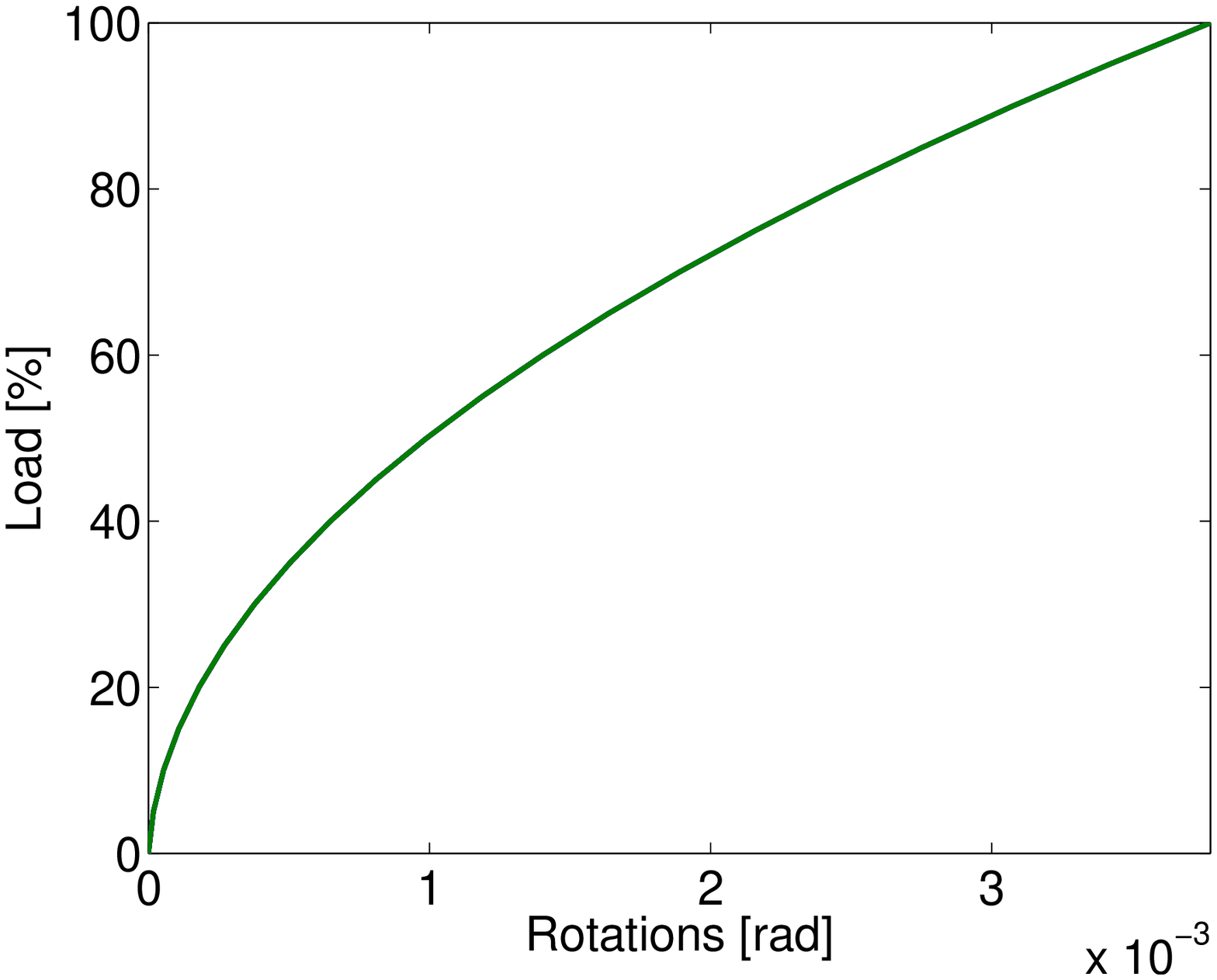}
    \caption{}
    \label{level2_spiderweb_square_quasi_rot}
  \end{subfigure}
  \begin{subfigure}[t]{.325\textwidth}
    \centering
    \includegraphics[height=3.5cm]{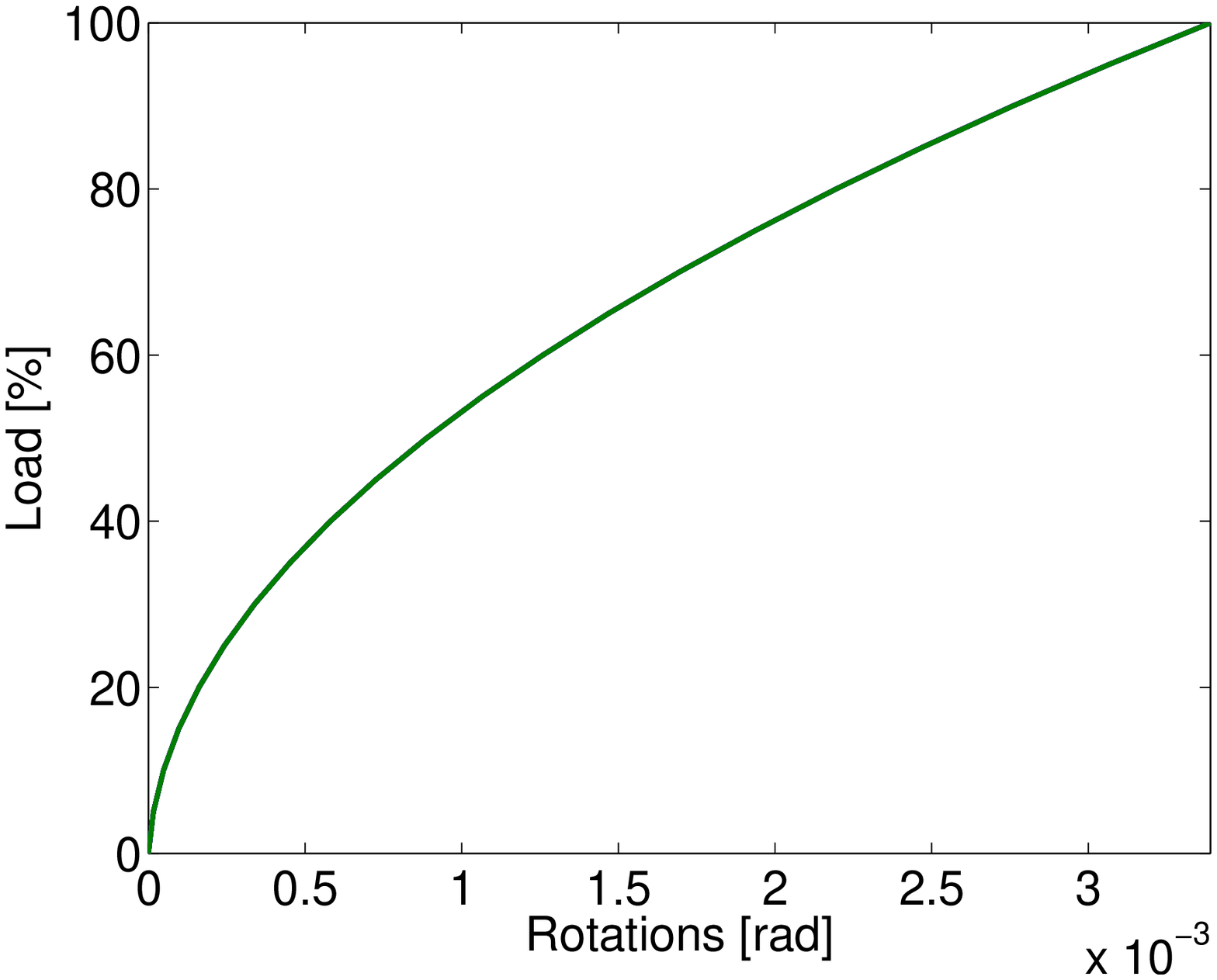}
    \caption{}
    \label{level2_spiderweb_hexagonal_quasi_rot}
  \end{subfigure}
  \begin{subfigure}[t]{.325\textwidth}
    \centering
    \includegraphics[height=3.5cm]{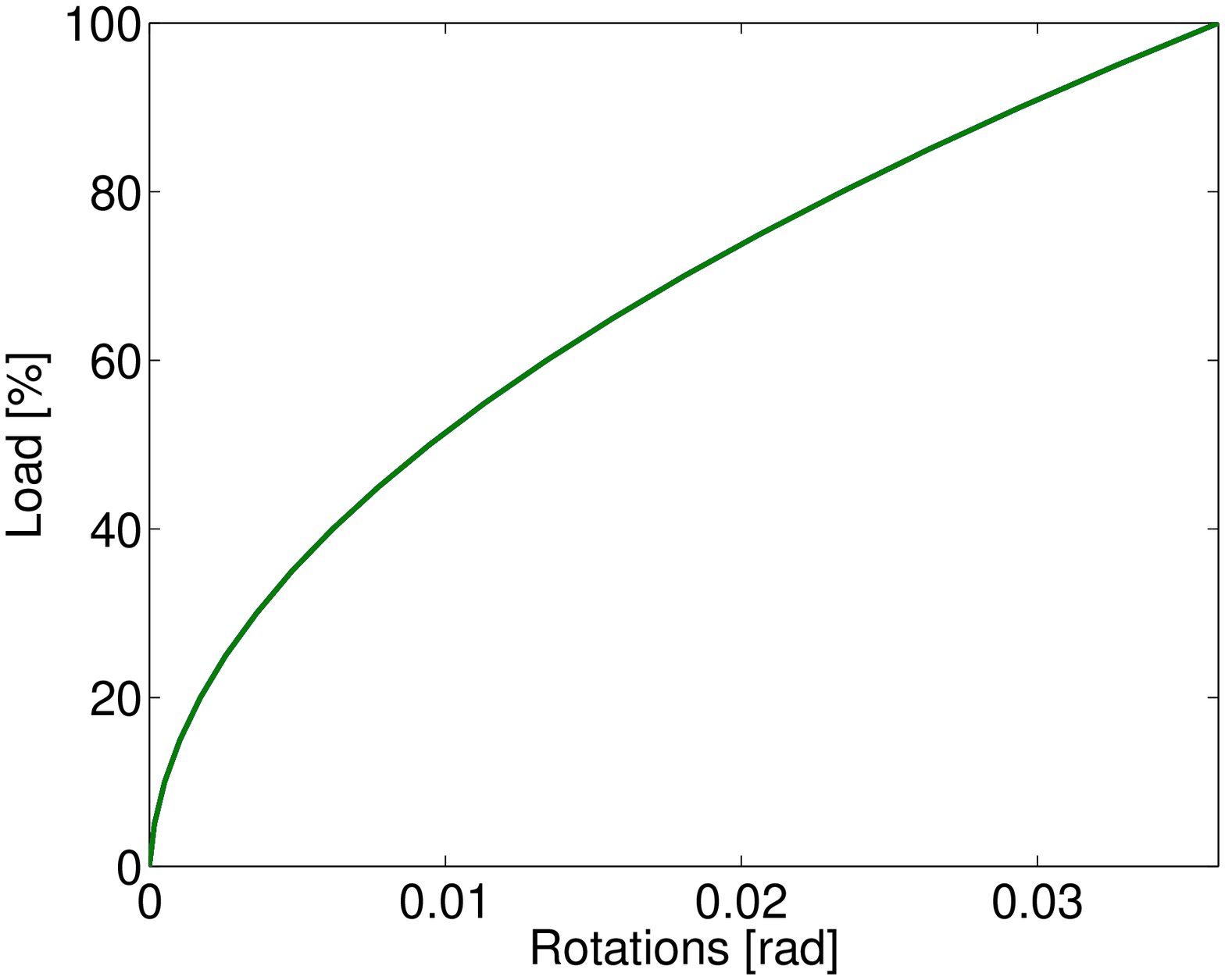}
    \caption{}
    \label{level2_spiderweb_square_inclusion_quasi_rot}
  \end{subfigure}
  }
  \caption{Stresses and load-displacement curves considering torsional behavior for (\subref{level2_square_inclusion_quasi_rot_defr}, \subref{level2_square_inclusion_quasi_rot}) non-hierarchical, (\subref{level2_spiderweb_square_quasi_rot_defr}, \subref{level2_spiderweb_square_quasi_rot}) square hierarchical, (\subref{level2_spiderweb_hexagonal_quasi_rot_defr}, \subref{level2_spiderweb_hexagonal_quasi_rot}) hexagonal hierarchical, and (\subref{level2_spiderweb_square_inclusion_quasi_rot_defr}, \subref{level2_spiderweb_square_inclusion_quasi_rot}) hierarchical square structures.}
  \label{level2_quasi_torsion}
\end{figure}

Results for both displacements and Von Mises stresses considering in-plane bending behavior are shown in Figure \ref{level2_quasi_shear}, and considering torsional behavior in Figure \ref{level2_quasi_torsion}.

Results from Figure \ref{level2_quasi_shear} indicate that for in-plane bending, the structure depicted in Figure \ref{level2_spiderweb_square_quasi_disp_defr} has the largest stiffness and shows no stress concentration. The structure depicted in Figure \ref{level2_spiderweb_hexagonal_quasi_disp_defr} shows an improvement in stiffness, but has the limitation of stress concentration at its junctions. The structure depicted in Figure \ref{level2_spiderweb_square_inclusion_quasi_disp_defr} has both disadvantages of decreased stiffness and stress concentration, which can be explained to the smaller radii of its elements, since all structures had their radii chosen to present the same mass as the structure in Figure \ref{level2_square_inclusion_quasi_disp_defr}.

Results from Figure \ref{level2_quasi_torsion} indicate that structures depicted in Figures \ref{level2_spiderweb_square_quasi_rot_defr} and \ref{level2_spiderweb_hexagonal_quasi_rot_defr} show increases in their effective torsional stiffness, but also show stress concentration at their junctions. Also confirming previous results, the structure shown in Figure \ref{level2_spiderweb_square_inclusion_quasi_rot_defr} presents a decrease in its torsional stiffness, and larger stress values at its elements, mainly due to their cross-section reduction.

\section{Band diagrams for simple two-dimensional structures} \label{app_simple_band_diagrams}

\begin{figure}[h!]
  \makebox[\textwidth]{
  \begin{subfigure}[h]{.2\textwidth}
    \centering
    \includegraphics[width=2cm]{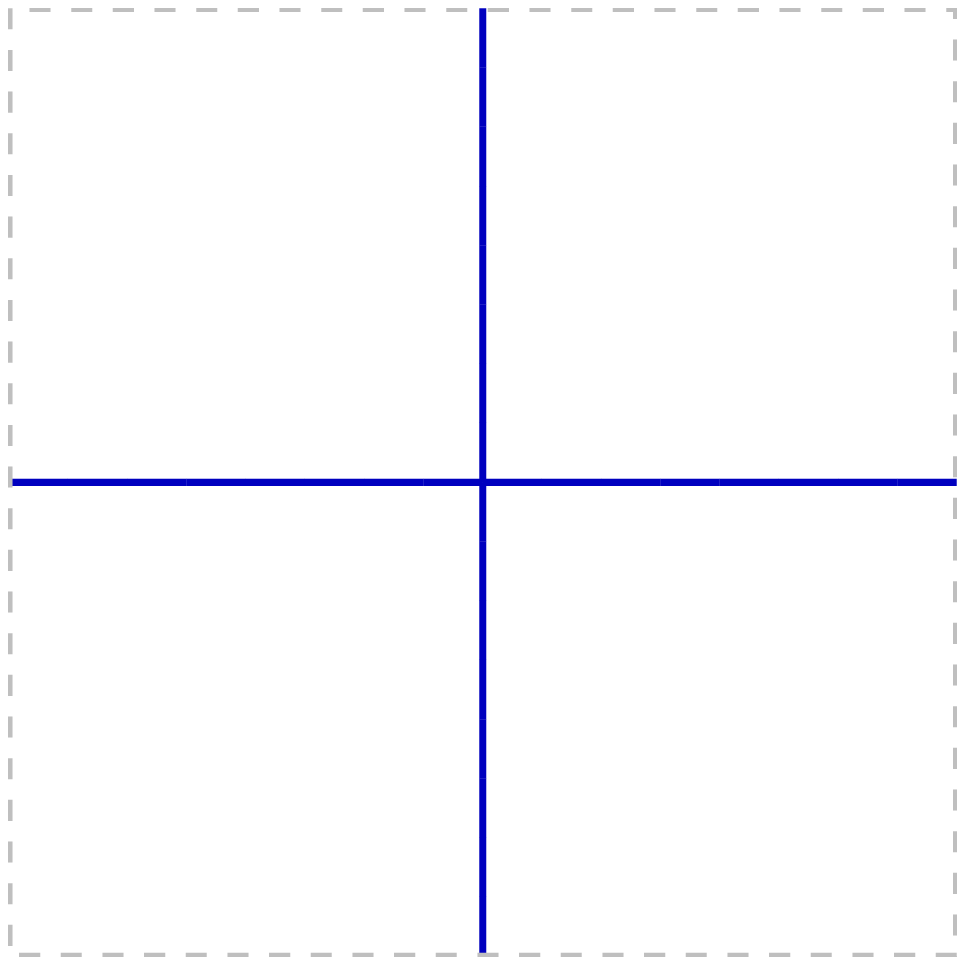}
  \end{subfigure}
  \begin{subfigure}[h]{.35\textwidth}
    \centering
    \includegraphics[width=5cm]{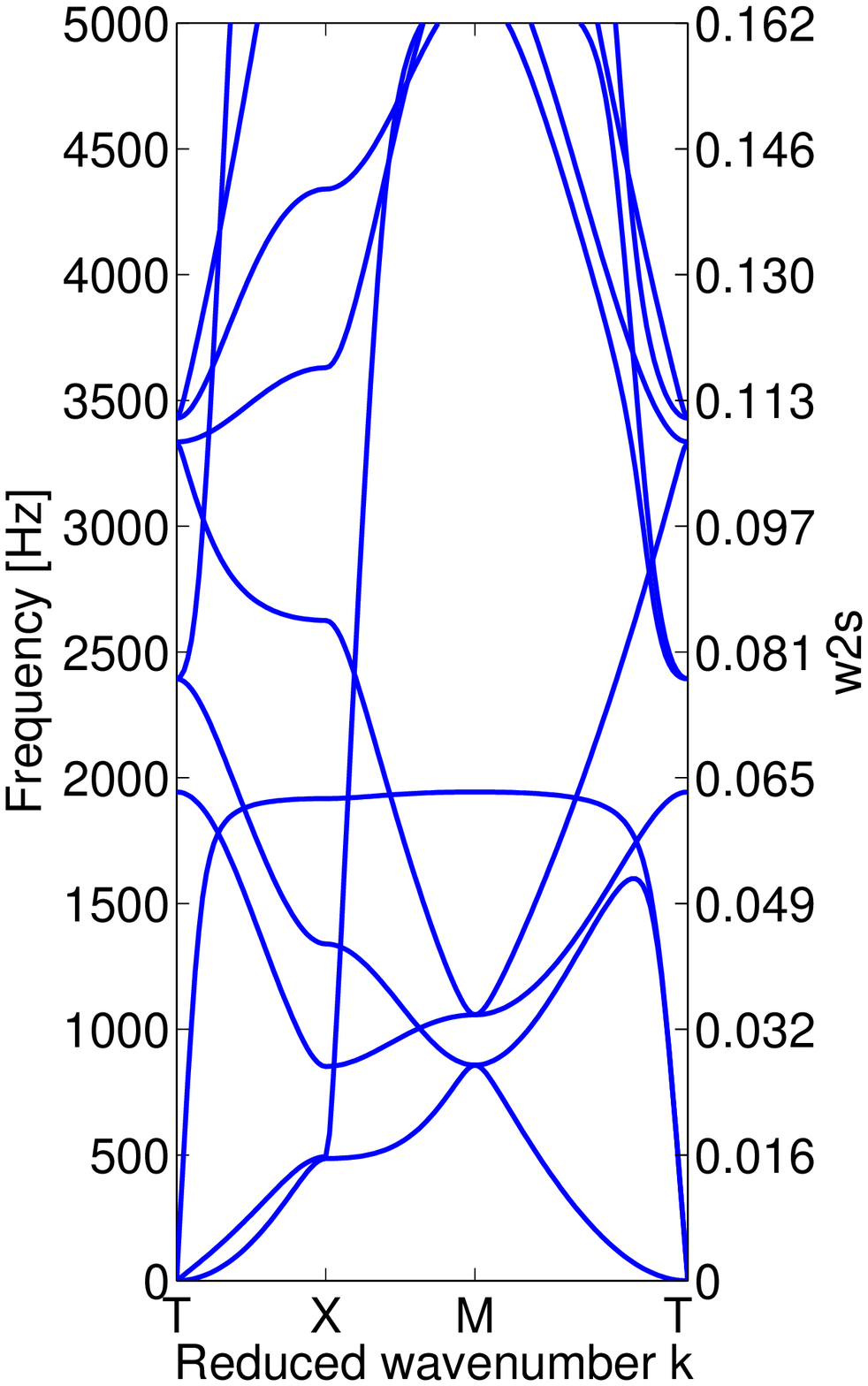}
    \caption{}
    \label{level2_square_band_diagram}
  \end{subfigure}
  \begin{subfigure}[h]{.25\textwidth}
    \centering
    \includegraphics[width=2cm]{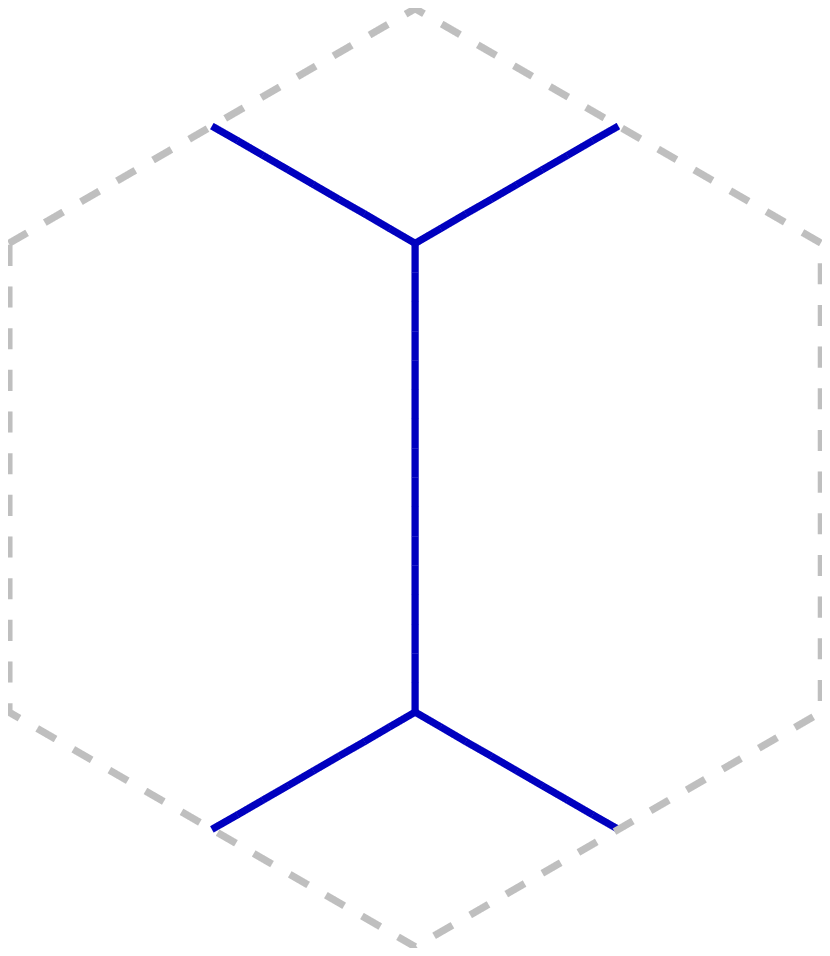}
  \end{subfigure}
  \begin{subfigure}[h]{.35\textwidth}
    \centering
    \includegraphics[width=5cm]{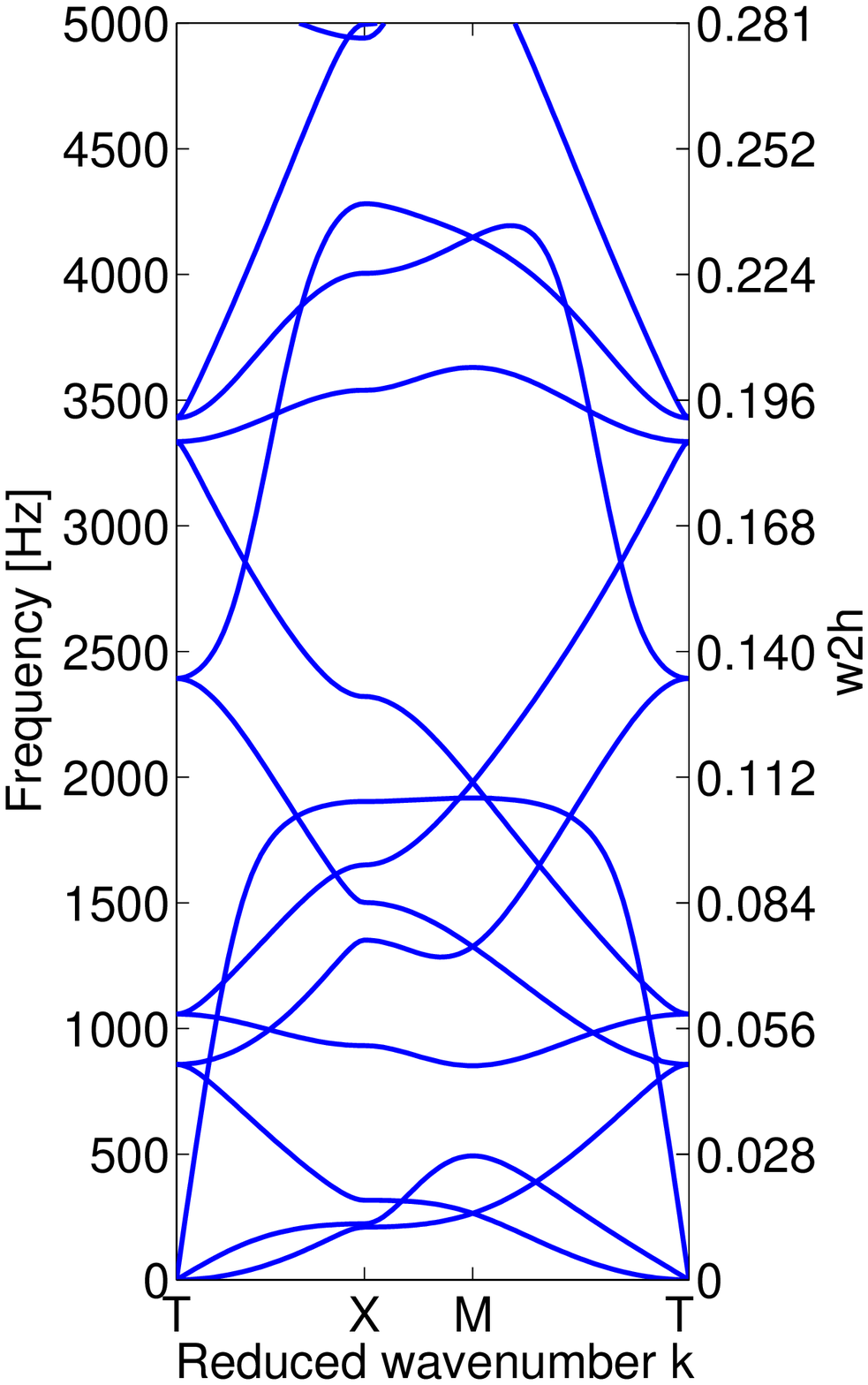}
    \caption{}
    \label{level2_hexagonal_band_diagram}
  \end{subfigure}
  }
  \caption{Band diagrams for the two-dimensional (\subref{level2_square_band_diagram}) square and (\subref{level2_hexagonal_band_diagram}) hexagonal structures. No remarkable features are noticed.}
  \label{level2_simple_band_diagrams}
\end{figure}

Figure \ref{level2_simple_band_diagrams} presents the band diagrams of the simple two-dimensional structures presented in Figures \ref{level2_square_top} and \ref{level2_hexagonal_top}, also indicated here with the corresponding contour of the periodic cell using dashed gray lines. The masses of the square and hexagonal periodic cells are, respectively, $1.15 \times 10^{-3}$ kg and $1.73 \times 10^{-3}$ kg. No distinguishable features (such as partial BGs) are noticed.

\section{Variations in FRF of the structure depicted in Figure \ref{level2_square_inclusion_top}} \label{app_square_inclusion_frf}

\begin{figure}[h!]
  \makebox[\textwidth]{
  \begin{minipage}[l]{.45\textwidth}
  \begin{subfigure}[h]{\textwidth}
    {\centering
    \includegraphics[width=6cm]{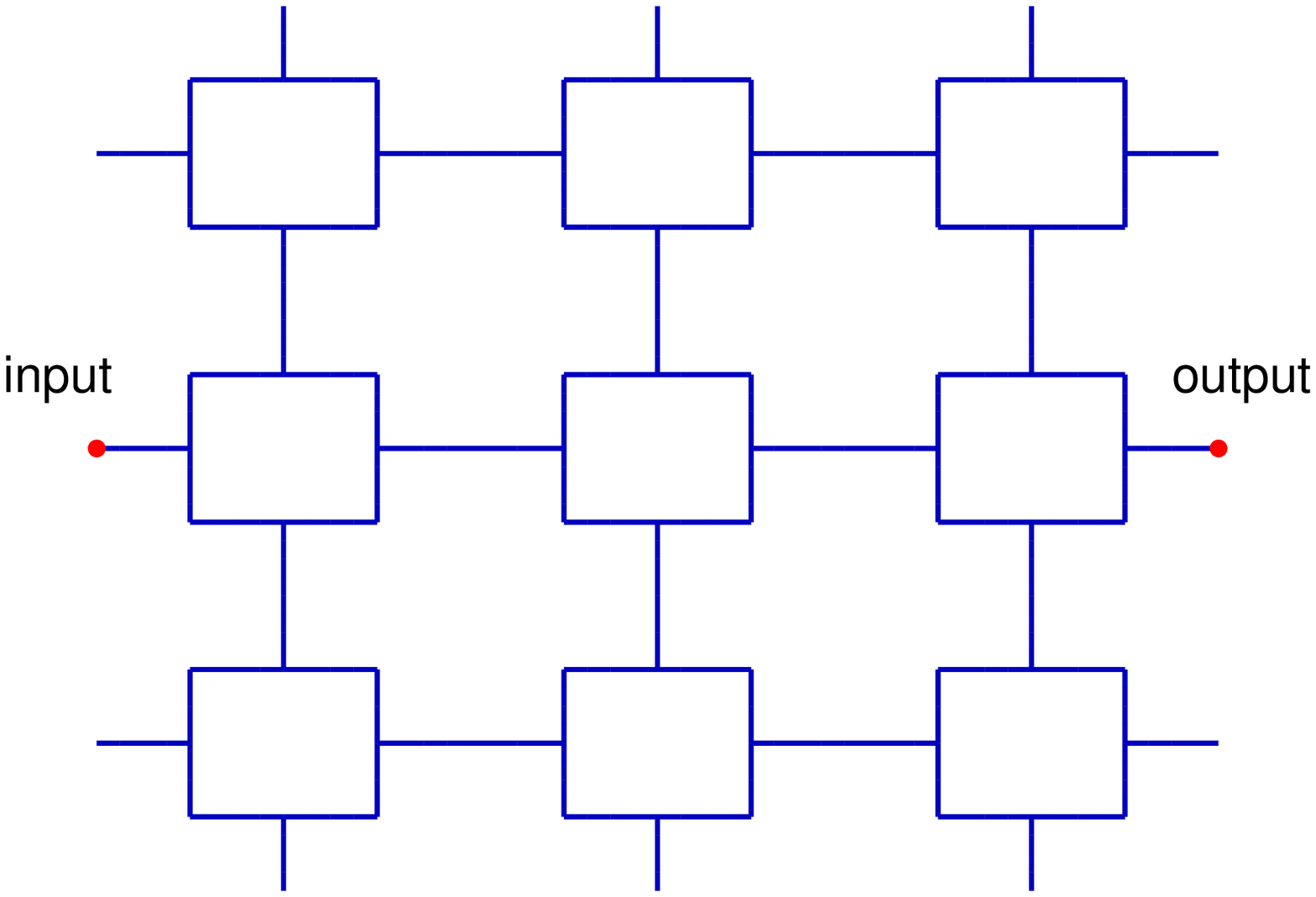}}
    \includegraphics[scale=0.5]{./xy_coordinates.eps}
    \caption{}
    \label{level2_square_inclusion_frf_top}
  \end{subfigure}
  \end{minipage}
  \begin{minipage}[l]{.6\textwidth}
    \centering
    \begin{subfigure}[h]{\textwidth}
      \centering
      \includegraphics[width=10cm]{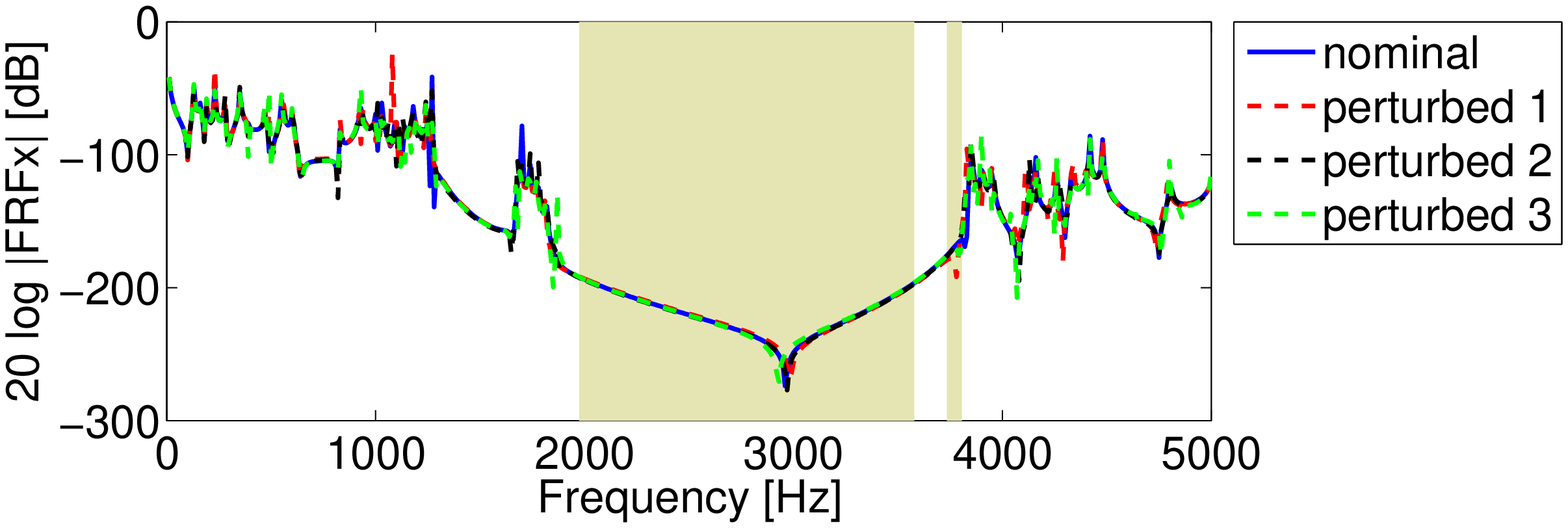}
      \caption{}
      \label{level2_square_inclusion_frf_x}
    \end{subfigure}

    \begin{subfigure}[h]{\textwidth}
      \centering
      \includegraphics[width=10cm]{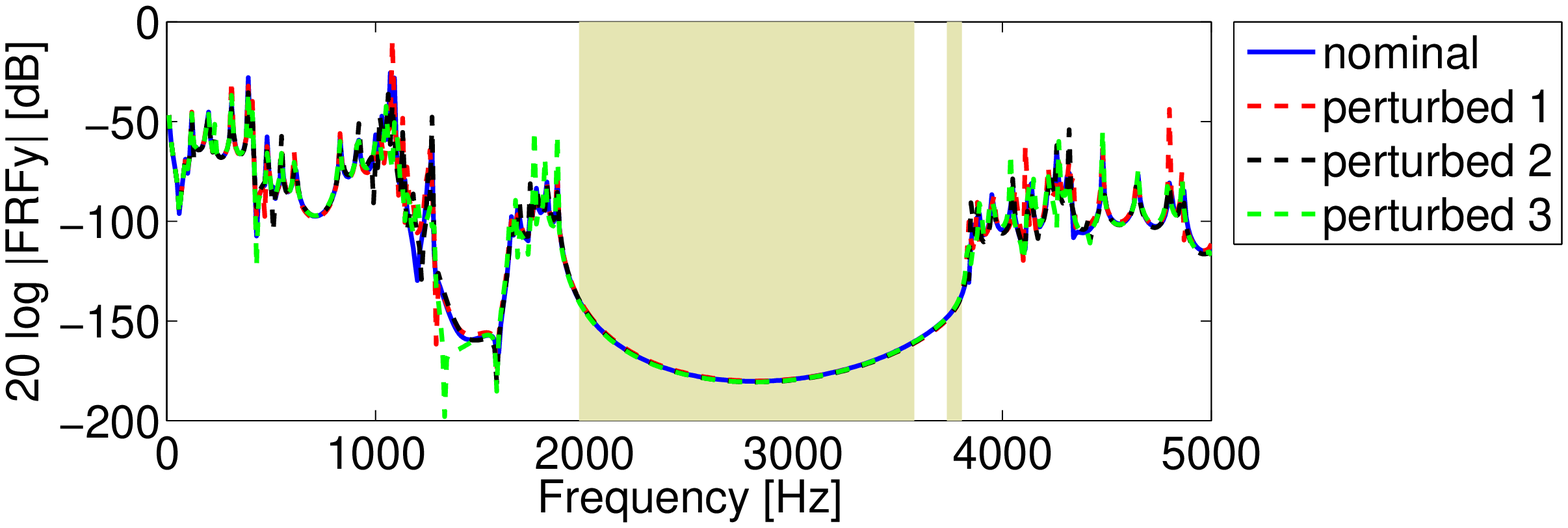}
      \caption{}
      \label{level2_square_inclusion_frf_y}
    \end{subfigure}
    
    \begin{subfigure}[h]{\textwidth}
      \centering
      \includegraphics[width=10cm]{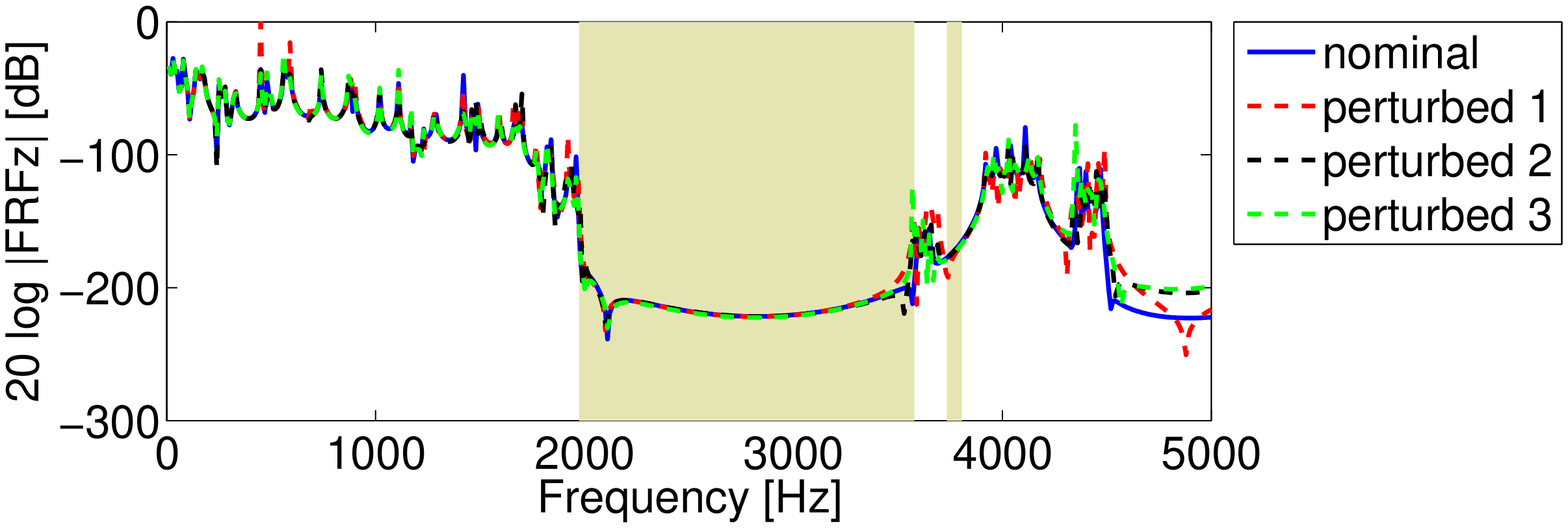}
      \caption{}
      \label{level2_square_inclusion_frf_z}
    \end{subfigure}
  \end{minipage}
  }
  \caption{Variations in FRFs considering (\subref{level2_square_inclusion_frf_top}) a force input at the center of the left side and outputs averaged at the center nodes of the right side with (\subref{level2_square_inclusion_frf_x}) $x$-, (\subref{level2_square_inclusion_frf_y}) $y$-, and (\subref{level2_square_inclusion_frf_z}) $z$-direction FRFs.}
  \label{level2_square_inclusion_frf}
\end{figure}

The results for the non-hierarchical two-dimensional finite structure (see Figure \ref{level2_square_inclusion_top}) are shown in Figure \ref{level2_square_inclusion_frf}. Frequency response functions computed using nominal radii values are marked in blue (\textcolor{blue}{--}, nominal), and those obtained using perturbed samples are marked in red (\textcolor{red}{- -}, perturbed 1), black (- -, perturbed 2), and green (\textcolor{green}{- -}, perturbed 3). The previously computed BGs are marked in yellow.

As depicted in Figure \ref{level2_square_inclusion_frf}, there is an excellent correlation between computed BGs and large decreases in FRFs for the structure depicted in Figure \ref{level2_square_inclusion_frf_top}. The first BG is mostly restricted by the $z$-direction attenuation (Figure \ref{level2_square_inclusion_frf_z}), and the second BG has its lower limit associated with the $z$-direction and upper limit associated with the $x$- and $y$-directions (Figures \ref{level2_square_inclusion_frf_x} and \ref{level2_square_inclusion_frf_y}). These correlations are weakly affected by the variation in the radii of the elements.

\clearpage
\bibliographystyle{elsarticle-harv}\biboptions{authoryear}
\bibliography{frame_hierarchy}

\end{document}